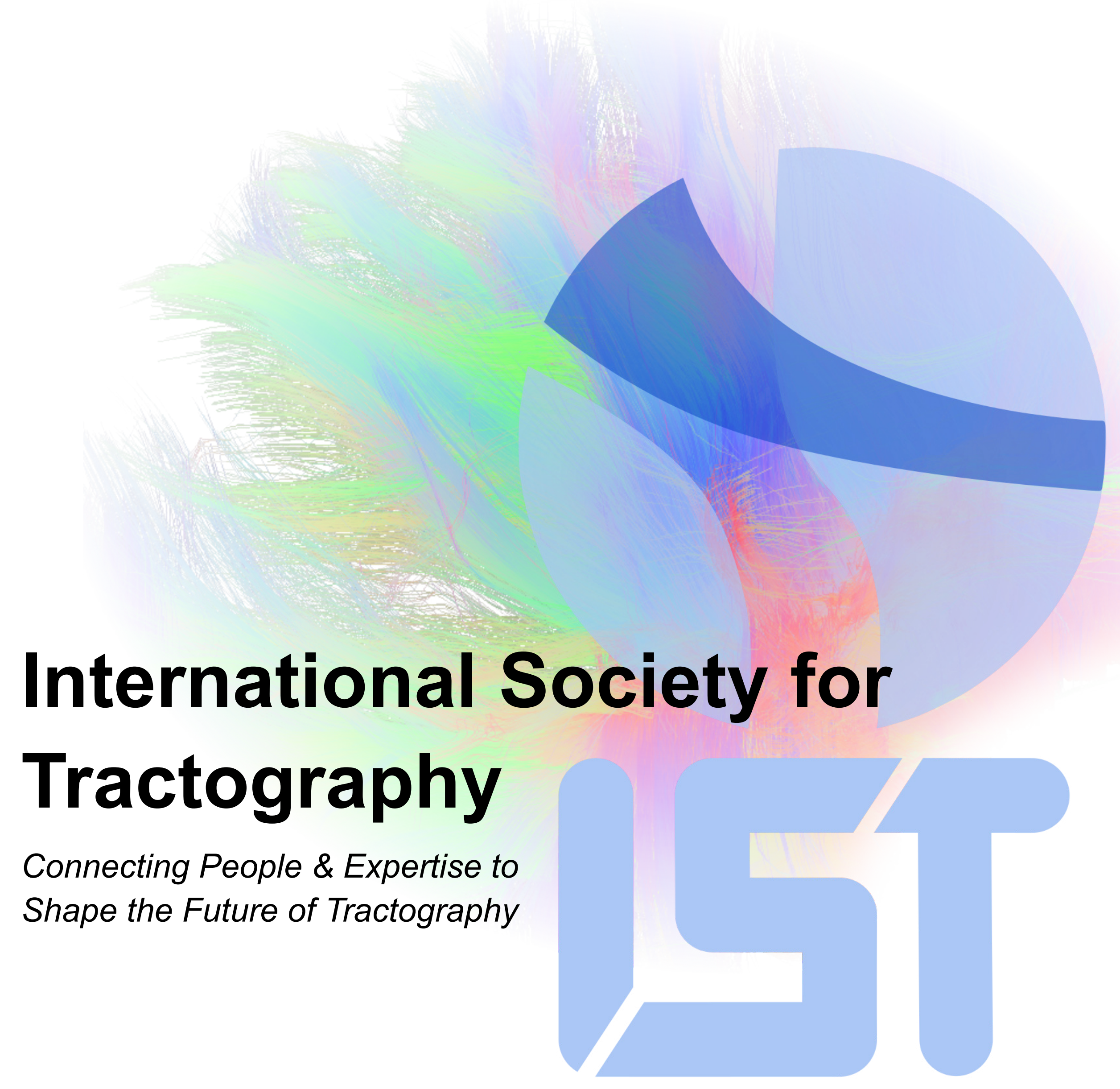

# International Society for Tractography

*Connecting People & Expertise to Shape the Future of Tractography*

## Conference Proceedings

# IST 2025 Bordeaux

*October 13th - 16th, 2025, Bordeaux, France*

**Title:** Proceedings for the Inaugural Meeting of the International Society for Tractography - IST 2025 Bordeaux

**IST 2025 Bordeaux - Scientific Organizing Committee**

**Proceedings Editors:** Flavio Dell'Acqua[1], Maxime Descoteaux[2,3,4], Graham Little[5], Laurent Petit[4,6]

**Proceedings Reviewers:** Dogu Baran Aydogan[7,8], Stephanie Forkel[9,10,11,12], Alexander Leemans[13], Simona Schiavi[14], Michel Thiebaut de Schotten[15,16]

**Affiliations:**

[1] Institute of Psychiatry, Psychology and Neuroscience, King's College London, London, United Kingdom

[2] Computer Sciences Department, Université de Sherbrooke

[3] Imeka Solutions Inc., Sherbrooke, Canada

[4] IRP OpTeam, CNRS Biologie, France and Université de Sherbrooke, Canada

[5] Jump Ship Labs, Edmonton, Canada

[6] Université Bordeaux, CNRS, CEA, IMN, GIN, UMR 5293, F-33000, Bordeaux, France

[7] Department of Neuroscience and Biomedical Engineering, and Advanced Magnetic Imaging Center, Aalto NeuroImaging, Aalto University School of Science, Espoo, Finland

[8] A.I. Virtanen Institute for Molecular Sciences, University of Eastern Finland, Kuopio, Finland

[9] Donders Institute for Brain and Cognition Behaviour, Radboud University, Nijmegen, the Netherlands

[10] Max Planck Institute for Psycholinguistics, Nijmegen, the Netherlands

[11] Brain Connectivity and Behaviour Laboratory, Sorbonne Universities, Paris, France

[12] Centre for Neuroimaging Sciences, Department of Neuroimaging, Institute of Psychiatry, Psychology and Neuroscience, King's College London, London, UK

[13] Image Sciences Institute, University Medical Center Utrecht, Utrecht, Netherlands

[14] ASG Superconductors S.p.A., Genoa, Italy

[15] Groupe d'Imagerie Neurofonctionnelle, Institut des Maladies Neurodégénératives 5293, Centre National de la Recherche Scientifique (CNRS), University of Bordeaux, Bordeaux 33076, France

[16] Brain Connectivity and Behaviour Laboratory, Sorbonne Universities, Paris 75006, France

**Description:** This collection comprises the abstracts presented during poster, power pitch and oral sessions at the Inaugural Conference of the International Society for Tractography (IST Conference 2025), held in Bordeaux, France, from October 13-16, 2025. The conference was designed to foster meaningful exchange and collaboration between disparate fields. The overall focus was on advancing research, innovation, and community in the common fields of interest: neuroanatomy, tractography methods and scientific/clinical applications of tractography. The included abstracts cover the latest advancements in tractography, Diffusion MRI, and related fields including new work on; neurological and psychiatric disorders, deep brain stimulation targeting, and brain development. This landmark event brought together world-leading experts to discuss critical challenges and chart the future direction of the field.

| | | | |
|---|---|---|---|
| 16 | Critical white matter tracts for bimanual motor skill learning: Exploring structural connectivity in acute stroke. | Coralie van Ravestyna | Yves Vandermeeren |
| 17 | ON-Harmony: A multi-site, multi-modal travelling-heads resource for brain MRI harmonisation with integration of UK Biobank scanners | Shaun Warrington | Stamatios N Sotiropoulos |
| 18 | Curvature properties on Estimating Asymmetric Fiber Orientation Distribution | Mojtaba Taherkhani | Tim B. Dyrby |
| 19 | Quantifying the white matter pathways supporting written word production using tractometry | Romi Sagi | Michal Ben-Shachar |
| 20 | Estimating microscopy-informed fibre orientations from dMRI data in the UK Biobank | Silei Zhu | Amy F.D. Howard |
| 21 | A microscopy-trained model to predict super-resolution fibre orientations from diffusion MRI | Silei Zhu | Amy F.D. Howard |
| 22 | Superficial white matter association with cognitive decline using UKBiobank database (N=13747) | Nabil Vindas | Jean-François Mangin |
| 23 | Mapping the Superior Longitudinal System: anatomical insights from BraDiPho | Laura Vavassori | Silvio Sarubbo |
| 24 | Sulcal morphology reflect the organization of short U-shape association fibers | Arnaud Le Troter | Olivier Coulon |
| 25 | Do current automated tractography methods hold up in tumour and epilepsy pathology? A comparison of six methods with expert manual tractography | Steven Greenstein | Joseph Yuan-Mou Yang |
| 26 | MouseFlow, a pipeline for diffusion MRI processing and tractogram generation in mouse brain validated using Allen Brain Atlas Connectivity with m2m. | Elise Cosenza | Laurent Petit |
| 27 | Inferior Frontal Projections of the Arcuate Fasciculus Tractography in Broca's Area: Validation against Direct Cortical Electrical stimulation Language Mapping performed in a Paediatric Epilepsy Surgery Case | Joseph Yuan-Mou Yang | Wirginia Maixner |
| 28 | Exploring the macaque precentral gyral white matter using 11.7T dMRI | Fanny Darrault | Frédéric Andersson |
| 29 | Micro- and macrostructural fiber tract changes during pediatric posterior fossa tumor surgery. | Pien E.J. Jellema | Jannie P. Wijnen |
| 30 | Multicenter approach for validation of white matter tracts involved in cognition | M.J.F. Landers | G.J.M. Rutten |
| 31 | BundleParc: automatic bundle parcellation in tumor data | Antoine Théberge | François Rheault |
| 32 | Linking Continuous and Interrupted Human Central Sulcus to the Variability of the Superficial White Matter using a β-Variational AutoEncoder | C. Mendoza | J.-F. Mangin |
| 33 | Can a reduced diffusion MRI protocol achieve optic radiation tractography comparable to those reconstructed using a multi-shell diffusion MRI acquisition for temporal lobe epilepsy surgery planning and surgical image-guidance? | Joseph Yuan-Mou Yang | Wirginia Maixner |
| 34 | Combining tractography and intracranial EEG to define the structural epileptic network | Arash Sarshoghi | Sami Obaid |
| 35 | Harmonizing Free Water Metrics in Aging: A Comparative Study of Single-Shell and Multi-Shell Diffusion MRI | Stanislas Thoumyre | François Rheault |
| 36 | Mapping the structural connectome of temporal lobe epilepsy variants to improve surgical outcomes | Emna Guibene | Sami Obaid |
| 37 | Implicit Neural Tractography: Mapping White Matter in Continuous Space | Ruben Vink | Maxime Chamberland |
| 38 | Unveiling the functional specialization of human circuits with naturalistic stimuli | Ovando-Tellez M. | Michel Thiebaut de Schotten |
| 39 | Local Spherical Deconvolution (LSD) for Tractography of High-Resolution Diffusion MRI of Chimpanzee Brains | A. Anwander | C. Eichner |
| 40 | Robust pipeline to bring automated tractography into neurosurgical practice | R. Bakker | G.J.M. Rutten |
| 41 | nf-pediatric: A robust and age-adaptable end-to-end connectomics pipeline for pediatric diffusion MRI | Anthony Gagnon | Maxime Descoteaux |
| 42 | Tractography in the Developing Knee Joint at Microscopic Resolution | Nian Wang | Nian Wang |
| 43 | Tract morphing: A novel 3D mesh-based voxel-wise framework for multimodal white matter tract analysis | Saludar C | V Shim |
| 44 | Tract-Specific DKI Reveals Early White Matter Microstructure Alterations in Alzheimer's Disease | Santiago Mezzano | Catherine A Morgan |
| 45 | Mapping Reward Processing Circuits- Cortico-Striatal White Matter Tracts and their Associations to Psychopathological Traits | Santiago Mezzano | Flavio Dell'acqua Acqua |
| 46 | High-resolution diffusion MRI and tractography in the in-vivo & ex-vivo non-human primate brain at 10.5T | Mohamed Kotb Selim | Stamatios N. Sotiropoulos |
| 47 | SBI Meets Tractography: A new approach for Bayesian inference in diffusion MRI models | J.P. Manzano-Patrón | Stamatios N. Sotiropoulos |
| 48 | A Deep Diffusion Model Approach for Diffusion MRI White Matter Fiber Tractography | Yijie Li | Fan Zhang |
| 49 | Deconstructing DTI-ALPS: Clarifying the biological interpretation in aging and cerebral small vessel diseases | Ami Tsuchida | Stephanie Debette |
| 50 | Supervised Learning for Tractogram Alignment | Gabriele Amorosino | Paolo Avesani |
| 51 | DeepDisco: A Deep Learning Framework for Rapid Brain Connectivity Estimation | Anna Matsulevits | Michel Thiebaut de Schotten |
| 52 | White matter bundle segmentation with deformation features in glioma patients | Chiara Riccardi | Paolo Avesani |
| 53 | Identifying the Microstructural Neurobiological Signature of Brain Lesions and Disconnected Tissue Using the UK Biobank | Anna Matsulevits | Mallar Chakravarty |
| 54 | Diffusion MRI tractography to reduce risks of postoperative neurological deficits: A systematic review and meta-analysis | Guido L. Guberman | Sami Obaid |
| 55 | Surface-based Tractography uncovers 'What' and 'Where' Pathways in Prefrontal Cortex | Marco Bedini | Daniel Baldauf |
| 56 | GPU tractography: What can we learn from half a trillion streamlines? | Yanis Aeschlimann | Romain Veltz |
| 57 | A principled mathematical study of the limit of fiber tractography | Samuel Deslauriers-Gauthier | Romain Veltz |
| 58 | Unbiased tractogram density optimisation for robust estimation of white matter connectivity differences | Philip Pruckner | Robert Smith |
| 59 | Intraoperative fast fibre tract segmentation in paediatric tumour patients | Dana Kanel | Jonathan D. Clayden |
| 60 | Integrating normative and patient models of tractography for accurate prognosis in human glioblastoma | Joan Falcó-Roget | Alberto Cacciola |
| 61 | Improving tractography reconstruction with asymmetric FOD tractography: preliminary evidence on the cortico-spinal tract | Richard Stones | Flavio Dell'Acqua |
| 62 | Tractography on Implicit Neural Representations of Diffusion MRI | Sanna Persson | Rodrigo Moreno |
| 63 | Connectivity Patterns across Bipolar Disorder Stages: a Tractography-based Graph Analysis | Serena Capelli | Annabella Di Giorgio |
| 64 | Predicting Facial Nerve Condition and Functional Outcome in Cerebellopontine Angle Tumours Using MRI-Based Models | Alberto Arrigoni | Simonetta Gerevini |
| 65 | Assessment of Different Tractography Methods for Superficial White Matter Reconstruction | Xi Zhu | Fan Zhang |
| 66 | Structural connectivity-based individual parcellations using various tractography algorithms | C. Langlet | J.-F. Mangin |
| 67 | Using Large Language Models to Inform Tractography | Elinor Thompson | Daniel C. Alexander |
| 68 | Towards Whole-Brain Tractography of the Mouse from Serial Optical Coherence Tomography | Charles Poirier | Maxime Descoteaux |
| 69 | Probing the clinical value of tractography reconstruction for glioma surgery | Ludovico Coletta | Silvio Sarubbo |
| 70 | Pre- versus postoperative tractography in patients with (supra)sellar tumors: correlations to optic pathway deformations and vision | Andrey Zhylka | Nico Sollmann |
| 71 | A comprehensive, high-resolution 7T atlas of structural brain connectivity in humans | Hélène Lajous | Patric Hagmann |
| 72 | Improved Riemannian FOD averaging for fiber bundle priors incorporation in FOD-based tractography algorithms | G. Ville | J. Coloigner |
| 73 | Multimodal Interactive White Matter Bundles Virtual Dissection | Garyfallidis E | Paolo Avesani |
| 74 | Multi-compartment tractometry approach for white matter neuroinflammation investigation in late-life depression | Nathan Decaux | Julie Coloigner |
| 75 | An Atlas of the augmented Corticospinal tract | Guillaume Mahey | Bertrand Michel |
| 76 | Extracting tract-specific neurodegeneration by differentiating converging fibers using fixel-based analysis | Lloyd Plumart | Frans W. Cornelissen |
| 77 | Quantifying the Impact of Probabilistic Streamline Turning Angle on Brainstem-Inclusive Whole-Brain Connectomes | Monica Duran | Brittany Travers |
| 78 | The Connectome Analysis for Pediatric Epilepsy Surgery (CAPES) Study: Leveraging Normative Disconnectome Mapping to Predict Seizure Outcomes | Sudarsan Packirisamy | Alexander G. Weil |
| 79 | SWM bundles segmentation using streamlines and voxel information in VAE latent space | S. Navarrete | P. Guevara |
| 80 | Nevrolens 2.0: Augmented Reality Atlas for Cross-Species Neuroanatomical Understanding | Thanh P. Doan | Ekaterina Prasolova-Førland |
| 81 | High-Resolution Tractography Shows Later Maturation of Superficial White Matter Across the Lifespan | J. Urbina-Alarcón | C Beaulieu |
| 82 | Structural Connectivity Mapping of the Central Amygdala | Vinod Kumar | Ivan de Araujo |
| 83 | Post-operative Clinical Outcome Is Not Correlated to Fronto-Striatal Tract Involvement by Diffuse Gliomas of the Supplementary Motor Area: Preliminary Results | Jahard Aliaga-Arias | Francesco Vergani |
| 84 | TractoSearch: a Faster Streamline Search for Scalable | Etienne St-Onge | Etienne St-Onge |

| | Tractography Analysis | | |
|---|---|---|---|
| 85 | Amount of white matter activation and microstructures explain depression recovery in subcallosal cingulate deep brain stimulation | Ha Neul Song | Helen S. Mayberg |
| 86 | Generation of synthetic data for validating tractography-based cortical parcellation and fiber clustering algorithms | Elida Poo | Pamela Guevara |
| 87 | Principal Component Analysis of Diffusion MRI and Magnetization Transfer Metrics Reveals Distinct Lesion Microstructure in Multiple Sclerosis | E. Hernandez-Gutierrez | A. Ramirez-Manzanares |
| 88 | Hybridization Strategies for Robust Brain Tractography | Jesús Martínez-Miranda | Alonso Ramírez-Manzanares |
| 89 | Clinical-ComBAT: Towards Flexible Harmonization of White Matter measures in Clinical Diffusion MRI | Manon Edde | Pierre-Marc Jodoin |
| 90 | Variability of white matter activation pathways for connectomic targeting in subcallosal cingulate deep brain stimulation | Ha Neul Song | Ki Sueng Choi |

**The emergence of consciousness from the brain connectome**

Denis Le Bihan[1]

[1] NeuroSpin, CEA, Paris-Saclay University, Saclay, France

Diffusion MRI, now a pillar of medical imaging, was introduced in 1985[1], demonstrating how the diffusive motion of water molecules could be spatially encoded with MRI to produce images revealing the underlying structure of biological tissues on a microscopic scale. A major subfield of diffusion MRI (Diffusion Tensor Imaging or DTI[2]) makes it possible to obtain non-invasively, from the anisotropic movement of water molecules in the brain's white matter, 3-dimensional images of the cerebral connections making up the Connectome. DTI has opened up new avenues to investigate brain diseases, recently extending to psychiatry, revealing how mental illnesses could be seen as disorders of the Connectome's spacetime, in a new framework which merges structural (anatomical) connectivity and functional connectivity between neural nodes. Given that there is a finite limit on the action potential propagation speed within the connectome, functional “distances” between neural nodes (geodesics), thus, depend on both the *spatial* distances between nodes and the *time* to propagate between them, through a *relativistic connectome spacetime* with four intricated dimensions[3]. Neural nodes act as “masses” which, depending on their degree of activation, “curve” the functional connectome spacetime and the flow of action potentials around them, as for gravity in the Universe. Notably, the propagation speed depends on axonal length: longer fibers carry faster speeds because they are surrounded by a thicker myelin sheath. In patients in minimally conscious states the density of functional connections is known to decrease, with the disappearance of the short (slow) connections to the benefit of long-range (fast) connections[4]. This suggests that conscious activity must be associated with a slowdown of the overall propagation speed, by mobilizing short range connections, in a similar way a gravitational field slows the speed of light according to general relativity. In short, *the equivalent of the gravitational field for the connectome must be consciousness,* which appears as both the source and the result of connected neural activity within the Connectome. Following recent development in theoretical physics[5], we conjecture that the Connectome should be considered to have five dimensions, the fifth dimension allowing the natural, *immaterial* emergence of consciousness as a dual form of the 4D spacetime activity embedded in our *material* cerebral cortex[3]. This curvature may be estimated from DTI (structural) and rs-fMRI (functional) data, providing a quantitative and dynamic signature of consciousness, as shown in subjects in different stages of sleep[6]. Advanced diffusion MRI methods and recent progress in gradient hardware will further allow to estimate axon diameters within the Connectome[7], which will allow maps of conduction velocities to be obtained, with implications to better understand psychiatric syndromes, such as in depression[3,8] and psychotic disorders[3,9].

# White-matter SEEG stimulation identifies a role for the IFOF in semantic control

*D Giampiccolo1,2, N Li3, A Granados4, L Binding1, E Jefferies5, R Jackson5, A O'Keeffe1, V Litvak1, U Vivekananda1, N Burgess1, F Xiao1, J Oliveira1, A McEvoy1,2, J Duncan1, B Diehl1#, A Miserocchi1,2#, F Chowdhury1#*

*1 UCL Queen Square IoN 2 National Hospital for Neurology and Neurosurgery, Queen Square, London 3 Charité – Universitätsmedizin Berlin, 4 King's College London, London, UK 5 University of York*

**Background:** We recently proposed that the inferior fronto-occipital fasciculus (IFOF) would support multimodal semantic control,[1] but this proposal has to be tested

**Methods:** We adopted a novel SEEG stimulation technique with stable white matter contacts in 16 patients implanted (7 left, 2 bilateral, 7 right) for drug-resistant epilepsy. Each patient's IFOF was identified on native preoperative high-resolution tractography and tested with a multimodal neuropsychological battery. We calculated volume of tissue activated (VTA) for each stimulation site in the native space to confirm IFOF stimulation location in LeadDBS. VTAs were then normalised and used for tract-wise two-sample t-test after permutation using a 760 µm ex-vivo normative connectome.

**Results:** 56 contacts pairs in the IFOF were tested (4320 trails, 1429 with stimulation). IFOF stimulation impaired picture naming ($p <0.001$), but also non-language domains such as visual semantics ($p <0.001$), face-perception ($p <0.001$), tool use ($p <0.001$), perseverations ($p <0.001$), hallucinations ($p <0.001$) within the same pair of contacts and among different IFOF contacts in the same patient. VTA analysis in LeadDBS confirmed a role of the IFOF and showed a graded dorso-ventral differentiation its streamlines.

**Conclusions:** Using this novel, powerful white matter stimulation technique, our results demonstrate that IFOF stimulation impacts multimodal semantic control. Besides, SEEG white matter stimulation may represent a powerful technique to disentangle contribution to function beyond language in both hemispheres.

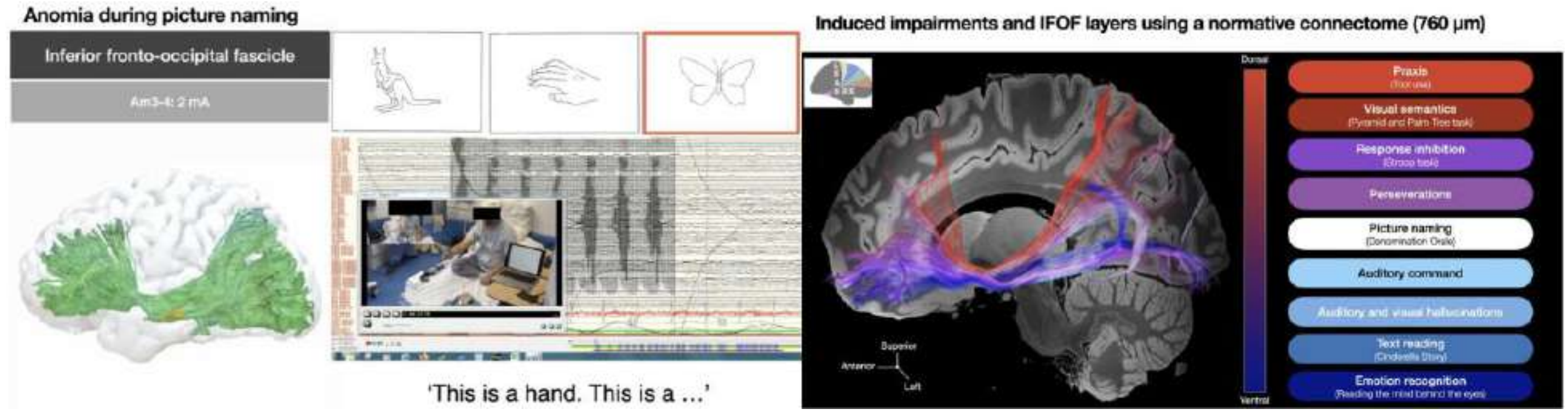

**Title:** White Matter Microstructural Alterations in Young Heavy Alcohol Users: Evidence from 7T Diffusion Imaging of the Anterior Thalamic Radiation.

**Authors:** Alan N. Francis, PhD; Ihsan M. Salloum, MD, MPH
Department of Neuroscience, School of Medicine, University of Texas Rio Grande Valley

**Abstract**

**Background:** Adolescence and young adulthood are critical periods of neurodevelopment during which alcohol use may exert long-lasting effects on brain structure and function. Diffusion tensor imaging (DTI) enables assessment of white matter microstructural integrity through metrics such as fractional anisotropy (FA) and radial diffusivity (RD), which provide insights into axonal organization and myelination.

**Methods:** This study examined white matter alterations in heavy alcohol users compared to light users using ultra-high-field 7 Tesla DTI data from the Human Connectome Project. Twenty-four heavy alcohol users were matched to Twenty-four light users based on age and sex. TRACULA (TRActs Constrained by UnderLying Anatomy) was employed to reconstruct major white matter tracts. Probabilistic tractographic analyses focused on the bilateral anterior thalamic radiation (ATR), a pathway implicated in executive function and cognitive control.

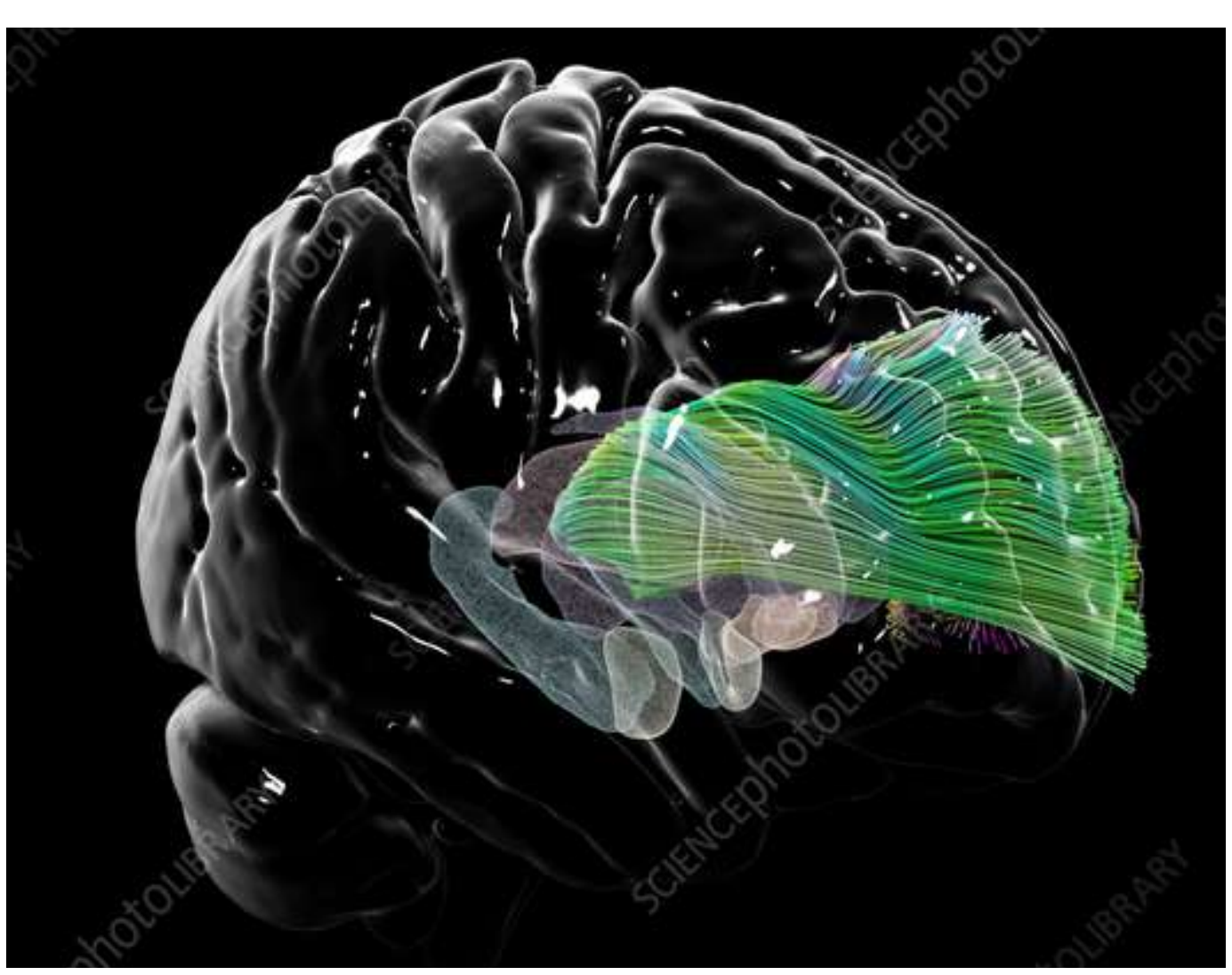

**Results:** Heavy alcohol users exhibited significantly higher FA in the left ATR compared to light users ($p = 0.033$), alongside significantly lower RD in the same region ($p = 0.017$). These findings suggest altered white matter microstructure, potentially reflecting aberrant myelination or compensatory reorganization in response to alcohol-related neurotoxicity.

**Conclusions:** The observed microstructural alterations in the left ATR may represent early biomarkers of alcohol-induced neuroadaptation, with potential implications for cognitive functioning. These results underscore the vulnerability of developing white matter to the effects of heavy alcohol use and highlight the need for early intervention. Future longitudinal studies are warranted to assess the functional consequences of these alterations and their potential reversibility with sustained abstinence.

**Title:** **Advanced reconstruction of white matter tracts in preterm neonates from clinical diffusion MRI data.**

**Authors and Affiliations:** **Laurie Devisscher[1,2,3], Yann Leprince[2], Nicolas Elbaz[4], Chloé Ghozland[5], Parvaneh Adibpour[1,2], Catherine Chiron[1,2], Sara Neumane[1,2], Aline Gonzalez-Carpinteiro[1,2], Lucie Hertz-Pannier[1,2], Marianne Barbu-Roth[3], Alice Heneau[5], Valérie Biran[1,5], Marianne Alison[1,4], Jessica Dubois[1,2]**

1.Université Paris-Cité, INSERM, NeuroDiderot, F-75019 Paris, France 2.Université Paris-Saclay, CEA, NeuroSpin, UNIACT, F-91191 Gif-sur-Yvette, France 3.Université Paris-Cité, CNRS, Integrative Neuroscience and Cognition Center, F-75005 Paris, France 4.Assistance Publique-Hôpitaux de Paris - APHP, Robert-Debré Hospital, Department of Pediatric Radiology, F-75019 Paris, France 5.Assistance Publique-Hôpitaux de Paris - APHP, Robert-Debré Hospital, Neonatal Intensive Care Unit, F-75019 Paris, France

**Introduction:** Our study aims to develop a robust pipeline with diffusion MRI (dMRI) and tractography to automatically extract a wide range of white matter bundles in the whole brain of preterm infants at term equivalent age with anatomical particularities such as cerebral lesions, increased volumes of cerebral ventricles and extracerebral cerebrospinal fluid (CSF). As a first step, we here focused on sensory and motor tracts (cortico-spinal tract, CST) to characterize the developing microstructural properties.

**Methods:** We collected and analyzed 3T-MRI clinical data of 105 very and extremely preterm babies (gestational age at birth: 24-32 weeks), scanned at term equivalent age (38-43 weeks of post-menstrual age–wPMA). We used the baby-XTRACT tool implemented in FSL that provides tractography protocols for mapping 42 white matter bundles (19 bilateral and 4 central tracts) defining seeding, stop and exclusion regions on the Schuh neonatal template [3]. A key point of this study was to obtain a robust registration of individual images to the template despite the brain anatomical specificities of our population. We optimized this registration by creating brain masks from a combination of iBEAT and drawEM segmentations [4][5] of super-resolved T2w images (0.8mm isotropic,obtained with NiftyMIC[6]), which enabled us to remove part of the CSF. Individual dMRI images without diffusion weighting (b=0) were then coregistered to T2-weighted images, which were themselves registered to the template [7]. The registrations were conducted using Ants 2.5.3 with finely tuned parameters. Besides, following the pre-processing of dMRI data (b=1000 s/mm$^2$ with 42 directions), multiple fibre orientations were estimated with BEDPOSTX with a two fibres model [8]. Tractography was performed with PROBTRACKX through the baby-XTRACT framework and tract reconstructions were visually checked for the CST. Maps of diffusion tensor imaging (DTI) metrics were estimated, allowing us to compute the tract-density-weighted averages of metrics in the CST. To further evaluate the approach robustness, we assessed the correlation between metrics and PMA at scan.

**Results:** Despite a wide range of brain anatomical features, we were able to achieve optimal registration for all infants (Figure a) as well as correct CST reconstructions (Figure b). The exploration of CST microstructure confirmed a decrease of mean, axial and radial diffusivities with PMA at scan, as well as an increase in fractional anisotropy (Figure c).Residual inter-individual variability might rely on other factors.

**Conclusion:** This study provides a proof of concept for applying the baby-XTRACT tool to clinical diffusion MRI data in very preterm infants with cerebral injury. This approach allowed us to obtain accurate extraction of the CST and will be generalized to other white matter tracts. By characterizing the tract microstructural properties, we will explore the effects of several clinical factors beyond PMA at scan (e.g., gestational age at birth, respiratory, digestive and infection complications, brain severity score proposed by Kidokoro et al [9]). This approach will also be applied to MRI data collected at 2 months of corrected age in a subgroup of 39 babies with longitudinal data to evaluate the effect of an early motor training [10].

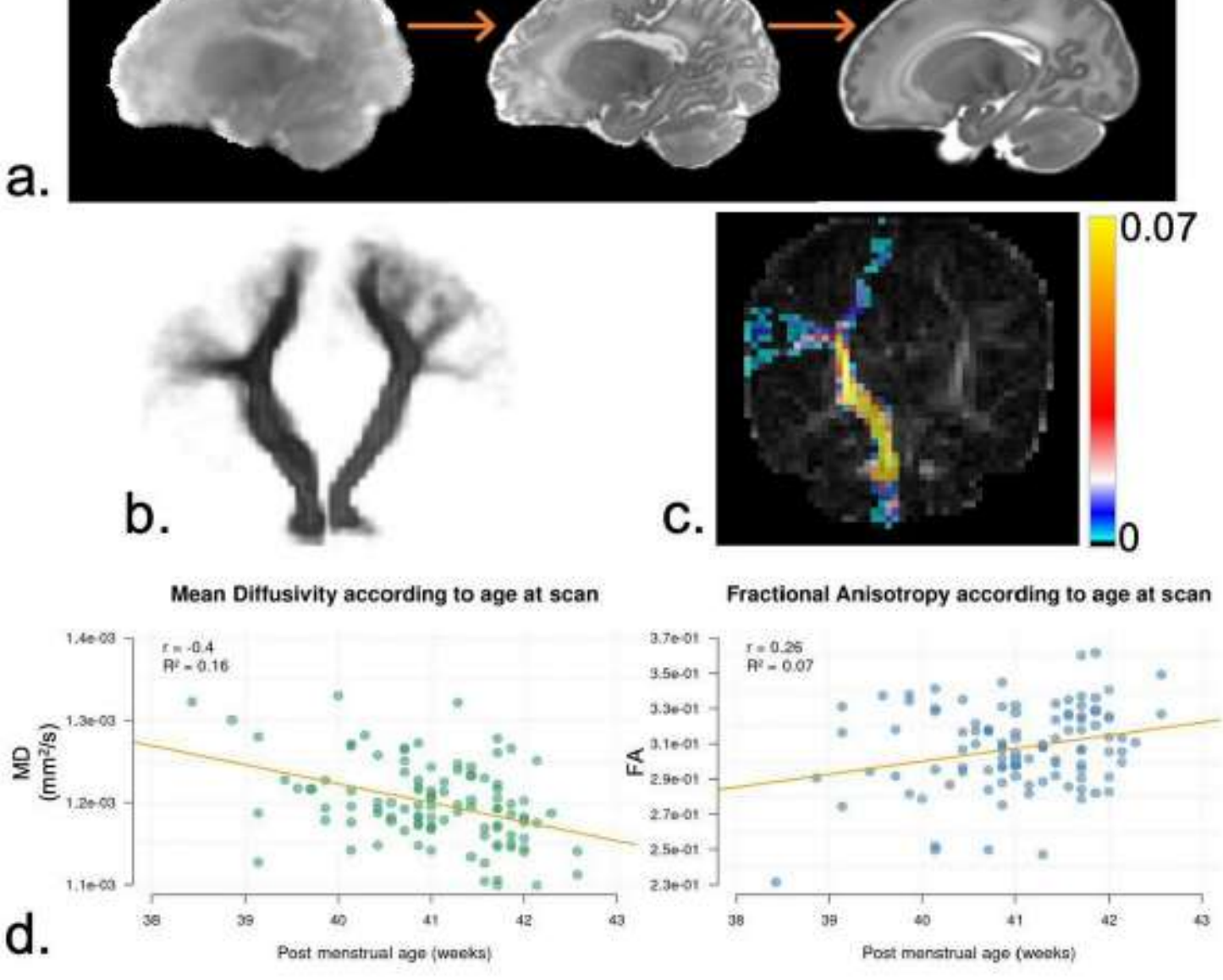

***Figure 1*** *a. Example of individual registrations, from left to right: dMRI images (b0) to T2w images, T2w images to Schuh template. b,c. Individual reconstructions of corticospinal tracts: 3D coronal views (b) and normalized density of right tract, projected on the FA map (c). d. Graphs of MD and FA according to post-menstrual age at MRI scan. Regression lines show statistically significant Pearson correlations (p<0.01).*

## Reduced White Matter Microstructure in Opiate Use Disorder: A Diffusion MRI Study

Richard Nkrumah[1], Lea Wetzel[2], Justin Böhmer[3], Katharina Eidenmüller[2], Wolfgang Sommer[2,4], Hendrik Walter[3], Gabriele Ende[1]

[1] Department of Neuroimaging, Central Institute of Mental Health, Heidelberg University, Germany
[2] Department of Addictive Behavior & Addiction Medicine, Central Institute of Mental Health, Heidelberg University, Germany
[3] Department of Psychiatry and Psychotherapy, Charité – Universitätsmedizin Berlin, Germany
[4] Department of Psychopharmacology, Central Institute of Mental Health, Heidelberg University, Germany

**Introduction:** Substance use disorders (SUD), particularly involving opiates, cannabis, and nicotine, are associated with widespread white matter (WM) alterations that may underlie cognitive and behavioral dysfunctions[1]. Traditional diffusion tensor imaging (DTI) provides insight into general WM microstructure but is limited in resolving crossing fibers. Fixel-based analysis (FBA), an advanced tractography-based method, offers higher specificity by quantifying fibre density (FD), fibre cross-section (FC), and their product (FDC), enabling a more accurate assessment of white matter pathology[2].

**Methods:** Participants = 143 adults (Controls = 51, Opiate = 17, Cannabis = 51, Nicotine = 24), recruited across two sites (Berlin and Mannheim). All substance users met DSM-5 criteria for addiction. Imaging & Analysis: Diffusion-weighted imaging acquired using single-shot spin-echo EPI sequence with the following parameters: TR/TE=3200/63ms, matrix = 110 × 110, 72 slices, $2 \times 2 \times 2\text{mm}^3$ and 60 directions with a b value of 1000 $\text{s/mm}^2$ and processed using MRtrix[3]. Measures: DTI= Fractional Anisotropy (FA) & Mean Diffusivity (MD). FBA= FD, FC & FDC. Statistics: Independent t-tests comparing each user group to controls, adjusting for age, sex, and site. FWE correction using threshold-free cluster enhancement (TFCE) with $p < 0.05$.

**Results:** DTI and FBA results highlighting reduced FA, increased MD, and decreased FDC in opiate users.

### Results I - DTI measures

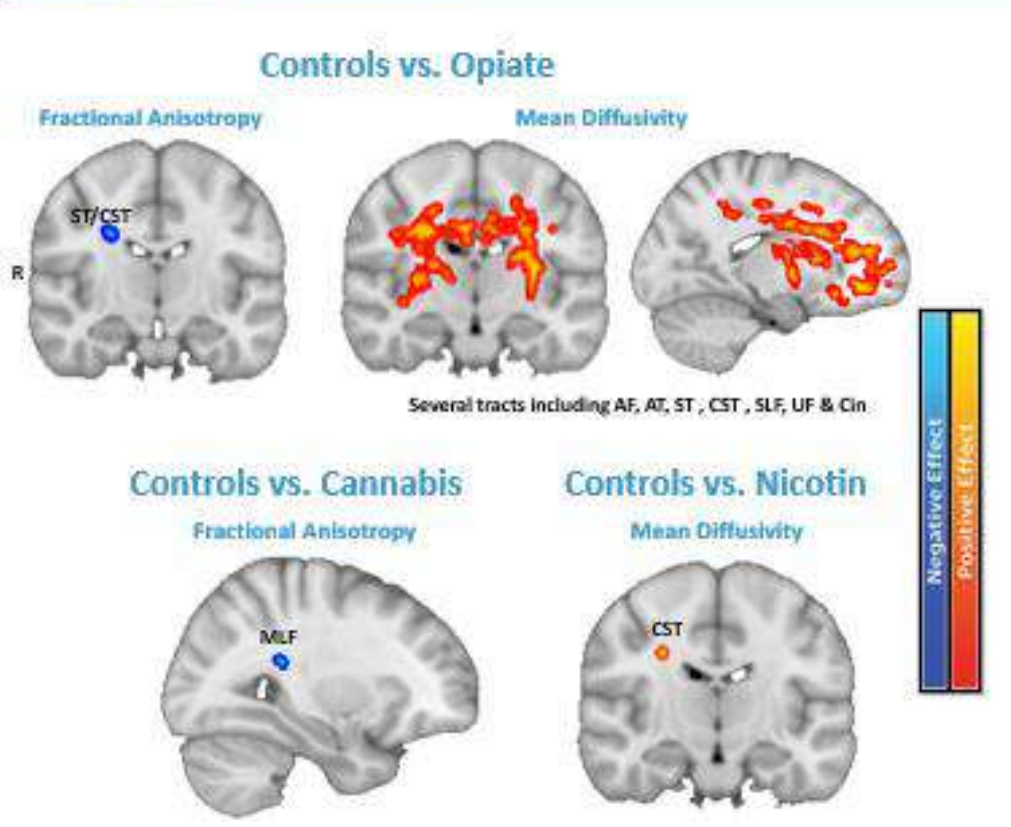

Differences between substance users and healthy control in DTI measures. AF: Arcuate fasciculus, AT: Anterior thalamic radiation, ST: Superior thalamic radiation, CST: Cortical spinal tract, MLF: Middle Longitudinal Fasciculus, Cin: Cingulum, SLF: Superior longitudinal fasciculus, UF: Uncinate fasciculus, R: right

### Results II- FBA measures

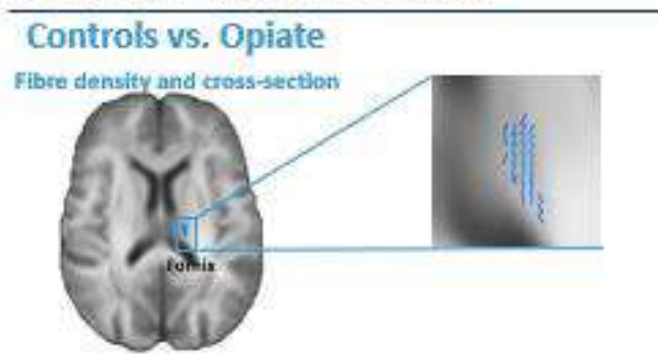

Reduced fibre density and cross-section in left fornix in opiate users compared to healthy controls. No significant results were found for either FD or FC measures in opiate users, nor between cannabis or nicotine users and healthy controls.

Opiate Users:

- DTI: Significantly reduced FA and increased MD across multiple WM tracts (e.g., corticospinal tract, arcuate fasciculus).
- FBA: Significant reduction in FDC in the left fornix, indicating disrupted limbic connectivity.

Cannabis Users:

- Reduced FA in the middle longitudinal fasciculus (MLF), suggesting localized WM alterations.

Nicotine Users:

- Increased MD in the corticospinal tract, possibly reflecting demyelination or reduced axonal packing.

**Conclusions:** Tractography-based FBA reveal that opiate use is associated with the most extensive WM micro- and macrostructural alterations, particularly in tracts supporting memory and emotion regulation (e.g., fornix)[4]. Cannabis and nicotine use show more localized changes[5]. These findings highlight the potential clinical utility of advanced tractography techniques like FBA in identifying substance-specific brain alterations, potentially informing tailored interventions and substance-specific biomarker development in SUD.

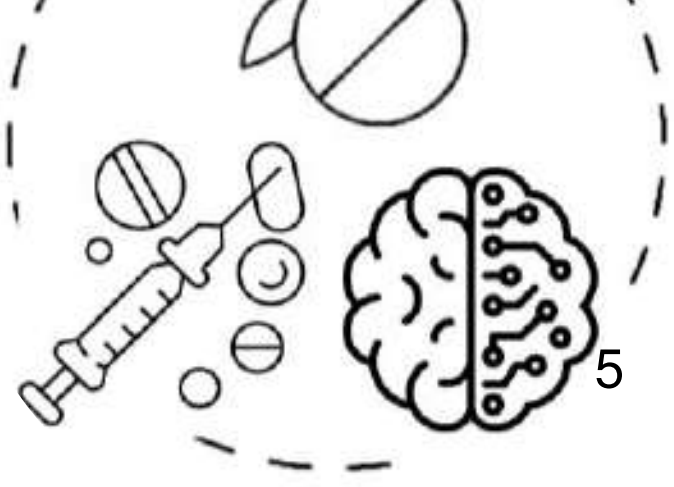

# Bimetric Invariants for Geodesic Tractometry and Machine Learning †‡

Luc Florack, Rick Sengers, Eindhoven University of Technology, The Netherlands

**Introduction.** Unlike streamline tractography, geodesic tractography [1, 2] must contend with the implications of the Hopf-Rinow theorem, which states that *any* two points in the brain can be geodesically connected. This requires additional criteria to distinguish anatomically plausible tracts from arbitrary geodesics. Specification of two side conditions will single out one (or at best a few) candidate(s), but this in itself begs the question of anatomical validity [3].

Given a candidate tract, we would therefore like to assess its anatomical plausibility. *Complete systems of bimetric tractometric invariants* provide data driven evidence for this purpose. By integration with pattern recognition or machine learning trained on expert feedback, geodesic redundancy enables optimization of specificity without loss of sensitivity.

**Theory.** A *tractometric invariant* of a putative tract $\gamma$ is a functional $\phi(\gamma) \in \mathbb{R}$ independent of curve parameterization and spatial coordinates. *Completeness* of a system of invariants entails that all data evidence is accounted for given some symmetry constraint. The system is said to be *irreducible* if there are no mutual dependencies among its elements. It is notoriously hard to construct such systems in general [4], but some simple known instances are relevant for our purpose.

**Application.** We focus on *global* invariants assigned to an entire tract, and impose the constraint that only zeroth order data confined to the tract's spatial locus are admitted. A trivial example is the complete irreducible system $\{\phi_i(\gamma) = \lambda_i(D)\}_{i=1,2,3}$ of DTI eigenvalue averages $\lambda_i(D)$ along a tract, which is equivalent to the set of DTI power traces $\{\psi_i(\gamma) = \operatorname{tr} D^i\}_{i=1,2,3}$ ('polynomial invariants'). Completeness implies that DTI matrix powers $D^k$ for $k \geq 4$ play no role (Cayley-Hamilton). Adding curve orientation requires an extension of this system with 2 additional invariants. To construct polynomial invariants there is a simple visual 'diagrammar', with the following elements:

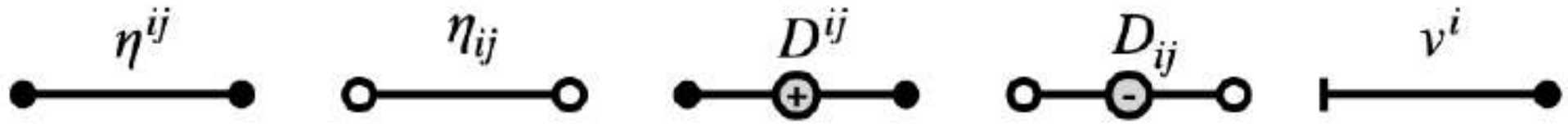

*Diagrammar.* $\eta_{ij}$ is the Euclidean metric, with dual $\eta^{ij}$ (identity matrices in Cartesian coordinates); $D^{ij}$ is the DTI tensor, with inverse $D_{ij}$ (acting as a space curving metric for geodesic tractography); $v^i$ is the geodesic unit tangent; endpoints $\bullet$ and $\circ$ symbolize free indices that may be connected ('contraction') and thereby resolved. Consecutive labels $\oplus$ and $\ominus$ annihilate pairwise. These rules make it easy to construct invariants, viz. any closed diagram (without dangling $\bullet$ or $\circ$) is a polynomial invariant, vice versa. The 'tadpole' diagram for the tangent vector $v^i$ has only one free index and could be replaced by the dyadic tensor $v^i v^j$ without loss of generality (with some superficial diagrammatic consequences).

*Example.* A complete irreducible system of global tractometric invariants based on lowest order DTI data and tract orientation is shown below. Replacing $\oplus$ by $\ominus$ yields an equivalent system. Trivial invariants have been omitted. The diagrams are insightful pictures corresponding to algebraic expressions. A simple, unsupervised and spatially unbiased clustering of the 3-dimensional scatter plot of the three loop diagram features only (equivalent to the DTI eigenvalue averages along tracts), ignoring the interplay with tract orientation, is consistent with the neuroanatomical branching of the CST into medial and lateral sub-bundles. The example also shows the idea of pruning using pattern recognition or machine learning algorithms, in this case an unsupervised isolation forest algorithm, increasing the internal micro-structural coherence by removing anomalous tracts (per cluster). Interestingly coherence outliers tend to be spatial outliers, despite the fact that no spatial information is encoded in our tractometric invariants (e.g., no exclude regions have been used in the CST figure below).

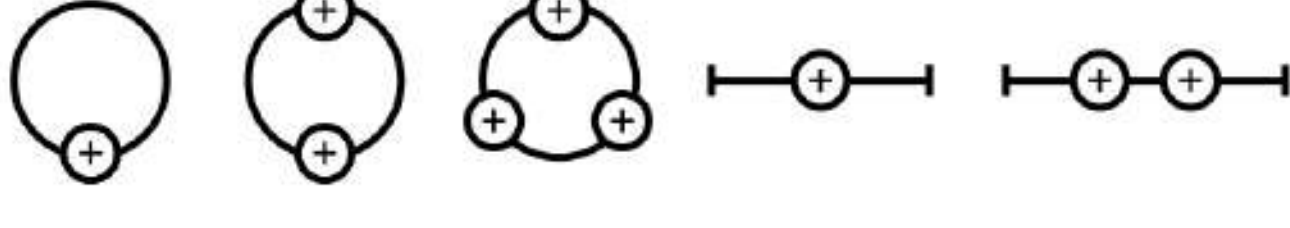

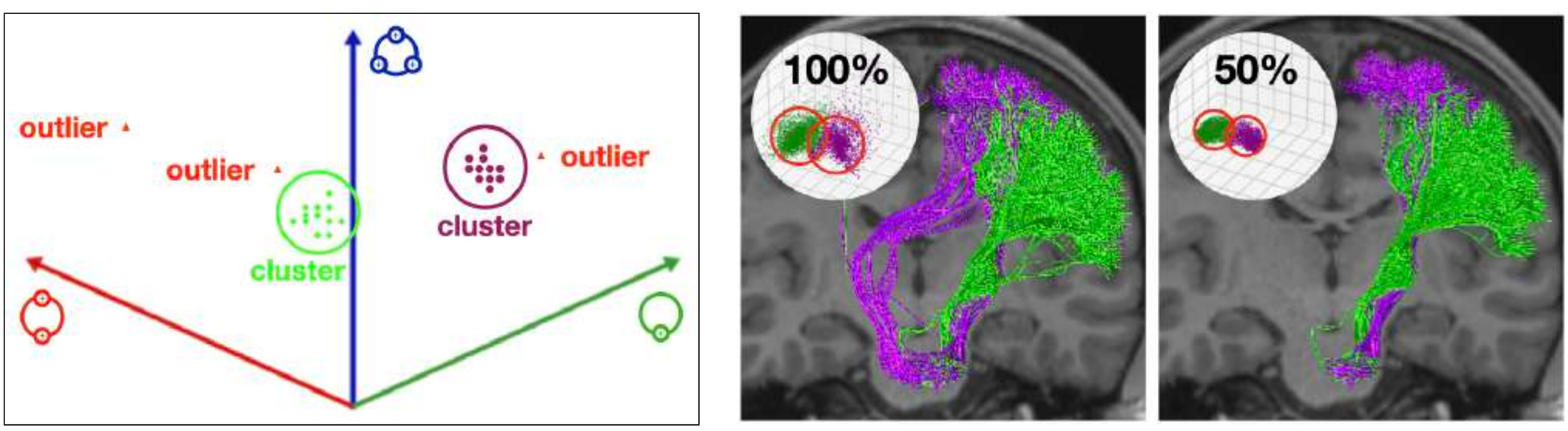

**Conclusion.** Geodesics are optimal-diffusion paths and as such form a natural yet highly redundant set of primitives for tractography. Anatomically plausible geodesics ('true positives') require disambiguating criteria in terms of meaningful tractometric invariants in a necessarily bimetric setting. Completeness ensures that, under specified symmetries, no information is a priori discarded. We have sketched the idea by constructing a small yet potentially powerful irreducible system to feed pattern recognition and machine learning methods for anatomical specificity. This system can be enriched with additional global invariants, or generalized to include local invariants, at a concomitant computational price.

† International Society for Tractography: Inaugural Conference, October 13–16, 2025, Centre Broca, Bordeaux, France.

‡ This publication is part of the project Bringing Tractography into Daily Neurosurgical Practice with the project number KICH1. ST03.21.004 of the research program Key Enabling Technologies for Minimally Invasive Interventions in Healthcare, which is (partly) financed by the Dutch Research Council (NWO). Elisabeth-TweeSteden Hospital (Tilburg), Erasmus Medical Center (Rotterdam), Amsterdam University Medical Center (Amsterdam) and Medtronic are gratefully acknowledged for their support.

# Normative Tract Profiles of White Matter Microstructure and Metabolite Ratios Along the Superior Longitudinal Fasciculus in Healthy Human Brain

Archith Rajan[1], Sourav Bhaduri[2], Subhanon Bera[2], Laiz Laura de Godoy[1], Mauro Hanaoka[1], Sulaiman Sheriff [3], Suyash Mohan[1], and Sanjeev Chawla[1]

[1]Department of Radiology, University of Pennsylvania, Philadelphia, PA, United States, [2]Institute for Advancing Intelligence (IAI), TCG CREST, Kolkata, India, [3]Department of Radiology, University of Miami, Miami, FL, United States,

**Purpose:** Superior longitudinal fasciculus (SLF) is the largest association tract in the brain. The SLF is critical in multiple normal functions like cognition, visuospatial attention and memory, and its microstructural integrity is known to be compromised in several neuropsychological conditions [1]. Several studies have also reported significant metabolite alterations in several neurological and neuropsychiatric disorders including schizophrenia [2]. Despite documenting significant differences in diffusion MRI derived parameters between normal controls and schizophrenia patients, no significant differences in metabolite patterns from brain regions encompassing SLF were observed between two groups in a study [3]. This might be due to the fact that conventional proton MR spectroscopy (1H-MRS) and diffusion MRI (dMRI) voxels were not co-registered together. With this limitation in mind, this methodological study was designed (i) to overlay metabolite maps on SLF-I and II segments (ii) to assess the metabolite distribution along the path of SLF-I and II segments across various regions and (iii) to evaluate the relationships among metabolite ratios and (dMRI) derived parameters from these segments in normal healthy adults.

**Methods:** Anatomical images, WBSI and high angular resolution diffusion imaging (HARDI) sequences were acquired from 10 healthy normal adults (age: 33.67 ± 2.52 years; 7M/3F) on a 3T MRI scanner. To evaluate the intrasubject variability, one subject underwent MRI scans three times. The preprocessing pipeline is illustrated in **Figure 1**. The WBSI data were analyzed using MIDAS package with the standard processing steps [4]. Quality assurance was evaluated by considering Cramer-Rao lower bounds (<20%), line shape, line width (2-12Hz), CSF contamination, and degree of residual water and lipid signals. Parametric maps of choline (Cho) / N-acetylaspartate (NAA) and choline (Cho) / creatine (Cr) were computed. Subsequently, only those voxels in the maps that had greater than 50% probability of white matter tissue were retained for further analysis. A three-shell diffusion imaging protocol with b-values of 300, 800 and 2000 s/mm$^2$ and a total of 109 unique diffusion encoding directions, with 9 interspersed $b_0$ images, was used to generate neurite orientation dispersion density imaging (NODDI) derived intra-cellular volume fraction (ficvf), isotropic volume fraction (fiso) and orientation density index (ODI) and diffusion tensor imaging (DTI) derived mean diffusivity (MD) and fractional anisotropy (FA) maps. The whole brain tractograms were generated and an automated tool [5] was used to delineate the segments I and II of SLF. White matter maps of Cho/Cr and Cho/NAA were then co-registered to non-diffusion weighted ($b_0$) image using the Water signal intensity maps. The SLF I and II segments were divided into 20 anatomically distinct sections and section-wise mean values of parameters (MD, FA, ficvf, fiso, ODI, Cho/NAA and Cho/Cr) for each tract profile were computed [6] from all the subjects. Repeatability was determined by calculating the coefficients of variation (CVs), intra-classcorrelation (ICC), and repeatability coefficient (RC) of WBSI and dMRI derived parameters for each SLF segments. To assess the covariance between microstructural and metabolite metrics, mixed effect linear models were assessed, with WBSI parameters as the dependent variable for mean tract profiles from 10 subjects. The significance level was set at $p<0.01$.

**Figure 1: Overview of the image processing pipeline to generate along tract profiles**

**Results:** The white matter metabolite maps of Cho/NAA and Cho/Cr were successfully overlaid over SLF segments I and II in all cases. Repeatability analyses revealed the intra-subject CVs in the acceptable range of 0.8% (ficvf: SLF I right) to 13.36% (Cho/NAA: SLF II right). Metabolite measures from the SLF-I right were the most reliable (ICC: 0.944 and 0.845; CV: 0.023 and 0.032; RC: 0.33 and 0.20 for Cho/Cr and Cho/NAA respectively). The mean inter-subject CVs across the sections of tract profiles were in the range of 3.2% (MD: SLF I right) to 23.1% (fiso: SLF II left). Strong linear associations were observed between the tract profiles of the metabolite ratio maps and the corresponding tract profiles of dMRI parameter maps, with the model including all the dMRI parameters showing a consistent association for the segments SLF I right, SLF II right and SLF II left:

**Cho/NAA ~ f (1 + ODI + fiso + ficvf + FA + MD)** Adjusted $R^2$ = 0.953, 0.954, 0.818, 0.811 for SLF I right, SLF I left, SLF II right and SLF II left respectively
**Cho/Cr ~ f (1 + ODI + fiso + ficvf + FA + MD)** Adjusted $R^2$ = 0.968, 0.945, 0.965, 0.804 for SLF I right, SLF I left, SLF II right and SLF II left respectively

**Discussion and Conclusion:** Our findings may be useful for simultaneously assessing metabolite alterations and white matter microstructure of SLF-I and II segments under multiple pathological conditions. The proposed approach may allow more objective and unbiased assessment of regional metabolite patterns along the path of SLF as opposed to the conventional single and multi-voxel MR spectroscopy methods of placing distinct large voxels along a specific section of the tract. Our study is small and cross-sectional involving only ten normal healthy individuals and would require further validation in larger cohorts. Such multi-parametric normative tract profiles of white matter microstructure and metabolism could serve as the basis for early detection of neurodegeneration in various pathologies.

# Cross-Species Cortical Parcellation via Homology Consensus Graph Representation Learning from Diffusion MRI Tractography

**Yazhe Zhai[1], Yifei He[1], Jiaolong Qin[1], Fan Zhang[2], and Ye Wu[1,*]**
[1] School of Computer Science and Technology, Nanjing University of Science and Technology, Nanjing, China
[2] School of Information and Communication Engineering, University of Electronic Science and Technology of China, Chengdu, China

## Introduction

Cross-species cortical parcellation seeks to identify conserved brain regions by analyzing human and macaque brains together[1]. However, accurate alignment is challenging due to differences in cortical folding and anatomy[2]. We propose a two-stage clustering framework that combines structural and geometric connectivity to better identify homologous parcels while maintaining species-specific features.

## Methods

We group cortical vertices into super-vertices by minimizing a weighted sum of geodesic and feature-space distances for consistency across species. Next, we create a vertex-cluster matrix by mapping streamline endpoints to cortical vertices and a vertex-atlas matrix by aggregating overlaps with the XTRACT atlas[3]. These form multi-view graphs optimized through low-rank tensor learning to create a shared low-dimensional embedding. Finally, we apply spectral clustering and align human and macaque parcels using pairwise matching via the Dice coefficient.

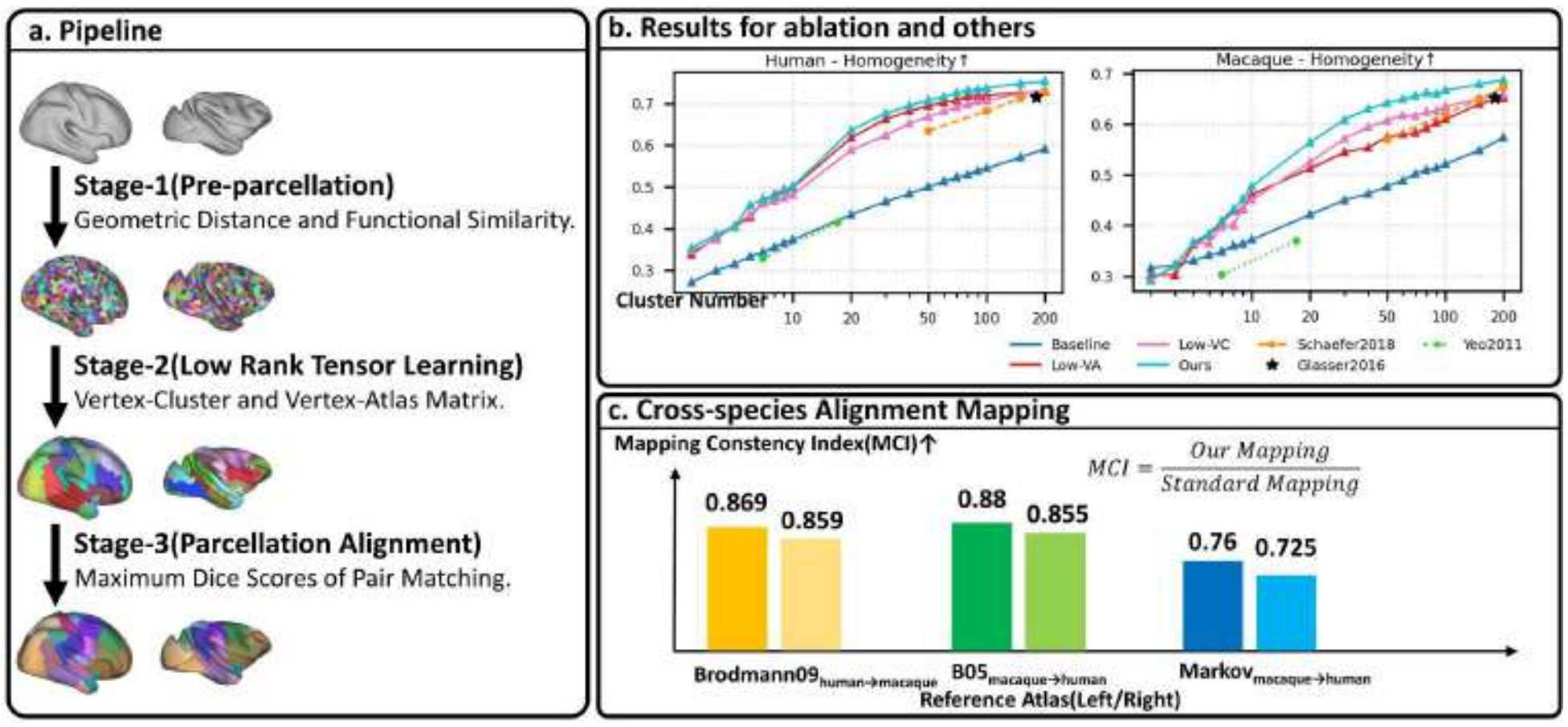

Fig.1 The pipeline of our method and results.

## Results

Our framework consistently produced biologically meaningful parcellations across a range of cluster counts, with optimal balance at 20–30 parcels. When benchmarked against established atlases (Yeo[4], Schaefer[5], Glasser[6]) and ablations using only vertex–cluster or vertex–tract features, our two-stage method achieved superior homogeneity in both species. Cross-species alignment, quantified by a novel Mapping Consistency Index against Brodmann, B05, and Markov references, alignment by Xu et al.[7], exceeded 0.85 for humans to macaque and 0.75 for macaque to humans in 20 parcels.

## Conclusion

We present a two-stage framework for cross-species cortical parcellation that combines geometric and connectivity features. We use multimodal fusion and low-rank tensor learning to identify homologous regions.

# Cross-species Standardised Cortico-Subcortical Tractography

Stephania Assimopoulos[1], Shaun Warrington[1], Davide Folloni[2], Katherine Bryant[3], Wei Tang[4], Saad Jbabdi[5], Sarah Heilbronner[6], Rogier B Mars[5], Stamatios N Sotiropoulos[1]

*[1]Sir Peter Mansfield Imaging Centre, School of Medicine, University of Nottingham, UK; [2]Nash Family Department of Neuroscience and Friedman Brain Institute, Icahn School of Medicine at Mount Sinai, New York, NY, USA; [3]Centre de Recherche en Psychologie et Neurosciences, UMR 7077, CNRS/Université Aix-Marseille, France; [4]Luddy School of Informatics, Computing and Engineering, Indiana University Bloomington, IN, USA; [5]Oxford Centre for Integrative Neuroimaging, University of Oxford, UK; [6]Baylor College of Medicine, Houston, TX, USA.*

**Introduction:** White matter (WM) bundles connecting cortical areas with subcortical nuclei are crucial for relaying and modulating cortical function [1]. Their disruption is linked to abnormal function and pathology in neurodegenerative and mental health (MH) disorders [2,3]. Diffusion MRI (dMRI) and tractography enable exploration and reconstruction of such WM bundles [4], but their relative size, the complexity and associated bottlenecks, make their estimation challenging [5]. As a result, cortico-subcortical WM tracts are under-represented in dMRI tractography studies. Here, we introduce a set of standardised tractography protocols for delineating tracts between the cortex and various deep subcortical structures, including the caudate, putamen, amygdala, thalamus and hippocampus. Our protocols were first devised in the macaque brain, guided by chemical tracer literature, and then extended to the human. We assessed the tract reconstructions against tracer studies patterns and their robustness to data quality. We subsequently incorporated these protocols into a common space framework [6,7] to assess their efficacy in connectivity-based cross-species prediction of homologous cortical structures and subcortical nuclei.

**Methods:** We built upon our previous cross-species tractography framework (FSL-XTRACT [8]). Using prior anatomical knowledge from macaque tracers, we defined new generalisable protocols in template space for the amygdalofugal tract (AMF), the Muratoff bundle (MB) and the striatal bundle (StB) (external capsule) for its frontal, sensorimotor, temporal and parietal parts, adding to our previous protocols for hippocampal and thalamic tracts [8]. Due to their close proximity, we also developed new protocols for the respective extreme capsule (EmC) parts (frontal, temporal, parietal) and revised previously protocols for the uncinate fasciculus (UF), the anterior commissure (AC) and the fornix (FX). Each protocol includes a unique combination of seed, target, waypoint, exclusion masks, delineated in standard macaque space (F99), and then correspondingly defined in human standard (MNI) space. Using dMRI data from 6 ex-vivo rhesus macaque brains [6] (from PRIME-DE) and 50 HCP subjects [9], we performed tractography in human and macaque.

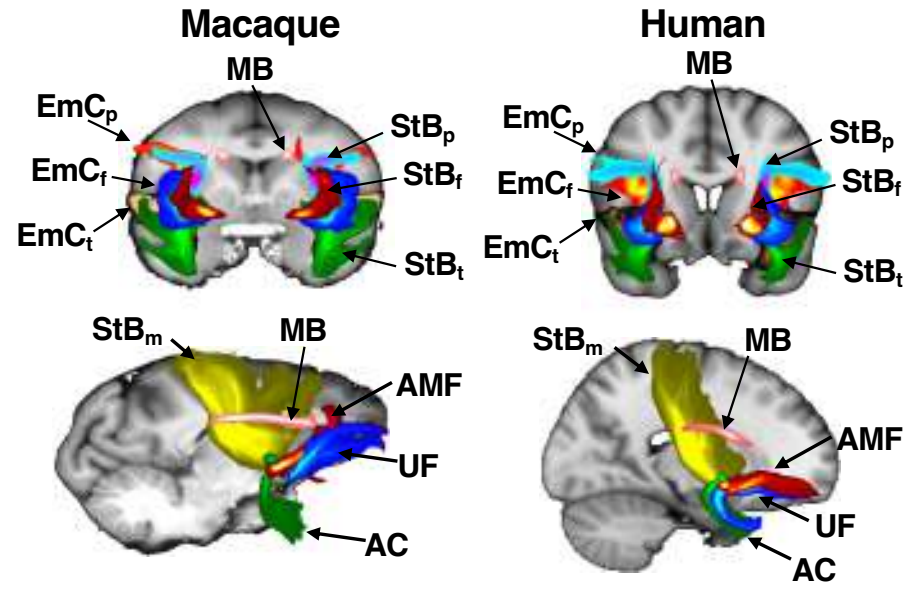

Figure 1: Cross-species cortico-subcortical tractography.

**Results:** Tractography reconstructions are shown in Fig 1. These were tested for robustness against data quality (HCP & UK Biobank data in the human) and template spaces (F99 & NMT in the macaque) – results not shown. Subsequently, we ensured that reconstructed tracts follow principles known from tracers in the macaque and how these translate to humans (Fig. 2). Fig. 2A demonstrates how white matter relative spatial organisation is preserved between the reconstructed bundles. Fig. 2B shows how different cortical areas connect to the putamen, as revealed by tracer injections (from superior to inferior: more connections to parietal, sensorimotor, frontal and temporal). Tractography of the StB parts reveals similar patterns in the putamen, both for the macaque and the human. Lastly, we used these species-matched tractography protocols (new & original XTRACT [8] – 59 tracts in total) to map homologous grey matter regions in humans and macaques. Fig. 3A shows how similar frontal cortical regions across macaques and humans connect similarly to the same set of bundles (blue and orange polar plots, respectively). The similarity of such tractography patterns can therefore be used to identify homologous regions between species [6, 7]. Fig. 3B demonstrates maps of (dis)similarity (KL divergence) between tractography patterns across the whole macaque/human cortex, when neighbouring seed frontal regions are selected in the human (first column)/macaque (second column) (dmPFC & vmPFC, OFC & FOp). Predictability of these regions based solely on their tractography patterns agrees well with reference atlas boundaries, showing the richness of information these patterns convey.

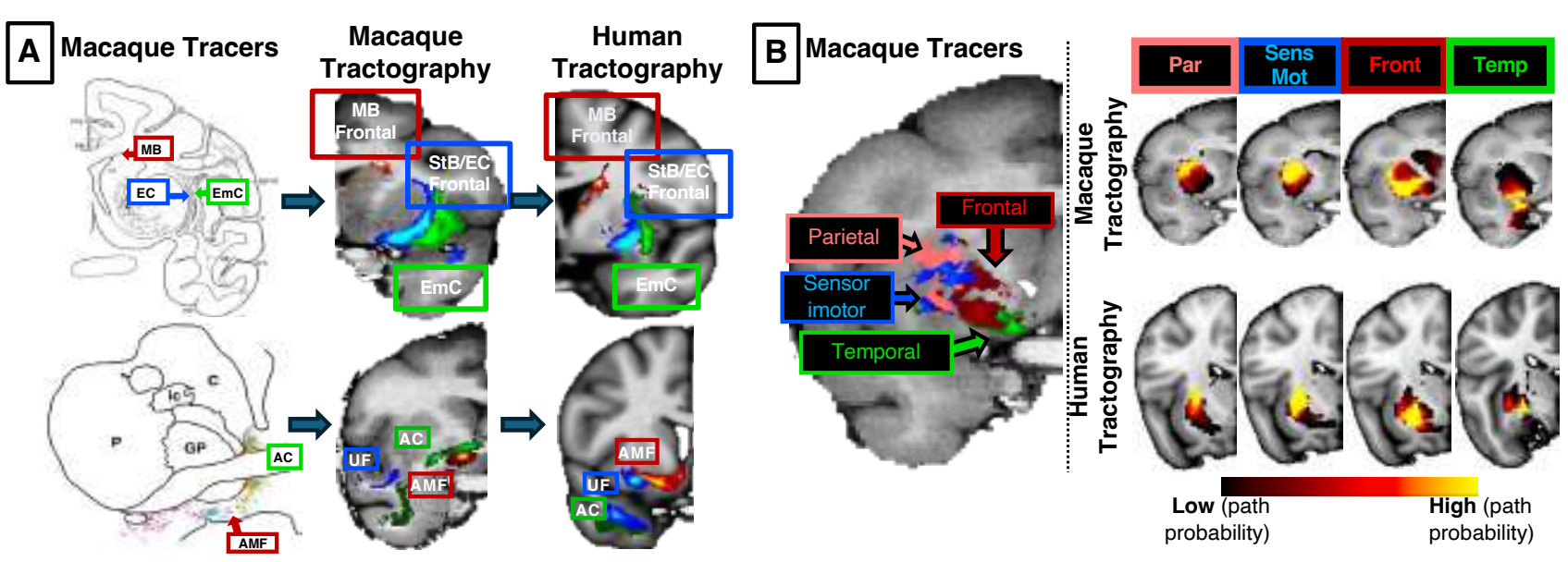

Figure 2: (A) Relative spatial organisation of WM bundles is preserved with tractography, following patterns of tracers. (B) Connectivity patterns between different cortical areas (parietal, sensorimotor, frontal, temporal) and the putamen as found with tractography, are in agreement with tracer patterns.

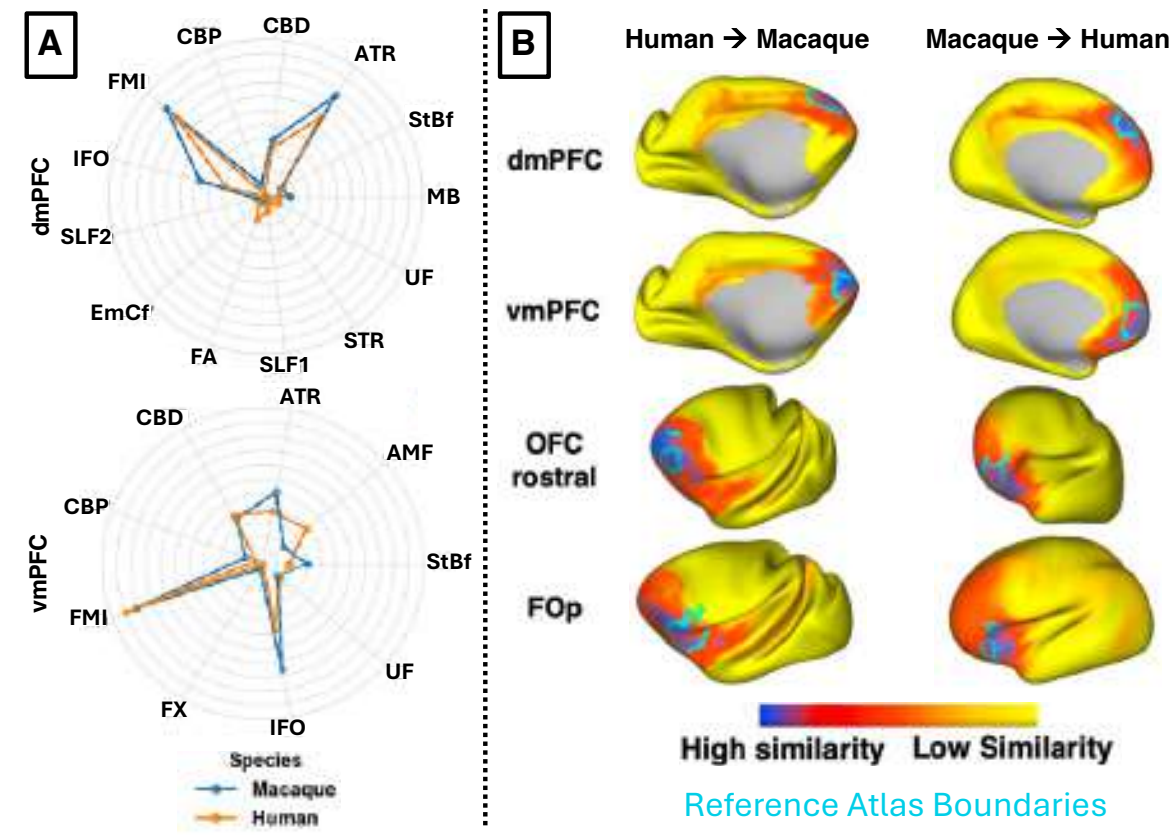

Figure 3: (A) Patterns of connectivity of cortical areas to WM bundles are similar across species for corresponding cortical areas. (B) Cross-species cortical ROI prediction based on tractography patterns.

**Conclusion:** Building upon our previous work on FSL-XTRACT [6, 8, 10], we introduced standardised protocols for automated cortico-subcortical tractography in the macaque and human brain. We demonstrate the robustness of the reconstructed patterns against tracers and their value in mapping homologous grey matter regions across species.

**Spatial Pointwise Orientation Tracking (SPOT): Resolving the spatial layout of fibre ODFs for super-resolution streamlining**

Saad Jbabdi[1], Amy Howard[1,2]
[1]*Oxford Centre for Integrative Neuroimaging, University of Oxford, UK*
[2]*Department of Bioengineering, Imperial College London, London, United Kingdom*

## Introduction

A fundamental and unsolved challenge in diffusion MRI tractography is to uncover the spatial distribution of orientations within an imaging voxel from the voxelwise fibre orientation distributions (FODs). Streamlining in tractography attempts to solve this problem through a combination of interpolation and heuristics [1], such as penalising sharp turns in the streamlining. Although this approach has been successful over the years in mapping large coherent bundles, tractography notoriously still suffers from "bottleneck" issues where the FOD can ambiguously represent different underlying orientation configurations [2]. The problem is that once we've built an FOD, it is too late to recover the underlying pointwise orientations. Here we propose a potential paradigm shift: rather than estimating pointwise orientations through streamlining *after* fitting FODs, we propose to model the underlying pointwise orientations directly in a forward model of the data that bypasses the need for fitting FODs (Fig 1A). Pooling the data across neighbourhoods of voxels makes this model tractable (invertible). We present a preliminary version of the model in simulated and real data, and suggest how this type of approach can be used for joint modelling of dMRI and polarised light imaging (PLI) data from the same tissue, where the PLI further constrains the model.

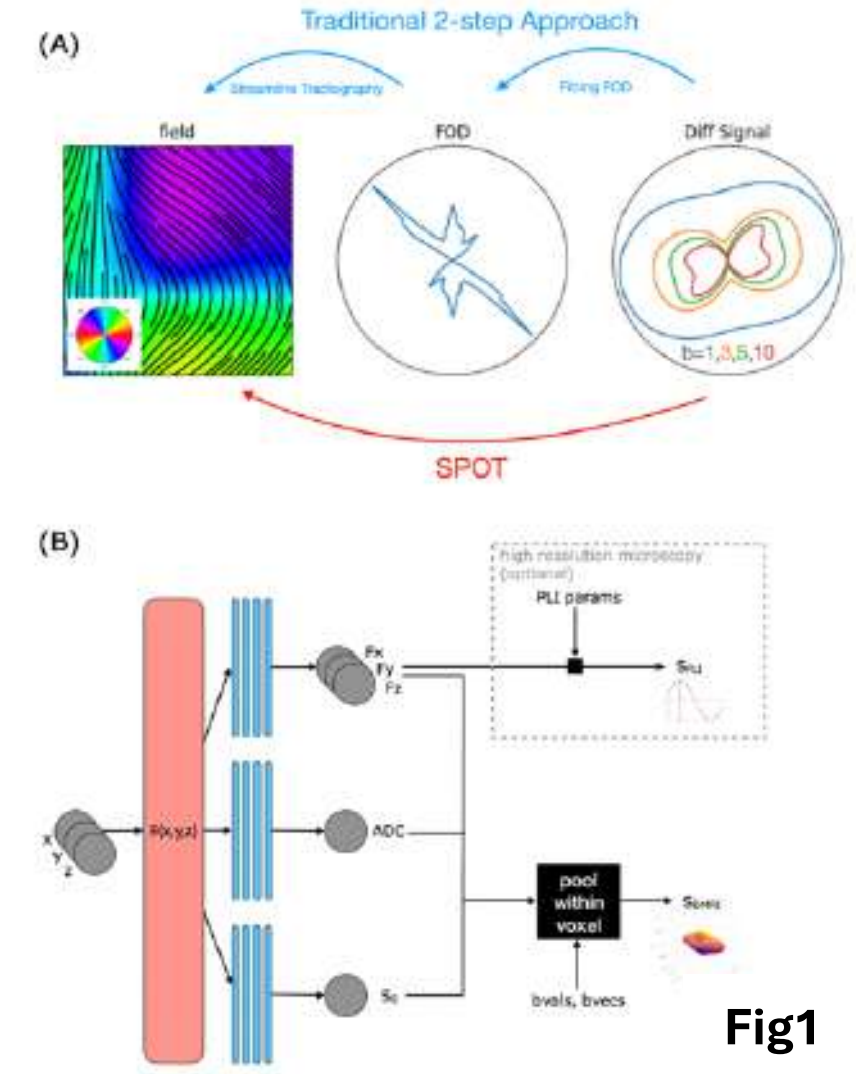

**Fig1**

## Methods

At the core of the forward model are simple multi-layer perceptrons (MLPs, which in these experiments have 4 layers, 256 neurons per layer, tanh activations, linear output). The MLPs take as input the (x,y,z) coordinates of any point in space (not necessarily at the centre of a voxel), and output either a 3D vector (for modelling the orientation vector field) or a scalar (for modelling e.g. the diffusion coefficient). To enable modelling of both high and low frequencies in the spatial domain, we insert a Fourier Feature embedding layer between the input and the MLPs [3]. The outputs of the MLPs (for the vector and scalar fields) are then used within forward models of the data (Fig 1B). For example, to predict diffusion data, sample positions from within a voxel are fed through the MLPs to generate orientations and scalars, then fed through the diffusion model equation for a stick given a set of bvals/bvecs, and then averaged over the voxel. The predicted signal is then compared to the measured diffusion data to calculate a mean squared error loss. This is summed over a neighbourhood of voxels to calculate the overall loss used for training the MLPs. Other data, such as polarised light imaging (PLI), can be similarly predicted from the orientation vector field, enabling the joint modelling of multiple modalities (optional extra). SPOT was evaluated using simulated diffusion data from an orientation vector field (here 2D) which we aimed to recover. For comparison with interpolation, the data was resampled onto a 100x100 grid using splines of order 3 (with Scipy's map_coordinates), and a tensor was fitted in each pixel to obtain a principal orientation per pixel. Next, we assessed the impact of jointly fitting a single dMRI voxel alongside one or two slides of high-resolution PLI intersecting the voxel. Finally, we show some early results fitting SPOT to in vivo dMRI data (60 directions, 2mm isotropic, b=1k).

## Results

Preliminary results are shown in figures 2-4. In 2D simulations, we compare SPOT to spline interpolation (Fig 2) for different grid sizes (1x1, 2x2 and 4x4) to demonstrate the effect of including neighbourhood information. SPOT outperforms interpolation in terms of the resolved pointwise orientations and the FODs. When jointly fitting to MRI and PLI (Fig 3), SPOT manages to recover the underlying pointwise orientations from a single voxel (no neighbourhood information) even with a small amount of PLI as additional constraint . Results comparing SPOT to constrained spherical deconvolution (CSD) in in vivo data (Fig 4) show how FODs with crossing fibres resolve into streamlines bending onto the cortex in SPOT.

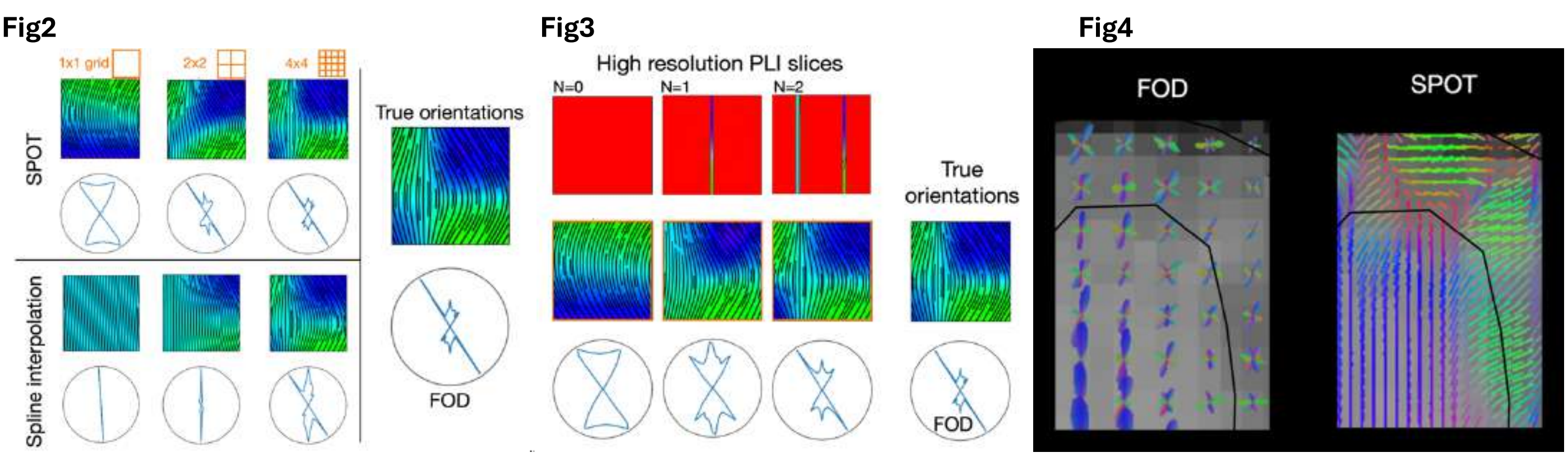

**Fig2** **Fig3** **Fig4**

# Estimating Brain Fibers via Volumetric Cortical Folding Deformation

## *Generating fibers without diffusion data or machine learning*

Besm Osman, Ruben Vink, Andrei Jalba and Maxime Chamberland
**Eindhoven University of Technology, The Netherlands**

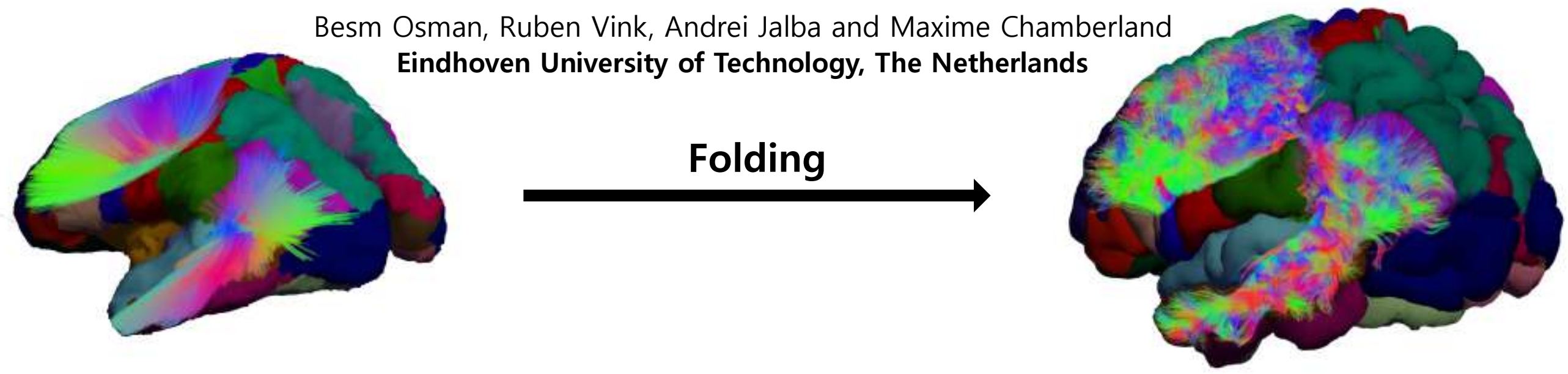

**Folding of generated fibers using our volumetric model. Video: [https://besm.dev/misc/ist-abstract](https://besm.dev/misc/ist-abstract)**

## Introduction

Diffusion data appear insufficient for distinguishing true fibers from spurious ones due to fundamental limitations of dMRI (Schilling et al. 2017), motivating the search for anatomical information within structural MRI that can inform fiber bundle geometry (St-Onge et al. 2018, Cai et al. 2024). Computational models have shown that cortical folding structure predicts various aspects of underlying fibers (Garcia et al. 2021). Unfortunately, current cortical folding models are unable to reproduce subject-specific gyrification patterns due to the complexity of the underlying mechanisms and high inter-subject variability (Alenya et al. 2022), making them unsuitable for subject-specific analysis in tractography.

In this work, we present a subject-specific computational model that simulates cortical fold deformations from the fetal to neonatal stage using anatomical features derived from T1-weighted MRI alone. We demonstrate potential use of the deformation field by generating and deforming simplified fetal-like fiber geometries from unfolded (fetal) to a folded (neonatal) configuration resulting in comparable fiber geometry to state-of-the-art tractography algorithm **without** the use of any diffusion data.

## Methods

We developed a **volumetric**, **subject-specific** cortical folding model based on reversed unfolding. The model estimates the deformation between a smooth fetal brain and the intricate folding pattern observed in an individual's T1-weighted MRI. This is achieved by simulating quasi-static cortical growth as a reversible process within a constraint-based system (Macklin et al. 2016). We begin with a T1w scan from a HCP Young Adults subject (Essen et al 2012), from which we extract a cortical surface and generate a volumetric mesh. We approximate the deformation by resolving constraints for surface area and volume, gradually scaled from neonatal to fetal values, yielding a deformation that approximates cortical development in reverse.

We use this deformation model to study how simplified fetal fibers are shaped by cortical folding. To isolate the contribution of deformation alone, we explicitly avoid using any diffusion data for fiber generation. The pipeline takes as input a set of Desikan cortical parcellation labels, from which we generate simple geometries (straight lines or quadratic Bézier curves) between labeled regions in the fetal model. Fiber points $\mathbf{p}(t)$ are deformed by maintaining barycentric coordinates $\lambda_i$ within linearly interpolated tetrahedra with vertices $\mathbf{v}_i(t)$, such that $\mathbf{p}(t) = \sum_{i=0}^{3} \lambda_i \mathbf{v}_i(t)$.

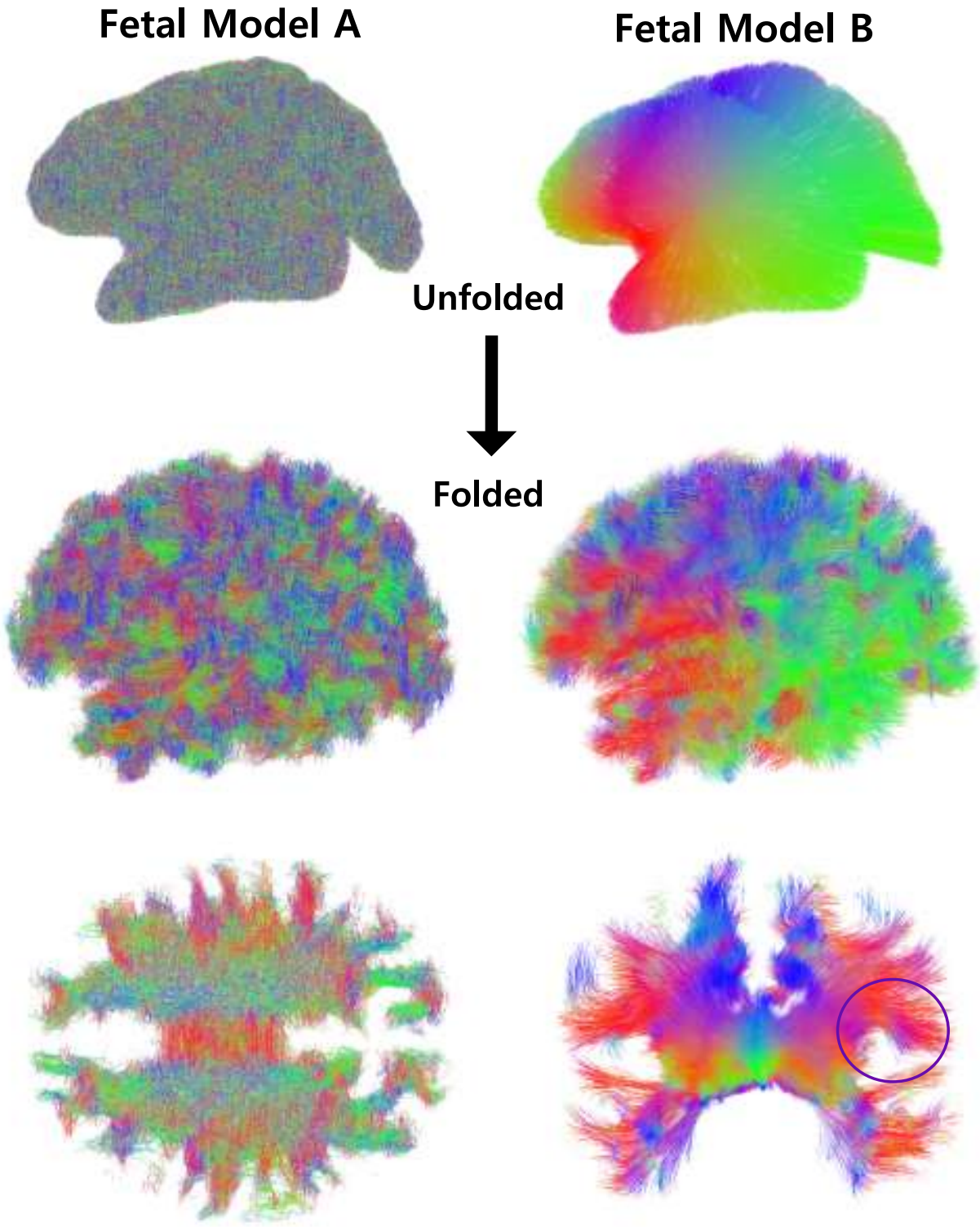

**Fig 1: Deformation of simplified fetal-like fiber models resulting in organized anatomically-aligned structures.**

## Results

We first demonstrate how our pipeline generates and deforms whole-brain fetal-like fiber configurations, as shown in Figure 1. Fetal Model A consists of short, randomly oriented fibers distributed throughout the brain volume. Despite the random initialization, the deformed fibers exhibit clear directionality, which is consistent with theories suggesting that tension-induced cortical growth contributes to fiber alignment. Expanding fibers during deformation tend to align with primary or secondary peaks found in FOD data, particularly near the surface. Fetal Model B consists of radial fibers extending from the cortical surface toward the center. The resulting deformation shows that fiber tips near the surface curve toward gyri and away from sulci (see purple circle), consistent with previous observations (Garcia et al. 2021).

We also show that fiber bundles can be synthesized using this deformation-based approach. The Middle Longitudinal Fascicle (MLF) is generated by connecting regions corresponding to four labels from the Desikan parcellation atlas via Bézier curves. In Figure 2, we compare our generated bundle to TractSeg (Wasserthal et al. 2018) output. The deformation-based bundle is similar in orientation and shape, despite using neither diffusion data nor filtering . The only specified inputs are the atlas labels (middle temporal, superior temporal, inferior parietal, and supramarginal) and a single control point defining the Bézier curve midpoints.

**In conclusion**, we present a subject-specific cortical folding pipeline that approximates volumetric deformations between fetal and neonatal configurations. By deforming simplified fetal-like fibers through this model, we generate anatomically plausible bundles **without** the use of diffusion data. Results align with known fiber bundles and geometry, demonstrating the potential of cortical deformation extracted from T1-weighted scans as a source of anatomical information to inform fiber structure. Future work will include validation using fetal MRI and refinement of the fetal fiber model to produce more anatomically realistic fetal geometries.

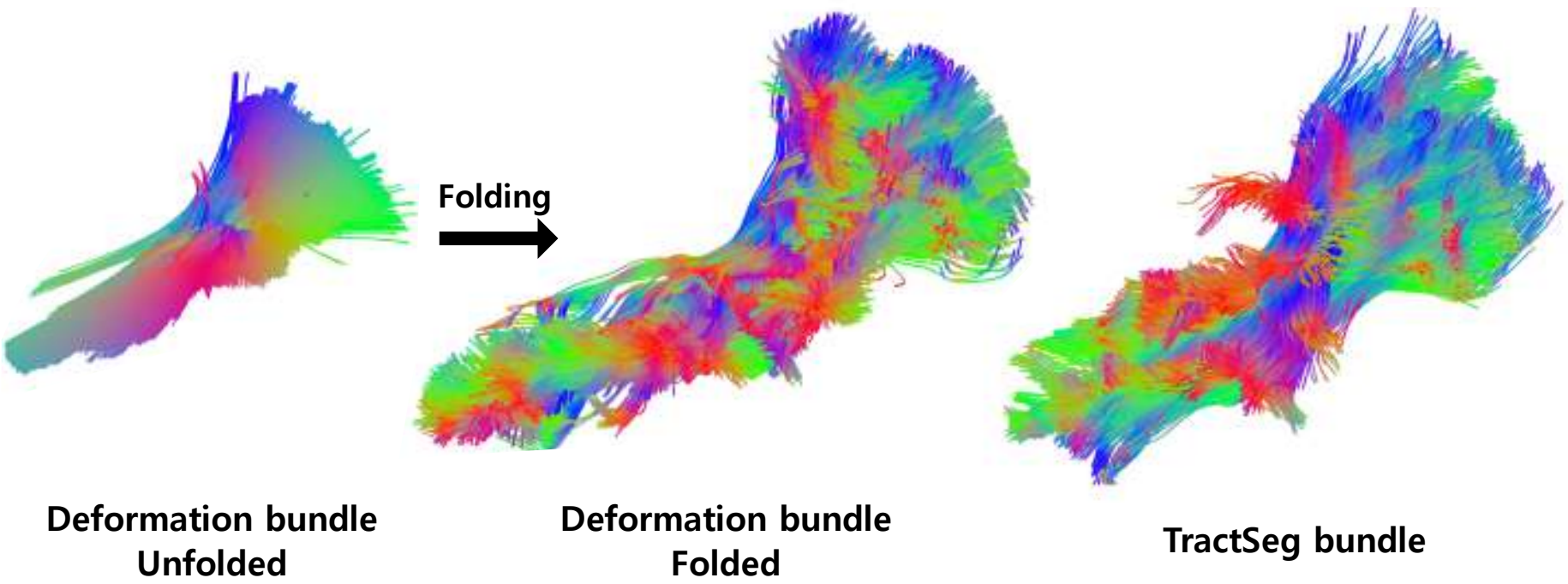

**Deformation bundle Unfolded** | **Deformation bundle Folded** | **TractSeg bundle**

**Fig 2: MLF bundle generated with only deformation data compared with TractSeg diffusion data-based MLF.**

**Background:**

ADHD is increasingly conceptualized as a disorder of disrupted structural connectivity (1). While cortical white matter has been extensively studied, subcortical networks—particularly the limbic system—remain underexplored, despite their central role in emotion regulation and behaviour (2). The current study uses advanced diffusion MRI acquisition and manual tractography to overcome prior limitations in resolving complex limbic pathways.

**Methods:**

We analysed multi-shell HARDI data from 169 participants (72 ADHD, 97 controls) scanned at three timepoints between ages 9 and 14 using a Siemens 3T system. Acquisition employed a multi-band accelerated sequence (b = 1000/2000/2800 s/mm²; 130 directions), allowing for high angular resolution. Preprocessing in ExploreDTI (3) included EPI and motion correction, B-matrix rotation, and robust tensor estimation via REKINDLE.

Deterministic Constrained Spherical Deconvolution (CSD) tractography was used to reconstruct whole-brain white matter fibres (CSD) (4). Limbic system white matter were isolated using manual tractography and anatomical ROIs. Microstructure was assessed using diffusion kurtosis imaging (DKI), specifically kurtosis anisotropy (KA), a sensitive marker of myelination and fibre complexity (5). Connectomes were built using Destrieux-based nodes, and macrostructural features were evaluated using graph theory metrics such as routing efficiency and network density (6).

**Results:**

Children with ADHD showed significantly lower KA in the bilateral cingulum bundle across all timepoints, suggesting reduced myelination. Within the ADHD group, higher symptom severity was associated with lower routing efficiency and network density in limbic circuits, highlighting altered macrostructural connectivity.

**Conclusion:**

Using dMRI and tractography techniques, this study captures longitudinal changes in subcortical limbic networks previously difficult to resolve. Findings provide compelling evidence for disrupted limbic connectivity in ADHD, broadening our understanding of the disorder's neural mechanisms and opening promising avenues for future exploration of subcortical brain networks.

# Integrating Jacobian Determinant and LDDMM for Evaluating White Matter

**Zhaoxian Ming[1], Zhijian Yao[2], Yifei He[1], Yu Xie[1], Ye Wu[1,*], and Jiaolong Qin[1,*]**

[1] School of Computer Science and Technology, Nanjing University of Science and Technology, Nanjing, China

[2] Department of Psychiatry, the Affiliated Brain Hospital of Nanjing Medical University, Nanjing, China

## Introduction

Diffusion Magnetic Resonance Imaging (dMRI) is key for studying white matter (WM) changes in depression by revealing microstructural alterations[1]. However, current metrics often miss broader structural alterations, hindering understanding of antidepressant-induced neuroplasticity. This study proposes a framework combining the Jacobian determinant and large deformation diffeomorphic metric mapping (LDDMM) to evaluate WM fiber bundle remodeling.

## Methods

The study investigated dMRI data from 51 patients with major depressive disorder (MDD) before and after 12-week antidepressant selective serotonin reuptake inhibitor (SSRI) treatment. Following standard preprocessing, a two-level method was used for fiber bundle segmentation, initially identifying inter-regional fibers with the Yeo atlas and MRtrix3[2, 3], which were then subdivided using K-means clustering. Centroid fibers representing these bundles were generated using QuickBundles clustering[4]. The LDDMM algorithm then registered each category's pre- and post-treatment centroid fibers to capture morphological changes. Momentum vectors were derived through LDDMM's iterative optimization, and the Jacobian determinant was calculated from perturbed point displacements. After quantifying WM longitudinal alterations by Jacobian determinant, a correlation analysis was conducted to explore the relationship between these alterations and clinical improvement, as measured by reduction in Hamilton Depression Rating Scale (HAMD) scores, in MDD patients. Specifically, Pearson's correlation analysis was used, and the false discovery rate (FDR) method with $q = 0.05$ was applied for multiple corrections.

## Results

FDR correction identified 32 WM fiber clusters with deformation significantly correlating with clinical improvement (Fig. 1b). Specifically, most significant positive results involve fibers connecting somatomotor-limbic regions ($r = 0.35 \sim 0.6$, $p_{corr}<0.02$), somatomotor-putamen ($r = 0.35 \sim 0.6$, $p_{corr}<0.03$), somatomotor-accumbens ($r = 0.6 \sim 0.9$, $p_{corr}<0.01$), limbic-putamen/pallidus ($r = 0.35 \sim 0.65$, $p_{corr}<0.03$), and amygdala-accumbens ($r = 0.8 \sim 0.95$, $p_{corr}<0.04$). Negative results involve fibers connecting limbic-default mode network/amygdala/brainstem ($r = -0.7 \sim -0.3$) and central executive network-accumbens ($r = -0.7 \sim -0.3$, $p_{corr}<0.04$). These findings are in line with prior research [5].

## Conclusion

This study integrates LDDMM and Jacobian determinant analysis to explore the effects of SSRI treatment on WM fiber structure. The results indicate that the Jacobian determinant is a valuable metric for quantifying WM changes related to SSRI treatment and aiding in the understanding of neuroplasticity.

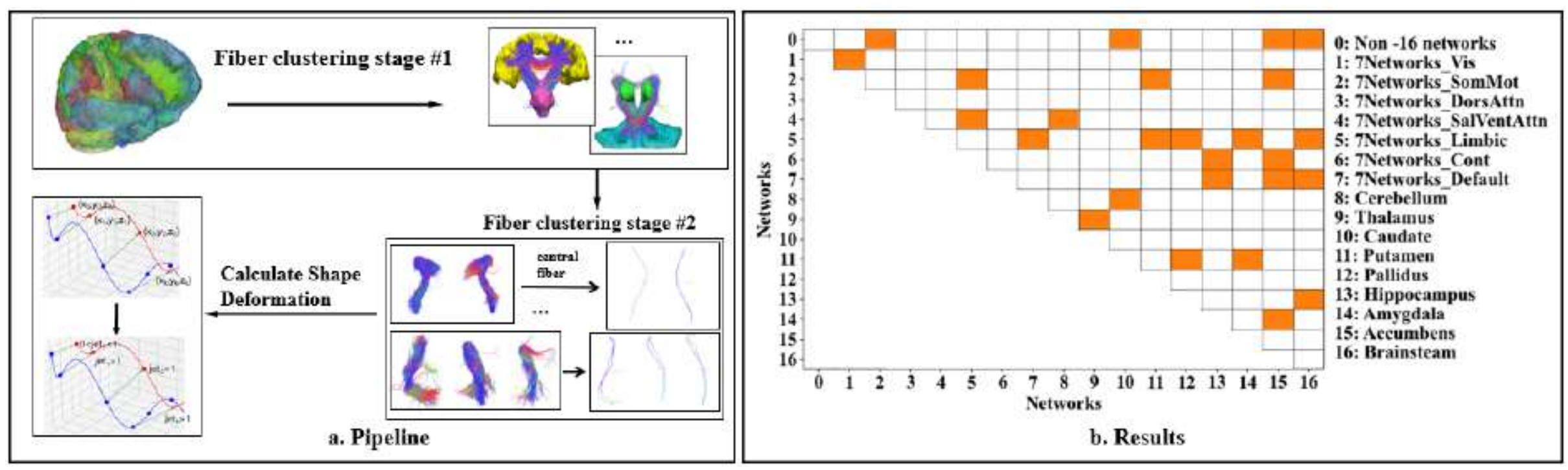

Fig.1 Pipeline and Results. (a) Methodological pipeline: Initial segmentation of WM fibers using the Yeo atlas and MRtrix3, followed by fiber clustering, central bundle extraction, and shape deformation calculation. (b) The color of the triangle elements in the result matrix indicates a significant correlation between the deformation of their corresponding fiber clusters and the reduction ratio of total HAMD scores.

# Connectome-based consensus graph learning for fine-scale subcortical parcellation

**Zhonghua Wan[1], Yu Xie[1], Yazhe Zhai[1], Lei Xie[2], Yifei He[1], and Ye Wu[1,*]**
[1] School of Computer Science and Technology, Nanjing University of Science and Technology, Nanjing, China
[2] College of Information Engineering, Zhejiang University of Technology, Hangzhou, China

## Introduction

Subcortical regions like the thalamus and amygdala are essential for sensory processing and memory[1]. Diffusion MRI maps their complex white matter pathways, making it a powerful tool for identifying structurally informed subcortical subdivisions. We present a novel framework for fine-scale subcortical parcellation by integrating tractography-based connectivity and voxel-level features within a multiscale consensus graph learning framework.

## Methods

We propose a multiscale subcortical parcellation framework combining tractography-based structural connectivity with voxel-level anatomical features using diffusion MRI data from 171 healthy participants in the Human Connectome Project (HCP). Whole-brain tractography was used with deterministic and probabilistic algorithms and registered to MNI152 space. Streamlines connecting brain regions were clustered into fiber bundles, allowing us to create sparse voxel-wise structural connectivity matrices. We utilized a 3D extension of the SLIC[2] supervoxel algorithm to group similar voxels and recalculated connectivity matrices at the supervoxel level (Fig. 1a-b). A consensus graph representation learning strategy[3] was employed to produce individualized and population-consistent parcellations, leading to a group-level atlas generated by averaging embeddings across subjects (Fig.1c).

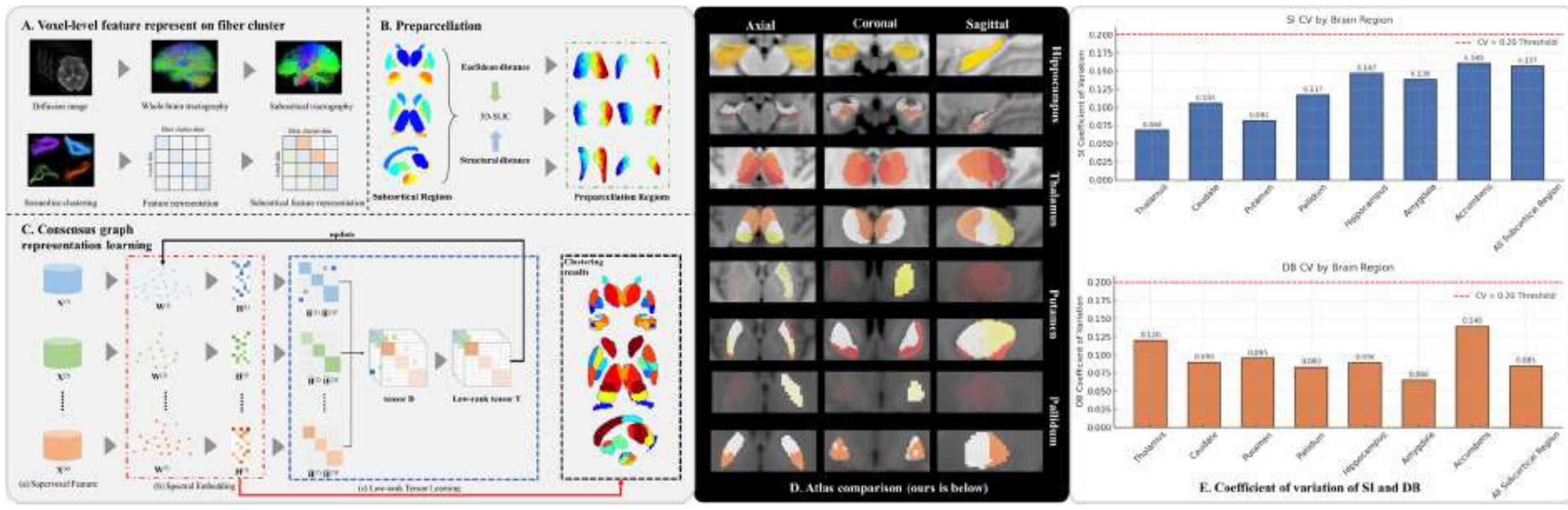

Fig.1 Overview of the proposed method. (a) Fiber cluster feature representation; (b) 3D-SLIC; (c) Consensus graph learning combining spectral embedding and low-rank tensor learning; (d) Atlas comparison; (e) Coefficient of variation.

## Results

Figure 1d shows strong alignment between our atlas and established subcortical atlases, including AAL3[4] for the thalamus, HOA2[5] for the hippocampus, and the Melbourne subcortical atlas[6] for the pallidus and putamen. We assessed the robustness of our parcellation by calculating the coefficient of variation (CV) for the Silhouette Index (SI) and Davies–Bouldin Index (DB) across 171 subjects, indicating low inter-subject variability and high intra-subject stability.

## Conclusion

This study introduces a multiscale framework for subcortical parcellation that combines diffusion MRI connectivity with anatomical features for improved segmentation and a population-level atlas.

## Automated Mapping of Cranial Nerves Pathways in Human Brain: A Multi-Parametric Multi-Stage Diffusion Tractography Atlas

**Lei Xie[1], Qingrun Zeng[1], Ye Wu[2], Yuanjing Feng[1,*]**

[1] Zhejiang University of Technology, China. [2] Nanjing University of Science and Technology, China

### Introduction

The cranial nerves (CNs) play a crucial role in various essential functions of the human brain, and mapping the pathways of them from diffusion MRI provides valuable preoperative insights into the spatial relationships between individual CNs and key tissues. In this work, we present what we believe to be the first study to develop a comprehensive diffusion tractography atlas for automated mapping of CN pathways in the human brain.

### Methods

In this work, we present what we believe to be the first study to develop a comprehensive diffusion tractography atlas for automated mapping of CN pathways in the human brain. The CN atlas is generated by fiber clustering by using the streamlines generated by multi-parametric fiber tractography (Fig.1(b)) for each pair of CNs, which can identify 8 fiber bundles associated with 5 pairs of CNs, including the optic nerve CN II, oculomotor nerve CN III, trigeminal nerve CN V and facial-vestibulocochlear nerve CN VII/VIII. Instead of disposable clustering, we explore a new strategy of multi-stage fiber clustering (Fig.1(c)) for multiple analysis of approximately 1M streamlines generated from the 50 subjects from the Human Connectome Project (HCP).

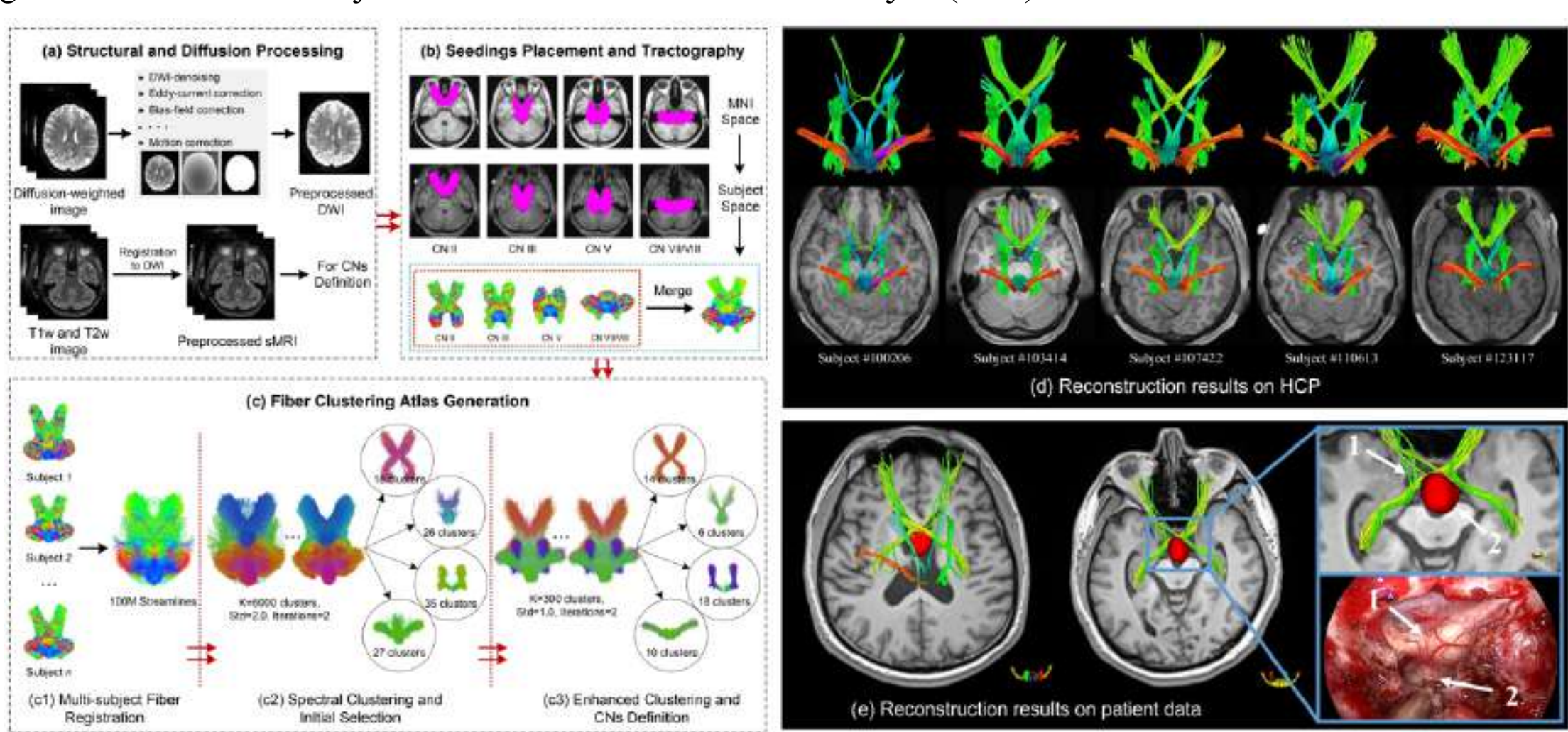

Fig.1 Overview of CN atlas generation pipelines. (a) Structural and diffusion MRI were passed through corresponding preprocessing steps for anatomical definition and tracking. (b) Fiber tractography for the reconstruction of CN pathways from the seeding images. (c) Multi-stage fiber clustering atlas generation using tractography data from HCP dataset. (d) Reconstruction results on HCP data. (e) Reconstruction results on patient data.

### Results

we demonstrate the application of the proposed CN atlas for automated mapping of the pathways in new subjects from different acquirement sites, including the HCP dataset, the multi-shell dMRI (MDM) dataset, PA patients, and CP patient. Qualitative and quantitative experimental results (Fig.1(d) and (Fig.1(e))) demonstrate that the proposed method has ideal colocalisation with expert manual identification.

### Conclusion

In this work, we propose a comprehensive multi-parametric diffusion tractography atlas for automated mapping the pathways of CNs in the human brain.

**Critical white matter tracts for bimanual motor skill learning: Exploring structural connectivity in acute stroke.**

**Authors:** Coralie van Ravestyn[a,b,c], Laurence Dricot[b], Nicolas Delinte[b,c], Benoit Bihin[d], Beatrijs De Coene[e], Nicolas Mulquin[e], Yves Vandermeeren[a,b,c]

**Affiliations:** [a]UCLouvain/CHU UCL Namur (Godinne), Neurology Department, Stroke Unit/Motor Learning Lab, Yvoir, Belgium; [b]UCLouvain, Institute of NeuroScience (IoNS), NEUR Division, Brussels, Belgium; [c]UCLouvain, Louvain Bionics, Louvain-la-Neuve, Belgium; [d]UCLouvain, CHU UCL Namur (Godinne), Scientific Support Unit (USS), Yvoir, Belgium; [e]UCLouvain/CHU UCL Namur (Godinne), Radiology Department, Yvoir, Belgium

**Introduction**. Bimanual motor skill learning (bim-MSkL) is central to most activities of daily life[1] and may be impaired after stroke. Since a bilateral network involving sensory-(pre)motor and higher-order areas supports bim-MSkL[2], the integrity of the white matter tracts (WMT) involved in this network is essential for the efficient transmission of information[3]. The exploration of WMT disrupted by acute stroke may reveal connections critical for bim-MSkL between motor and/or cognitive areas.

**Objectives**. To identify microstructural degradations or disconnections within WMT associated with impaired bim-MSkL.

**Methods**. 90 (sub)acute stroke patients and 62 age-matched healthy individuals (HI) trained during three consecutive days with a bim-MSkL robotic cooperation task (REA$^2$Plan, Axinesis)[4]. Patients were scanned using multimodal MRI for performing WMT microstructural integrity and tractography analyses.

**Results**. Bim-MSkL quantified through speed-accuracy trade-off was impaired in patients compared to HI (-0.52 [a.u.], $p<0.001$). Altered WM integrity indexed by tract-averaged diffusion-tensor imaging (DTI) and neurite orientation dispersion and density imaging (NODDI) metrics including fractional anisotropy, axial and radial diffusivity and orientation dispersion index in the corticospinal tract (CST), corpus callosum (CC) and superior longitudinal fasciculus (SLF) was associated with poorer bim-MSkL. Further tractography analyses are still ongoing and will be presented at the Conference. These include: (1) subdividing these main tracts of interest into equidistant segments to enhance the spatial resolution of microstructural integrity assessments and to characterize the variation of integrity metrics along each tract length; (2) investigating the relationships between these variations and stroke-related injury by quantifying the distance from each tract segment to the lesion, under the hypothesis that segments closer to the acutely infarcted zone exhibit greater structural compromise; (3) performing whole brain analyses to quantify the inter-hemispheric asymetry of WMT integrity induced by an acuter stroke; and (4) correlating tract-specific damage and asymmetry measures with bim-MSkL.

**Conclusion**. Acutely damaged microstructural integrity and/or disconnections of key WMT correlate with poorer bim-MSkL, providing a fine dissection of the network underlying bim-MSkL and paving the way for developing biomarkers that could allow personalising rehabilitation.

# ON-Harmony: A multi-site, multi-modal travelling-heads resource for brain MRI harmonisation with integration of UK Biobank scanners

Shaun Warrington[1], Andrea Torchi[1], Olivier Mougin[2], Jon Campbell[3], Asante Ntata[1,4], Martin Craig[1], Stephania Assimopoulos[1], Fidel Alfaro-Almagro[3], Stephen M Smith[3], Adam J Lewandowski[5,6], Karla L Miller[3], Mark Jenkinson[3,7], Paul S Morgan[1], and Stamatios N Sotiropoulos[1]

[1]Sir Peter Mansfield Imaging Centre, School of Medicine, University of Nottingham, UK. [2]Sir Peter Mansfield Imaging Centre, School of Physics, University of Nottingham, UK. [3]Oxford Centre for Integrative Neuroimaging, University of Oxford, UK. [4]National Physical Laboratory, UK. [5]Nuffield Department of Population Health, University of Oxford, UK. [6]UK Biobank Ltd, UK. [7]Australian Institute for Machine Learning (AIML), School of Computer and Mathematical Sciences, The University of Adelaide, Australia.

**Introduction:** MRI quantifiability is hindered by non-biological sources of variability, e.g scanner hardware/software[1–3]. Several approaches aim to standardise/harmonise acquisition and processing[4–6], but lack of harmonisation is an open challenge. We ran one of the most comprehensive, freely accessible, multi-modal travelling heads studies, ON-Harmony[7–9] (Fig1): 20 subjects, each scanned in up to 8 3T scanners, 3 vendors (Siemens/Philips/GE) and 5 modalities (T1w/T2w/dMRI/swMRI/fMRI), plus within-scanner/within-subject repeats, enabling within-scanner, between-scanner and between-subject variability to be mapped across multi-modal imaging-derived phenotypes (IDPs). We have recently scanned 11 of the subjects at the Stockport UK Biobank (UKB) imaging centre (Reading UKB centre also planned), enabling linkage of UKB population-level imaging data with various clinical scanners. Here, we showcase the data and reuse scenarios for assessing harmonisation efficacy.

**Figure 1: ON-Harmony: a multi-modal travelling-heads harmonisation resource**

**Figure 2: a) Example of a single subject's data. b) Between-session similarity for different pools of variability**

**Methods:** ON-Harmony consists of 2 primary phases (10 subjects each), plus the UKB extension. Acquisition protocols were aligned with the UKB imaging study[10], while respecting best practices and hardware limitations (i.e. parameters not simply nominally-matched)[8]. Each subject was scanned in at least 6 different scanners (out of a collection of 8 scanners) and 9 subjects had 5 additional within-scanner repeats. 11 participants were re-scanned at the UKB centre (1 subject with 5 within-scanner repeats). All data underwent quality control through visual inspection and then using MRIQC[11] (T1w/T2w/fMRI) and eddyQC[12] (dMRI). Data were processed with a modified version[8] of the UKB pipeline[13]. Hundreds of IDPs were derived for each session, allowing us to quantify IDP variability.

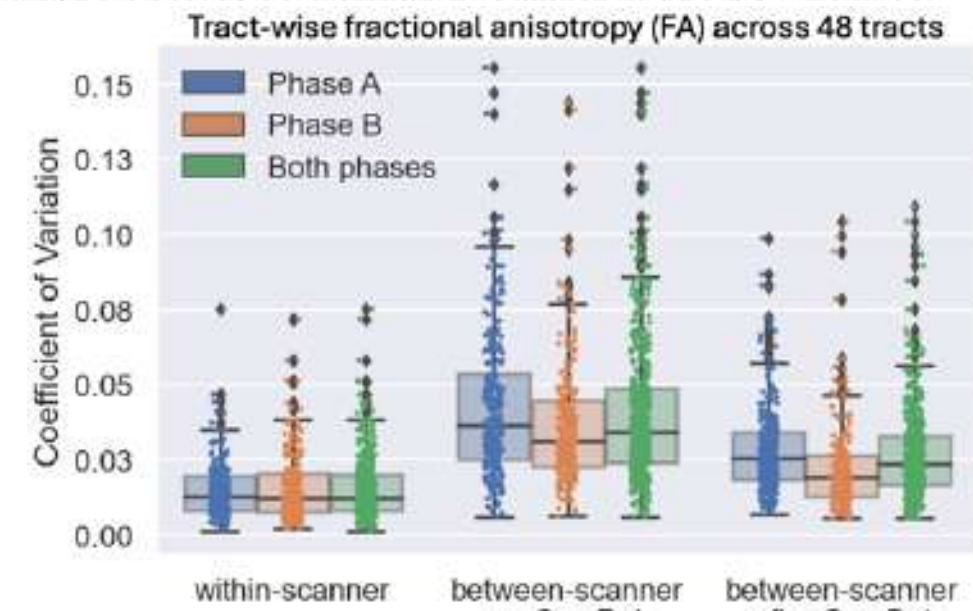

**Figure 3: Example of assessing explicit harmonisation efficacy**

**Results:** Fig2a shows a subject's raw data, depicting sessions across scanners (columns) and modalities (rows). We assessed between-session IDP similarity for within-scanner, between-scanner, between-subject pools (Fig2b). Between-subject variability has little overlap with scan-rescan variability, however, it overlaps quite substantially with between-scanner variability. We explored how ON-Harmony can be used to assess harmonisation efficacy, e.g. ComBat[14,15] (explicit harmonisation). Pre/post harmonisation between-scanner variability was compared to within-scanner variability (Fig3) for dMRI tract-wise FA measures. Reductions in between-scanner variability following harmonisation were revealed but did not match the within-scanner baseline. ON-Harmony can also be used to assess pipeline/tool generalisability across scanners (implicit harmonisation). Fig4 shows generalisability of tractography for FSL-XTRACT[16]. For each scanner, subject-averaged tract maps were obtained and correlated against a UKB atlas, with moderate-high correlation values across all scanners and generally consistent trends across vendors/phases, although with some exceptions (GE-A was a low gradient system with a single-shell protocol). Such comparisons showcase how ON-Harmony can be used to assess susceptibility of tools/pipelines to between-scanner effects.

**Figure 4: a) Tractography results for ON-Harmony phases and the UKB reference. b) Tract-wise correlations across vendors/phases**

**Conclusion:** We have presented a comprehensive harmonisation resource (ON-Harmony) for multimodal neuroimaging data, based on a travelling-heads paradigm. ON-Harmony is openly released, freely available[9] and can be used to assess harmonisation efficacy and to develop new vendor agnostic tools/pipelines. The novel UK Biobank extension will enable direct linkage of a representative sets of scanners from all vendors with one of the largest population-level studies.

# Curvature properties on Estimating Asymmetric Fiber Orientation Distribution

Mojtaba Taherkhani [1,2], Marco Pizzolato [2,3], Morten Mørup [2] and Tim B. Dyrby [2,3]

[1] Department of Information Engineering, University of Pisa, Pisa, Italy

[2] Department of Applied Mathematics and Computer Science, Technical University of Denmark, Kgs. Lyngby, Denmark

[3] Danish Research Centre for Magnetic Resonance, Copenhagen University Hospital Amager and Hvidovre, Copenhagen, Denmark

***Introduction.*** In diffusion magnetic resonance imaging (dMRI), symmetric fiber orientation distributions (FODs) can detect symmetric single and crossing fibers. However, in regions where fibers are bending, branching, or fanning, the FODs may appear asymmetric—often forming T-shaped, Y-shaped, or uneven crossing patterns. To address this limitation, the use of asymmetric FODs (A-FODs) has been proposed [1,2]. A-FODs have been shown to be aligned with the expected underlying anatomy [1] and have also been reported to improve the accuracy of fiber tractography modeling [3]. The derivation of A-FODs is based on incorporating local neighborhood information of each voxel [2]. A central assumption in modeling the intervoxel information is the fiber continuity principle, which posits that the radius of curvature of white matter fibers typically exceeds the dimensions of a voxel. Consequently, it is assumed that a fiber entering a voxel with a given orientation is more likely to exit along a similar orientation, reflecting local directional coherence. In [1], the general filtering-based formulation to model A-FODs from a weighted sum of symmetric FODs inside a neighborhood has been proposed.

$$\tilde{\psi}_x(u) = \frac{1}{W} \sum_{y \in N_u(x)} \sum_{v \in \Omega} R \cdot \psi_y(v) \tag{1}$$

Let $\tilde{\psi}_x(u)$ denote the A-FOD at voxel position $x$ along direction $u$, and let $N_u(x)$ represent the set of neighboring voxels surrounding the central voxel $x$. The set $\Omega$ corresponds to the unit sphere directions over which the orientation distribution function is projected. The term $R$ denotes the regularization weight applied to the original FOD $\psi_y(v)$ at voxel $y$ along direction $v$, while $W$ serves as a normalization factor. Furthermore, the displacement vector from voxel $x$ to voxel $y$ can be expressed as the product of its magnitude $I_{xy}$ by its unit direction $D_{xy}$. When the distance between neighboring voxels is relatively large, the assumption that fibers propagate as straight trajectories along directions $U$ and $-U$ across adjacent voxels becomes invalid [4] (Fig. 1a). To address this issue, the proposed method aims to incorporate the curvature characteristics of the underlying fiber architecture, thereby generalizing the fiber continuity assumption.

***Methods***. Different regularization weights have been formulated for the estimation of A-FODs [5], and these can be incorporated into the general formulation presented in Eq. (1). In this study, a novel curvature-aware regularization weight is introduced to better account for fiber curvature during A-FOD estimation. To incorporate curvature information, the direction $v$ at neighboring voxels is rotated along the vector $D_{xy}$ where the degree of rotation depends on the distance $I_{xy}$

$$\varphi = \left(e^{\log_{\frac{3}{\max(|I_{xy}-r_f|)}}(|I_{xy}-r_f|)} - 1\right) \times acos(-U \cdot -D_{xy}) \tag{2}$$

$$V_{rot} = v\cos(\varphi) + (D_{xy} \times v)\sin(\varphi) + D_{xy}(D_{xy} \cdot v)(1-\cos(\varphi)) \tag{3}$$

The rotated direction $V_{rot}$ is computed using Rodrigues' Rotation Formula, and $\varphi$ is the angular deviation introduced to align with the presumed fiber curvature. The parameter $r_f$ denotes the fiber's radius of curvature, and the rotation becomes significant when $I_{xy} > r_f$, thus enabling the model to capture fiber trajectories. The novel regularization weight is integrated with other well-known weights in Eq. (4). $G_\sigma$ is a Gaussian distribution with standard deviation $\sigma$.

$$\tilde{\psi}_x(u) = \frac{1}{W} \sum_{y \in N_u(x)} \sum_{v \in \Omega} \left(G_{\sigma_{spatial}}(I_{xy}) \times G_{\sigma_{fiber}}(U \cdot V_{rot}) \times G_{\sigma_{orient}}(U \cdot D_{xy})\right) \cdot \psi_y(v) \tag{4}$$

The first weighting function, $G_{\sigma_{spatial}}$, assigns weights to neighboring voxels based on their Euclidean distance from the voxel $x$ under consideration. The function $G_{\sigma_{orient}}$ accounts for the alignment between the currently processed direction $u$ and direction $D_{xy}$. Similarly, the $G_{\sigma_{fiber}}$ models the alignment between the current direction $u$ and the rotated direction $V_{rot}$ derived from Eq. (3). A visual interpretation of Eq. (4) is presented in Fig. 1a and Fig. 1b. By minimizing the MSE between the input symmetric FODs and the symmetric portion of the estimated A-FODs, we find the best parameter combination.

***Results.*** To demonstrate the advantages of our method, we tested the method on the HCP dataset and DiSCo phantom [6] that were simulated for different complex configurations. The Constrained Spherical Deconvolution (CSD) method was used to estimate FODs per voxel. Fig. 1c illustrates the ground truth streamlines and A-FODs extracted from multiple regions of the DiSCo phantom that exhibit complex configurations.

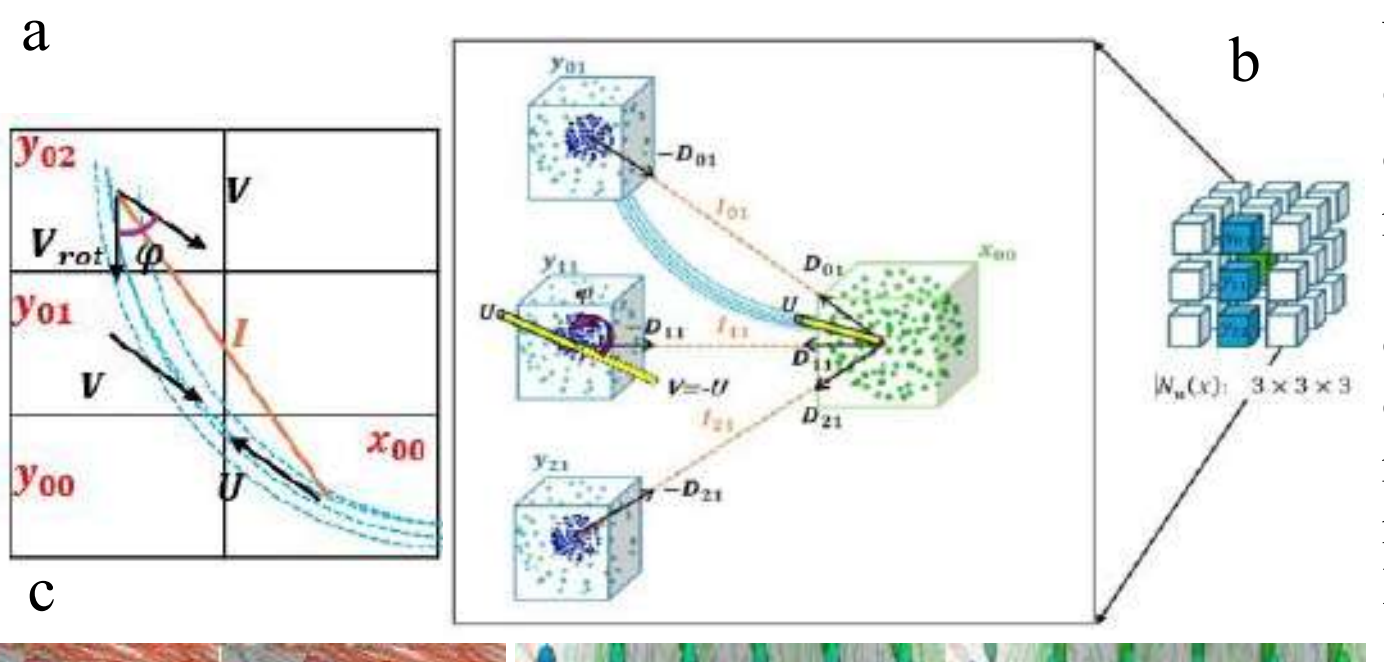

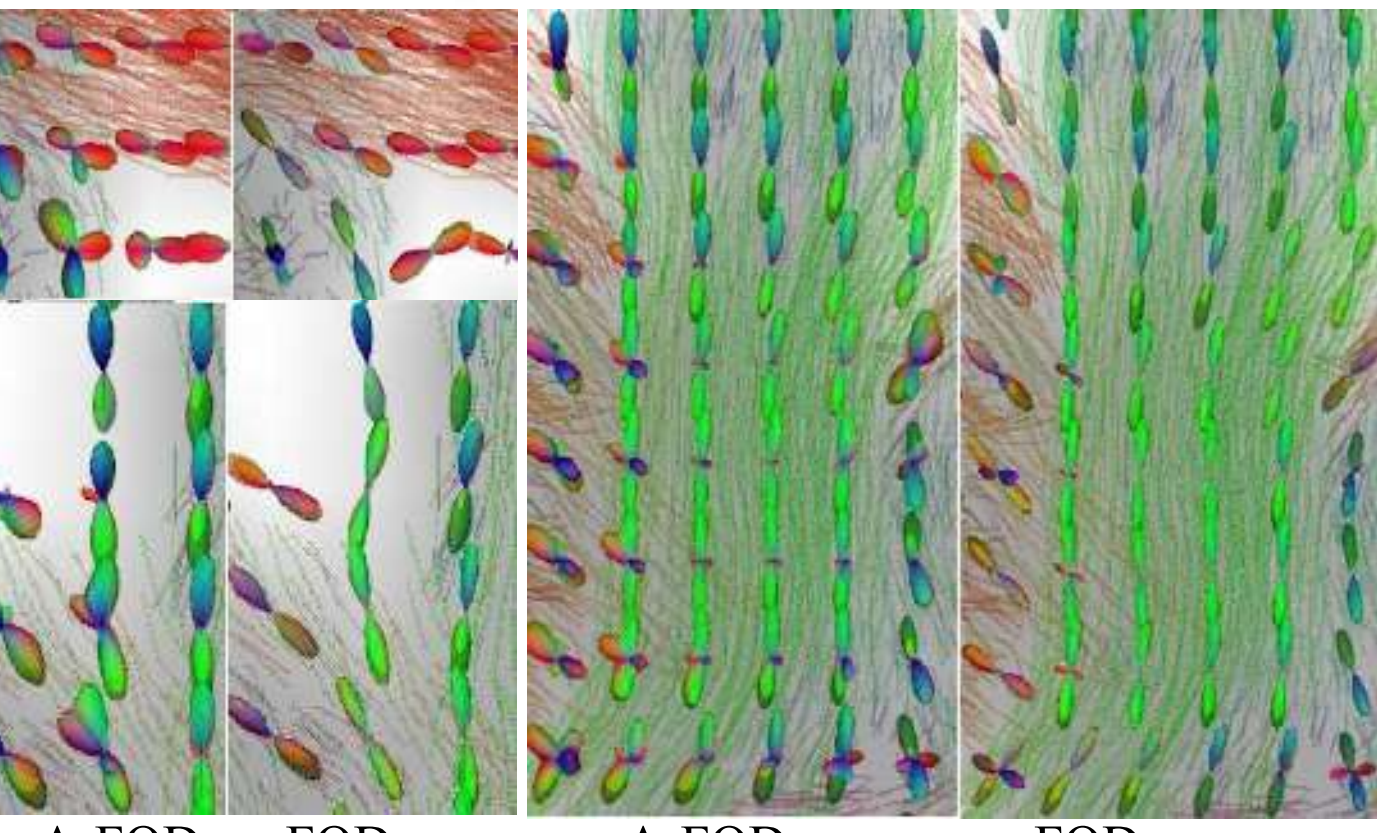

***Discussion.*** by excluding the center voxel $x$ from neighboring voxels $N_u(x)$, the A-FODs is estimated solely based on its surrounding voxels, allowing for the estimation of complex subvoxel fiber configurations. The proposed model both supports and generalizes the fiber continuity assumption to infer asymmetric orientations. Notably, the resulting asymmetries are comparable to those produced by optimization methods, while allowing sharper fiber turns than conventional FODs [2]. Incorporation of fiber curvature in the modeling promotes smoother fiber segments and enhancing connectivity. ***Conclusion.*** we generalized the filtering-based methods by proposing a novel regularization weight for estimating A-FODs, integrating fiber curvature properties through a local information of surrounding voxels and directional alignments within a single formulation. The results demonstrate that A-FODs effectively resolve subvoxel fiber configurations, including bending, fanning, and asymmetric kissing. Incorporating curvature information enables smoother fiber segments in continuous space compared to conventional FODs. Future work will assess if the curvature-aware regularization improves tractography in complex regions.

**Quantifying the white matter pathways supporting written word production using tractometry**

Romi Sagi[1], J.S.H. Taylor[2], Kyriaki Neophytou[3,4], Sivan Jossinger[1,5,6], Brenda Rapp[3], Kathleen Rastle[7] and Michal Ben-Shachar[1]

[1]The Gonda Multidisciplinary Brain Research Center, Bar-Ilan University, Ramat-Gan, Israel; [2]Division of Psychology and Language Sciences, University College London, London, UK; [3]Department of Cognitive Science, Johns Hopkins University, Baltimore, USA; [4]Department of Neurology, Johns Hopkins Medicine, Baltimore, USA; [5]Western Centre for Brain and Mind, Western University, London, Ontario, Canada; [6]Department of Computer Science, Western University, London, Ontario, Canada; [7]Department of Psychology, Royal Holloway, University of London, UK

**Introduction.** Producing written language is a fundamental aspect of everyday communication, but its neurobiological bases remain understudied. fMRI studies have implicated a distributed network of cortical, subcortical, and cerebellar brain regions in spelling tasks[1,2]. Still, the structural connectivity supporting these processes is less characterized. We leveraged probabilistic tractography methods and automated tract segmentation and quantification tools, coupled with sensitive behavioral assessment, to identify the white matter pathways that support spelling performance in a large cohort of neurotypical adults. **Methods.** Native English speakers completed a difficult spelling-to-dictation task and underwent MRI scanning ($N = 73$, mean age = $21 \pm 3$ years). Diffusion-weighted images were acquired on a 3T Siemens scanner using a single-shot EPI sequence (64 diffusion-weighted volumes at $b = 1000$ s/mm² and one reference volume at $b = 0$ s/mm², voxel size $\approx 2\times2\times2$ mm³). Constrained spherical deconvolution (CSD) modeling and probabilistic tractography yielded whole-brain tractograms. Language-related dorsal, ventral and cerebellar pathways were reconstructed automatically in each participant's native space, using a multiple-region-of-interest approach. Specifically, we utilized the AFQ[3] package together with tools developed in our lab[4,5,6] to segment bilaterally the three branches of the superior longitudinal fasciculus (SLF I, II, III), as well as the arcuate fasciculus, inferior longitudinal fasciculus (ILF), frontal aslant tract (FAT), and cerebello-thalamo-cortical pathways (CTC). Fractional anisotropy (FA) values were extracted from multiple nodes along each tract, and associations with spelling performance were assessed while controlling for additional behavioral factors. **Results.** Spelling accuracy varied widely across participants (7%–97%) and was bimodally distributed. Tractometry results revealed both dorsal and ventral stream associations with spelling: High-performing spellers showed a significant positive correlation between spelling accuracy and FA in the left ILF ($r_s = .53$, $p < .05$, FWE-corrected), implicating efficient lexical-orthographic spelling processes. In contrast, low-performing spellers showed a significant negative correlation between spelling accuracy and FA in the right SLF III ($r_s = -.47$, $p < .05$, FWE-corrected), suggesting reliance on phonological, fronto-parietal systems[4]. Additionally, FA in the left CTC tract was significantly correlated with spelling performance across the full sample ($r_s = .38$, $p < .05$, FWE-corrected), highlighting the cerebellum's contribution to higher-order cognitive-linguistic processing via cerebello-cortical loops[7]. No significant associations were found between spelling performance and FA in the bilateral FAT, which may be relevant for more peripheral-motor aspects of written word production. **Conclusion.** Our findings demonstrate that spelling is supported by a distributed set of white matter pathways, including ventral occipitotemporal, dorsal frontoparietal, and cerebello-cortical connections. Moreover, high- and low-performing spellers appear to rely on different cognitive processes, supported by distinct white matter pathways. High-performing spellers may rely more heavily on whole-word-form retrieval, whereas low-performing spellers appear to depend on analytic sound-to-letter mapping. These results underscore the value of tractometry methods for elucidating the highly complex neurocognitive architecture of written word production.

# Estimating microscopy-informed fibre orientations from dMRI data in the UK Biobank

Silei Zhu[a], Nicola K. Dinsdale[a,b], Saad Jbabdi[a], Karla L. Miller*[a], Amy F.D. Howard*[a,c]
a Wellcome Centre for Integrative Neuroimaging, FMRIB Centre, Nuffield Department of Clinical Neurosciences, University of Oxford, UK b Oxford Machine Learning in NeuroImaging Lab (OMNI), Department of Computer Science, University of Oxford, UK c Department of Bioengineering, Imperial College London, UK

**Introduction** There is an unmet need for non-invasive methods to map complex neuroanatomy at the meso-scale. Diffusion MRI (dMRI) tractography is effective but limited by inaccuracies in fibre orientation estimates. We present a deep-learning model that reconstructs microscopy-informed fibre orientations from in vivo human dMRI data, and investigate whether this tractography can better capture individual brain connectivity differences compared to conventional methods.

## Methods

**Data for model training:**
The network was trained on macaque data from BigMac with in-vivo dMRI (b=1000 ms/mm$^2$ at 1mm), postmortem dMRI (b=4000 ms/mm$^2$ at 0.6mm) and microscopy (polarised light imaging, 4 μm/pixel) in a single brain. The network was then fine-tuned and applied to human data from the UK BioBank (UKB) with pre-processed single-shell dMRI (b=1000 ms/mm$^2$, 2mm, 50 directions), alongside 2.4mm rsfMRI.

**Microscopy-informed fibre orientation distributions (FODs) estimation:**
We developed a domain adaptation network with 3 components: a feature extractor, predictor, and domain classifier (Fig1a). The feature extractor and predictor generate predicted FODs from dMRI, while the domain classifier ensures invariance between postmortem and in vivo data. The training involved comparing predicted FODs to "ground truth" hybrid dMRI-microscopy FODs using mean squared error. Hybrid FODs were created by combining 2D microscopy with through-plane dMRI orientations to obtain 3D FODs that are maximally informed by microscopy. The network was fine-tuned for human data in two steps (Fig1b) during which both human and macaque data were input to the network. First, it was trained on macaque FODs while the discriminator aimed to differentiate species. Second, the discriminator was fixed, and the feature extractor learned species-invariant features. This two-stage approach is crucial because "ground truth" FODs are not available for human data. Finally, the trained network was applied to UKB subjects. Network-derived FODs were compared to constrained spherical deconvolution (CSD). Tractography was performed using MRtrix3 and tract termination maps (TTMs) were generated by mapping normalised streamline counts onto the cortical surface.

**Relating tract termination masks to rsfMRI data:**
We analysed TTMs from both network-derived and CSD FODs to determine which of these two "parcellations" is in better agreement with rsfMRI (Fig2). Two analyses evaluated how well TTMs capture inter-individual variability in rsfMRI. Focusing on homotopic connections between the left and right visual cortex, we seeded tractography from the left visual cortex and mapped streamlines to the right hemisphere. The left visual cortex mask was also used in rsfMRI dual regression to identify each individual's visual network, which was compared to the TTM in the right hemisphere.

## Results

Our network can successfully reconstruct biologically plausible FODs and tractography from in vivo human data (Fig3a). It produces noticeable differences in tractography across subjects and compared to CSD (Fig3b). Figure 4 asks whether these differences are meaningful by examining tract-rsfMRI alignment in the visual cortex. Compared to CSD, the network TTM defines a smaller, more localized region that excludes the anticorrelated resting-state region (blue). This follows neuroanatomical expectations, as anticorrelated networks are typically not connected by direct structural pathways.

## Conclusion & Future work

We demonstrate the efficacy of microscopy-informed FODs estimated from conventional in vivo human MRI for meaningful fibre tracking. Future work will evaluate whether the individual differences of TTMs are associated with non-imaging phenotypes.

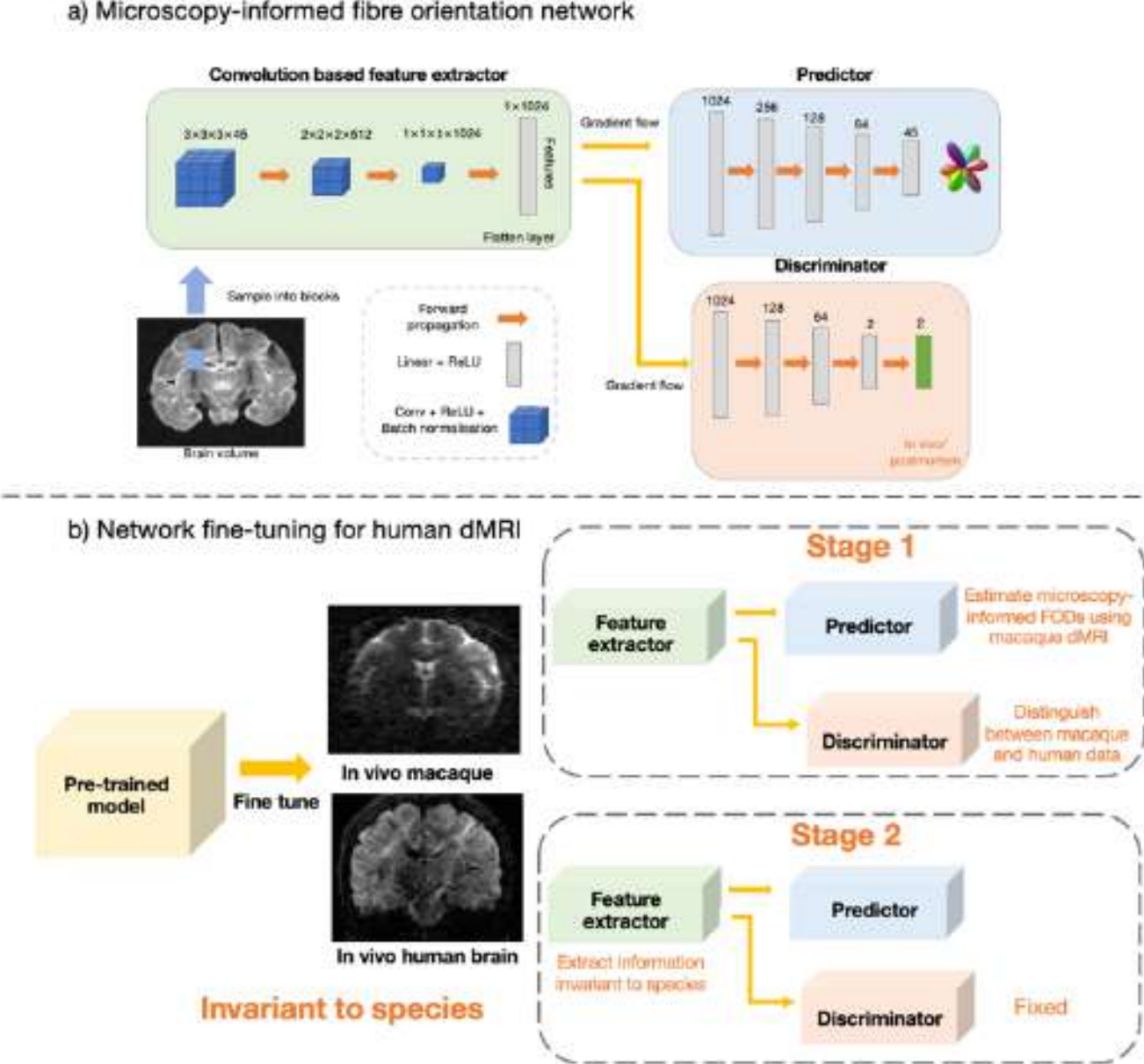

**Fig 1: Microscopy-informed fibre orientation network**
(a) The network employs a domain adaptation architecture to achieve invariance between postmortem and in-vivo tissue. (b) Fine-tuning on in-vivo human data establishes species invariance (macaque vs. human)

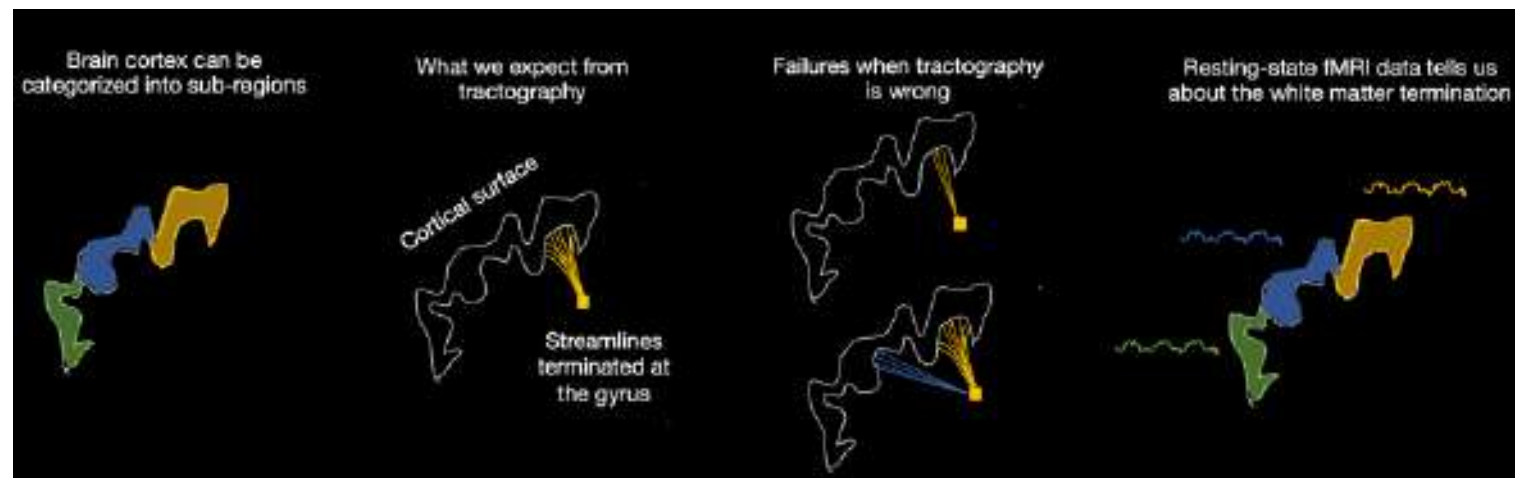

**Fig 2: White matter termination and rsfMRI network**
Brain regions consist of distinct functional subregions. White matter tract terminations, defined by tractography, help delineate these boundaries, though is prone to errors. rsfMRI may improve boundary precision, as regions with similar tract terminations show similar rsfMRI activity.

**Fig 3: Microscopy-informed FODs and tractography in UKB**
**(a)** Comparison of network FODs with CSD FODs in UKB data, showing improved, less noisy FODs near the gray-white matter boundary. **(b)** Reconstruction of the corticospinal tract reveals individual differences in tract terminations and noticeable differences between network and CSD.

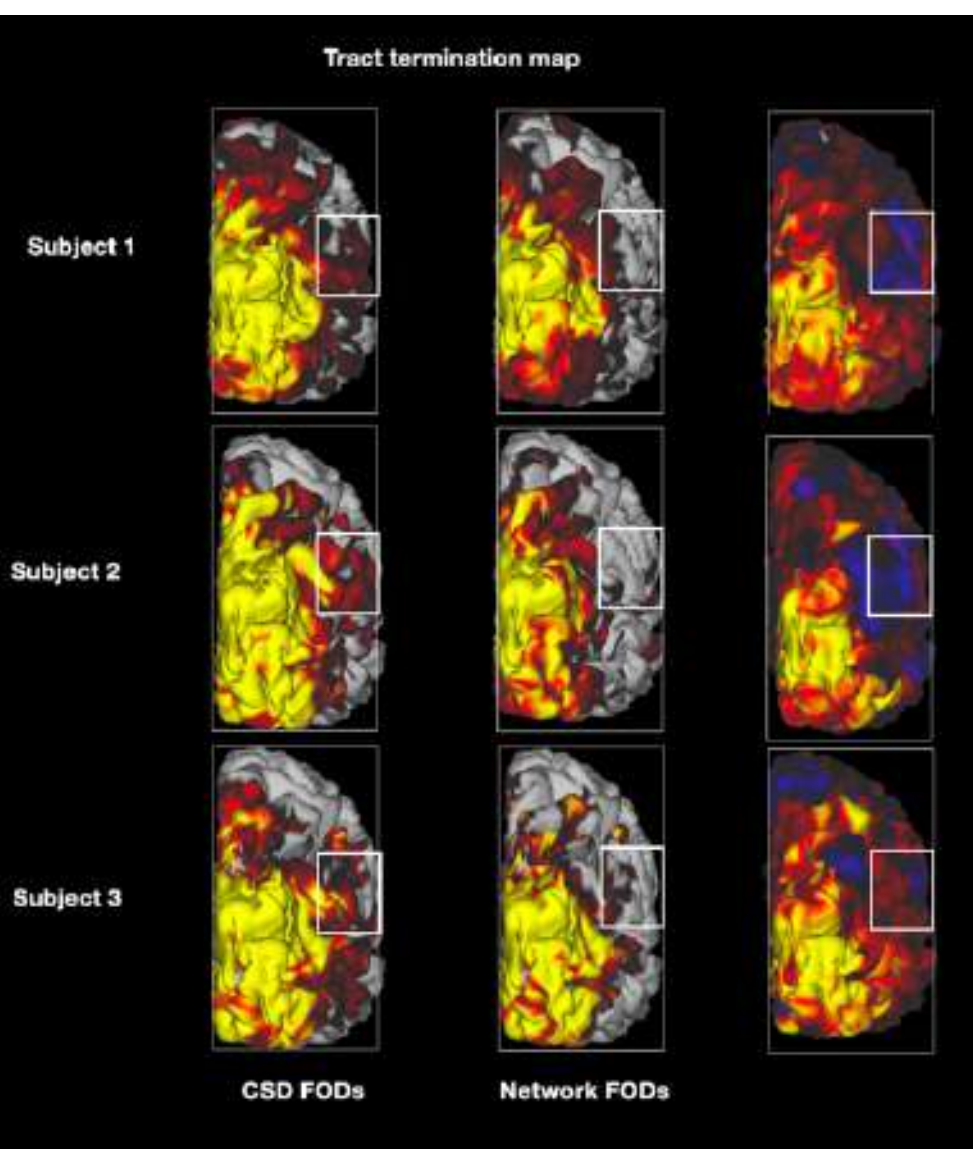

**Fig 4: Alignment of white matter tract termination and rsfMRI Network**
To assess alignment, tractography was seeded from the left visual cortex to generate right hemisphere white matter tract termination maps. The right hemisphere's rsfMRI network was identified by applying the left visual cortex mask in dual regression. Network FODs show better alignment with the rsfMRI network than CSD (N=3).

# A microscopy-trained model to predict super-resolution fibre orientations from diffusion MRI

Silei Zhu[a], Karla L. Miller[a], Nicola K. Dinsdale[a,b], Saad Jbabdi[a], Amy F.D. Howard[a,c]
a Wellcome Centre for Integrative Neuroimaging, FMRIB Centre, Nuffield Department of Clinical Neurosciences, University of Oxford, UK b Oxford Machine Learning in NeuroImaging Lab (OMNI), Department of Computer Science, University of Oxford, UK c Department of Bioengineering, Imperial College London, UK

**Introduction:** Microscopy provides fibre orientations at much higher resolutions than MRI. Diffusion MRI (dMRI) and microscopy in the same brain offer the opportunity to train a machine-learning model to super-resolve fibre orientation distributions (FODs). We develop a microscopy-informed network that provides super-resolved FODs from single-shell dMRI, doubling the resolution. Notably, through domain adaptation, our network can be applied to in-vivo MRI where microscopy is unavailable, including in humans, and offers the possibility of more precise fibre tracking in widespread applications.

**Methods**

**Training data:** We used the BigMac dataset with in-vivo dMRI (b=1 ms/$\mu$m$^2$ at 1mm), in-vivo structural MRI (0.5mm), postmortem dMRI (b=4 ms/$\mu$m$^2$ at 0.6mm), postmortem structural MRI (0.3mm) and microscopy data (polarised light imaging, 4 µm/pixel) from a single, whole macaque brain.

**Hybrid FODs:** As BigMac microscopy is 2D, we combined it with dMRI to reconstruct 3D “hybrid” FODs at twice the dMRI resolution. Microscopy provided orientations within the microscopy plane, whilst dMRI provided through-plane information (drawn from a posterior distribution of possible orientations). These hybrid FODs were considered “ground truth” during training.

**Network:** Super-resolution was achieved by i) leveraging high resolution (HR) microscopy (hybrid FODs) during training and ii) inputting dMRI alongside HR structural MRI from the same subject. Our network has 3 components:

1) **“Feature-fusion” module:** This combines HR structural MRI (twice dMRI resolution) with directional information from dMRI (Fig1A). The dMRI is spatially up-sampled to the structural resolution and the structural MRI up-sampled to 45 channels using transposed convolutions. The modalities are concatenated and put through a convolutional layer to create fused features. Fused features are generated for both postmortem and in-vivo MRI.

2) **Main network:** The network has a feature extractor and predictor that take a 3x3x3 cube of fused features and output a 2x2x2 cube of FODs at double the dMRI resolution. During training, these FODs were compared to hybrid FODs at high resolution using the mean squared error.

3) **Domain adaptation framework:** A domain classifier was designed to ensure that the feature extractor outputs were domain-invariant to in-vivo or postmortem MRI. This meant we could benefit from training on postmortem data while enabling in-vivo MRI usage.

**Validation:** Network FODs from unseen postmortem (1mm) macaque MRI were compared to constrained spherical deconvolution (CSD) at low-resolution, and CSD FODs up-sampled to HR using linear interpolation. The network was also applied to UK Biobank[6] in-vivo human MRI (2mm, b=1000 s/mm$^2$). The network was “fine-tuned” using the domain classifier to ensure the feature extractor’s last layer was common to both species.

**Figure 1: Microscopy-Informed Network.**
(a) **Feature Fusion:** dMRI provides low-resolution orientation information, while structural MRI supplies high-resolution spatial details. (b) **Main Network:** These fused features are processed by a domain-adaptation network to predict microscopy-informed fibre orientations.

**Results**

In postmortem macaque dMRI, our super-resolved FODs performed well on lower-resolution data, estimating 0.5mm FODs from 1mm dMRI follow microscopy anatomical expectations even in complex regions like the occipital lobe (Fig 2). Applied to in-vivo human data from the UK BioBank, the network successfully predicted 1mm FODs from 2mm dMRI (Fig 3). The network FODs could resolve more complex neuroanatomy in this small subcortical structure.

**Conclusion & Future work**

We demonstrate a method facilitating microscopy-informed super-resolved FOD reconstruction in datasets where microscopy is unavailable. The trained network is applicable to conventional MRI (single-shell in-vivo diffusion) facilitating widespread application. Future work will include comparisons to other super-resolution methods, investigate higher super-resolution factors (3-fold), and demonstrate benefits for in-vivo fibre tracking.

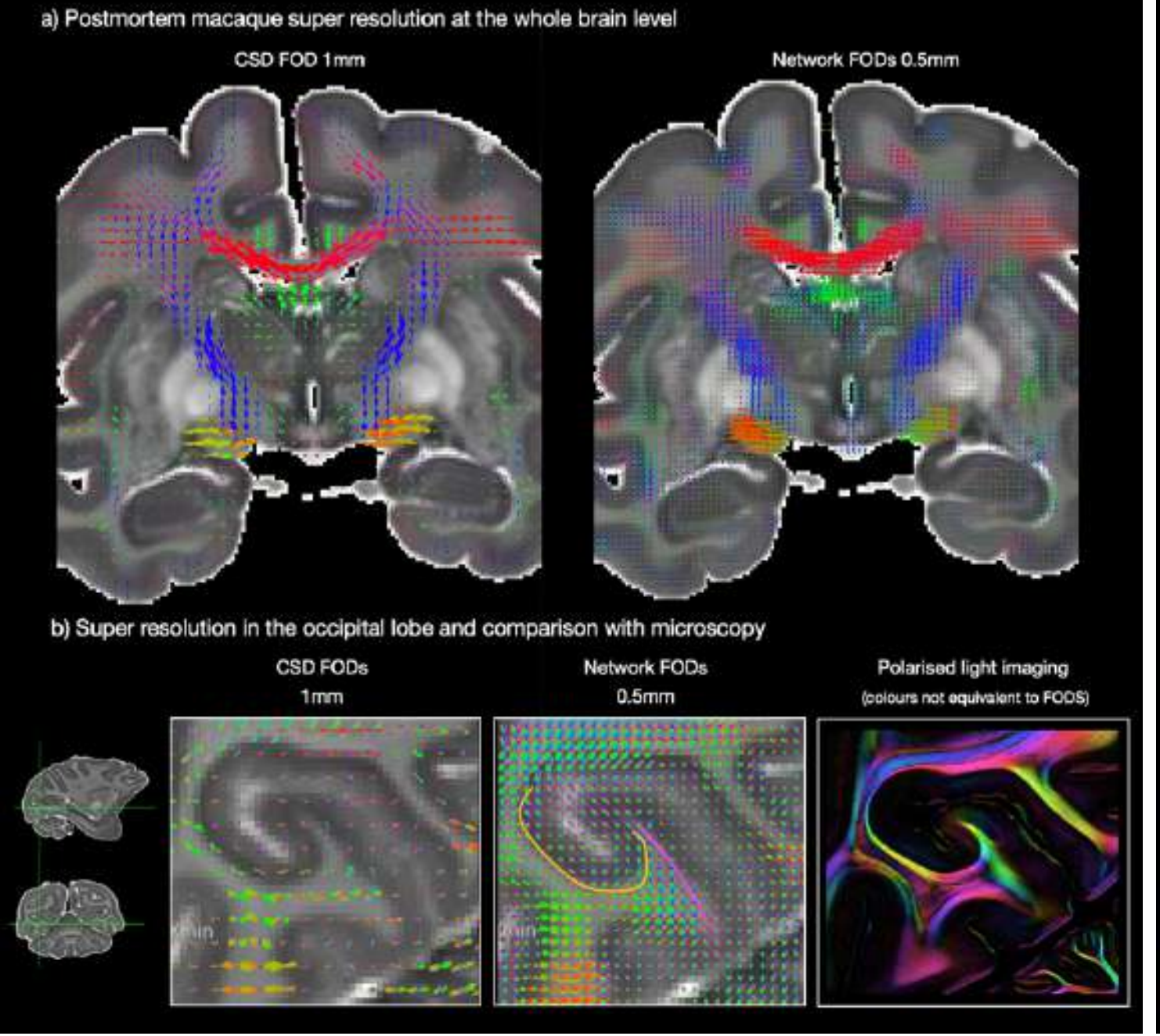

**Fig2: Super-Resolved FODs in Postmortem Macaque at 1 mm.** Postmortem dMRI (1 mm) and structural MRI (0.5 mm) were used to enhance FODs.

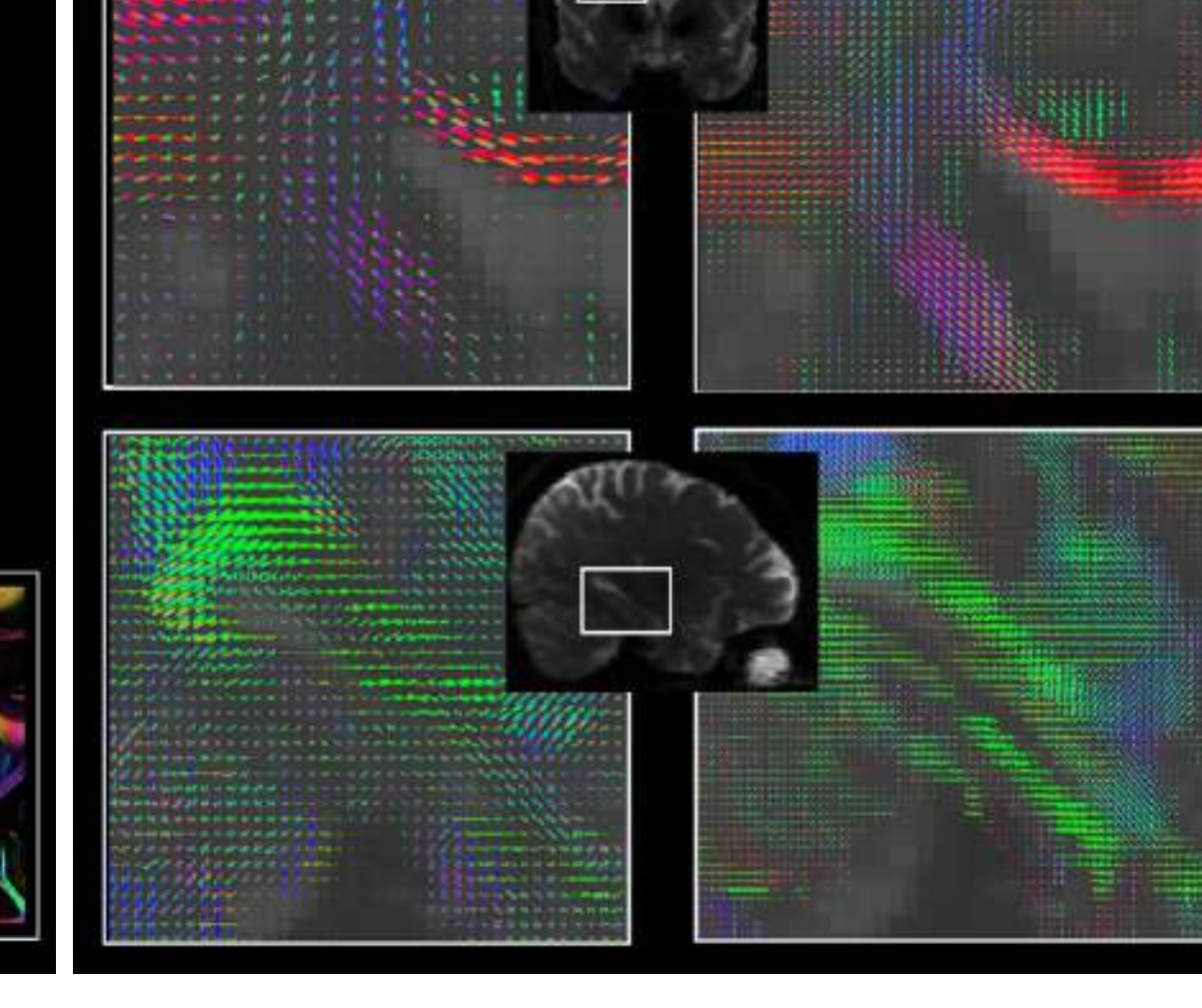

**Fig3: Super-Resolved FODs in In-Vivo Human.** Single-shell in-vivo dMRI (2 mm) and structural imaging (1 mm) were used to estimate microscopy-informed FODs.

# Superficial white matter association with cognitive decline using UKBiobank database (N=13747)

*Nabil Vindas[1,2], Nicole Labra[1,3], Vincent Frouin[1,2] ,Jean-François Mangin[1,2].*
[1]NeuroSpin CEA Saclay, Gif-sur-Yvette, France.
[2]Univeristé Paris-Saclay, Gif-sur-Yvette, France.
[3]University College London, London, United Kingdom

## I – Introduction

Aging of white matter (WM) leads to both structural and functional alterations at the synaptic level, which are vital for effective neuronal communication. Although changes in deep white matter (DWM) during aging are well characterized, superficial white matter (SWM) has been less studied due to its intricate location near the cortex and the small size of its fiber bundles [5]. However, recent improvements in neuroimaging have made it possible to examine SWM in greater detail [1]. In this study, we leveraged diffusion MRI (dMRI) and cognitive assessment data from 13,747 participants in the UK Biobank database. Using Partial Least Squares Path Modeling (PLS-PM), we explored the relationship between cognitive performance and SWM microstructure, analyzing 444 short-range bundles from the ESBA atlas and 35 long-range bundles from the LNAO-DWM12.

## II - Materials and Methods

We analyzed data from 13,747 healthy UK Biobank participants (aged 45–82, 52% female). Linear and non-linear transforms from diffusion space to T1, as well as from T1 to MNI space, were computed with FSL's FLIRT tool and the ANTs toolbox. We used the multi-shell, multi-tissue constrained spherical deconvolution model [2] with anatomically constrained tractography [6] and a second-order integration over the fiber orientation distribution [9] to produce 10 million streamlines. Spherical-deconvolution Informed Filtering of Tractograms [7] was used to filter and reduce the tractography to 5 million streamlines. The result was registered to the MNI152 space using the transforms mentioned above, and it was segmented using GeoLab [10] coupled with the ESBA atlas (SWM atlas) [3]. The segmentation of DWM was done using BundleSeg [8]. Finally, five diffusion MRI measures (FA, MD, ICVF, ISOVF, and OD) were mapped to the segmented bundles using MRtrix.

To investigate links between SWM/DWM microstructure and cognitive performance, we applied PLS-PM, a multivariate technique capable of capturing complex interactions between observed and latent variables without assuming input data normality. While DWM bundles were modeled individually, SWM bundles were grouped based on the Desikan-Killiany atlas, depending on the cortical regions that their endpoints connected and whether they were intra-regional or inter-regional connections. For each diffusion metric, reflective latent variables were constructed for the grouped SWM bundles. In the path model, age was connected to all variables, including SWM latent constructs, DWM bundles, and the derived cognitive component. We also linked white matter measures (SWM constructs and DWM bundles) to the cognitive component to evaluate the indirect influence of age on cognition via WM microstructure. Statistical significance was assessed through bootstrapping (1,000 samples), using 95% bias-corrected confidence intervals; effects were considered significant if the interval excluded zero. All PLS-PM analyses were conducted in R v4.4.1 using the plspm package v0.5.1.

## III - Results

In the PLS-PM analysis, the SWM blocks demonstrated good model quality, with communalities exceeding 0.5—indicating that a substantial portion of the variance within each block was captured by its corresponding latent variable—and unidimensionality, as evidenced by second eigenvalues below or near 1 and significantly smaller than the first. After isolating statistically significant paths, we found that SWM microstructure consistently played a more prominent mediating role in cognitive aging compared to DWM, across all diffusion metrics except for OD, which did not exhibit statistically significant results (Table 1).

*Table 1: DWM bundles (orange) and SWM latent factors (green) showing statistically significant association with the cognitive component. Abbreviations**:** LH: left hemisphere; RH: right hemisphere; Inter: inter-regional; Intra: intra-regional; Cu: cuneus; LO: lateral occipital; PoC: post-central; PrC: precentral; SF: superior frontal; CMF: caudal middle frontal; IP: inferior parietal; IT: inferior temporal; PoCi: posterior cingulate; PrCu: precuneus; SP: superior parietal; ST: superior temporal; IL: inferior longitudinal fasciculus; POST AR: posterior arcuate fasciculus.*

| | FA | MD | ICVF | ISOVF |
|---|---|---|---|---|
| **LH Cu inter** | - | - | -0.13 | - |
| **LH LO inter** | - | -0.14 | 0.11 | - |
| **LH PoC inter** | -0.16 | - | 0.21 | - |
| **LH PrC inter** | - | - | -0.18 | -0.26 |
| **LH PrC intra** | - | 0.11 | - | - |
| **LH SF intra** | 0.10 | - | - | - |
| **RH CMF inter** | 0.13 | - | - | - |
| **RH IP inter** | - | 0.21 | - | - |
| **RH IT intra** | - | 0.11 | 0.15 | - |
| **RH LO intra** | - | - | -0.12 | - |
| **RH PoCi inter** | - | 0.14 | - | - |
| **RH PrCu inter** | - | - | - | -0.11 |
| **RH SP inter** | - | - | - | -0.15 |
| **RH ST inter** | - | - | -0.15 | - |
| **RH IL** | - | -0.12 | - | - |
| **RH POST AR** | - | - | -0.1 | - |

## IV - Conclusion

Our analysis of data from over 13,747 UK Biobank participants indicates that superficial white matter (SWM) has a more pronounced influence on cognitive aging than deep white matter (DWM). Building on previous findings by [4], which focused on fractional anisotropy (FA), we extended the investigation to include additional diffusion metrics—particularly NODDI-based measures that offer a more detailed and biologically informed characterization of white matter microstructure through multi-compartment modeling. Despite these advances, several limitations should be acknowledged. The use of mean values to summarize diffusion metrics across bundles may introduce bias, potentially oversimplifying within-bundle variability. Additionally, for computational efficiency, each streamline was resampled to 15 points—a resolution that may sufficiently capture short-range connections but could limit accuracy in characterizing longer tracts. Future work will aim to explore the genetic underpinnings of SWM by leveraging the FUMA platform to identify genomic regions associated with SWM bundles that show strong associations with cognitive performance.

## V - References

# Mapping the Superior Longitudinal System: anatomical insights from BraDiPho

Laura Vavassori[1,2], François Rheault[3], Paolo Avesani[4], Alessandro De Benedictis[5], Francesco Corsini[1], Luciano Annicchiarico[1], Luca Zigiotto[1], Umberto Rozzanigo[6], Mattia Barbareschi[2,7], Laurent Petit[8] and Silvio Sarubbo[1,2].

*[1]Department of Neurosurgery, "S. Chiara" Hospital, APSS, Trento, Italy | [2]Department of Cellular, Computational and Integrative Biology (CIBIO), Center for Medical Sciences (CISMed), Center for Mind and Brain Sciences (CIMeC), University of Trento, Trento, Italy | [3]Sherbrooke Connectivity Imaging Laboratory (SCIL), Université de Sherbrooke, Sherbrooke, QC, Canada | [4]Neuroinformatics Laboratory (NiLab), Bruno Kessler Foundation (FBK), Trento, Italy | [5]Neurosurgery Unit, Bambino Gesù Children's Hospital, IRCCS, Rome, Italy | [6]Department of Radiology, "S. Chiara" Hospital, APSS, Trento, Italy | [7]Department of Laboratory Medicine - Pathology Unit, "S. Chiara Hospital, APSS, Trento, Italy | [8]Groupe d'Imagerie Neurofonctionnelle (GIN-IMN), University of Bordeaux, Bordeaux, France.*

**Introduction.** Charting the organization of white matter (WM) pathways is essential for understanding the functioning of the human brain[1]. Currently, there is no consensus about the anatomical extent, course and terminations of many WM bundles[2]. Indeed, descriptions of the cortical terminations of a bundle are generally bound to the specific aim of each study and rely on the assumption of prior anatomical definitions for bundle segmentation[3]. Concurrently, while tractography and ex-vivo WM dissection provide complementary insights, their integration remains a longstanding challenge[4]. This study provides a comprehensive, anatomically enhanced characterization of the superior longitudinal system (SLS) by integrating in-vivo tractography with ex-vivo dissection within the same radiological space with the BraDiPho approach (https://bradipho.eu)[5].

**Methods.** Constrained spherical deconvolution and particle-filtering tractography with anatomical priors[6] was computed for 39 healthy participants of the BIL&GIN database[7]. Using a data-driven cortex-to-cortex pairing approach leveraging gyral-sulcal anatomical landmarks, we reconstructed the dorsal associational connectivity of the frontal cortex, defined by a set of patterns of connectivity between ipsilateral gyri (*i.e.,* sub-SLS). Each sub-SLS was non-linearly registered to 10 photogrammetric models of ex-vivo Klingler WM dissection for a multimodal evaluation of its anatomical reliability[5] (Fig. 1).

**Results.** We identified 45 sub-SLS, of which (i) 22 were validated and refined through ex-vivo dissection, (ii) 17 were deemed anatomically plausible despite lacking ex-vivo confirmation, and (iii) 6 were classified as anatomically implausible. The anatomical description of plausible sub-SLS templates revealed fundamental organizational principles of the system: (i) a medio-lateral and dorso-ventral hierarchy, where dorsal regions connect dorsally and ventral regions ventrally, and (ii) a depth-dependent organization, with shorter, superficial fibers linking proximal areas and longer, deeper fibers connecting distal regions. Pearson's correlation confirmed a significant positive relationship between streamline length and distance from the cortex ($r=0.689$, $p<0.001$).

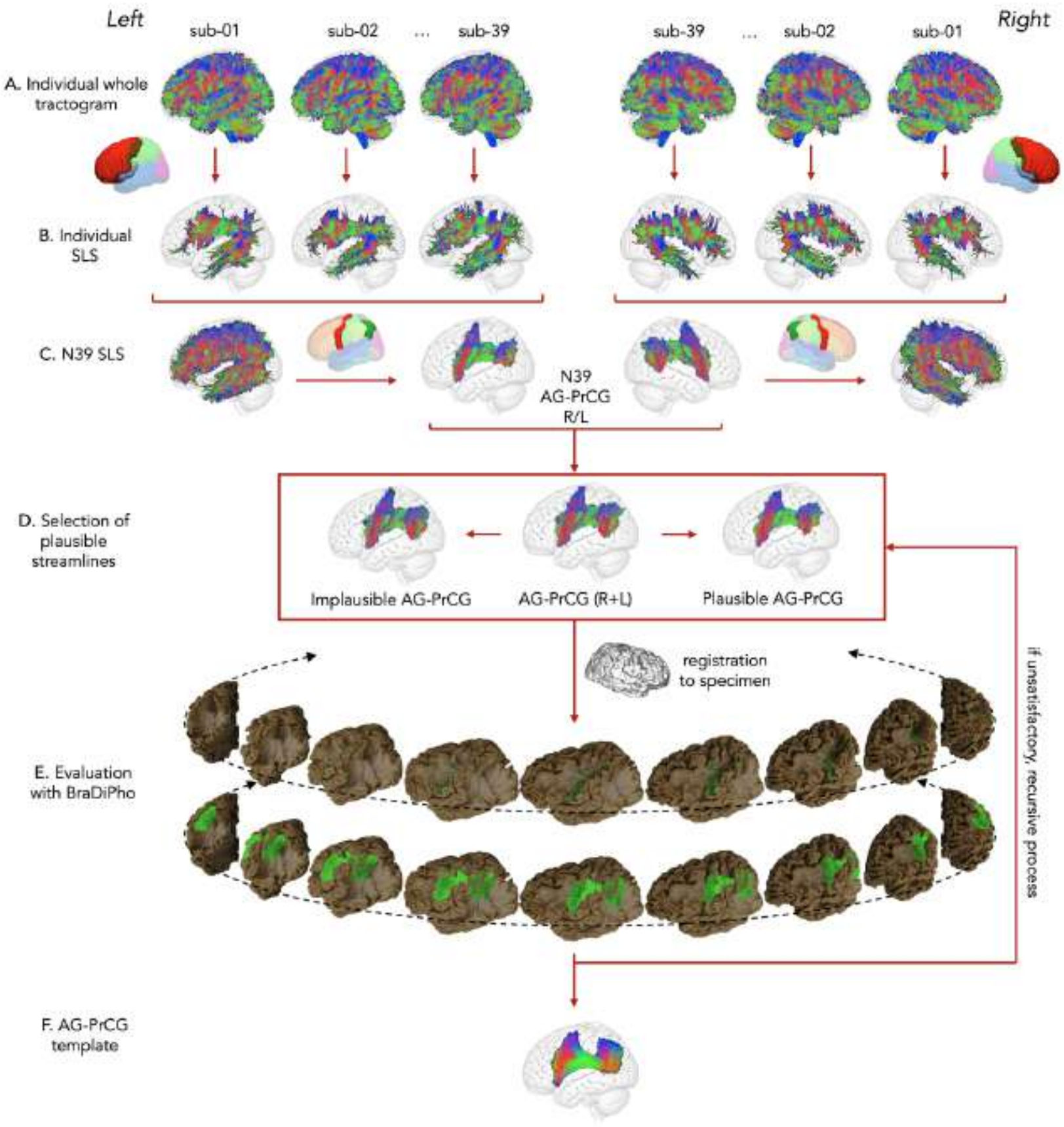

***Figure 1.*** Schematic representation of the pipeline for creating sub-SLS templates. Starting from the whole-brain tractograms of 39 individuals in MNI space (**A**), individual SLS tractograms are extracted (**B**). The individual SLS tractograms are concatenated, and the cortex-to-cortex pairing approach is applied to parcel the system into sub-SLS (**C**). sub-SLS from the right hemisphere are flipped to the left hemisphere and concatenated to their contralateral homologs (**C**) Broad clustering is performed, and streamlines of each sub-SLS are classified as plausible or implausible based on geometrical features (**D**). Filtered sub-SLS undergo anatomical evaluation using the BraDiPho framework (**E**), followed by smoothing (**F**). If anatomical evaluation identifies gaps in spatial coverage, the process from **D** to **F** is repeated, with finer clustering and refinement based on the information from ex-vivo dissection.

**Conclusions.** This study provides a robust anatomical foundation for future population-based WM atlases using bundle-specific tractography[8] and emphasizes the need for a distributed and integrated understanding of brain connectivity beyond classical bundle definition.

**References.**

## Sulcal morphology reflect the organization of short U-shape association fibers

Arnaud Le Troter, Olivier Coulon

*Institut de Neurosciences de la Timone, Aix-Marseille Univ, UMR CNRS 7289, Marseille, France*

### Introduction

Short Association Fibers (SAF) in superficial white matter constitute a large portion of the overall WM connectivity and have been associated to various pathologies. Despite this, SAF connectivity is understudied, mostly because in-vivo tractography for SAF using diffusion MRI is notoriously difficult [1,2]. It has been hypothesized from Klingler dissections [3] or MRI studies [4] that sulcal morphology is linked to U-shape SAF connectivity. In particular, some sulcal morphological landmarks called three-way junctions [3] or wall pinches (WP) [4] are thought to be associated to higher densities of U-shape fiber bundles. In this context, it has been shown [2] that the resolution of diffusion MRI is a determinant factor for SAF tractography. In this work we use a very high-resolution dataset [5] to infer U-shape SAF connectivity around two sulci and investigate its link with sulcal morphology

### Methods

The pipeline consisted of five main stages, as illustrated in Fig1, and was applied to the MGH single-subject dataset [5], offering DWI data with a high spatial resolution of $(0.76\text{mm})^3$ acquired in 9 sessions and 18 volumes. The T1w volume was registered to the T2w volume, that in turn was registered to the mean B0 volume computed from the 18 DWI volumes. The resulting warped T1w was resampled in the diffusion space. Cortical surface extraction and segmentation were performed using FreeSurfer 7.4 on the T1w volume warped on DWI (Fig1.A). ROIs were selected on the white matter surface, based on the Destrieux parcellation (aparc2009), targeting the central sulcus (CS: precentral and postcentral gyri and central sulcus labels) and the superior frontal sulcus (SFS: superior and middle frontal gyri and sulcus). These ROIs were then projected into the diffusion volume space and the resulting binary masks were morphologically dilated (kernel size: 5×5×5 voxels), and a bounding box was defined around the union of all dilated ROIs to crop the full dMRI dataset accordingly (Fig1.B). The cropped diffusion volumes were then processed using unringing and denoising filters (MRtrix3).

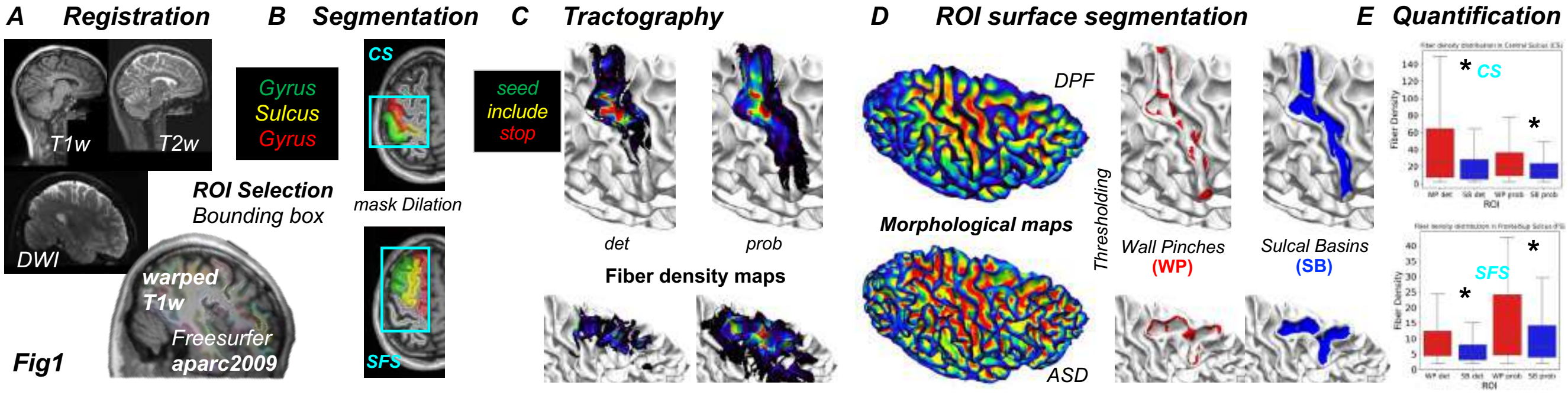

Tractography was performed as follow (Fig1.C). For the short U-shaped fiber extraction, we estimated the response function using the *Dhollander* method and computed fiber orientation distributions (FODs) via multi-shell multi-tissue constrained spherical deconvolution (MSMT-CSD), constrained within the dilated ROI volume. For each sulcus (SFS and CS), we generated 5000 streamlines from one gyrus to the other via the sulcus, in both directions using both the probabilistic (iFOD2) and deterministic (SD_STREAM) algorithms, with common parameters: minimum length of 10 mm, step size of 0.2 mm, maximum curvature angle of 30°, and unidirectional seeding. Streamlines were used to generate fiber density maps at diffusion resolution using tckmap. These maps were then projected onto the white matter cortical surface into the cortical ribbon using mri_vol2surf (trilinear interpolation of voxel 5 mm under the surface and over to 2 mm).

Sulcal morphology was characterized on the white cortical surface using two complementary morphological maps (Fig1.D): the DPF map (Depth Potential Function [6]), sensitive to sulcal depth variations, and the ASD map (Average Sample Distance [7]), enabling detection of WPs along the gyrus walls [7]. Sulcal basins were delineated by thresholding the DPF (DPF > -0.75). WPs within the basins were extracted by thresholding the ASD map (-1.5 < ASD < -0.1). To assess the relationship between SAF and WP, we compared the distribution of fiber densities within WPs versus the rest of the sulcal basin (SB) for SFS and CS and for both tractography algorithms. Statistical differences were assessed using a Mann–Whitney U test.

### Results and Discussion

Our analysis revealed a significantly higher fiber density beneath WPs compared to the rest of the sulcal basin, across both CS and SFS, for both tractography algorithms. Mann–Whitney tests revealed highly significant differences ($p < 0.001$) between WPs and SB for both deterministic and probabilistic approaches in SC and SFS (Fig1.E). This finding highlights the fact that short U-shaped association fibers are more frequently located beneath white matter protrusions such as WPs than elsewhere in the sulcus, suggesting a link between sulcal morphology and the local organization of superficial white matter connectivity.

**Title: Do current automated tractography methods hold up in tumour and epilepsy pathology? A comparison of six methods with expert manual tractography**

Steven Greenstein[3], Sila Genc[1,3,4], Francois Rheault[5], Maxime Descoteaux[5], Alison Wray[6], Wirginia Maixner[6], and Joseph Yuan-Mou Yang[1,2,4]

*[1]Neuroscience Advanced Clinical Imaging Service (NACIS), Department of Neurosurgery, Royal Children's Hospital, Melbourne, Australia. [2]Neuroscience Research, Murdoch Children's Research Institute, Melbourne, Australia. [3]Developmental Imaging, Murdoch Children's Research Institute, Melbourne, Australia. [4]Department of Paediatrics, University of Melbourne, Melbourne, Australia. [5]Sherbrook Connectivity Imaging Laboratory, University of Sherbrooke, Quebec, Canada. [6]Department of Neurosurgery, Royal Children's Hospital, Melbourne, Australia.*

**Introduction**: Diffusion MRI tractography is a valuable imaging adjunct for pre-surgical planning in neurosurgery[1]. Although expert-based manual tractography is considered the clinical silver-standard, its time-intensive nature and requirement for neuroanatomical expertise limits its use, especially in emergency settings[2]. Automated tractography methods are being developed to address these challenges[3], but their performance remains uncertain in cases with pathology[4]. Large lesions can cause significant anatomical distortion and peri-lesional white matter oedema can introduce additional technical complexities. This study aims to evaluate the performance of six current state-of-the-art automated tractography methods against expert-based manual tractography in a cohort of paediatric brain tumour and epilepsy surgery patients with challenging lesion-related imaging characteristics.

**Methods**: Cohort: Fourteen patients who had brain tumour or epilepsy surgery at the Royal Children's Hospital were selected based on having peri-lesional oedema or diffuse infiltrative tumours on imaging and received expert-based manual tractography for pre-surgical planning (8 males, median age=11.46 years [interquartile range, IQR 8.41-13.62], median lesion volume=81.89 $cm^3$ [IQR 57.87-131.77]).

MRI: Pre-surgical MRI scans were conducted on clinical 3T Siemens systems with 40 or 80mT/m gradients, acquiring T1-weighted data (voxel-sizes=0.8-1.0 mm) and multi-shell diffusion MRI (dMRI) data with multi-band accelerated EPI sequence (2.3 $mm^3$ isotropic voxels, TE/TR=77/3500 ms,11 interleaved b0s, b=1000 $s/mm^2$ applied over 30 directions, and b=3000 $s/mm^2$ applied over 60 directions). dMRI data were pre-processed with MRtrix3[5], FSL[6], ANTs[7] to correct for imaging noise[8], Gibbs ringing artefacts[9], motion and susceptibility distortion [10,11], and bias field inhomogeneity[12].

Manual Tractography: Tractography was performed using multi-tissue constrained spherical deconvolution[13] and iFOD-2 probabilistic tracking algorithm in MRtrix3[5]. Tracking ROIs were manually delineated by a neuroanatomy expert based on known anatomical priors. We evaluated the arcuate fasciculus (AF), corticospinal tract (CST), and optic radiation (OR) (n=84) bilaterally for all cases due to their common implications in surgical planning. Manual tracts were converted to binarised tract masks and acted as references to evaluate automated tractography outputs. Between 5,000-15,000 streamlines were retained per tract, depending on the tract type.

Automated Tractography: We assessed six automated methods: *TractSeg*[14], *BundleSeg*[15], *Classifyber*[16], *White Matter Analysis (WMA)*[17], *DeepWMA*[18], and *TRACULA*[19]. Each method was performed with default settings to generate automated tract outputs for each patient. For methods required the generation of a whole-brain tractogram, one million streamlines were used.

Statistics: As tract anatomy definitions varied between manual and automated methods, we first computed Dice similarity coefficients (DSC) between manual tract masks and each automated method's template-based tract ROIs[14-17] using scilpy (v1.10.1). Each automated method was then compared to the manual tracts by computing DSC, percentages of false-positive volumes (FP) for automated-only segments, and false-negative volumes (FN) for manual-only segments. FP and FN voxel distances from the lesion were plotted in histograms. Lesion- and non-lesion side DSC, FP and FN were compared using Wilcoxon rank sum test, with statistical significance set at $p<0.05$.

**Results**: All automated methods successfully reconstructed tracts on the non-lesion side, apart from the OR using *BundleSeg*. On the lesional side, reconstruction success varied: *TractSeg* (83.3%), *BundleSeg* (71.4%), *Classifyber* (92.9%), *WMA* (90.5%), *DeepWMA* (64.3%), and *TRACULA* (78.6%). Median Dice similarity coefficients (DSC) between manual tracts and template-based ROIs were low (0.20–0.37), indicating significant discrepancies in tract anatomy definitions. Visual inspection confirmed substantial variation across automated methods compared to manual tracts (Figure 1). All automated approaches showed low DSC (median 0.40, IQR 0.22–0.49), high FP (28.3%, IQR 15.12–40.49), and high FN (44.9%, IQR 26.48–62.65). Lesion-side tracts had significantly lower DSC and higher FN than non-lesion tracts ($p < 0.01$), while FP was only slightly higher ($p = 0.079$). Most FP and FN voxels were near lesions, suggesting that pathology and peri-lesional oedema negatively impact automated tractography (Figure 2).

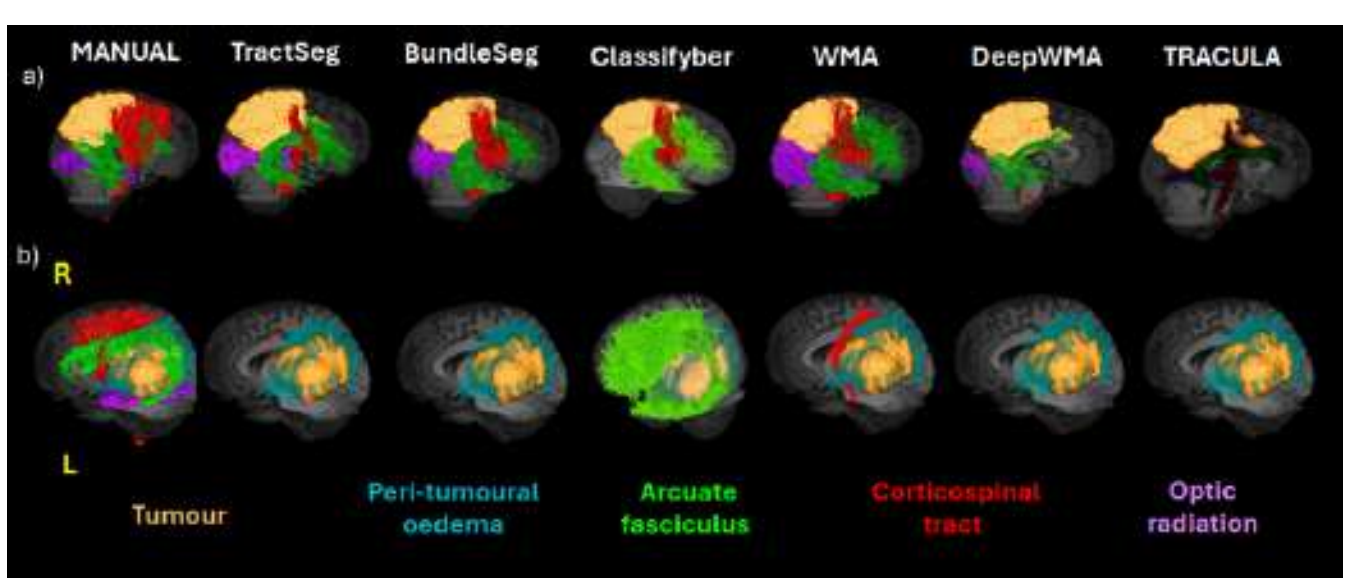

***Figure 1.*** *Manual and automated tractography outputs for two cases, a) a right superior parietal tumour, and b) a left temporal-parietal tumour with marked peri-tumoural oedema. The four automated tractography outputs demonstrate high variability in tract appearance compared to manual tractography results, that were used for pre-surgical planning.*

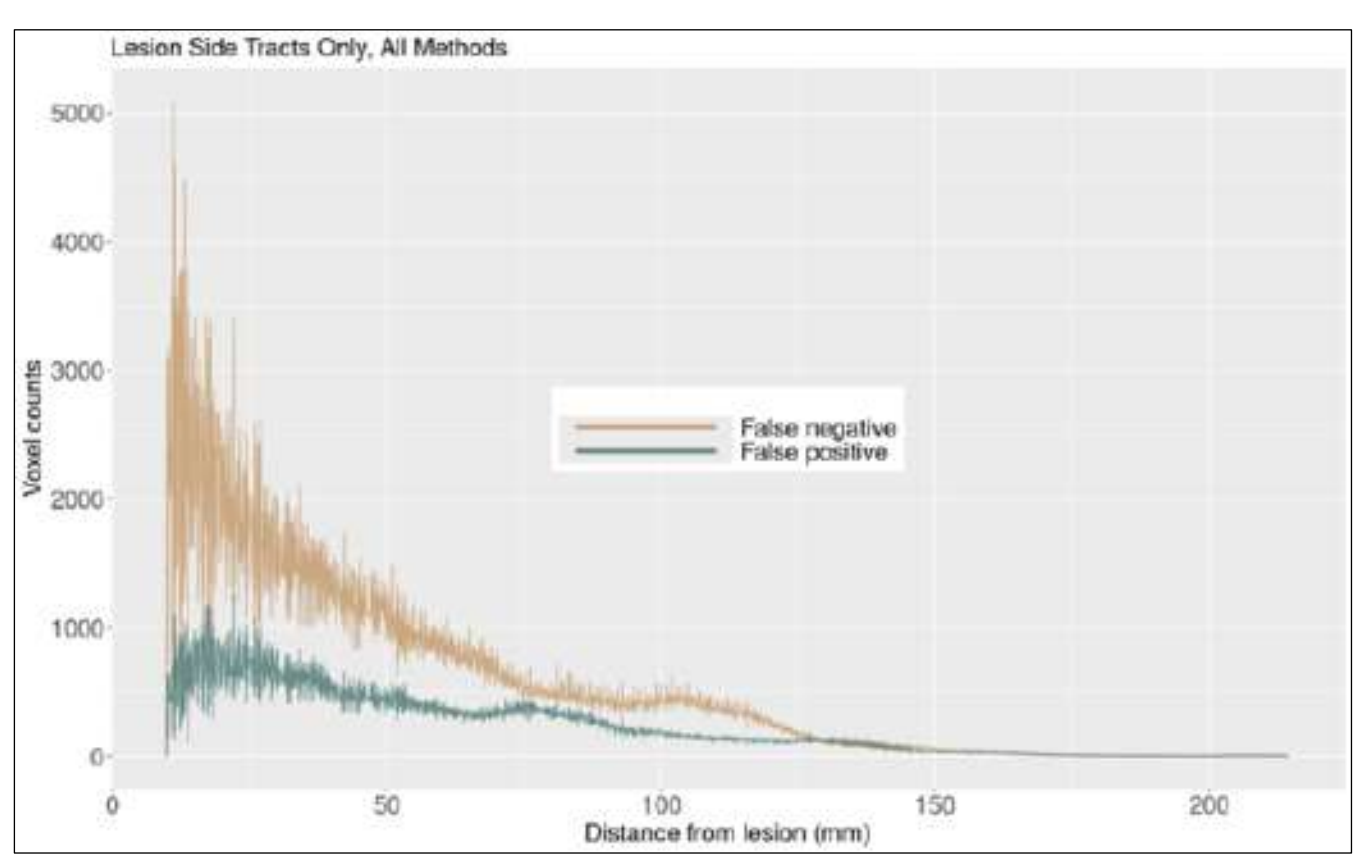

***Figure 2.*** *Frequency plots for the number of voxels within false-positives (FP) and false-negatives (FN) regions of automated tract outputs for lesion side tracts, expressed as a function of distance from the lesion. Gold lines indicate FPs results, and green lines indicate FNs results. Each subplot shows results from a different automated method. Only results from the lesion side tracts are shown. Binwidth was set to 0.1mm. The plots showed most FP and FN are located close to lesions.*

**Conclusion**: Despite high tract completion rates, automated methods showed substantial variability, especially on the lesion side, due to differing tract definitions. False positives and negatives often occurred near lesions, reflecting the challenge of modelling pathology-driven distorted white matter. These inaccuracies may increase surgical risk, highlighting the importance of standardised tract definitions and improved training of automated models using appropriate surgical cases. Our findings urge caution in relying on current automated tractography for complex paediatric tumour and epilepsy surgeries.

## *MouseFlow*, a pipeline for diffusion MRI processing and tractogram generation in mouse brain validated using Allen Brain Atlas Connectivity with *m2m*.

**Elise Cosenza[1], Arnaud Boré[2,4], Julien Fouilloux[1,3], Joël Lefebvre[3,4], Maxime Descoteaux[2,4], Sylvain Miraux[4,5,6], Laurent Petit[1,4]**

[1]Groupe d'Imagerie Neurofonctionnelle (GIN) IMN, UMR5293 CNRS, U. Bordeaux, France ; [2]Sherbrooke Connectivity Imaging Laboratory, U. Sherbrooke, QC, Canada ; [3]Université du Québec à Montréal, Montréal (QC), Canada ; [4]International Research Project OpTeam, CNRS Biologie, France – U. Sherbrooke, Canada ; [5]Centre de Résonance Magnétique des Systèmes Biologiques (CRMSB), UMR5536 CNRS, U. Bordeaux, France ; [6]Plateforme d'Imagerie Biomédicale (pIBIO), UAR3767 CNRS U. Bordeaux, France

In this work, we introduce **MouseFlow**, an open-source and standardised Nextflow-DSL2 pipeline specifically designed to address the current heterogeneity in processing diffusion magnetic resonance imaging (dMRI) data in rodent studies. While dMRI provides valuable insights into brain network organisation and white matter structure, the absence of unified processing protocols for rodents, unlike in human imaging, has led to variability in results and limited reproducibility. MouseFlow fills this gap by offering a robust, automated workflow that includes essential preprocessing steps such as denoising, eddy current and bias correction, brain extraction, and the reconstruction of both diffusion tensor imaging (DTI) and q-ball imaging. Importantly, it incorporates the Allen Mouse Brain Atlas (AMBA) [1] to enable anatomically guided extraction of specific diffusion metrics and tractography bundles. Built on the Nextflow DSL2 framework, the pipeline is highly adaptable to different dataset types, configurable via JSON, and optimised for reproducibility, transparency, and ease of use, making it a valuable tool for preclinical imaging research.

In addition to MouseFlow, we also present **m2m** [2], a complementary suite of Python-based tools developed to extend the analytical possibilities offered by the AMBA. The AMBA provides a uniquely detailed, high-resolution map of axonal projections in the mouse brain derived from viral tracer experiments, representing a powerful ground truth resource for studying structural connectivity. However, access to and analysis of AMBA connectivity maps have traditionally been confined to static visualisations on the Allen Institute website, without the ability to interactively combine it with user-acquired datasets. To overcome this limitation, m2m enables interoperability between Allen and user data spaces by leveraging the Allen Software Development Kit (AllenSDK) for data import and ANTsPyX for high-precision image registration. Using this framework, we compute a transformation matrix to map AMBA-derived projection density maps onto the native space of user dMRI data, facilitating direct visual and analytical comparison such as overlaying tractography with viral tracer data. Conversely, the inverse transformation allows researchers to trace any given white matter location in their own data back to its corresponding region and experimental context within the **Allen Mouse Brain Common Coordinate Framework**.

This bidirectional mapping supports rich data interaction, including visualisation through platforms like MI-Brain [3], and opens new avenues for multimodal integration of tractography and tracer-based connectivity data.

Together, MouseFlow and m2m offer a comprehensive and interoperable toolkit that significantly advances rodent neuroimaging research. By enabling standardised preprocessing and reproducible analysis workflows, along with interactive and bidirectional mapping between diffusion and tracer-based datasets, they open new opportunities for exploring structural connectivity. This integration not only enhances tractography interpretation but also supports its validation against tracer-derived ground truth (Figure), fostering deeper insights into white matter organisation in the mouse brain.

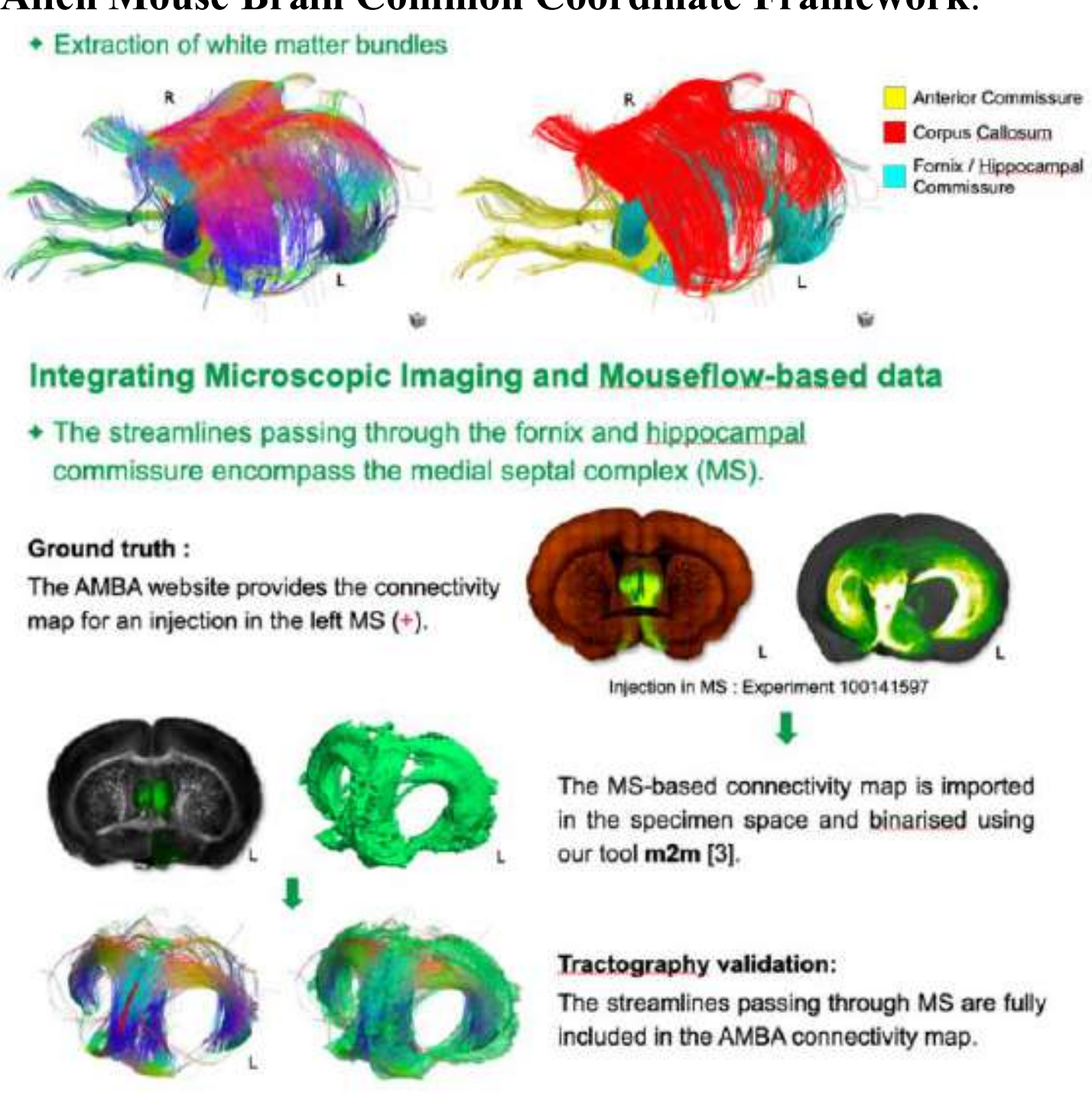

**Title: Inferior Frontal Projections of the Arcuate Fasciculus Tractography in Broca's Area: Validation against Direct Cortical Electrical stimulation Language Mapping performed in a Paediatric Epilepsy Surgery Case**

Joseph Yuan-Mou Yang[1,3,5], Sila Genc[1,4,5], Bonnie Alexander[1], Sarah Barton[2], Laura Moylan[2], Emma Macdonald-Laurs[2,3], A Simon Harvey[2], Wirginia Maixner[6]

*[1]Neuroscience Advanced Clinical Imaging Service (NACIS), Department of Neurosurgery, Royal Children's Hospital, Melbourne, Australia. [2]Department of Neurology, Royal Children's Hospital, Melbourne, Australia. [3]Neuroscience Research, Murdoch Children's Research Institute, Melbourne, Australia. [4]Developmental Imaging, Murdoch Children's Research Institute, Melbourne, Australia. [5]Department of Paediatrics, University of Melbourne, Melbourne, Australia. [6]Department of Neurosurgery, Royal Children's Hospital, Melbourne, Australia.*

**Introduction**

Accurate localisation of Broca's area and its connecting white matter tract, particularly the arcuate fasciculus (AF), is essential for safe neurosurgical planning near the language-dominant inferior frontal lobe.[1] However, tractography results can vary significantly depending on the reconstruction method used.[2] This study compares the accuracy of expert-guided manual versus automated AF tractography in delineating inferior frontal cortical projections, validated against direct cortical electrical stimulation (DCES) findings in a 15-year-old boy with drug-resistant, MRI-negative focal epilepsy undergoing extra-operative language mapping with subdural grid electrodes for evaluation of planned surgical resection.

**Methods**

MRI: Pre-surgical 3T MRI was acquired including 3D T1-weighted (voxel-sizes = 0.8mm) and multi-shell diffusion MRI (dMRI) data with multi-band accelerated EPI (2.3 $mm^3$ isotropic voxels, TE/TR=77/3500 ms,11 interleaved b0s, b = 1000 $s/mm^2$, 30 directions, and b=3000 $s/mm^2$, 60 directions). dMRI data were pre-processed with MRtrix3[3], FSL[4], ANTs[5] to correct for imaging noise[6], Gibbs ringing artefacts[7], motion and susceptibility distortion[8,9], and bias field inhomogeneity.[10]

Manual Tractography: Tractography was performed using multi-tissue constrained spherical deconvolution[11] and iFOD-2 probabilistic algorithm in MRtrix3. Tracking ROIs for AF were manually delineated based on expert anatomical priors[2,12]. Automated Tractography: was performed using TractSeg[13], a deep-learning method based on learning peak FOD of tract bundle segmentation. Both AF tracts were reconstructed using 3000 streamlines and both tracts were converted to thresholded tract density images as binarised tract masks.

DCES: Cortical stimulation was performed using a biphasic current stimulator, delivering bipolar stimulation to adjacent electrode pairs at 50 Hz with 0.5 msec pulse duration, 5 seconds trains, and currents up to 10 mA. EEG was continuously monitored for after discharges and seizures, noting their threshold, distribution, and any language disruption. Stimulation-positive sites were electrode contacts where stimulation at low currents (3 mA) induced reproducible expressive language deficits (anomia, impaired sentence repetition) with retained receptive language (following verbal commands). Stimulation-negative sites were electrode contacts where stimulation at high currents (8-10 mA), above after discharge threshold, showed no expressive language deficits. These stimulation results served as the ground truth for evaluating tractography accuracy.

Points along the grey matter-white matter interface corresponding to the stimulation positive and negative electrode locations were located on T1-weighted MRI co-registered with the dMRI. Each point consists of 3 T1-weighted MR voxel cubes (i.e. 2.4 mm in dimension), approximating the electrode contact size. These points were used as the stimulation locations for the tractography analysis.

Analysis: Volume overlaps between stimulation-positive, stimulation-negative sites along the grey matter-white matter interface and the manual or automated AF tract masks were computed. Tract overlap with stimulation-positive sites defines true positives (TP), while non-overlap defines false negatives (FN). Tract overlap with stimulation-negative sites defines false positives (FP), and non-overlap defines true negatives (TN). We then calculated: Positive Predictive Value (PPV) = TP/(TP+FP); Negative Predictive Value (NPV) = TN/(TN+FN), False-positive rate (FPR) = FP/(FP+TN); False-negative rate (FNR) = FN/(FN+TP), from both AF tractography methods.

**Results**: Surgical resection carried out including stimulation-negative inferior frontal lobe (*pars orbitalis* and anterior *pars triangularis*; black dotted line, Fig.1A), with no post-operative expressive language deficits observed. Using DCES as the functional reference for identifying AF projections to Broca's area, manual AF tractography achieved 100% PPV, 80.2% NPV, a 0% False-positive rate, and a 84.0% False-negative rate. *TractSeg*-based automated AF tractography achieved 66.7% PPV, 83.0% NPV, a 4.9% False-positive rate, and a 66.7% False-negative rate.

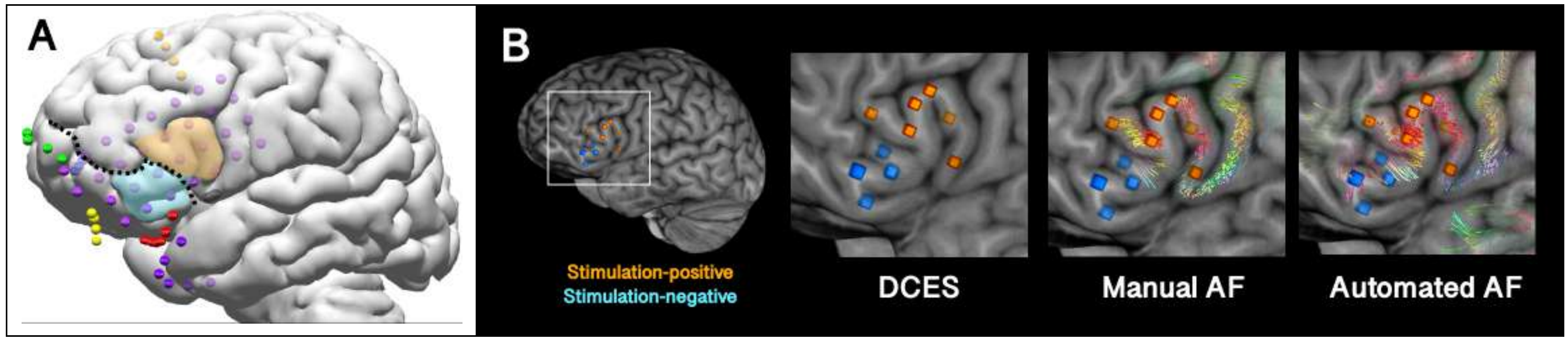

***Figure 1.*** *Language mapping and tractography in this patient.* ***A:*** *3D render displaying subdural and strip electrodes, with stimulation (expressive language)-positive (orange) and -negative (blue) regions in the inferior frontal lobe. The black dotted line indicates the surgical resection margin.* ***B:*** *Corresponding cortical electrical stimulation sites shown on 3D render, with close-ups of the left inferior frontal lobe overlaid with both manual and automated arcuate fasciculus (AF) tractography.*

**Conclusion**

Expert-guided manual tractography demonstrated superior specificity, with a 100% PPV and 0% FPR, making it highly reliable with remarkable precision for identifying language-negative cortex. Conversely, *TractSeg*-based automated tractography showed a much lower PPV and a higher FPR suggesting reduced precision in identifying language-negative cortex. Both methods showed high FNRs, suggesting limited sensitivity in detecting the full extent of DCES language-positive cortex. These findings support our practice of combining expert-based manual AF tractography with functional brain mapping to guide safer surgical resections near the language-dominant inferior frontal lobe. The elevated FPRs with *TractSeg*-based automated method suggests the need to refine the AF tract definition underlying its development.

# Exploring the macaque precentral gyral white matter using 11.7T dMRI

[a*]Fanny Darrault; [a,b*]Guillaume Dannhoff; [c]Maëlig Chauvel; [c]Cyril Poupon; [c]Ivy Uszynski; [a,d]Christophe Destrieux; [a,d]Igor Lima Maldonado; [a]Frédéric Andersson

[a] Université de Tours - INSERM - Imaging Brain & Neuropsychiatry iBraiN U1253
[b] Centre Hospitalier Régional Universitaire de Strasbourg, Strasbourg, France
[c] BAOBAB, NeuroSpin, Paris-Saclay University, CNRS, CEA, France
[d] CHRU de Tours, Tours, France
[*] These authors contributed equally

*Introduction*: Brain connectivity is being extensively studied and is crucial to comprehend its considerable complexity. Understanding the organization of connections will help decipher the function of the cortical areas. Since Penfield's homunculus, the precentral gyrus has been reconceptualized as more than the primary motor cortex[1,2].
*Methods*: We analyzed its connective profile by quantifying association, projection, commissural, and striatal fibers and their distribution pattern at the grey matter/white matter interface (GMWMI). We used a single-subject ultra-high field 11.7T MRI dataset[3] from an ex vivo Macaca fascicularis brain.
*Results*: Results revealed a **heterogeneous fiber distribution** at the GMWMI, with dense foci and lower-density areas.
<u>Association</u> fibers were **predominant** and distinctly distributed, with high-density areas in the ventral two-thirds of the gyrus, on its convexity, and to a lesser extent, in the paracentral lobule. **Other categories** were found in the medial first third on the convexity of the gyrus, the midline and on the ventral gyrus crest. We **noted left/right asymmetries** with more commissural fibers in the right precentral gyrus and less striatal and projection than in the left precentral gyrus.
We focused <u>on the association fibers,</u> which showed that fibers going through the frontal adjacent areas are found rostrally on the precentral gyrus, whereas those going through the parietal adjacent areas are found on the caudal part of the gyrus. The remaining fibers (going through distant areas) are found rostrally and on the paracentral lobule.
*Conclusion*: These findings suggest that the precentral gyrus has a strong associative component that intermingles with motor regions[4]. Ultra-high resolution MRI's ability to reconstruct fiber pathways highlights its potential to bridge macaque study data to human neuroanatomy.

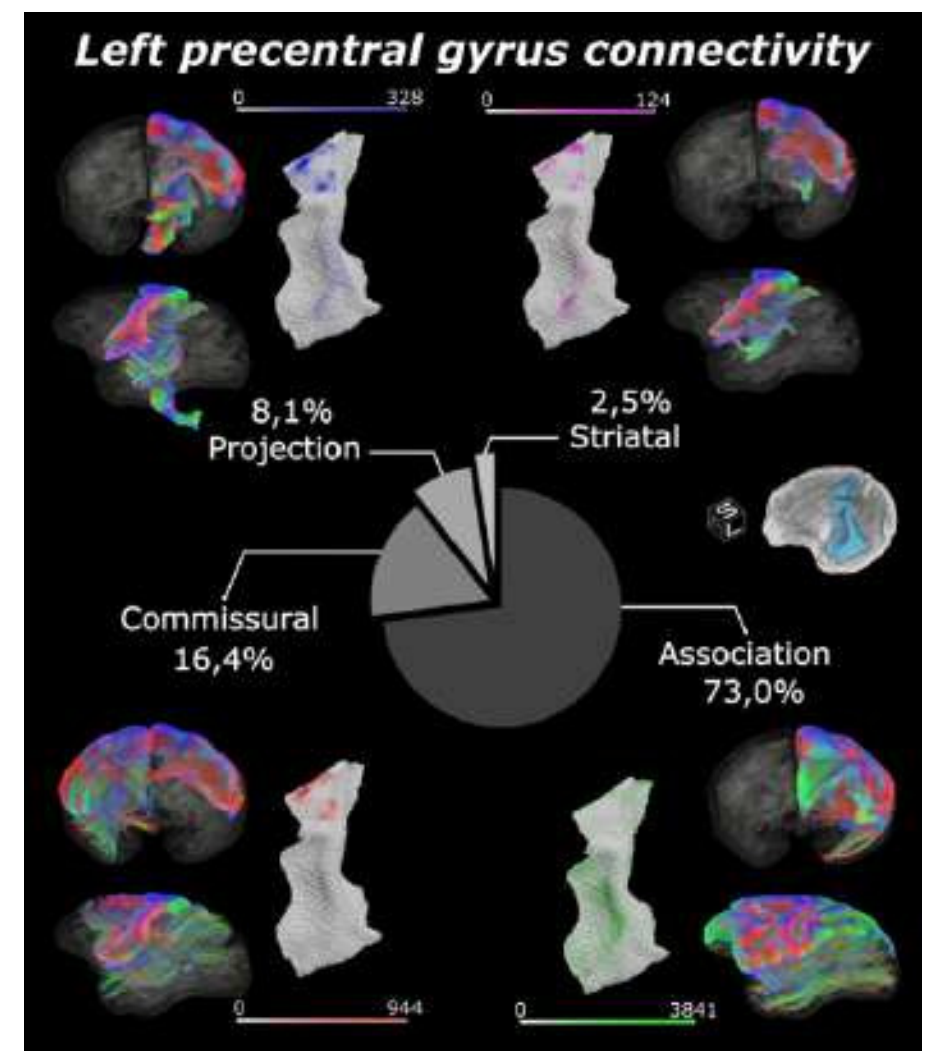

*Figure 1: Tractography of the four categories of fibers of the left precentral gyrus (association, striatal, commissural and projection), with their distribution on the white matter/grey matter interface and their relative amount in percentages*

## Micro- and macrostructural fiber tract changes during pediatric posterior fossa tumor surgery.

**Authors**: Pien E.J. Jellema (1,2), Karina J. Kersbergen (1), Eelco W. Hoving (1), Maarten H. Lequin (3), Wouter (P.) Nieuwenhuis (1,4) Alberto De Luca (2,5)*, Jannie P. Wijnen (2)*. **Affiliations**: 1) Department of Pediatric Neuro-Oncology, Princess Máxima Center for Pediatric Oncology, Utrecht, the Netherlands. 2) Centre for Image Sciences, University Medical Center Utrecht, Utrecht, the Netherlands. 3) Edward B. Singleton Department of Radiology, Texas Children's Hospital, Austin, Texas, USA. 4) Department of Radiology, University Medical Center Utrecht, Utrecht, the Netherlands. 5) Department of Neurology, University Medical Center Utrecht, Utrecht, The Netherlands. *Shared last authors

### Introduction

Intraoperative fiber tractography enables visualization of eloquent white matter tracts such as the dentate-rubro-thalamic tract (DRTT) during pediatric posterior fossa tumor (pPFT) surgery. This allows assessment of intraoperative physiological changes relative to preoperative data while minimizing the confounding effect of brain shift that results from craniotomy. This study investigates pre- to intraoperative changes in DRTT macro- and microstructure.

### Methods

MR-images of 30 pPFT patients (average age: 7.6±4.6 years, Fig. 1a & b) were acquired using a 3-Tesla intraoperative MRI scanner (Philips Ingenia ElitionX, Philips Healthcare, Best, The Netherlands) with two single channel RF coils (Fig 1d). Both pre- and intraoperative images were acquired on the day of surgery with patients positioned prone in the DORO head clamp (Fig. 1c). Preoperative imaging was performed prior to craniotomy, and intraoperative imaging after tumor resection, with an open skull. The imaging protocol included a T1-weighted scan and diffusion MRI (dMRI) (2.5 mm isotropic, 6 b = 0 s/mm$^2$, 20 b=1000 s/mm$^2$, 32 b=2000 s/mm$^2$ images). Whole brain deterministic fiber tractography was performed after GRL deconvolution (1) with a 1 mm step size, 60° angle threshold, and FOD threshold of 0.01. Inclusion regions of interest (ROIs) for DRTT reconstruction were the dentate nucleus (2), thalamus (3), and primary motor cortex (4). Exclusion ROIs were semi-manual segmentations of the tumor (Smart Brush, Brainlab, Munich, Germany) and the corpus callosum (5). The arcuate fasciculus (AF), an anatomically distant tract from the pPFT resection cavity, serves as a control to assess measurement consistency across timepoints. The AF were segmented from whole-brain tractography (1) using the white matter analysis pipeline (6). Macrostructural metrics included tract volume and diameter (7). DRTT and AF volumes were normalized to supratentorial white matter volume. Paired Wilcoxon signed-rank tests were used to compare pre- to intraoperative data. Microstructural metrics included fractional anisotropy (FA) and mean diffusivity (MD) sampled along the tract centerline (8). Signal-to-noise ratios were calculated by dividing the mean signal intensity of all b0 volumes by their standard deviation.

### Results

Both the DRTT and AF (Fig. 2) exhibited a reduction in volume and diameter during surgery. This decrease was statistically significant for the left contralateral DRTT in terms of both volume (45% reduction, $p = 0.002$) and diameter (25% reduction, $p = 0.002$), as well as for the right AF volume (10% reduction, $p = 0.019$) and diameter (5.2% reduction, $p = 0.020$). SNR decreased by 9.5% ($p = 0.004$). Comparing pre- to intraoperative data, the along tract analysis of FA and MD remained stable along the AF, with minimal variation (Fig. 3b). In contrast, FA measurements along the DRTT were characterized by high fluctuations (data not shown). The DRTT showed greater variability in MD within the cerebellar segment compared to the cortical segment, particularly seen in the contralateral DRTT (Fig. 3a).

### Discussion

Microstructural metrics of the AF remained stable across surgery, as expected given its distance from the resection cavity. In contrast, the DRTT showed increased variability in MD, especially in its cerebellar segment near the resection cavity, suggesting localized microstructural changes. The macrostructural decrease was more pronounced in the DRTT than in the AF, suggesting an effect of pPFT surgery on the DRTT. The reductions observed in both tracts, including the anatomically distant AF, may reflect global effects such as decreased intraoperative SNR or non-linear brain shift. This could result from cerebrospinal fluid evacuation and altered brain consistency due to gravity on the brain in the prone position. Overall, our results highlight the partly unsolved challenge of achieving reliable intraoperative tractography, especially in the context of an open skull, even when using consistent tractography parameters. Taken together, the primary effects of pediatric posterior fossa surgery were observed mainly in microstructural rather than macrostructural measures, with localized changes in the DRTT near the resection cavity and stability in anatomically distant AF tracts. Improving the reliability of intraoperative reconstructions will require technical advances to improve SNR and data quality.

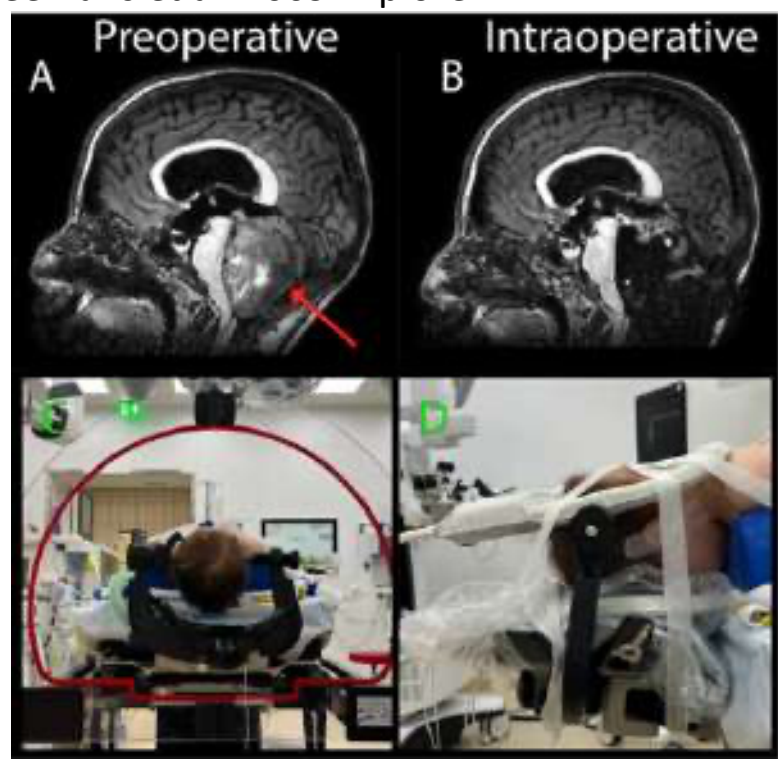

**Fig.1 Study set-up.** T1-weighted (T1w) images of a 14 year old girl with a 4th ventricle tumor before resection (A, preoperative) and after surgical resection but before closure of the skull (B, intraoperative). The red arrow in A indicates the tumor in-situ. The patient is prone positioned in the DORO surgical head-frame (C). The two RF coils are then placed anterior and posterior of the patients' head (D).

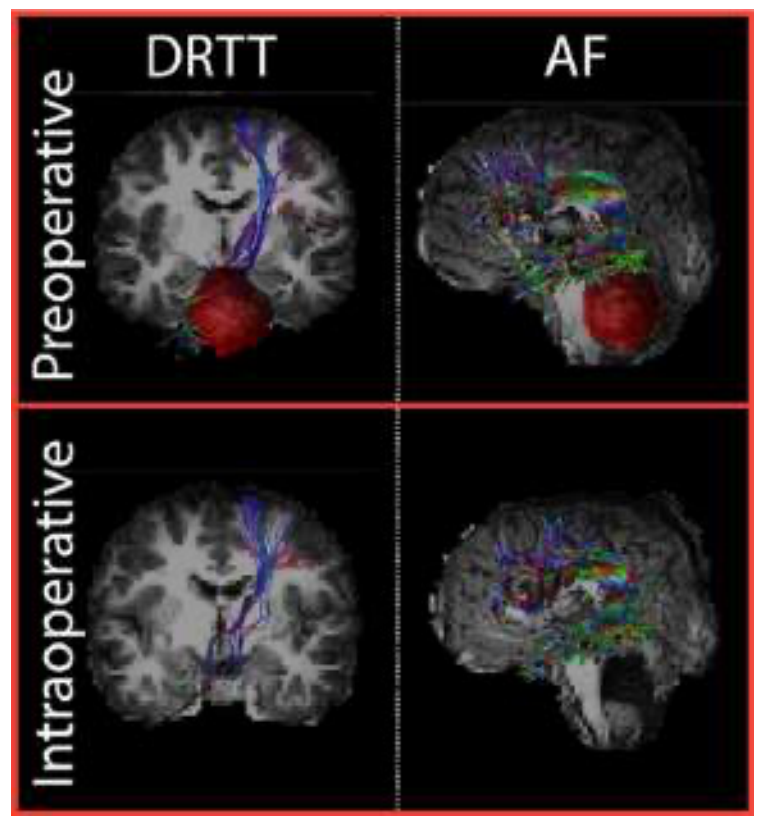

**Fig. 2. Examples of tract reconstructions.** Coronal and sagittal views of fiber reconstructions in a 9-year-old boy with a pilocytic astrocytoma. The left column shows part of the dentate-rubro-thalamic tract (DRTT) originating in the left dentate nucleus and terminating in the contralateral hemisphere. The right column shows the left arcuate fasciculus (AF).

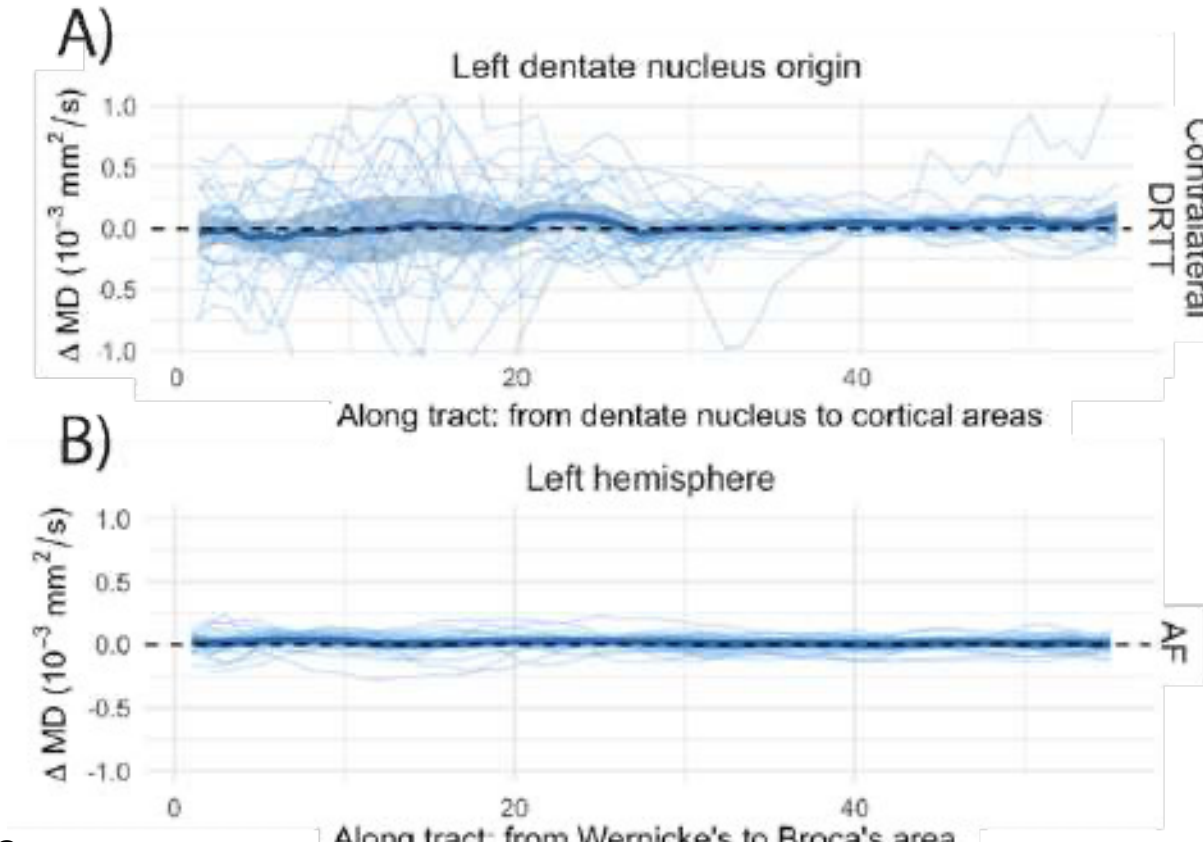

**Fig. 3. Along tract analysis.** Mean diffusivity changes (Δ MD, intraoperative-preoperative) are shown along the contralateral DRTT (A) and AF (B) originating in the left hemisphere. Population average (thick line), 95% confidence interval (ribbon), and individual subject data (thin lines) are shown.

# Multicenter approach for validation of white matter tracts involved in cognition

M.J.F. Landers[1], M.M.G.H. van de Veerdonk[1,2], R. Bakker[1,3], E. Madonnet[4], E.M. Bos[5], A.P.J.E. Vincent[5], D. Satoer[5], P.A. Robe[6], M.J.E. van Zandvoort[6], H.B. Brouwers[1], G.J.M. Rutten[1]

[1] Department of Neurosurgery, Elisabeth-Tweesteden Hospital Tilburg, The Netherlands
[2] Department of Radiology & Nuclear Medicine, Erasmus MC Cancer Institute, Erasmus University Medical Center, Rotterdam, the Netherlands
[3] Department of Mathematics and Computer Science, Technical University, Eindhoven, the Netherlands
[4] Department of Neurosurgery, Lariboisière Hospital, Paris, France
[5] Department of Neurosurgery, Erasmus MC Cancer Institute, Erasmus University Medical Center, Rotterdam, the Netherlands.
[6] Department of Neurology & Neurosurgery, University Medical Centre Utrecht, Utrecht, The Netherlands

**Introduction**
White matter architecture is increasingly acknowledged in neurosurgical decision-making[1], especially in glioma surgery. MR-tractography enables visualization of white matter tracts[2] and is routinely used to plan surgery and guide intraoperative (extent of) resection[3]. However, tractography results need further validation, particularly for tracts involved in cognitive functions. During awake brain surgery, direct electrical stimulation (DES) of the brain offers a unique opportunity to validate tractography results[4]. However, no studies have systematically categorized DES responses and related them to specific cognitive tracts. In a previous multicenter study[5], we developed a standardized intraoperative protocol that was used to align procedures and documentation practices between neurosurgical centers. In the current study we investigate if multicenter DES-data can be used for the clinical validation of tractography.

**Methods**
In this, prospective, observational, multicenter study, we included adult patients that underwent awake primary glioma surgery from four neurosurgical centers: ETZ Tilburg, Lariboisière Paris, EMC Rotterdam, and UMC Utrecht. All patients underwent monitoring of cognitive functions with a standardized set of tasks according to the previously developed protocol[5,6]. Diffusion-weighted MRI scans were processed on a protected automatic tractography pipeline (CSD-based) that was developed in collaboration with the Technical University Eindhoven[7]. The pipeline automatically generates the following white matter tracts: Frontal Aslant Tract, Superior Longitudinal Fasciculus II and II, Arcuate Fasciculus and Inferior Fronto-Occipital Fasciculus. Positive and negative DES response sites were recorded and co-registered with the patient-specific tractograms and shortest distances between response site and tracts were quantified. Figure 1 demonstrates a case example. Postoperative DWI scans were made for verification of positive and negative mapping sites.

**Preliminary results**
Currently, 75 patients are included in this multicenter study. The results demonstrated that surgical teams all adhere to the standardized protocol, which allowed for pooling of data. Preliminary analyses demonstrate that intraoperative DES findings can be used to validate white matter tracts involved in cognitive functions. We expect to be able to present the final results by October 2025.

**Conclusion**
This study represents an important step toward systematic clinical validation of white matter tracts involved in cognition. Standardized DES data collection during awake brain tumor surgery using a standardized intraoperative monitoring protocol is feasible and provides a framework to address current gaps in evidence. Moreover, it can serve as a scalable and adaptable framework for future refinement and clinical validation.

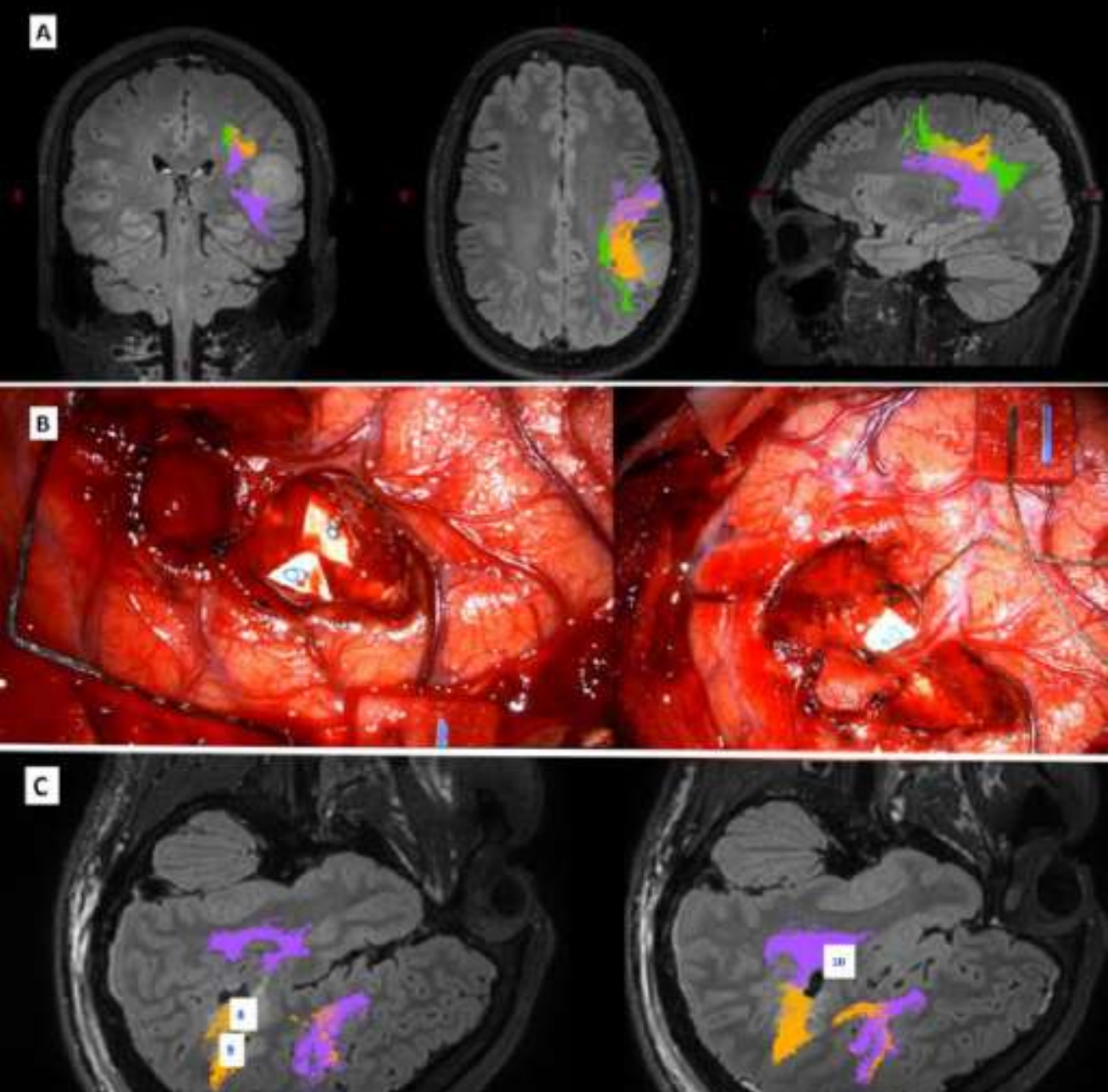

Figure 1: Case example[8]. A. 32-year-old male patient was operated on a left supramarginal low-grade astrocytoma. A) Preoperative coronal, axial and sagittal view of MRI with tractography output from the pipeline of the SLS II (green), SLS-III (orange) and AF (purple); B) intraoperative photo, tag 8=speech arrest; tag 9=multiple order errors during digit span forward and backward; tag 10=anomia during picture naming; C) intraoperative stimulation points transferred to postoperative MRI images, indicating that tags 8 and 9 correlate with SLS-III (minimal distance 6.2mm and 8.6mm for SLF and 11.9 and 16.5mm for AF), and tag 10 with AF (minimal distance 8.9mm for AF and 9.6mm for SLF).

# BundleParc: automatic bundle parcellation in tumor data

Antoine Théberge[1], Pierre-Marc Jodoin[1], Joseph Yuan-Mou Yang[2], Maxime Descoteaux[1], François Rheault[1]

[1] Sherbrooke Connectivity Imaging Lab (SCIL)/VitaLab, Université de Sherbrooke, Québec, Canada

[2] Neuroscience Advanced Clinical Imaging Service (NACIS), Royal Children's Hospital/Neuroscience Research, Murdoch Children's Research Institute/Department of Paediatrics, University of Melbourne, Melbourne, Australia

**Introduction** Bundle parcellation (i.e. splitting white matter bundles into multiple sections, or regions) is a critical step in tractometry [1, 2], the study of the microstructural properties of white matter bundles. More broadly, bundle delineation is a crucial step in preoperative planning for brain tumours [3]. While machine learning methods are gaining popularity for these tasks, most are trained and tested on high-quality research-grade data, and their applicability in challenging scenarios, such as pathological data, is often unknown. In this work, we assess the robustness of a recent method for bundle parcellation, *BundleParc* (formerly known as *LabelSeg*) [4] on a cohort of preoperative patients with brain tumours. We assess the similarity between BundleParc and two other well established procedures for bundle parcellation and delineation : tractography and atlas registration. We observe that all three methods offer very different bundle delineations and parcellation, each with their own advantages and pitfalls.

**Methods** BundleParc is a neural network-based tool that directly maps from fODFs to bundle parcellations in voxel space, i.e. $N$ regions representing different parts of a fiber bundle (also sometimes referred to as *label maps*). The neural network was trained on 105 subjects from Human connectome project, processed with *TractoFlow*, and their curated reference sets of fiber bundles [5]. Bundle parcellations on the reference set of streamlines (to train BundleParc) were computed as follows: we first registered a centerline from a population-averaged atlas in MNI space to the subject's space. We then resampled the centerline to $N$ points, the number of desired labels. Then, we assigned the nearest centerline's point (ie target label) to every streamlines' points, and trained a support vector machine to perform classification of the streamline points. The resulting separation hyperplanes were used to assign the proper labels to each voxel in the bundle's volume. Once trained, BundleParc can map directly from fODF to bundle parcellations, in the subject's space, in a few seconds on a consumer-grade GPU. We assessed the robustness of BundleParc on a publicly available set of preoperative subjects with brain tumours (11 glioma, 14 meningioma patients) [6], which includes T1w and diffusion MRI acquisitions (b=700,1200,2800 at 2.5mm iso) as well as manually delineated tumour masks. Subjects were processed using TractoFlow. We directly predicted bundle parcellations using BundleParc with 10 regions and compared with : (1) tractography and (2) atlas registration. Tractography (1) was performed using an ensemble [7] of local probabilistic and Particle Filtering Tractography. Tracking masks were computed by thresholding the Fractional Anisotropy maps at 0.10 and seeding was performed at 10 seeds per voxel in the tracking mask. Resulting tractograms were segmented using BundleSeg [8]. Label maps were then computed as described above using scilpy (specifying the "hyperplane" method). For atlas registration (2), we computed label maps in MNI space from the population average bundle atlas for BundleSeg (TractSeg definitions) [9] and then registered the label maps to each subject's space. Per subject, BundleParc took approximately 30 seconds to predict all bundles, tractography took around 3 hours (including label computation) and atlas registration took around 10 minutes (excluding label computation in MNI space).

**Experiments** We compute the bundle-wise (c.f. tab. 1) and label-wise (c.f fig. 2) Dice coefficient for all bundles and subjects between the three considered methods to assess overall delineation and parcellation agreement. We additionally measure tumor distance for segments that are less than 15mm away from the tumor (c.f fig 3) to assess the impact of tumor deformation on the methods. To quantify subject-specificness, we measure bundle-wise overreach wrt white matter tissue segmentation (c.f. table 2). Finally, we visually inspect two bundles from one patient to assess the similarity and differences between the reconstructions from our three methods (c.f. fig. 1). For all quantitative experiments, we perform a one-way ANOVA test to enquire if the scores are statistically distinct between all three methods and comparisons.

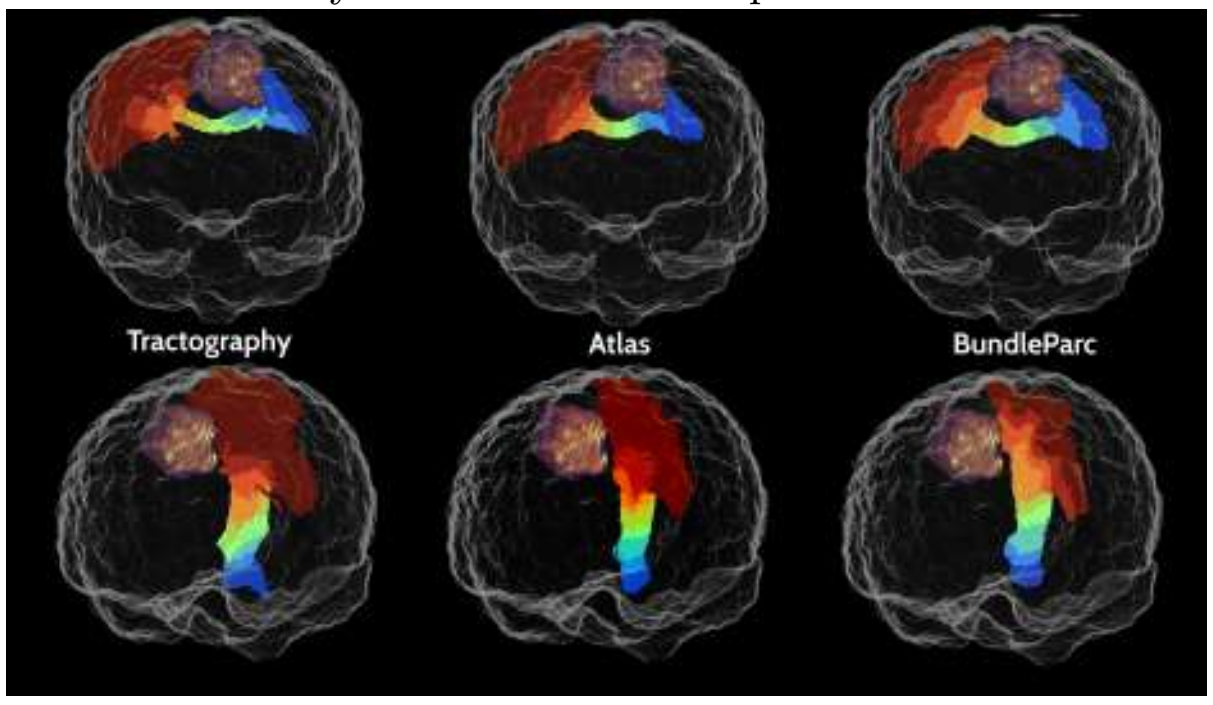

Figure 1: Single subject with a meningioma in the frontal lobe along with bundle parcellation from all three methods. Different colors represent different segments. Upper row: anterior mid-body segment of the corpus callosum. Lower row: left cortico-spinal tracts.

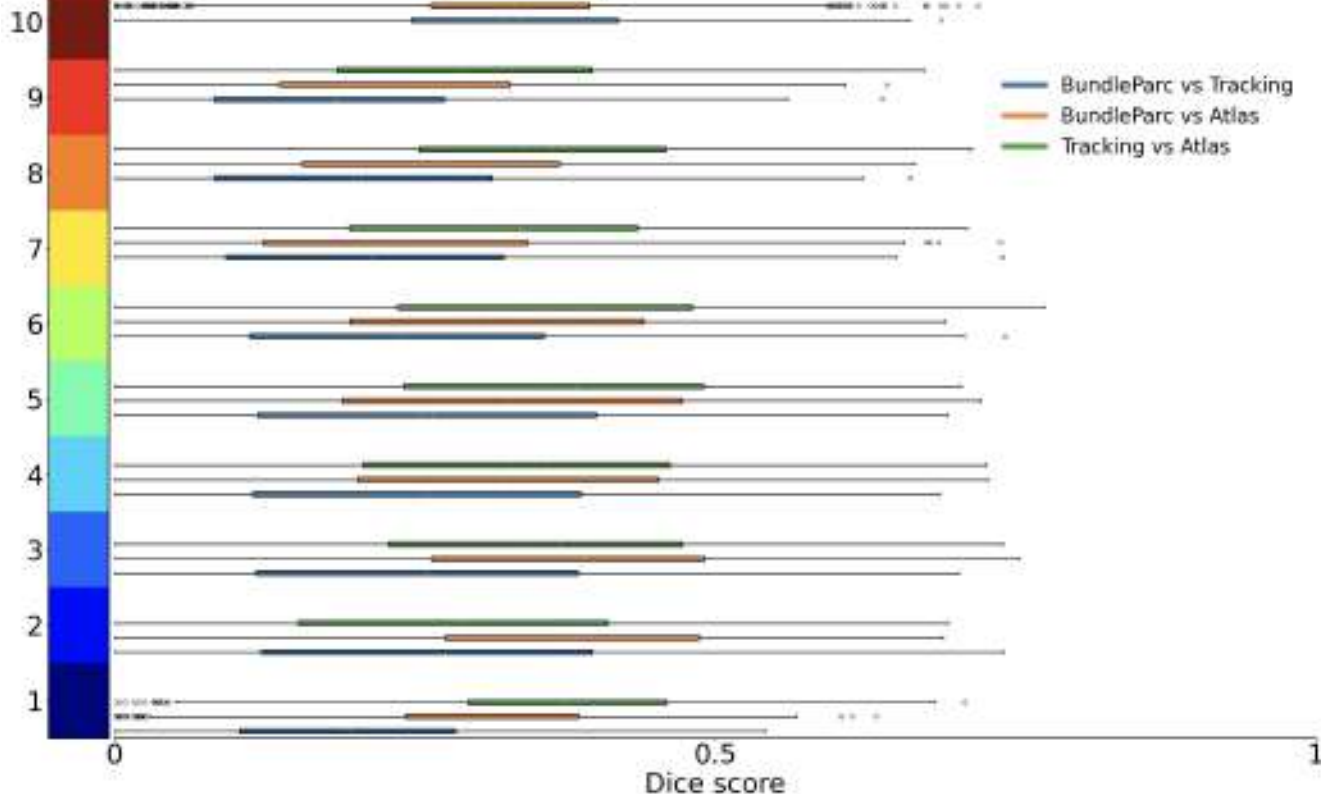

Figure 2: Label-wise Dice coefficients between pairs of methods, for all bundles and patients.

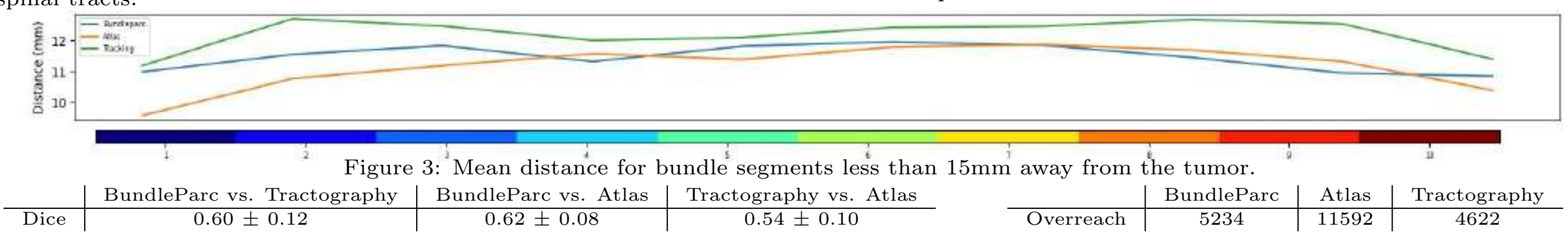

Figure 3: Mean distance for bundle segments less than 15mm away from the tumor.

| | BundleParc vs. Tractography | BundleParc vs. Atlas | Tractography vs. Atlas |
|---|---|---|---|
| Dice | 0.60 ± 0.12 | 0.62 ± 0.08 | 0.54 ± 0.10 |

Table 1: Dice coefficients between whole-bundle volume masks obtained from all three bundle parcellation methods for all patients and bundles.

| | BundleParc | Atlas | Tractography |
|---|---|---|---|
| Overreach | 5234 | 11592 | 4622 |

Table 2: Mean overreach of the white matter mask for all bundles and patients, in $mm^3$.

**Results** Bundle-wise, the volume of the bundles reconstructed by BundleParc is in fair agreement with both tractography and atlas registration, more so than between tractography and registration ($p < 0.005$). Label-wise, however, tractography is most similar with atlas registration; however the agreement is fairly low (below 0.5), highly variable and statistically distinct between all three methods ($p < 0.005$). In this case, BundleParc's labels are more similar to registration than tractography. In all cases, most outliers (i.e. highly similar or dissimilar parcellations) happen at the extremities (label 1 or 10) of bundles. Distance-to-tumor wise, all three methods perform similarly; endpoints (i.e. labels 1 or 10) tend to be closest to the tumor ($p > 0.005$), compared to the trunk segments of the tract. Atlas registration is more prone to produce bundle segments outside of the white matter mask than BundleParc or Tractography ($p < 0.005$). Visually, while no method provided "degenerate" reconstructions due to the displacement of tissues caused by the tumour, tractography seems to provide "fuller" bundles compared to BundleParc, which in return seems to produce "neater" parcellations.

**Discussion and conclusion** Overall, we report a fair agreement between our newly proposed method, BundleParc, and well-established bundle delineation methods. Overall, BundleParc was not more in disagreement with tractography or atlas registration than tractography is with atlas registration. While discrepancies were noted between methods (i.e. most ANOVA tests reported statistically significant differences between scores), no method exhibit catastrophic failures and every method was in fair (but not total) agreement. It should be noted that a large portion of the scores were obtained on bundles possibly far and unimpacted by tumoral deformation. While these may act as "control" cases, assessing the reconstruction similarity between methods in "normal" white matter fibres, future work should directly focus on clinically relevant bundles, located for example in the same hemisphere as the tumor. All in all, this study demonstrates that machine learning-based tract segmentation methods for diffusion MRI can achieve robust reproducibility, even in technically challenging brain tumour cases.

# Linking Continuous and Interrupted Human Central Sulcus to the Variability of the Superficial White Matter using a β-Variational AutoEncoder

C. Mendoza[1], J. Chavas[1], A. Dufournet[1] , J. Laval[1], M. Guevara[1], Z. Y. Sun[1], A. Grigis[1], P. Guevara[2], D. Riviere[1], J.-F. Mangin[1]
[1]Université Paris-Saclay, CEA, CNRS, BAOBAB, Neurospin, Gif-sur-Yvette, France
[2]Faculty of Engineering, Universidad de Concepción, Concepción, Chile

## Introduction

Superficial White Matter (SWM) and cortical folding can be seen as two endophenotypes of brain development, reflecting a deep structural interdependence that shapes the architecture of the cerebral cortex. Previous works have studied the relationship between the SWM and sulcal anatomy [1–3]; however, their morphological and functional interplay remains poorly understood. Here, we show for the first time a relationship between changes in short association fibers, Central Sulcus (CS) morphology and brain function in the right hemisphere. We demonstrate this by focusing on the transition from a single- to a double-knob configuration, an anatomical variation previously associated with differences in hand sensorimotor activation [4]. For this purpose, we leverage a β-Variational AutoEncoder (β-VAE) to obtain meaningful representations of short-range cortical connectivity. Additionally, to illustrate rearrangements in short fiber bundle organization, we study the rare case of an interrupted CS, in which hand sensorimotor activation is mapped to a bridging gyrus connecting the precentral and postcentral regions [5].

## Methods

We analyzed 333 subjects from the HCP database [6], including seven subjects with an interrupted CS in the right hemisphere. First, we calculated the SWM tractography [7] based on MRtrix3 software [8]. Then, we segmented the fibers surrounding the right CS using gyral crown lines, derived from the white matter mesh with the ABLE software [9] (Fig. 1-A). Fiber connectivity profiles, based on vertex-wise endpoints density, were computed on the white mesh and projected onto the FreeSurfer registered sphere (Fig. 1-B) [10]. Then, these profiles were transformed into a 2D angular grid and used to train a β-VAE with all the subjects (Fig. 1-C). Next, the β-VAE embeddings were used to compute a pairwise distance matrix, which served as input to the ISOMAP algorithm as in [4]. The β-VAE and ISOMAP parameters were selected based on a maximum correlation with a CS-shape-based embedding describing a transition between single to double knob configuration ($r=-0.34$, $p=2.31e-05$) [4]. To analyze inter-subject differences, moving averages of the connectivity profiles were computed along the ISOMAP axis. Regions of interest were defined based on differences between the first and last averages (Fig. 1-D), and were subsequently used to segment dorsal and ventral fiber bundles (Fig. 1-E). Finally, to assess fiber reorganization in subjects with an interrupted CS, we used a U-shapedness index, defined as the ratio of fiber length to the Euclidean distance between its endpoints. Lower values of this index, approaching 1, indicate straighter fiber trajectories. To obtain a single value per bundle, we computed the average U-shapedness across fibers.

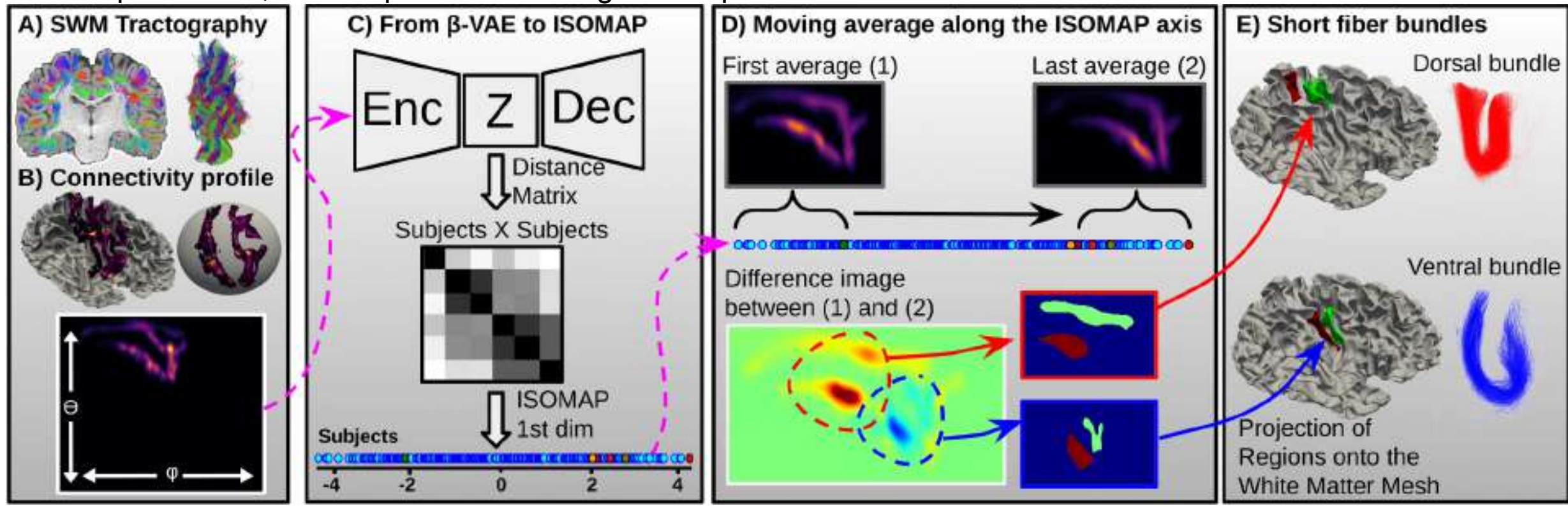

**Fig 1.** Methodology to study short fiber bundles driving inter-subject differences. In the ISOMAP axis: cyan (continuous CS), different colors (interrupted CS).

## Results

We found a negative correlation between the ISOMAP coordinates and the number of fibers in the dorsal bundle ($r = -0.57$, $p = 1.53e-30$) (Fig. 2-A). Since this bundle was located near the hand-knob, average fMRI left-hand motor maps are shown for subjects at both extremes of the ISOMAP axis, along with the corresponding average CS (Fig. 2-B). Notably, the group at the rightmost end of the ISOMAP axis exhibited a more upper positioned knob—typical of the double-knob configuration—and a reduced global activation volume. A positive correlation between ISOMAP coordinates and the number of fibers in the ventral bundle was found ($r = 0.45$, $p = 2.86e-18$) (Fig. 2-C). Finally, the dorsal bundle mapped directly or near the interruption of the CS, with a high proportion of fibers having a more straight shape for most interruptions ($p = 0.021$, Mann-Whitney U), as shown from the left shift of the U-shapedness distribution and in the zoomed view of subject 138231 (Fig. 2-D).

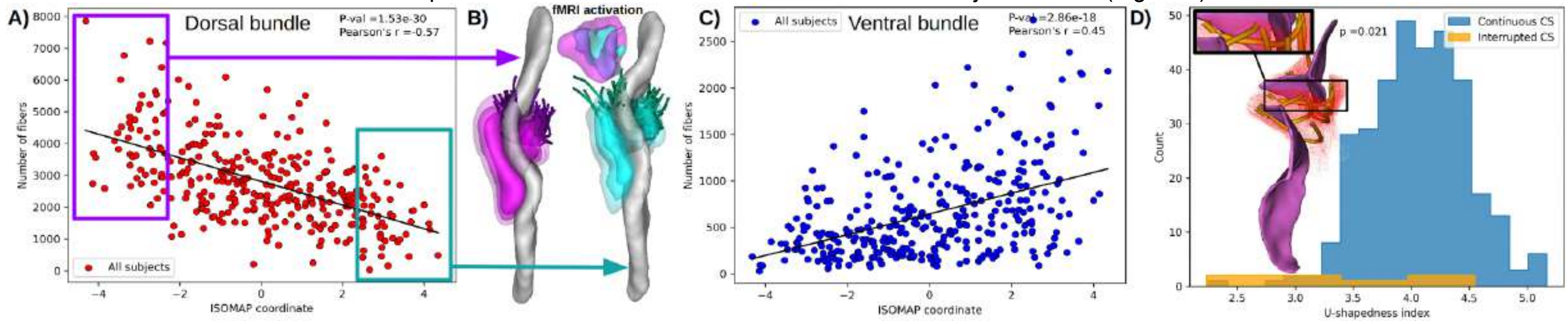

**Fig 2**. (A) Results for the dorsal bundle. (B) Average CS and fMRI left-hand motor activation, along with an average streamline per subject representing the dorsal bundle. (C) Results for the ventral bundle. (D) Histogram of dorsal bundle average U-shapedness for continuous (light blue) and interrupted (orange) CS.

## Conclusions

We demonstrate for the first time a link between changes in short fibers bundles, CS morphology and extent of functional areas in the right hemisphere. Our findings align with a previous work showing reduced left-hand motor activation when the right CS transitions from a single- to a double-knob configuration [4], which may correlate with the lower number of fibers observed in the dorsal bundle. Conversely, the increased number of fibers in the ventral bundle may reflect the emergence of a second knob in the ventral portion of the CS, although this hypothesis requires further investigation. Subjects with an interrupted CS had straighter fiber trajectories in the dorsal bundle, which is interesting given prior findings that sensorimotor fMRI contrast localizes in the bridging gyrus, suggesting that such interruptions may drive a reorganization of the underlying fiber architecture. While CS interruptions are rare (~1% prevalence [5]), they occur more frequently in variable regions such as the temporal lobe [3], showing the potential of our framework to disentangle the anatomo-functional relationship between SWM and cortical folding.

## Acknowledgements

This work was funded by the National Agency for Research and Development (ANID) / Scholarship Program / DOCTORADO BECAS CHILE / 2024 - 72240205. HCP Data were provided by the Human Connectome Project, WUMinn Consortium (Principal Investigators: David Van Essen and Kamil Ugurbil; 1U54MH091657) funded by the 16 NIH Institutes and Centers that support the NIH Blueprint for Neuroscience Research; and by the McDonnell Center for Systems Neuroscience at Washington University.

**Title: Can a reduced diffusion MRI protocol achieve optic radiation tractography comparable to those reconstructed using a multi-shell diffusion MRI acquisition for temporal lobe epilepsy surgery planning and surgical image-guidance?**

Joseph Yuan-Mou Yang[1,2,4], Jian Chen[3], Kurt Schilling[5], Alison Wray[6], Wirginia Maixner[6]

*[1]Neuroscience Advanced Clinical Imaging Service (NACIS), Department of Neurosurgery, Royal Children's Hospital, Melbourne, Australia.[2]Neuroscience Research, Murdoch Children's Research Institute, Melbourne, Australia. [3]Developmental Imaging, Murdoch Children's Research Institute, Melbourne, Australia. [4]Department of Paediatrics, University of Melbourne, Melbourne, Australia. [5]Department of Radiology and Radiological Sciences, Vanderbilt University Medical Center, Nashville, TN, United States. [6]Department of Neurosurgery, Royal Children's Hospital, Melbourne, Australia.*

**Introduction**

Diffusion MRI (dMRI) tractography is a valuable tool for pre-surgical planning, helping to avoid surgical white matter tract injuries.[1,2] At our hospital, expert-guided manual ROI-based tractography using multi-shell dMRI has been instrumental in visualising the optic radiations (OR) for temporal lobe epilepsy surgery, supporting both pre-surgical planning and intraoperative guidance to preserve visual function.[3] However, multi-shell acquisition poses a barrier to widespread adoption due to scanner hardware limitations and restricted clinical scan time at many centres. This study investigates whether reduced dMRI protocols, such as halved acquisition time or single-shell sampling at low or high b-values, can produce OR tractography of comparable quality to those reconstructed using a multi-shell protocol, for surgical use.

**Methods**

Patients/MRI: MRI data from five children with drug-resistant focal epilepsy who underwent temporal lobe epilepsy resective surgery at the Royal Children's hospital were included. All underwent OR tractography for pre-surgical planning and intra-operative image guidance assisting surgical execution. Tractography was based on using 3T multi-shell dMRI acquired using multi-band EPI (2.3 mm$^3$ isotropic voxels, TE/TR=77/3500 ms,11 interleaved b0s, b=1000 s/mm$^2$, 30 directions, and b=3000 s/mm$^2$, 60 directions), plus a reverse-phase b0 pair. TA =7.5 minutes. This original dataset is referred to as **dMRI scheme A** (original-2-90).

Four experimental dMRI schemes: with reduced protocols were derived from scheme A:

- **dMRI scheme B** (halved-2-45): 5 b0s, b=1000 s/mm$^2$, 15 directions, and b=3000 s/mm$^2$, 30 directions, TA = 3.75 minutes
- **dMRI scheme C** (high-b-60): 5 b0s, b=3000 s/mm$^2$, 60 directions, TA = 5 minutes.
- **dMRI scheme D** (high-b-45): 5 b0s, b=3000 s/mm$^2$, 45 directions, TA = 3.75 minutes.
- **dMRI scheme E** (low-b-30): 5 b0s, b=1000 s/mm$^2$, 30 directions, TA = 2.5 minutes.

All dMRI data were pre-processed with MRtrix3[4], FSL[5], ANTs[6] to correct for imaging noise[7], Gibbs ringing artefacts[8], motion and susceptibility distortion[9,10], and bias field inhomogeneity[11].

Tractography: Optic radiation tractography from the operating hemisphere was performed using the iFOD-2 probabilistic algorithm in MRtrix3.[4] FODs were estimated using multi-shell multi-tissue CSD[12] for schemes A and B, and single-shell 3-tissue CSD[13] for schemes C, D and E. Expert-guided ROIs defined on the original data were linearly transformed to each reduced dMRI space for consistent tracking. All other tractography parameters were held constant. For each experimental dMRI scheme, OR tractography was repeated 10 times. All tract masks were linearly transformed back to the original imaging space for comparison.

Analysis: Tracts from the original 2-shell scheme (A), used in surgical planning, served as the reference. Dice Similarity (DS) scores were calculated to assess spatial overlap between reference OR tracts and those generated from each reduced scheme (B–D). ANOVA comparisons of DS scores were conducted using a linear mixed model with subjects as a random effect, followed by pairwise t-tests. A p-value < 0.05 was considered statistically significant.

**Results**

Figure 1 summarises the mean DS scores for each dMRI scheme across all patients, and shows example OR tractography for one patient. Across all five cases, the median DS scores comparing the full protocol (Scheme A) to the reduced protocols were: A vs B = 0.81 [IQR 0.85–0.90], A vs C = 0.82 [0.86–0.90], A vs D = 0.81 [0.84–0.89], and A vs E = 0.64 [0.74–0.85]. ANOVA of linear mixed model revealed significant differences among the four reduced schemes ($p < 2.2e^{-16}$). All pairwise t-tests were statistically significant, with the overall trend: high-b-60 > high-b-45 > halved-2-45 > low-b-30, (p-values < $2.2e^{-16}$ to $1.9e^{-05}$). Among high b-value schemes, more diffusion directions yielded higher DS scores. Between schemes with the same total diffusion directions, the single shell high b-value scheme outperformed the halved 2-shell scheme.

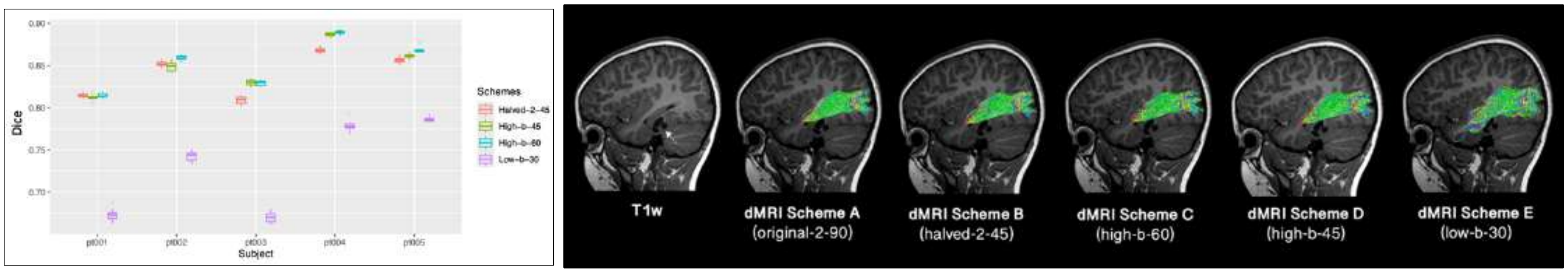

***Figure 1. Left:*** *Mean Dice Similarity scores comparing optic radiation tract masks from the full diffusion MRI scheme (A) with four reduced schemes [B (red), C (blue), D (green), E (purple)] across five patients.* ***Right:*** *Example case showing optic radiation tractography differences across schemes. Schemes B, C and D closely resemble A, while scheme E shows noisier, less reliable reconstructions. The white arrow points to the basal temporal tumour location. T1w = T1-weighted; dMRI = diffusion MRI.*

**Conclusion**

The high b-value diffusion MRI schemes yielded OR tractography most comparable to the full two-shell protocol, outperforming both the halved two-shell and low b-value schemes. Higher directional scheme (i.e. 60 > 45 directions) performs better for high b-value data. These findings offer preliminary insights for designing time-efficient dMRI protocols that maintain OR tractography quality for temporal lobe epilepsy surgery planning. Further studies with larger surgical cohorts are needed to validate these findings. Studies are also needed to assess whether observed differences in DS scores translate into differences in tractography utility for surgical decision-making, and evaluate performance across other clinically relevant tracts.

# Combining tractography and intracranial EEG to define the structural epileptic network

Arash Sarshoghi[1,2], Alexis Robin[8], Denahin Toffa[1], Guido Guberman[7], Emna Guibene[1,2], Arman Sarshoghi[4], Albert Guillemette[4], Maxime Descoteaux[3], François Rheault[3], Elie Bou Assi[2], Dang K. Nguyen[2,5], Guillaume Theaud[1], Sami Obaid[1,2,6]

[1]Neuroscience Research Axis, University of Montreal Hospital Research Center (CRCHUM), Montreal, Quebec, Canada, [2]Department of Neuroscience, University of Montreal, Montreal, Quebec, Canada, [3]Department of Informatics, Sherbrooke University, Sherbrooke, Quebec, Canada, [4]Department of Medicine, University of Montreal, Montreal, Quebec, Canada, [5]Department of Neurology, University of Montreal Hospital Center (CHUM), Montreal, Quebec, Canada [6]Division of Neurosurgery, Department of Surgery, University of Montreal Hospital Center (CHUM), Montreal, Quebec, Canada [7]Department of Neurology and Neurosurgery, Faculty of Medicine, McGill University, Montreal, Quebec, Canada, [8]Neurology Department, CHU Grenoble Alpes, INSERM, U1216, Grenoble Institut Neurosciences, Grenoble, France

## Introduction

One-third of epileptic patients are resistant to medication, with surgery as a potential solution[1,2]. However, 25–35% of operated patients continue to experience disabling seizures[3], showing the need for better surgical targeting. A network-based perspective has improved our understanding of seizure generation, yet the role of white matter (WM) connections in these networks remains underexplored. While intracranial EEG (icEEG) localizes cortical nodes, it omits WM pathways[4]. Here, we integrate diffusion MRI tractography with icEEG to visualize and quantify the WM edges of epileptic networks.

## Methods

High-resolution T1 and diffusion MRI scans were acquired from 20 patients having undergone an icEEG and from 49 healthy controls (HC). A comprehensive structural connectivity pipeline (Tractoflow[5], Surface-Enhanced Tractography[6], and Connectoflow[7]) generated tractograms while correcting for the gyral bias, parcellated the cortex into 246 Brainnetome atlas regions[8], and computed connectivity strength (CS) using streamline counts and COMMIT weights (a microstructure-informed CS metric). Expert epileptologists classified icEEG contacts of patients into seizure onset (SOZ), propagation (SPZ), irritative (IZ), or non-involved zones (NIZ). Each parcel was assigned to the same zone as the contact within it. Tracts connecting parcels were then grouped into intrazonal (SOZ↔SOZ…IZ↔IZ) and interzonal (SOZ↔SPZ…IZ↔NIZ) networks. CS of analogous networks in HCs were also computed. Across-group, within-network and within-group, across-network differences were assessed using FDR-corrected nonparametric tests.

## Results

Across-group FDR-corrected comparisons revealed that patients exhibited increased streamline counts in IZ↔NIZ and decreased COMMIT weights in NIZ↔NIZ. Across-network FDR-corrected comparisons revealed higher streamline counts in SOZ↔SOZ compared to SOZ↔SPZ, SOZ↔NIZ, SPZ↔IZ, SPZ↔NIZ, and IZ↔NIZ. Although COMMIT weights showed overall network differences, no comparisons remained significant after FDR.

## Conclusions

These findings underscore the potential of diffusion MRI tractography to complement icEEG by delineating the WM architecture of epileptic networks. Combining icEEG-derived zone definitions with tractography metrics may refine surgical targets and improve seizure outcomes.

# Harmonizing Free Water Metrics in Aging: A Comparative Study of Single-Shell and Multi-Shell Diffusion MRI

Stanislas Thoumyre[1,2], Arnaud Boré[1], Laurent Petit[2,3], and François Rheault[1,3]

[1]Sherbrooke Connectivity Imaging Lab (SCIL), Université de Sherbrooke, Canada
[2]Groupe d'Imagerie Neurofonctionnelle (GIN), Institut des Maladies Neurodégénératives-UMR 5293, CNRS, CEA Univ. Bordeaux, France
[3]IRP OpTeam, CNRS Biologie, France - Université de Sherbrooke, Canada

## Introduction

Diffusion-weighted MRI (dMRI) evaluates brain microstructure, but traditional models such as diffusion tensor (DTI) and kurtosis imaging lack specificity [1]. Biophysical models, such as the free water (FW) model, isolate extracellular water diffusion from tissue signals, providing corrected diffusion metrics and estimating the free water fraction (FWF), a marker for inflammation or tissue loss [2]. Multi-shell (MS) acquisitions enhance FW modeling but single-shell (SS) protocols remain common clinically due to shorter scan times. Although FW estimation from SS data is not optimal, it is highly correlated with MS data and can serve as an acceptable and trustworthy proxy [3]. This study compares FW metrics from SS and MS acquisitions within the same subjects globally and in white matter bundles, aiming to validate SS FW estimates, particularly for aging research using ADNI-3 data. This study also aims to assess whether harmonization of FW metrics between MS and SS acquisitions from the same subjects is feasible to improve longitudinal multi-protocol follow-up.

## Methods

We included 26 cognitively normal participants (mean age: 75.15 ± 7.2 years, 4 men, 22 women) from the ADNI-3 cohort [4], scanned approximately two years apart. dMRI used SS (b=1000 s/mm$^2$, 48 directions) initially, and MS (b=500, 1000, 2000 s/mm$^2$, 112 directions) later, both at 2 mm$^3$ isotropic resolution. Structural MRI included T1-weighted (1 mm$^3$ resolution) and FLAIR sequences. White matter lesions were segmented using SHIVA-WMH [5]; tractography and FW metrics were generated using TractoFlow [6], BundleSeg [7], and freewater_flow [8] pipelines. FW metric were extracted from "safe tissue" masks, which excluded lesions and eroded regions, ensuring analysis within intact regions, free from partial volume effects. Each subject were normalized to ensure their median were standardized. To assess agreement between SS and MS acquisitions, FWF estimates were compared using Pearson correlations across tissue types, and mixed-effects linear regression across white matter bundles. Additionally, to harmonize SS and MS data, the freewater_flow pipeline was rerun on the MS dataset using only the $b \leq 1200$ s/mm$^2$ shell (DTI shell), enabling generation of FW metrics comparable to those from SS acquisitions.

## Results

FWF values showed strong correlations between SS and MS acquisitions across brain tissue types: cerebrospinal fluid (CSF, $r = 0.815$), grey matter (GM, $r = 0.820$), and white matter (WM, $r = 0.819$), all with $p$-value $< 0.005$ (fig.1a). A mixed-effects linear regression performed across seven white matter bundles (AF, FAT, IFOF, MdLF, OR_ML, SCP, and SLF) revealed a significant fixed effect ($p$-value $< 0.001$, coefficient = 0.625), indicating a strong but sub-proportional relationship between FW values from SS and MS data (fig.1b). Random variance across bundles was minimal (Group Var = 0.001), suggesting the relationship is consistent across regions. Finally, harmonization of free-water corrected metrics (FA-t and MD-t) using only the DTI shell ($b \leq 1200$ s/mm$^2$) from MS data improved inter-protocol agreement but did not fully eliminate residual differences (fig.1c).

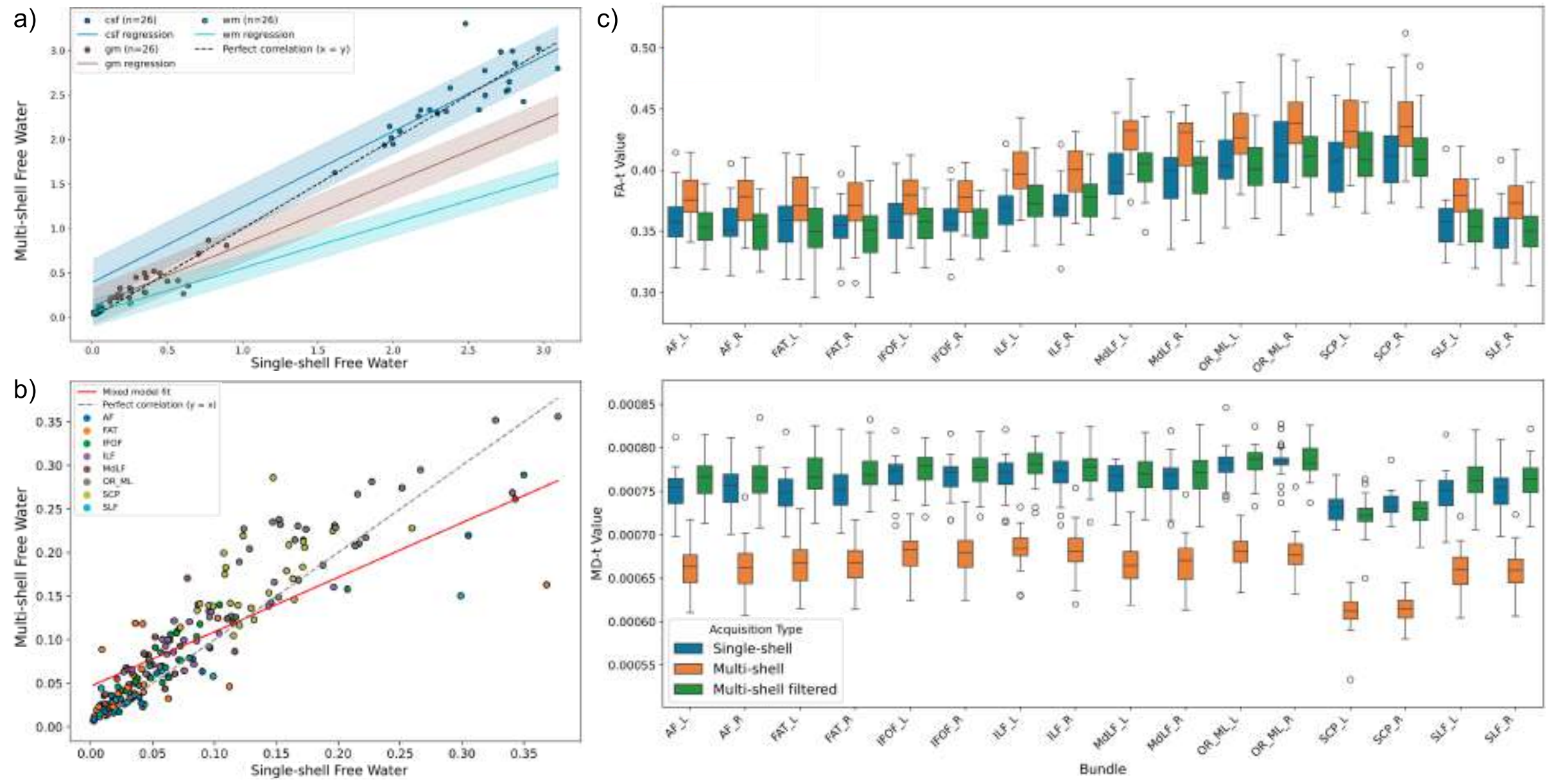

Figure 1: **Comparison of Free Water (FW) metrics between single-shell (SS) and multi-shell (MS) diffusion MRI acquisitions.** **(a)** Correlation of FWF values across cerebrospinal fluid (CSF), grey matter (GM), and white matter (WM) between SS and MS acquisitions. Each dot represents a subject (n = 26); lines show the correlation per tissue with 95% confidence intervals. **(b)** Mixed-effects linear regression of mean FWF values across seven white matter bundles: AF (Arcuate Fasciculus), FAT (Frontal Aslant Tract), IFOF (Inferior Fronto-Occipital Fasciculus), MdLF (Middle Longitudinal Fasciculus), OR_ML (Optic Radiation Meyer's Loop), SCP (Superior Cerebellar Peduncle), and SLF (Superior Longitudinal Fasciculus). The fixed effect is highly significant (p < 0.001, coefficient = 0.625). Random variance across bundles is low (Group Var = 0.001). **(c)** Harmonization of free-water corrected metrics (FA-t and MD-t) between SS and MS acquisitions, across right (R) and left (L) white matter bundles. We use DTI shells from MS (Multi-Shell filtered) to harmonize values with those from single-shell data.

## Conclusions

FWF estimates from SS and MS acquisitions show strong correlations across tissues and bundles, supporting the use of SS data as a reliable proxy in the ADNI-3 dataset. Regression analyses confirmed stable tract-wise relationships, and partial harmonization using the DTI shell improved inter-protocol agreement.Although this approach is specific to ADNI-3, applying the methodology to other datasets and acquisition protocols would be valuable for verifying its broader effectiveness. A key limitation of this study is the two-year gap between SS and MS scans, which may introduce aging-related changes; nonetheless, the results remain significant, highlighting the robustness of these findings.

**Title**: Mapping the structural connectome of temporal lobe epilepsy variants to improve surgical outcomes

**Authors**: Emna Guibene[1,2], Émile Lemoine[1,2,5], Maxime Descoteaux[3], François Rheault[4], Arash Sarshoghi[1,2], Dang K. Nguyen[1,2,5], Guillaume Theaud[1], Sami Obaid[1,2,6]

[1] University of Montreal Hospital Center Research Center (CRCHUM), Montreal, Quebec, Canada
[2] Department of Neurosciences, Faculty of Medicine, University of Montreal, Quebec, Canada
[3] Sherbrooke Connectivity Imaging Lab (SCIL), University of Sherbrooke, Sherbrooke, Quebec, Canada
[4] Medical Imaging and Neuroimaging (MINi) Lab, University of Sherbrooke, Sherbrooke, Quebec, Canada
[5] Department of Neurology, University of Montreal Hospital Center (CHUM), Montreal, Quebec, Canada
[6] Department of Surgery, Faculty of Medicine, University of Montreal, Quebec, Canada.

Temporal lobe epilepsy (TLE) is the most common type of focal epilepsy. Unfortunately, one-third of patients are drug-resistant. While temporal lobe (TL) surgery is an option for these refractory cases, ~30% of patients continue to experience seizures postoperatively[1], possibly due to epileptogenic networks extending beyond the TL, encompassing the contralateral TL (bitemporal epilepsy - BTE) or extratemporal regions (temporal plus epilepsy - TPE)[2]. Diffusion MRI tractography offers a non-invasive method to map epileptic networks, providing structural insights to differentiate TLE from BTE and TPE, thus improving surgical candidate selection[3].

We included 27 patients with TLE, 12 with BTE, 15 with TPE and 49 healthy controls (HC). All underwent high-resolution T1- and diffusion-weighted MRI sequences. Images were processed using TractoFlow[4], Surface-Enhanced-Tractography[5], and Connectoflow[6] to generate COMMIT-weighted connectivity matrices quantifying microstructural connectivity strength[7]. Matrices from patients with left-sided TLE/TPE were side-flipped. Between-group comparisons were performed using FDR-corrected two-sample t-tests. Graph-theoretical analyses (GTA) were also conducted, assessing nodal strength, betweenness centrality, clustering coefficient and local efficiency[8].

Comparisons revealed significant connectivity differences between all patient groups and HC, with BTE and TPE demonstrating more extensive alterations than TLE. BTE showed a more diffuse pattern of increased connectivity strength than TLE, including strong connections in the left TL. Compared to TLE, TPE patients revealed one increased ipsilateral connection between the medial amygdala and the rostral thalamus. GTA showed increased network metrics in BTE and TPE compared to TLE. BTE exhibited pronounced network alterations in the left TL and bilateral limbic regions, whereas TPE showed more widespread changes in bilateral subcortical-limbic regions and association cortices. These findings highlight more widespread alterations in BTE and TPE compared to TLE.

We highlight potential structural signatures of TLE, BTE, and TPE, enabling non-invasive differentiation of these variants to improve surgical candidate selection and postoperative outcomes.

# Implicit Neural Tractography: Mapping White Matter in Continuous Space

Ruben Vink[1,*], Tom Hendriks[1,*], Bram Kraaijeveld[1], Anna Vilanova[1], and Maxime Chamberland[1]

[1] *Department of Computer Science & Mathematics, Eindhoven University of Technology, Eindhoven, The Netherlands*

[*] *Authors contributed equally*

**Introduction** Tractography represents a continuous problem by nature [3], while diffusion MRI provides discrete measurements at the voxel level. Typically, linear interpolation of the FODs is used, as (probabilistic) tractography requires many FODs at arbitrary coordinates. This can often result in inaccuracies (e.g., create artificial peaks), such as in regions where the underlying fibres have high curvature [2, 4]. Implicit Neural Representations (INRs) have been shown to provide an accurate noise-robust mapping of **continuous** coordinates to FODs [2] and show great promise in downstream tasks such as tractography. This study evaluates tractography results using FODs sampled from a continuous INR.

**Methods** We first fit an INR on the DWI signal of two datasets; 1) the synthetic Tractometer dataset [1] and 2) the CDMRI 2018 challenge dataset [5]. The Tractometer dataset provides a quantitative comparison to other tractography methods, whereas the CDMRI 2018 dataset is used for in-vivo qualitative visual inspection. We perform whole-brain tracking using an in-house probabilistic method [8] that uses the fitted INRs to sample the FODs during the tracking process. We compare the result with a 'baseline' method that uses Constrained Spherical Deconvolution (CSD) [7] and samples the FODs with linear interpolation. The only difference between both methods is how the FODs are generated (i.e., INR vs. linear interpolation), with all other tractography parameters kept fixed.

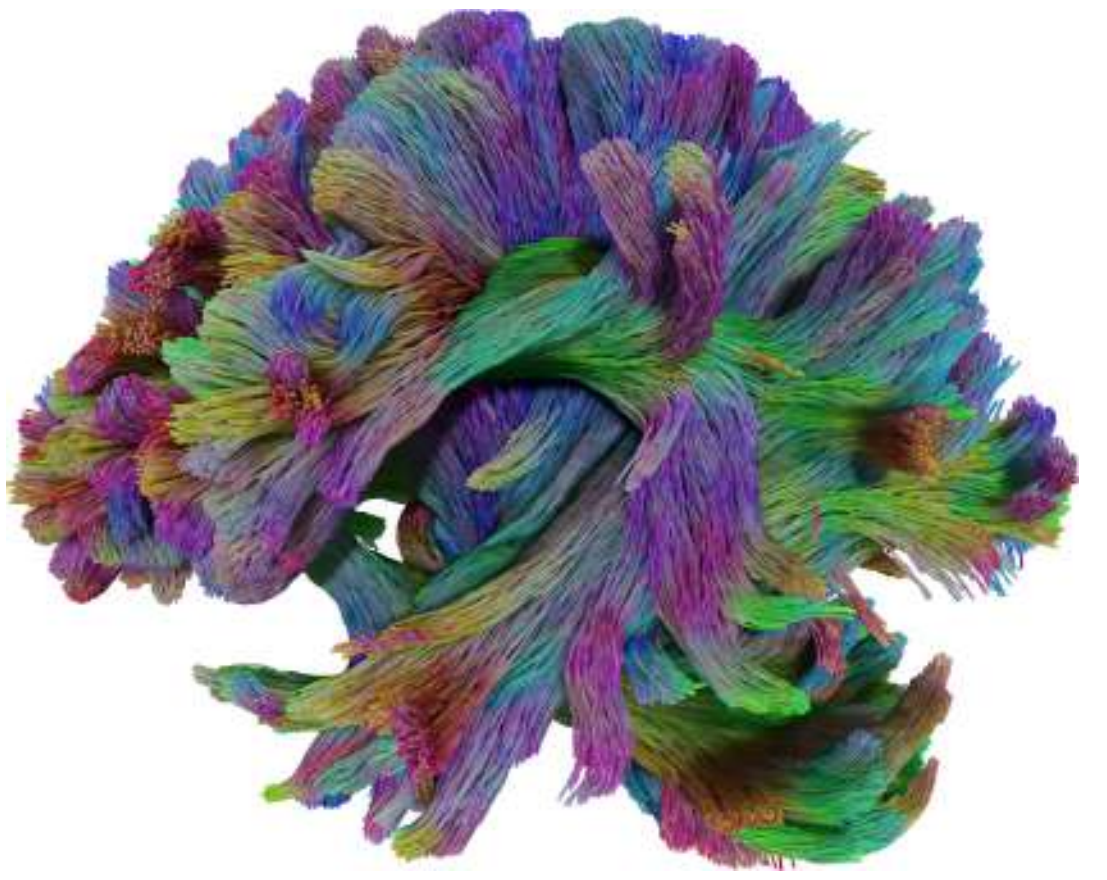

Figure 1: Continuous probabilistic tracking of the ISMRM 2015 challenge dataset.

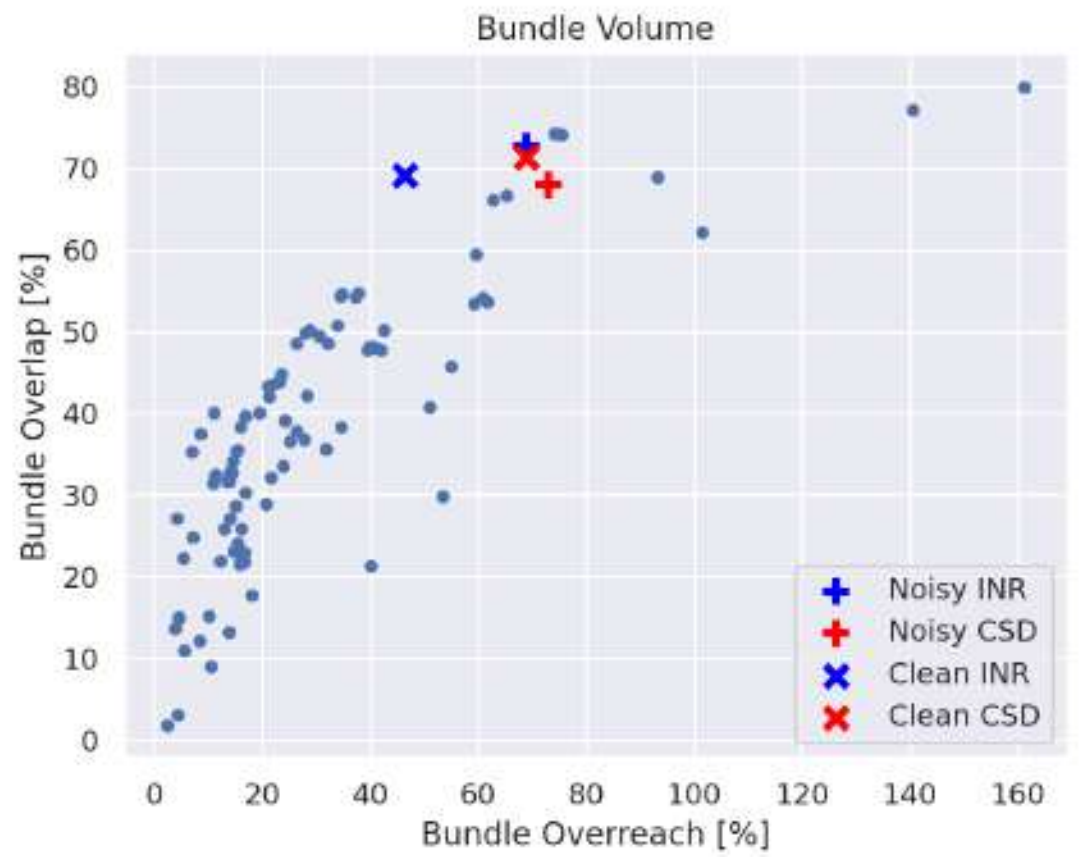

Figure 2: Tractometer scores for the INR (blue) and CSD (red) approaches with respect to noisy and ground truth datasets.

**Results** Figures 1 and 2 show results on the Tractometer dataset [6]. Notably, the INR method achieves 20 VB, 37.4% VS, 72.7% OL, 68.8% OR, and 58.5% F1 for the noisy dataset and 21 VB, 53% VS, 69.2% OL, 46.5% OR, and 62.7% F1 when fit on the ground truth data. Figure 3 shows a sagittal view of our tracking on the CDMRI 2018 challenge dataset [5]. One can observe a general smoother tracking and better coverage of the gyri when using INR- based tracking (right) compared to CSD based tracking (left). In Figure 4 we see a more clearly defined separation of the cerebellar lobes, without spurious fibers ending in the cerebellar sulcus. Finally, in the close-up of the complex optic nerve region, INR tractography provides a denser optic chiasm when compared to traditional tracking.

**Discussion & Conclusion** First, our findings on the synthetic Tractometer dataset indicate that using the INR for tracking achieves scores similar to the best performers, whereas linear interpolation of FODs with the same tractography configuration does not reach such levels. The qualitative analysis of the CMRI 2018 dataset shows smoother and more evenly distributed streamlines achieved by simply changing the FOD sampling from a discrete interpolation method to a continuous representation. All in all, this shows early merit in using continuous representations in fiber tracking, but further comparisons with ground truth datasets and clinical quality scans are necessary. This framework also opens up avenues to incorporate continuous microstructural information during tracking (i.e., microstructure-informed tractography). Finally, the INR can also provide fast tracking, since the model can be inferred directly using the GPU thus removing the load of sampling large FOD volumes from static memory.

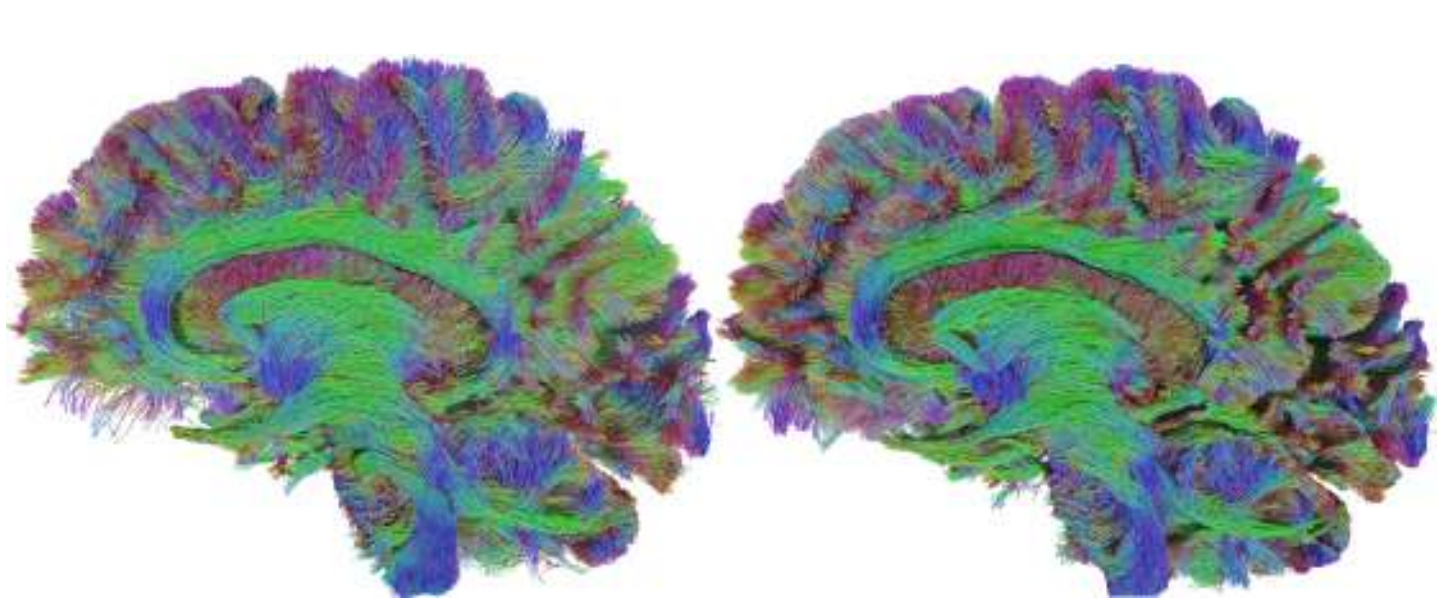

Figure 3: Saggital slice. Left: CSD based tracking. Right: INR based tracking.

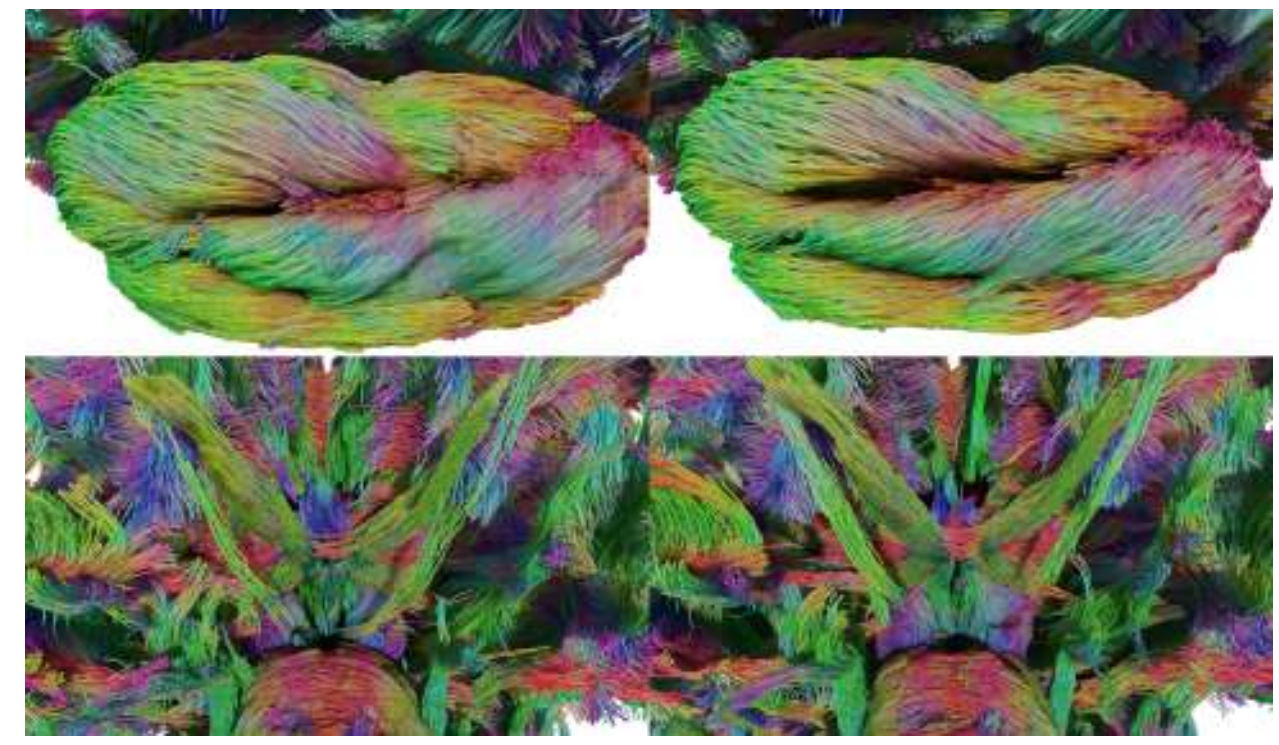

Figure 4: Left: CSD based tracking. Right: INR based tracking. Top: Lateral view of the cerbellum. Bottom: Inferior view of the optical nerve.

**Acknowlegements** The authors acknowledge funding from the Dutch Research Council (NWO) grant numbers OCENW.M.22.352 and KICH1.ST03.21.004.

## Unveiling the functional specialization of human circuits with naturalistic stimuli

Ovando-Tellez M.[1,2], Foulon C. [1,2], Nozais V.[1], Pacella V.[1,3], Thiebaut de Schotten M. [1,2]

[1] Brain Connectivity and Behaviour Laboratory, Paris, France
[2] Groupe d'imaginerie fonctionelle (GIN), Institut des maladies Neurodegeneratives (IMN) – UMR 5293, CNRS, Bordeaux, France
[3]IUSS Cognitive Neuroscience (ICON) Center, Scuola Universitaria Superiore IUSS, Pavia, Italy

**Introduction**: White matter has long been studied for its structural organization, yet its functional properties remain poorly understood. Leveraging the functionnectome framework [1], which maps functional MRI (fMRI) data onto white matter pathways using anatomical priors, we aimed to identify functional subdivisions of white matter fibers. We hypothesized that naturalistic video watching could reveal distinct patterns of involvement of the white matter[2], enabling a novel parcellation driven by function.

**Methods :** We analyzed 7T fMRI data from 110 Human Connectome Project participants during video watching. Functional signals were projected onto white matter using anatomical priors for association, commissural, and projection fibers[3]. Group-level temporal variation maps were generated via one-sample t-tests, followed by UMAP and HDBscan for unsupervised clustering. Parcels were validated using a separate 20-subject dataset, and functional homogeneity was compared to null models. We used machine learning to explore cognitive terms by aligning them with semantic video content[4] and validated the associations using two external datasets.

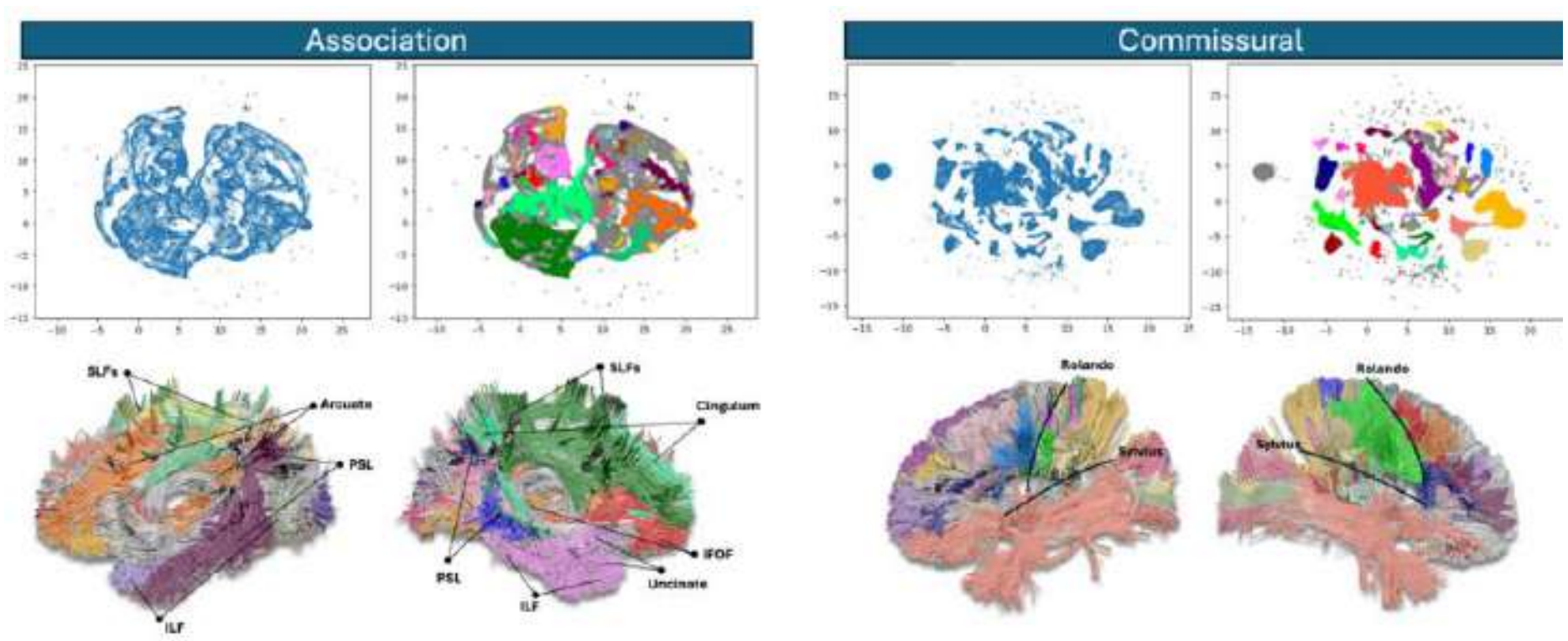

**Figure 1**. Top: UMAP plots show voxel distribution across groups; right panel colors indicate HDBscan-identified clusters. Bottom: Brain parcellation with white matter tracts showing color-coded parcels. SLFs : Superior longitudinal fasciculi, ILF : Inferior longitudinal fasciculus, PSL : Posterior segment of arcuate fasciculus, IFOF : Inferior fronto-occipital fasciculus

**Results :** We identified 36 association and 40 commissural fibre clusters with distinct functional profiles (Figure 1 - top); projection fibres lacked clear functional organization in our dataset. Results were projected onto the brain (Figure 1 - bottom), revealing parcels with high functional homogeneity ($Z > 3.5$) and reproducible activation patterns across datasets. External validation confirmed consistent associations with cognitive domains. Parcel-specific video frames are available at http://cognipact.bcblab.com

**Conclusion :** We present the first functionally driven parcellation of association and commissural white matter using naturalistic fMRI. These reproducible, semantically meaningful circuits suggest that white matter supports functionally specific roles[5], expanding our understanding of brain organization beyond the cortex and providing a novel functionally relevant parcellation of the white matter to our community.

**Local Spherical Deconvolution (LSD) for Tractography of High-Resolution Diffusion MRI of Chimpanzee Brains**

A. Anwander[1], M. Paquette[1], Y. Becker[1], F. Wermter[2], C. Bock[2], EBC Consortium, R. Wittig[3], C. Crockford[3], A. D. Friederici[1], and C. Eichner[1]
[1]Department of Neuropsychology, Max Planck Institute for Human Cognitive and Brain Sciences, Leipzig, Germany
[2]Alfred Wegener Institute, Helmholtz Centre for Polar and Marine Research, Bremerhaven, Germany
[3]Institut des Sciences Cognitives Marc Jeannerod, Lyon, France;

**Introduction:**

Human abilities such as cognitive skills and language use a network of interconnected cortical regions [1]. However, the evolutionary development of this structure remains unclear. Comparing the brains of non-human primates with their cognitive abilities provides indirect insight into possible evolutionary pathways [2].

Advances in diffusion MRI (dMRI) enable the microstructure of white matter (WM) to be imaged and WM fiber tracts to be reconstructed. Post-mortem MRI in preclinical systems enables data acquisition at a very high resolution and excellent quality [3, 4].

This study provides a high-resolution atlas of chimpanzee brain WM pathways using advanced dMRI techniques on a chimpanzee brain. The data collection was conducted in the 'Evolution of Brain Connectivity Project' studying brains and behavior of wild and captive primates [5].

**Methods:**

We acquired high-resolution dMRI data of a 47-year-old female chimpanzee using a 9.4T Bruker Biospec 94/30 MRI system with a Gmax=300 mT/m gradient system and a 154 mm coil [6]. The brain was positioned in a left-right orientation to optimize coil sensitivity. The data was acquired using a segmented 3D echo-planar imaging (EPI) spin-echo sequence with two refocusing pulses to minimize the effects of B1+ inhomogeneity. The scan had a resolution of 500 µm and used 55 diffusion directions with b = 5,000 s/mm² (32 segments). Dummy scans (13 hours) were performed to maintain a steady temperature. Three interspersed b = 0 images were used for field drift correction. Total acquisition time: ~90 h.

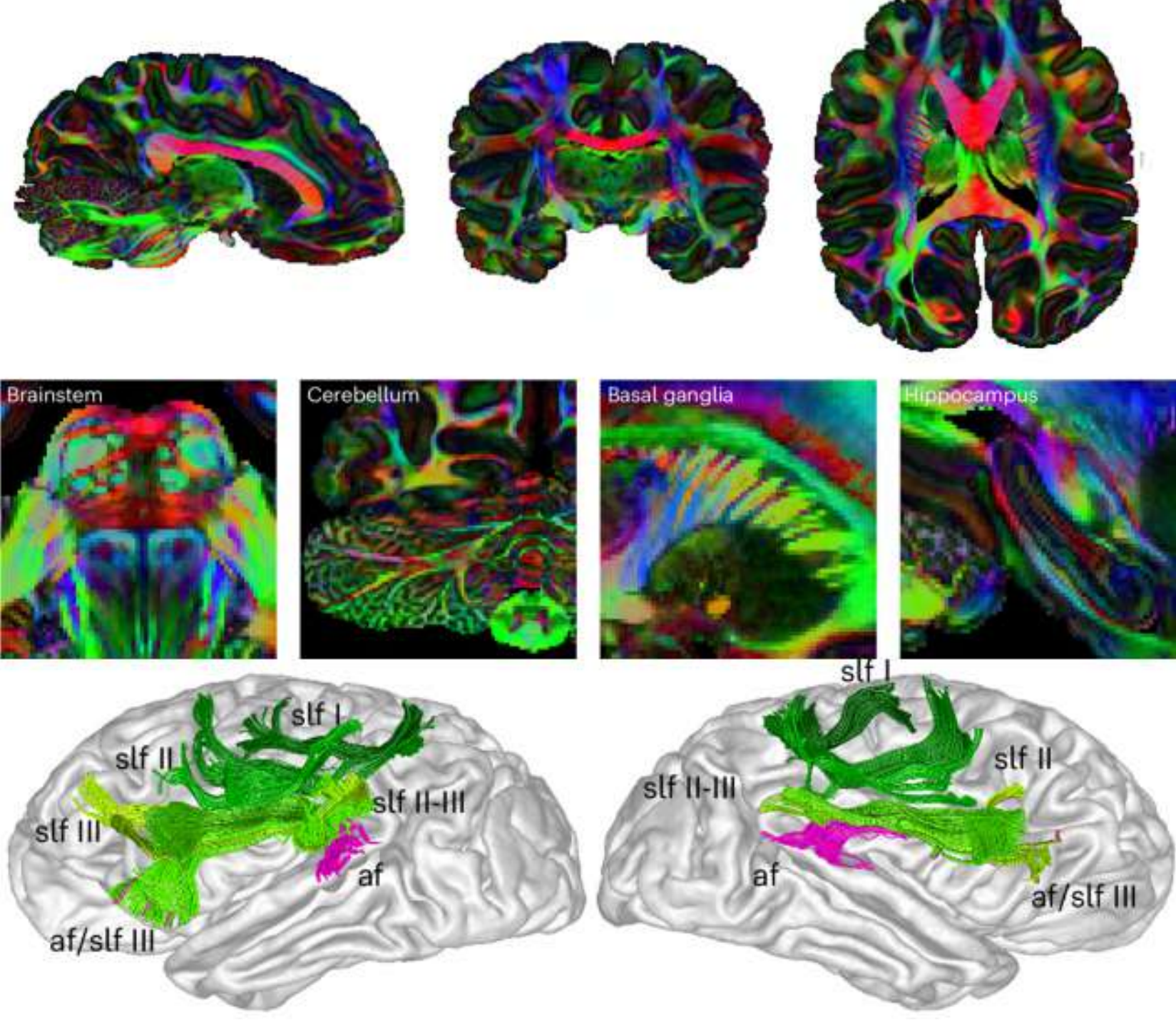

**Figure 1:** High-resolution MRI data quality. Top: Whole-brain color-coded FA reconstruction of the acquired high-resolution (500 µm isotropic) postmortem chimpanzee dMRI dataset. Middle: Zoomed regions for the brainstem, cerebellum, basal ganglia, and hippocampus. Bottom: Tractographic reconstruction of fronto-parietal association tracts. Figure adapted from [6].

The preprocessing included debiasing, denoising, Gibbs ringing correction, field drift correction, eddy current correction, and intensity normalization. We developed a new local spherical deconvolution method (LSD) to estimate the fODF [6]. Briefly, this method computes an optimized sharpening deconvolution transformation [7] by selecting the best underlying sharpening ratio from a series of predefined ratios ranging from 1.1 to 6. These ratios encode the diffusivity ratio between the main and secondary axes of the deconvolution kernel. We used the AIC to select the optimal deconvolution ratio for each voxel. This resulted in a robust fODF estimation and eliminated false positive peak directions. The fODF was used to compute whole-brain deterministic streamline tractography using MRtrix, followed by interactive fascicle segmentation (Fig. 1).

**Results and Discussion:**

The high-resolution dMRI data revealed anatomical details previously seen only in histologic data, but hidden in earlier MRI scans [6]. Key structures such as the pontine tract, corticospinal tract and lemnisci in the brainstem, as well as the finely branched cerebellar foliate, were clearly visible. In the striatum, Edinger's comb and internal capsule fibers intersecting with striatal cell bridges were well defined. The rolled dentate gyrus of the hippocampus was also clearly resolved. This level of detail surpasses current human in-vivo dMRI and previous chimpanzee MR studies [8] with higher resolution, improved microstructural contrast, and more accurate fiber tract reconstructions (Fig. 1). As a result, fine details of the fiber tracts, including the precise location of the endpoints, were revealed. Resource: https://openscience.cbs.mpdl.mpg.de/ebc

**Conclusion:**

This ultra-high-resolution dataset allows detailed studies of brain organization in great apes. In particular, the high quality data allowed revealing unknown anatomical details of the equivalent of the language network in a large group of chimpanzees [9]. This enables now cross-species comparisons of the brain connectivity which are critical for tracing brain evolution.

# Robust pipeline to bring automated tractography into neurosurgical practice

R. Bakker[1,3], M.M.G.H. van de Veerdonk[1,2], H.B. Brouwers[1], G.J.M. Rutten[1]

[1] Department of Neurosurgery, Elisabeth-Tweesteden Hospital Tilburg, The Netherlands
[2] Department of Radiology & Nuclear Medicine, Erasmus MC Cancer Institute, Erasmus University Medical Center, Rotterdam, the Netherlands
[3] Department of Mathematics and Computer Science, Technical University, Eindhoven, the Netherlands

**INTRODUCTION** In 2023 we released a pipeline for fully automated tractography in brain tumor patients [1]. It reconstructs seven white matter tracts, each of which plays an essential role in sensorimotor or cognitive functions. The pipeline is run in parallel to a clinically approved workflow (Medtronic Stealth™ Tractography) in which the same set of tracts is computed by manual selection of seed- and target regions. In comparison, we find the automated results more consistent, less prone to human error, and cost-saving. We therefore aim to bring the pipeline into clinical practice and are running a validation study with data from glioma patients in five hospitals [2]. The focus of the project is 1) to make the pipeline robust to the presence of large tumors; 2) to make it compatible with a variety of Diffusion MRI (DWI) protocols; 3) to make it modular to enable rapid prototyping of tractography algorithms; 4) to create a set of validation tools for quick assessment of the seven tracts; and 5) to integrate the tracks into the neuronavigational system used in the operating room.

**METHODS** The core of the pipeline consists of a) Constrained Spherical Deconvolution (CSD)-based preprocessing of the diffusion data, b) coregistration of the anatomical scans, c) cortical and subcortical parcellation of the brain to generate seed- and target regions for tractography, d) probabilistic tractography, where a) and d) rely mainly on MRtrix3 and FSL software [3,4]. Our main effort goes into making the co-registration and brain parcellation work robustly in the presence of large tumors. The parcellation workhorse has become the deep learning based SLANT segmentation algorithm [5], which chops the brain into a 3x3 grid of overlapping cubes and merges the 27 submodels by a voting strategy that results in a 133-area parcellation. This distributed approach performs remarkably well in the presence of large tumors, without having to first segment the tumor itself.

Two stumbling blocks hinder the use of pipeline results in clinical practice: 1) software to be installed on a hospital's intranet must meet very strict requirements that research tools such as MRtrix3 do not meet, and 2) to use the tracks during a tumor-resection procedure the neuronavigational system must load and display them. For the first issue we turn to fully web-based solutions that do not require software to be installed. For the second issue we are in touch with device vendors. Medtronic for example uses the Trackvis .trk format to store tracks generated on its Stealth™ Station, but cannot yet import these files from external sources.

**RESULTS** Validation of the pipeline with data from five hospitals revealed several issues with the initial release [1]: (a) Non-isotropic T1 data may cause issues with the resampling of seed regions in DWI space. (b) The Brain Extraction Tool of FSL does not always converge if the T1 covers large parts of the brain stem. (c) Some DWI protocols have a high level of susceptibility-induced distortion that can cause the T1-to-DWI registration to fail.

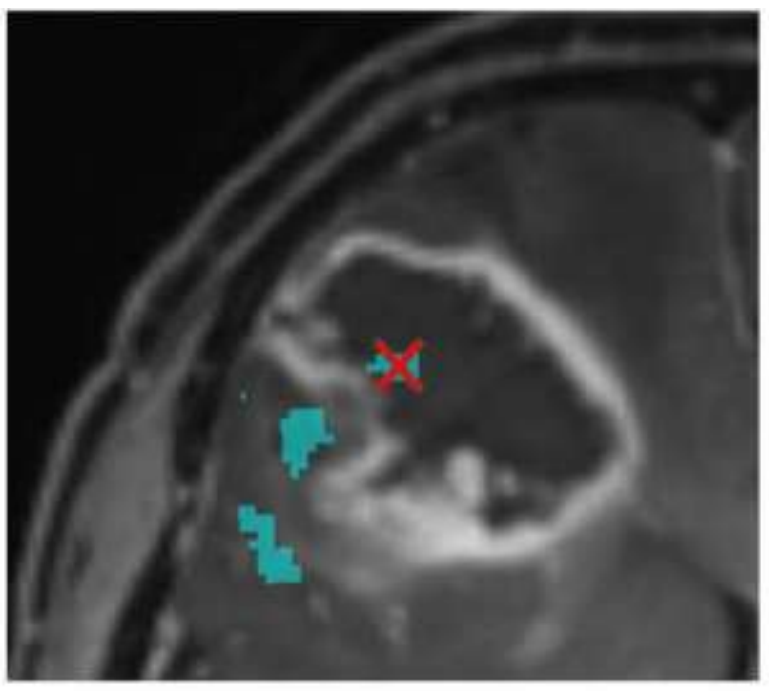

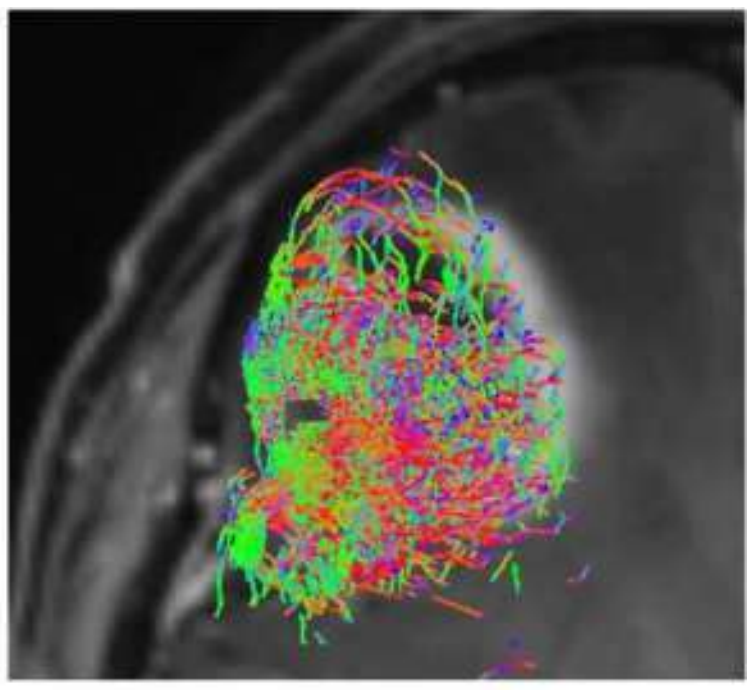

These problems can mostly be solved by using the feature-rich 'qsi-prep' [6] DWI preprocessing pipeline which we are currently testing. We also had cases where the SLANT parcellation method placed small, isolated patches of brain areas inside tumor zones. The above figure illustrates this for the AF-tract, which runs inside a tumor due to a wrongly assigned piece (marked by red cross) of 'triangular part of inferior frontal gyrus'. A cleanup routine now removes these patches if they are small and well separated from the core part of the brain area.

To aid in the clinical validation we released pilot versions of three web-based apps:

1. bvec-viewer (rbakker.github.io/bvec-viewer) shows the distribution of gradient directions on a sphere.
2. medtronic-tracts-extractor (rbakker.github.io/medtronic-tracts-extractor) reads tracks from a Medtronic export file.
3. tck-viewer (rbakker.github.io/tck-viewer) displays .tck or .trk tracks in a .obj mesh glass brain.

**CONCLUSION** We take an automated tractography pipeline further towards clinical practice by increasing its robustness against different diffusion scan protocols and presence of large tumors, implement web-based evaluation tools, and make the output ready for ingestion by neuronavigational systems.

Supported by grant "Bringing Tractography into Daily Neurosurgical Practice" no. KICH1.ST03.21.004 of the research program *Key Enabling Technologies for Minimally Invasive Interventions in Healthcare*, which is (partly) financed the Dutch Research Council (NWO).

# nf-pediatric: A robust and age-adaptable end-to-end connectomics pipeline for pediatric diffusion MRI

Anthony Gagnon[1,2], Arnaud Boré[2], Alex Valcourt Caron[2], Manon Edde[2], Stanislas Thoumyre[2,3], François Rheault[2], Marie A. Brunet[1], Larissa Takser[1], Maxime Descoteaux[2]

[1] Département de pédiatrie, Université de Sherbrooke, Qc, Canada. [2] Sherbrooke Connectivity Imaging Laboratory, Université de Sherbrooke, Qc, Canada. [3] Groupe d'imagerie Neurofonctionnelle (GIN), UMR 5293, Université de Bordeaux, France.

**Introduction.** New longitudinal pediatric initiatives, such as ABCD[1] and HBCD[2], aim to study the impact of various factors on early brain maturation/development by leveraging more than 10,000 multisite diffusion MRI acquisitions. These initiatives require scalable and reproducible pipelines, such as TractoFlow[3] or QSIPrep[4], for processing dMRI data. However, most dMRI processing pipelines are based on strong priors established in adults and, therefore, are often inadequate for pediatric/infant data. The few pipelines developed for that purpose[5] rely on outdated technologies, making their scalability to large cohorts difficult. Pediatric/infant data require specific approaches to account for the rapid neurophysiological changes in the developing brain[6]. Conventional methods for registration, tissue segmentation, and even brain extraction are unstable at these ages and require alternative approaches. To handle such large datasets, pipeline ecosystems such as Nextflow, recently deemed the optimal solution for neuroimaging pipelines[7], offer easy deployment on any computing environment. Therefore, we propose **nf-pediatric,** a robust, modular, and age-adaptable end-to-end connectomics pipeline for diffusion MRI written in Nextflow, specifically designed for individuals aged 0 to 18 years.

**Methods.** The **nf-pediatric** pipeline is built upon the nf-neuro[8], a community hosting a collection of thoroughly tested, containerized, and reproducible Nextflow modules covering a myriad of tools from state-of-the-art neuroimaging libraries. **nf-pediatric** follows a modular architecture, grouped into three profiles that enable users to select specific parts of the pipeline based on their needs (Figure 1). By leveraging the BIDS conventions, **nf-pediatric** extracts participants' age at scan and groups them in specific processing sequences (e.g., < 6 months, between 6-18 months, between 18-30 months, or > 30 months for the tracking profile). Groups below 30 months are age-matched to their closest infant template to extract tissue maps, while older participants undergo native tissue segmentation. The derivatives adhere to the BIDS-Derivatives specifications to ensure compatibility with other processing tools. Quality control (QC) reports are generated using MultiQC[9], a sophisticated QC reporting tool, which produces per-subject and across-subject QC reports.

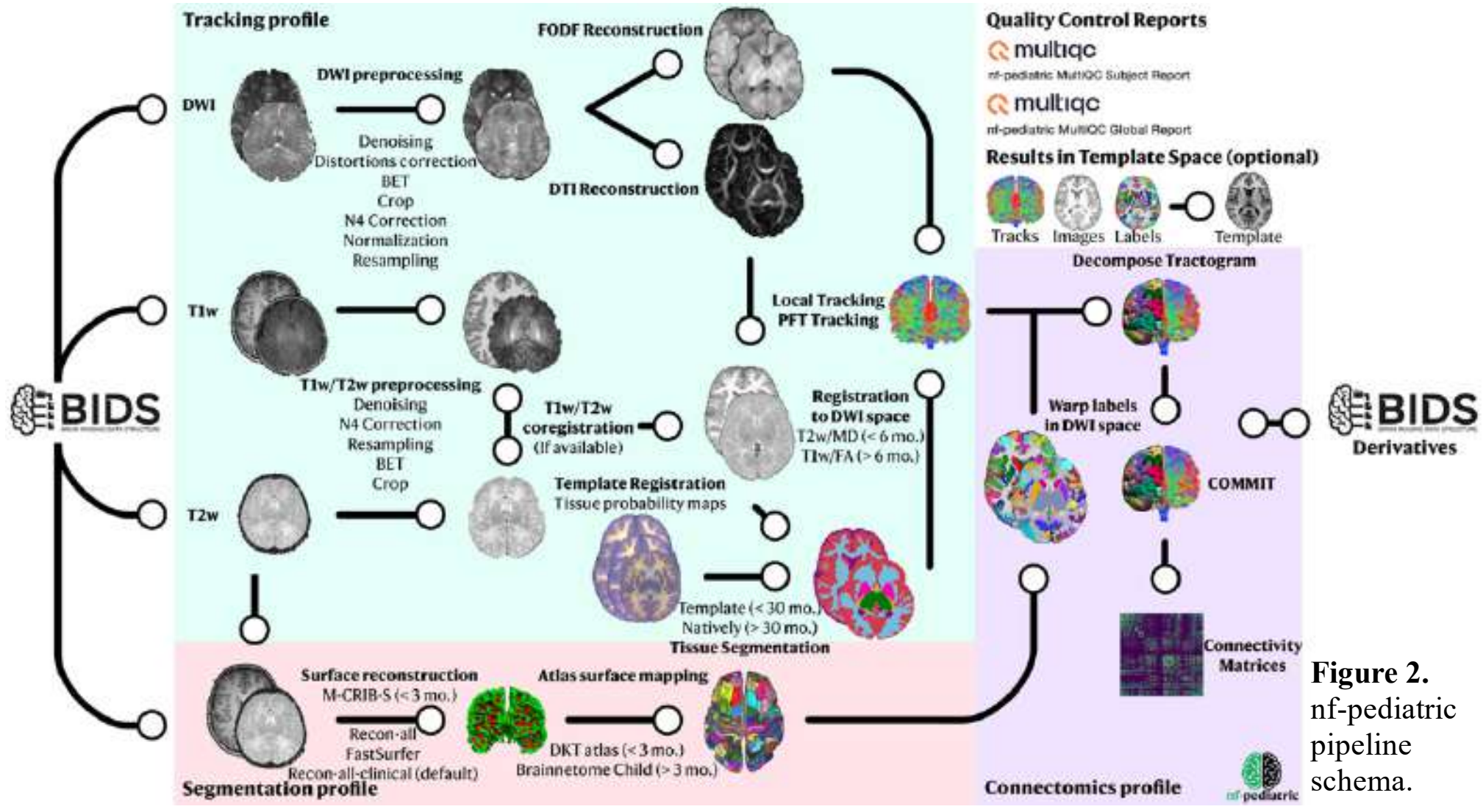

**Figure 2.** nf-pediatric pipeline schema.

**Results.** The major strength of **nf-pediatric** is its ability to automatically adapt based on the subject's age, particularly in multiple key processing steps: **(1)** age-adaptable brain extraction using tailored machine learning models, **(2)** age-adaptable moving and target images for multimodal registration maximizing similar image contrast and intensities, **(3)** age-adaptable tissue segmentation methods, leveraging TemplateFlow[10] to fetch age-matched templates followed by template registration in subject-space, **(4)** a combination of both local and particle filter tracking using either deterministic or probabilistic methods, enabling a larger range of downstream analyses, **(5)** age-adaptable cortical surface reconstruction and segmentation with pediatric atlases, and **(6)** integrated QC report, showcasing screenshots of key processing steps for visual assessment on the subject-level, but quantitative plots for assessment of outliers on the population-level. The final output contains essential files (metric maps, tractograms, processed volumes, etc.) in BIDS format, as well as statistic-ready connectivity matrices. To date, **nf-pediatric** has been tested on approximately 10,000 subjects, ranging in age from a few weeks to 17 years (Figure 2). The end-to-end design, spanning from raw data to connectivity matrices, within a single pipeline enables significantly faster processing times compared to traditional approaches, which often require manual data manipulation and the use of multiple pipelines to achieve the same results. Those key advantages make **nf-pediatric** a strong candidate for processing large pediatric datasets.

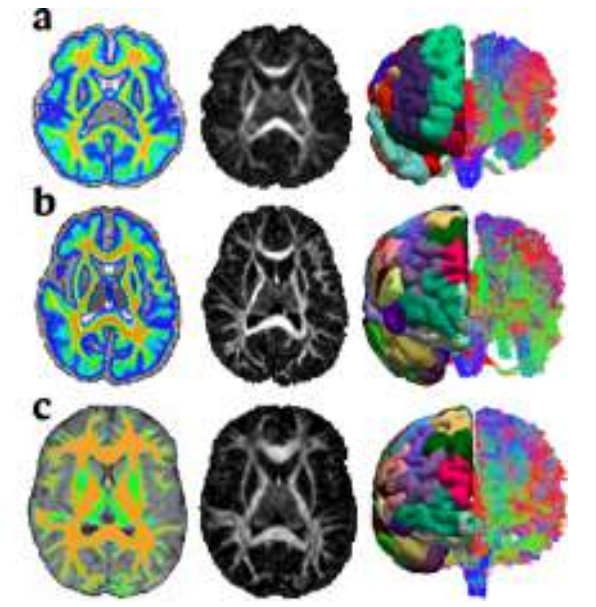

**Figure 1.** Outputs from nf-pediatric. Left to right: Tissue probability maps, FA maps, and filtered tractogram with cortical parcellation. **a.** 1 month. **b.** 16 months. **c.** 10 years.

**Conclusions. nf-pediatric** provides, for the first time, an end-to-end connectomics pipeline with age-tailored processes that adapt to each subject's age, allowing users to perform state-of-the-art diffusion MRI processing on infant/pediatric data. Due to its reliance on nf-neuro, **nf-pediatric** integrates a strong testing infrastructure and strict programming guidelines, ensuring its maintainability and improvability over time. To broaden the scope of **nf-pediatric**, new profiles will be added to support the automatic extraction of major white matter bundles using infant-tailored bundle atlases.

# Tractography in the Developing Knee Joint at Microscopic Resolution

**Nian Wang**[1,2]
[1]Advanced Imaging Research Center, UT Southwestern Medical Center, Dallas, Texas, USA
[2] Department of Biomedical Engineering, UT Southwestern Medical Center, Dallas, Texas, USA

**Introduction:** The knee joint relies on a variety of ligaments, muscles, tendons, bones, and cartilage to maintain flexibility, stability, and strength, and it is the largest and one of the most complex joints in the human body. The knee joint is susceptible to many types of injuries, including fractures, sprains, tears, and dislocations. Diffusion-based tractography has been widely used in identifying anatomical connections in human and animal brains. However, application of DTI to map the complex collagen fiber structures in the knee joint is still rare, probably because of limited spatial resolution, relatively low FA values, and relatively low signal-to-noise (SNR). Recently, we have developed high-resolution DTI method to probe the 3D collagen fiber architectures in the knee joint, its application to the developing knee joint has not been explored.

**Methods:** Animal experiments were carried out in compliance with the local Institutional Animal Care and Use Committee. The rats were sacrificed at 4, 8, 12, 16, and 25 weeks and the knee joints were harvested for MRI scans. Scans were performed using a 3D Stejskal-Tanner diffusion-weighted spin-echo pulse sequence at 9.4 T with parameters as follows: matrix size = 400x256x256, FOV = 18x11.52x11.52 $mm^3$, TE = 9.1 ms, TR = 100 ms, 31 unique diffusion directions with a b value of 1000 $s/mm^2$ and 4 non-diffusion-weighted (b0) measurements. All the diffusion-weighted images (DWIs) were registered to the baseline images (b0) to correct for eddy currents. The DTI and GQI models were used to characterize the collagen fiber directions in different connective tissues of knee joints.

**Results:** Figure 1 shows the representative b0 images of developing rat knee joints at 45 µm isotropic resolution. The growth plate and cartilage become thinner with development. It is interesting to note that laminar appearance is evident in the growth plate and articular cartilage (red arrows) at week 4 but gradually disappears at later ages (week 8 and week 16, data not shown). To explore the origin of laminar appearance, we calculated the ODFs and found the distinct collagen fiber orientations of the growth plate during knee development (Figure 2). The tractography further demonstrated that the collagen fiber architecture is rephased during the development. Compared to the basic DTI model, GQI can resolve more complicated collagen fiber architecture at the superficial zone of the growth plate (Figure 3). Besides the growth plate, distinct Collagen fiber orientations of articular cartilage during knee development were resolved by ODFs (Figure 4).

**Discussion and Conclusion:** This study demonstrates that both FA and MD values are sensitive biomarkers for knee development. High-resolution dMRI can nondestructively characterize the complex collagen fiber orientations and architectures of the knee joint. The diffusion tractography further helps to visualize the ultrastructure and quantify the integrity of the fibrillar collagen network. This capacity can provide unique insight into animal studies of developmental and degenerative joint disease.

**Acknowledgements:** This work was supported by NIH/NINDS NS125020.

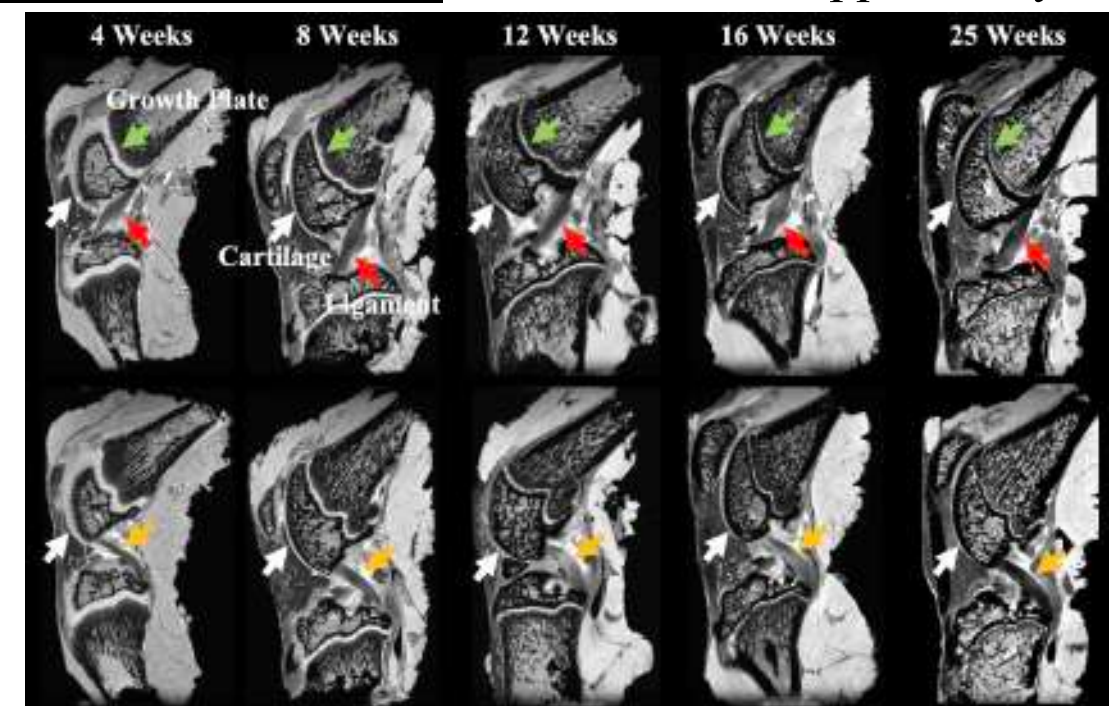

**Figure 1.** The b0 images of developing rat knee joints at 45 µm isotropic resolution. The growth plate and cartilage become thinner with development.

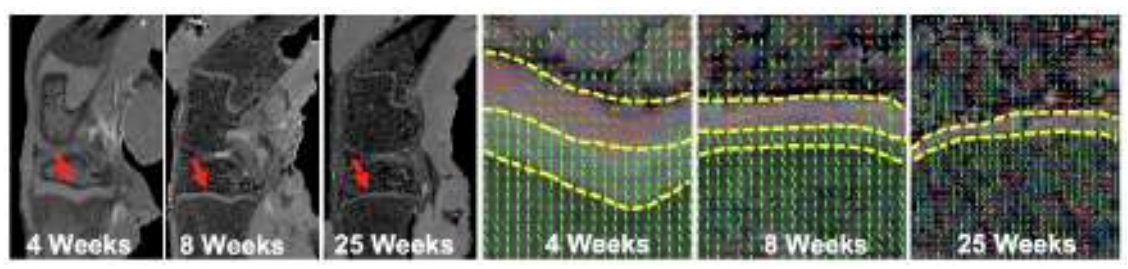

**Figure 2.** Distinct Collagen fiber orientations of the growth plate detected by ODFs.

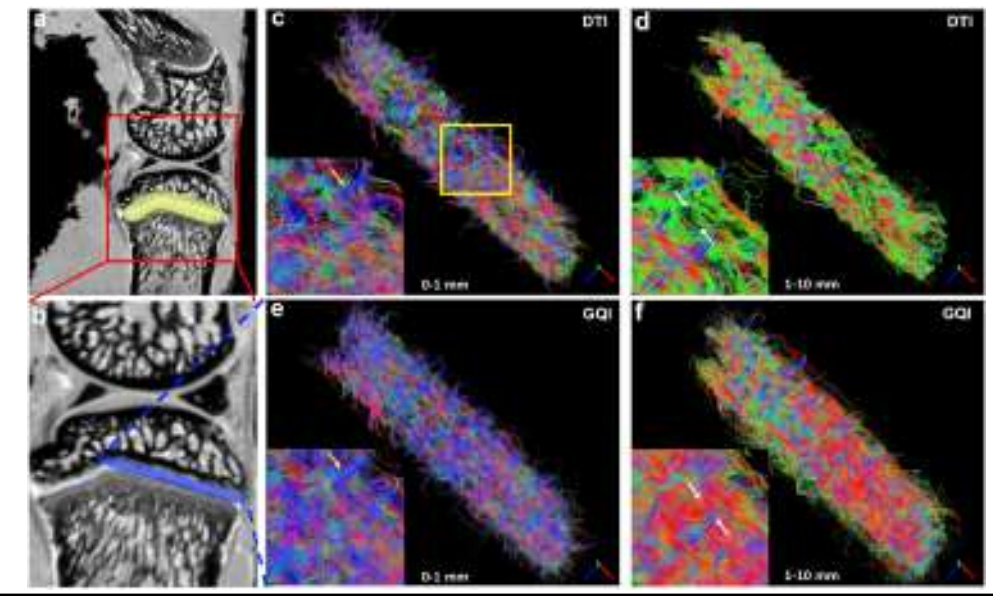

**Figure 3.** Compared to DTI, GQI resolves more complicated fiber architecture at the superficial zone of the growth plate.

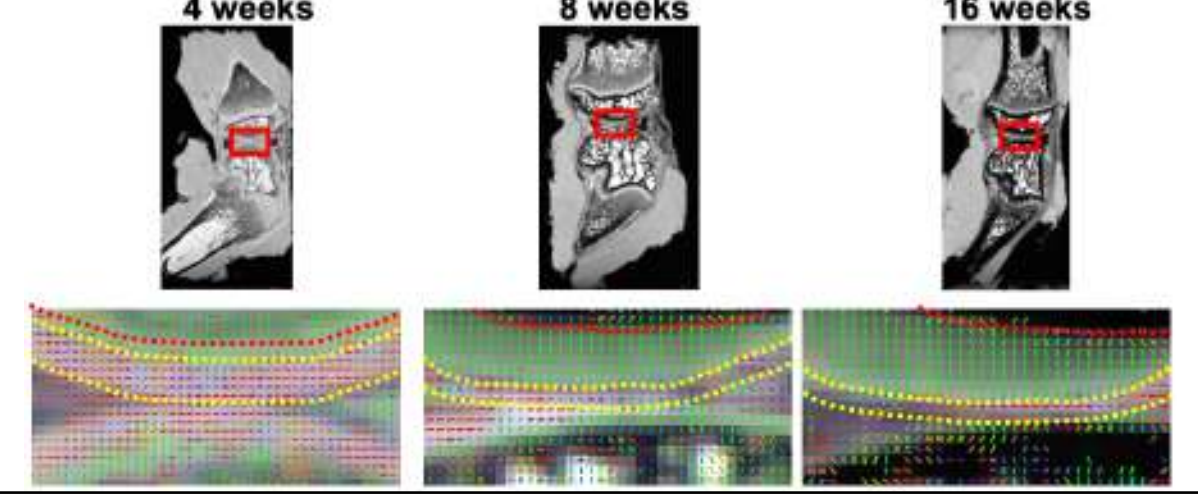

**Figure 4.** Distinct Collagen fiber orientations of articular cartilage during knee development resolved by ODFs.

## Tract morphing: A novel 3D mesh–based voxel-wise framework for multimodal white matter tract analysis

Saludar C [1], Tayebi M [1,2], Kwon E [1,2], McGeown J [2,3,4], Nepe-Apatu T [2], Condron P [2,3], Schierding W[2,5], Potter L [2], Mātai mTBI Research Group [2], Scadeng M [2,3], Wang A [1,2,3], Fernandez J [1,2], Holdsworth S [2,3], Shim V [1,2]
1.Auckland Bioengineering Institute, University of Auckland, Auckland, New Zealand
2.Mātai Medical Research Institute, Gisborne, New Zealand
3.Faculty of Medical and Health Sciences & Centre for Brain Research, University of Auckland, Auckland, New Zealand
4.TBI Network, Auckland University of Technology, Auckland, New Zealand
5. Vision Research Foundation, University of Auckland, Auckland, New Zealand

**INTRODUCTION:** Diffusion MRI (dMRI) is a widely used quantitative technique that generates a rich array of microstructural metrics (1), but this abundance of values from multiple imaging sequences creates a complex, high-dimensional analysis challenge. By coupling dMRI with tractography we can map voxel-wise tissue properties inside a single white-matter bundle, but common pipelines still compress this detail. Whole-tract averaging reduces everything to one mean number, tract profiling keeps only a few samples along the centreline, and TBSS strips each 3-D tract down to a skeleton. These shortcuts simplify statistics at the cost of valuable information. To preserve that detail, we introduce tract morphing, which projects the full voxel-wise dMRI metrics onto a subject-specific 3-D mesh that follows the tract's geometry and fibre orientations, retaining far more of the original data for downstream analysis (2). Principal component analysis (PCA) was then used to characterise patterns and variation of diffusion metrics along the WM tract, reducing the multi-dimensional data into a lower-dimensional set of principal components that capture the greatest variation in the original data. We applied this method to assess the effects of head acceleration events (HAE) exposure or a diagnosis of mild traumatic brain injury (mTBI) in collision-sport athletes relative to non-collision-sport athlete controls.

**METHOD:**
Thirty-three male high-school collision-sport athletes (HAE group) were scanned on a 3.0 T (GE Signa Premier) with T1-MPRAGE (0.5 mm isotropic) and multi-shell diffusion MRI (b=0,1000,2000,3000 s/mm²; 2 mm isotropic) at early, mid, and post-season. Twenty matched non-contact-sport controls underwent a single scan. Ten collision-sport athletes with in-season mTBI were imaged at approximately 1 week (Timepoint A), 3 weeks (B), and 8 weeks (C) post-injury. Diffusion data were preprocessed in FSL, and DTI/DKI metrics were extracted using MRtrix3 and Pydesigner. Fibre orientation distributions were estimated via multi-shell, multi-tissue constrained spherical deconvolution in MRtrix3, and bilateral corticospinal tracts (CST) were automatically segmented with TractSeg. A subject-specific CST mesh was created by morphing a tetrahedral template (generated from a single CST mask via TetGen in FEBIO) to each subject's CST segmentation using free-form deformation (3). Nearest-neighbour mapping in Python assigned voxel-wise diffusion metrics (FA, MD, AD, RD, KFA, MKT, MK, AK, RK) to mesh nodes. Permutation testing with independent-sample t-tests identified mesh elements showing significant group differences (controls vs. HAE; controls vs. mTBI). Finally, PCA reduced the significant-element metrics into principal components, and PC1–PC2 plots visualised clustering across groups and timepoints.

**RESULT:** Permutation testing revealed bilateral CST regions with significant differences between controls and the HAE group from early-season to post-season, characterized by an increasing number of significant elements concentrated along the CST spine. In the controls vs. mTBI comparison, a region-specific trend was observed in the CST: from Timepoint A to C, significant elements increased along the spine but decreased in the crown.

Principal component analysis (PC1 and PC2) revealed distinguishable clusters for both control vs. HAE and control vs. mTBI comparisons. Clustering of the PCA data enabled discrimination between groups, with the following contributions from DTI and DKI metrics: For control vs. HAE [Left CST: 46.87% (DTI), 53.13% (DKI); Right CST: 43.54% (DTI), 56.46% (DKI)] and for control vs. MTBI [Left CST: 50.79% (DTI), 49.21% (DKI); Right CST: 37.82% (DTI), 62.18% (DKI)]. Notably, while the HAE clusters overlapped across timepoints, the mTBI clusters at Timepoints A, B, and C showed partial separation, suggesting temporal divergence in the underlying diffusion profiles.

**DISCUSSION AND CONCLUSION:**
Our tract morphing framework preserves the full voxel-wise dMRI metrics within a subject-specific 3D mesh of the CST, capturing microstructural detail that conventional pipelines discard. By linking each mesh node to its nearest dMRI voxel, we maintain spatial fidelity across nine diffusion parameters and visualise localised alterations via permutation testing and PCA clustering. When applied to collision-sport athletes, this method detected both cumulative HAE-induced sub-concussive changes over a season and dynamic post-mTBI variations in distinct CST regions, highlighting its sensitivity. While we demonstrated PCA for dimensionality reduction and visualisation, our framework readily accommodates advanced machine-learning techniques - clustering algorithms, support-vector machines, or deep-learning classifiers - to refine subgroup identification and prognostic modelling. More importantly, the approach's real strength lies in its flexibility; multiple quantitative modalities - such as finite-element–derived strain, hemodynamic measures from functional MRI, axonal diameter estimates, myelin water fraction or various tissue properties - can be projected onto the same morph model. Integrating these diverse imaging and non-imaging parameters yields a comprehensive feature set that greatly enhances group differentiation, supports more robust clustering, and opens the door to predictive analytics that consider all relevant tissue and biomechanical metrics at once.

This extensible pipeline offers a powerful platform for multimodal data fusion along anatomically accurate tract geometries. By uniting microstructural, biomechanical, functional, and molecular information in a single framework, it facilitates the way for precise monitoring of brain health, individualised risk stratification, and real-time prediction of disease progression or treatment response across a wide range of neurological disorders.

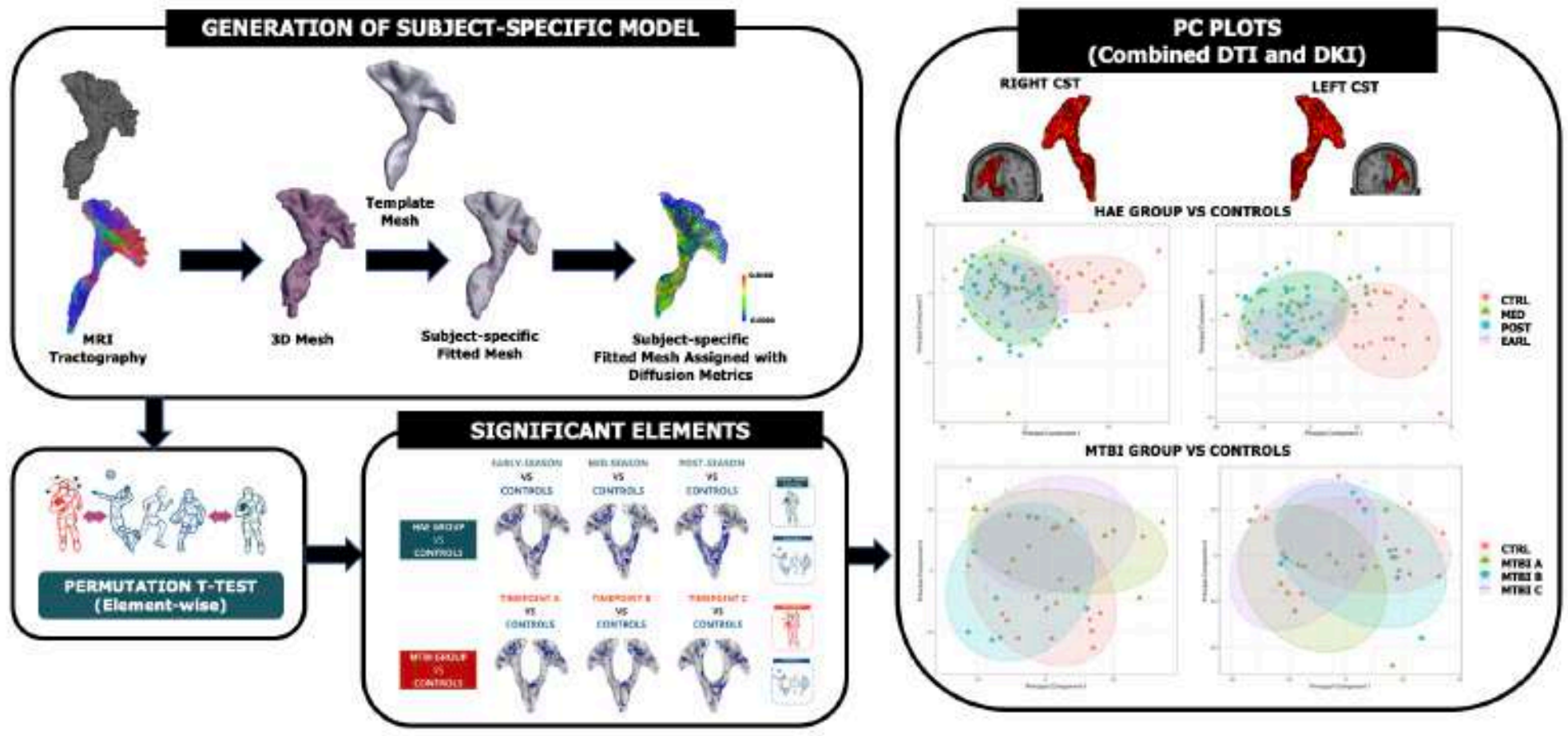

**Figure 1.** Schematic diagram of the methodology, analysis, and results from the generation of the subject-specific model from MRI tractography to tract morphing, element-wise permutation t-testing, identification of significantly different elements between groups compared, and plotting of principal components 1 and 2 from PCA to visualise clustering. Collision sport athletes (HAE group)

**Tittle**- Tract-Specific DKI Reveals Early White Matter Microstructure Alterations in Alzheimer's Disease

**Authors**- Santiago Mezzano[1,2], Reece P Roberts[1,2], Flavio Dell A'qua[3], Ian J Kirk[1,2], Tracy R Melzer[4,5,6], Campbell J Le Heron[5,6,] Kiri L Brickell[2,7], John C Dalrymple Alford[4,5,6], Tim J Anderson[6,] Nick J Cutfield[8], Lynette J Tippett[1,2] Catherine A Morgan[1,2] and the NZ-DPRC

[1]School of Psychology, University of Auckland, Auckland, New Zealand. [2]Centre for Brain Research, University of Auckland, Auckland, New Zealand. [3]Institute of Psychiatry, Psychology, and Neuroscience, Kings College London. [4]Te Kura Mahi ā-Hirikapo | School of Psychology, Speech and Hearing, University of Canterbury, New Zealand. [5]New Zealand Brain Research Institute, Christchurch, NewZealand. [6]Department of Medicine, University of Otago, Christchurch, New Zealand. [7]School of Medicine, University of Auckland, Auckland, New Zealand. [8]Department of Medicine, University of Otago, Dunedin, New Zealand.

People with Subjective Cognitive Decline (SCD) and amnestic Mild Cognitive Impairment (aMCI) are at increasing risk of developing Alzheimer's Disease (AD) dementia. Identifying neurobiological changes that underpin this progression is critical for early diagnosis and intervention. The parahippocampal cingulum within the Papez circuit connects the posterior cingulate cortex to the hippocampus, and has been shown to undergo early changes in dementia using MRI[5][4].

**Methods**

We analyzed diffusion MRI data from the NZ-Dementia Prevention Research Clinic (DPRC) cohort, which included healthy controls (n = 35), SCD (n = 57), MCI (n = 103), and AD participants (n = 26). Diffusion tensor and kurtosis parameters were estimated using DESIGNER2[6]. and FA, MD, Mean Kurtosis (MK), and Radial Kurtosis (RK) were computed. Whole-brain tractograms were then generated using a spherical deconvolution algorithm in StarTrack [1], providing additional Hindrance Modulated Orientational Anisotropy (HMOA) metrics. MegaTrack [3] was used to align all individual tractograms to a standard space, allowing a single manual dissection of the parahippocampal cingulum(see left panel in figure), then warped back to each subject's native space (see middle panel), where microstructural metrics were extracted for each individual.

**Results**

A two-way ANCOVA (Group x Hemisphere) controlling for age and sex revealed significant group differences in MD, MK, RK, and HMOA (all FDR-corrected $p < 0.001$). FA was not significant after correction ($F = 2.35$, $p = 0.071$) but showed a downward trend towards MCI and AD. No hemispheric differences were observed across metrics. Post-hoc pairwise ANCOVA showed significant increases in MD and decreases in HMOA in MCI and AD. Notably, MK and RK were significantly higher in the SCD group compared to Controls (MK: FDR $p = 0.0008$; RK: FDR $p = 0.0001$) but then reduced in the MCI and AD groups (right panel figure).

**Conclusion**

Our findings support the emerging view that DKI metrics are sensitive to early WM changes in preclinical AD[7]. Elevated MK and RK values that we find in SCD may reflect neuroinflammatory processes such as microgliosis and cytokine release, consistent with increased diffusion restriction[7]. Later-stage pathology such as myelin breakdown, axonal loss and apoptosis [9] may be detected by DTI metrics (increased MD, reduced FA, and HMOA ). In summary, our findings suggest fitting DKI in the ROI based tractography of the parahippocampal cingulum might be a sensitive biomarker for neural changes in SCD. The underlying mechanism may be neuroinflammatory responses that precede WM degeneration measured by DTI metrics.

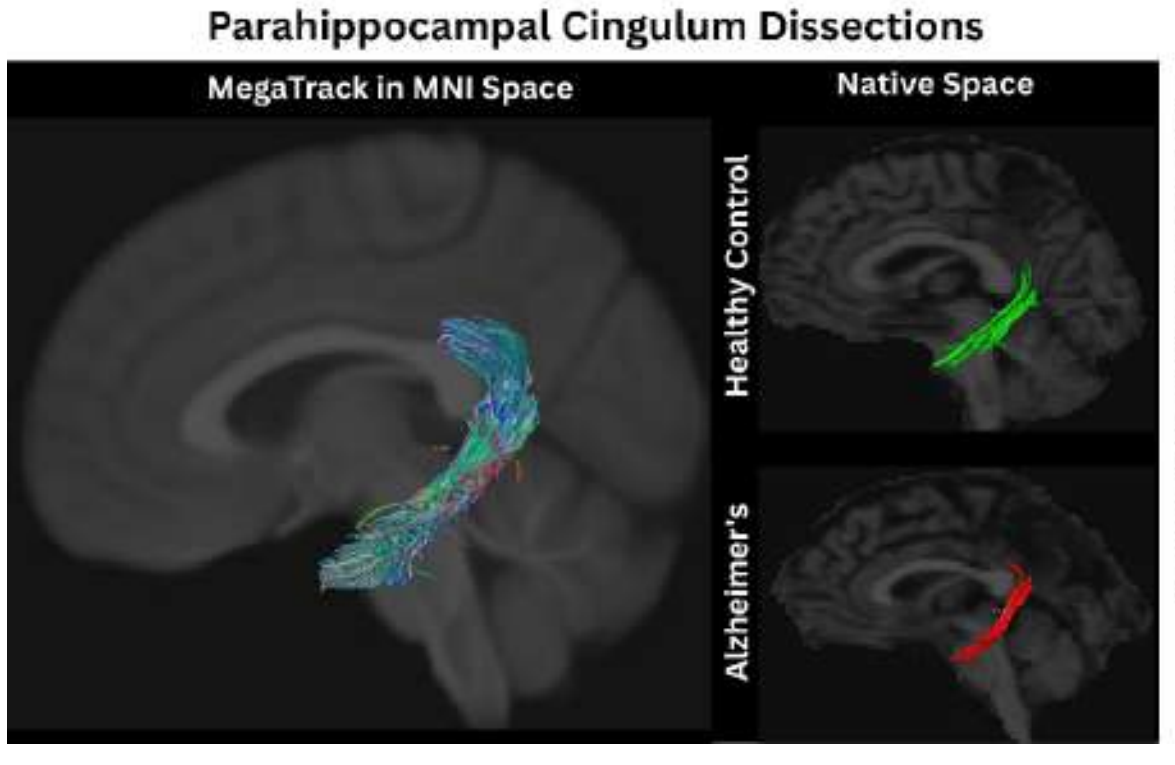

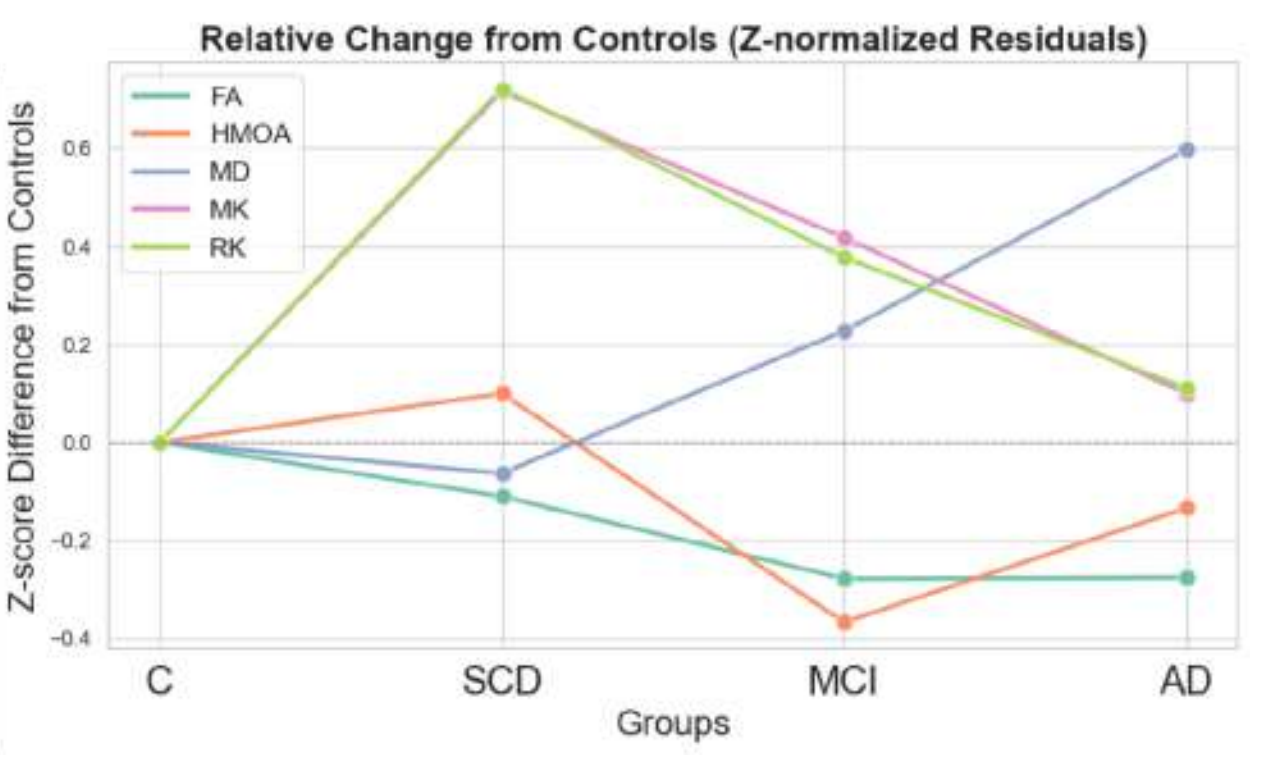

*Tittle*- Mapping Reward Processing Circuits- Cortico-Striatal White Matter Tracts and their Associations to Psychopathological Traits

*Authors*- Santiago Mezzano[1], Tobias Fernandez Borkel[1], Veronica Orbecchi[1], Manu Raghavan , Ahmad Beyh, Flavio Dell'acqua Acqua[1],

*Affiliations:* [1]Institute of Psychiatry, Psychology, and Neuroscience, King's College London

**Introduction**

Dysfunction in reward-related brain circuits, particularly involving the orbitofrontal cortex (OFC) and its projections to the nucleus accumbens (NAcc), have been implicated in various psychopathologies, including depression, anxiety, and attention-deficit/hyperactivity disorder (ADHD) ( Kujawa, 2019). The OFC is structurally, and functionally divided into the medial OFC (mOFC) and lateral OFC (lOFC), with the mOFC primarily involved in processing rewards and the lOFC associated with non-reward or punishment sensitivity (Rolls, 2019). These divisions influence goal-directed behavior and emotional regulation through their White Matter (WM) connectivity with the NAcc (Haber, 2012). Despite extensive functional imaging evidence, the structural underpinnings of these pathways remains unexplored. This study investigates the structural associations of lOFC-NAcc and mOFC-NAcc tracts to psychopathological scores, using diffusion MRI data from healthy subjects from the Human Connectome Project (HCP) (Van Essen, 2013).

**Methods**

Tractography for 166 participants form the HCP was processed with a spherical deconvolution algorithm using Startrack (Dell'Acqua, 2010). MegaTrack, a novel tractography segmentation tool (Dell'Acqua, 2025) enabled efficient and unbiased tract segmentation by aligning tractograms into a common standard space for a single manual dissection. This dissection was then mapped back to native space, allowing for precise diffusion and tractography metrics extraction for each subject. Virtual dissection identified four bilateral pathways connecting the lOFC and mOFC to the NAcc, with ROIs derived from the DKT atlas (Desikan, 2006)

Microstructural metrics extracted were fractional anisotropy (FA), hindered modulus of anisotropy (HMOA) (Dell'Acqua 2013), and tract volume. An mOFC/lOFC ratio was computed. Psychopathological traits were measured using HCP Achenbach Adult Self-Report DSM-oriented scores (Achenbach, 2003). Correlation analyses and multiple linear regression, adjusted for age and gender, evaluated associations between tract metrics and psychopathological scores.

**Results and Discussion**

The FA of lOFC-NAcc pathways positively correlated with total psychopathological traits, while mOFC-NAcc FA showed a negative correlation. A lOFC/mOFC ratio emerged as a stronger predictor of psychopathology than individual tracts. Multiple linear regression analyses and random split-half testing further supported the significant association of the OFC ratio with psychopathological traits.

To gain greater specificity into psychopathological traits, we analyzed their DSM derived subtypes and found that the FA ratio was significantly associated with depression, somatic problems, avoidant personality traits, and ADHD, with all associations surviving FDR correction (Figure 1). Additionally, other analyses revealed specific links between the lOFC/mOFC ratio and inattention traits in ADHD ($R = 0.226$, $p = 0.0036$), but not hyperactivity.

**Conclusion**

This study provides some promising evidence of structural distinctions between lOFC-NAcc and mOFC-NAcc pathways in relation to psychopathology. Particularly, findings align with the non-reward attractor theory of depression, and could suggest that the lOFC/mOFC ratio serves as a sensitive marker of reward and punishment dysfunction. The observed lateralization effects underscore the need to consider hemispheric differences in reward processing networks. Future investigations should validate these findings in clinical populations and explore their utility for psychiatric translation.

# High-resolution diffusion MRI and tractography in the in-vivo & ex-vivo non-human primate brain at 10.5T

Mohamed Kotb Selim[1], Shaun Warrington[1], Benjamin C. Tendler[2], Wenchuan Wu[2], Steen Moeller[3], Hamza Farooq[3], Pramod Pisharady[3], Edward J. Auerbach[3], Gregor Adriany[3], Franco Pestilli[4], Sarah Heilbronner[3,5], Essa Yacoub[3], Kamil Ugurbil[3], Christophe Lenglet[3], Karla Miller[2], Saad Jbabdi[2], Jan Zimmermann[3,6], Stamatios N. Sotiropoulos[1]

*[1]Sir Peter Mansfield Imaging Centre, School of Medicine, University of Nottingham, UK. [2]Oxford Centre for Integrative Neuroimaging, University of Oxford, UK. [3]Centre for Magnetic Resonance Research, University of Minnesota, MN, USA. [4]Department of Psychology & Neuroscience, University of Texas at Austin, TX, USA. [5]Baylor College of Medicine, TX, USA. [6]Department of Neuroscience, University of Minnesota, MN, USA.*

**Introduction:** A key challenge in diffusion MRI (dMRI) tractography stems from ambiguities in resolving white matter (WM) microscopic fibre patterns from macroscale data of limited spatial resolution. One approach to push resolution is to take advantage of the greater baseline signal available when scanning at ultra-high field[1]. However, ultra-high field MRI has its own set of challenges[2], for instance a shorter T2 and increased susceptibility-induced distortions. In this work, we present high resolution dMRI data for the non-human primate (macaque) brain obtained using a 10.5T whole-body human MRI scanner. We propose in-vivo and ex-vivo acquisition protocols for pushing resolution towards the mesoscale, along with a pipeline from image reconstruction to data quality assessment and biophysical modelling. This enables tractography reconstructions at ultra-high field, which we demonstrate here. The presented developments are part of the Centre for Mesoscoscale Connectomics (https://mesoscale-connectivity.org/), a large-scale programme that performs whole-brain tracing, microscopy and MRI for the same human and macaque brains, with the aim to map axons directly to dMRI signals and tractography.

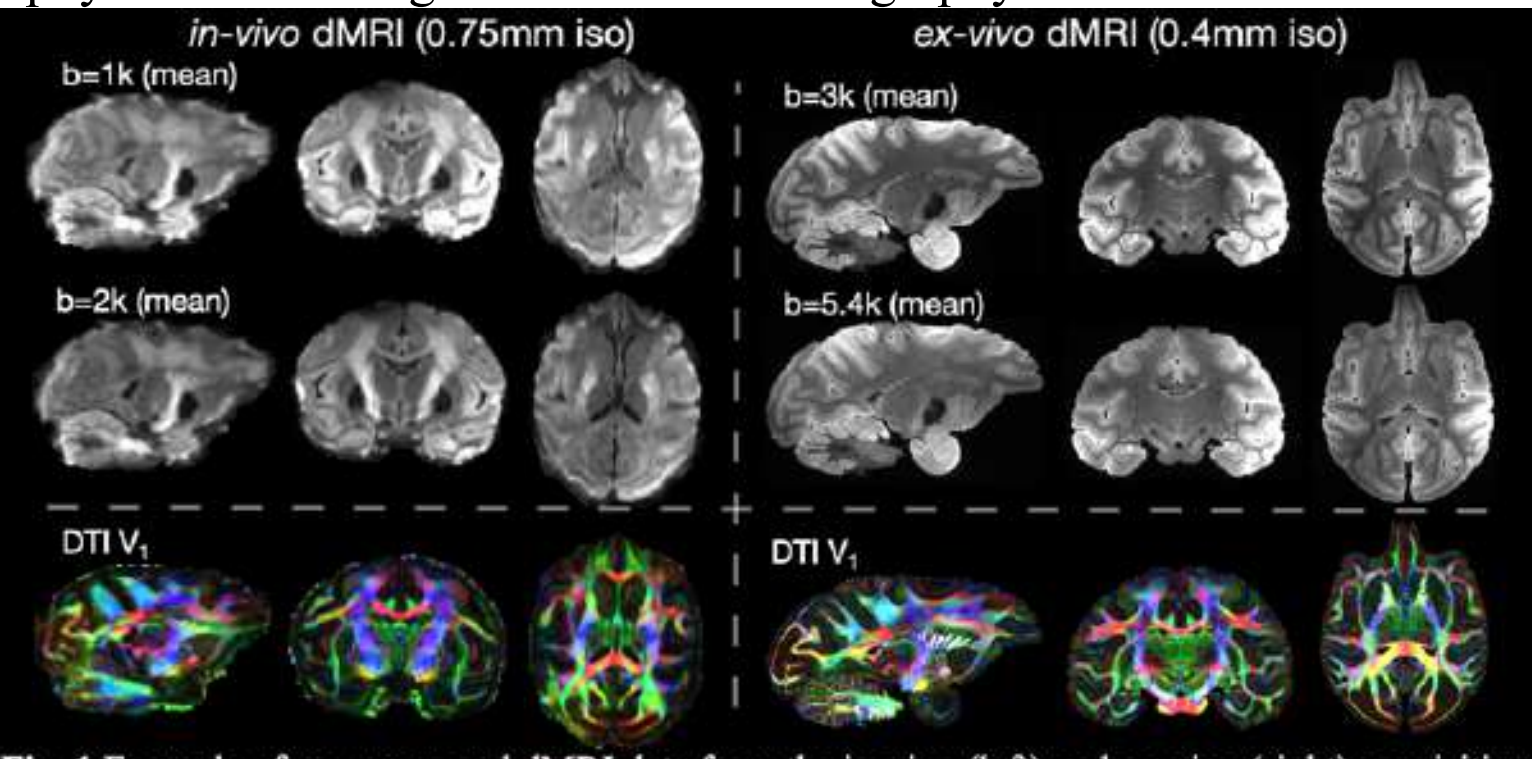

**Fig. 1** Example of pre-processed dMRI data from the in-vivo (left) and ex-vivo (right) acquisitions.

**Methods:** We developed dMRI protocols for the in-vivo and ex-vivo macaque brain using a Siemens MAGNETOM 10.5T human scanner (CMRR, University of Minnesota), fitted with SC72D gradients (70 mT/m) and a custom-built 32-channel receive coil. In-vivo acquisitions from an anesthetized macaque used a PGSE-EPI sequence at an isotropic resolution of 0.75mm (TE=66ms, TR=7.35s, GRAPPA=3, 75% partial Fourier) with b=1000, 2000 s/mm$^2$, 54 volumes per shell and 4 repeats for each of the two phase encoding directions AP/PA, scan time≅2hrs). Ex-vivo data were acquired using a diffusion-weighted steady-state free precession (DW-SSFP) protocol[3,4] (TE=16ms, TR=21ms, flip angle=14$^o$, single-line readout) with an isotropic spatial resolution of 0.4mm, 2 q-values (150 & 225 cm$^{-1}$ equivalent[5] to b=3,040 & 5,430 s/mm$^2$) and 121 volumes (scan time≅28hrs). Data were reconstructed offline using a SENSE1 coil combination[6] and complex-domain denoising (NORDIC[7]). In-vivo EPI data were motion and distortion corrected following standard practices[8]. Non-linear registration was performed to standard NMT macaque space[9]. Fiber orientations were estimated (up to 3 per voxel) using SE (for in-vivo) and SSFP (for ex-vivo) models of parametric spherical deconvolution, followed by landmark-based tractography (XTRACT)[10].

**Fig. 2** Fibre orientations within the centrum semiovale (up to 3 fibres per voxel).

**Results:** Fig1 displays preprocessed data (shell-wise averages and DTI-V$_1$ RGB-colour coded maps), with uniform signal and high diffusion contrast. The achieved spatial resolutions (0.75 mm in-vivo, 0.4 mm ex-vivo) are amongst the highest reported to date for the macaque brain from a human whole-body scanner with a conventional gradient insert. Fig2 displays crossing-fibre estimates in the centrum semiovale, revealing sensitivity to detect 2-way and 3-way crossings both in-vivo and ex-vivo. Tractography performance for a number of major WM bundles is demonstrated in Fig3, revealing good correspondence between in-vivo and ex-vivo data, but also thinner projections arising from the higher-resolution ex-vivo data. Finally, we explored whole-brain probabilistic tractography using the ex-vivo data, reconstructing a tract density image (TDI). Fig4 shows the spatial distribution of streamlines, RGB colour-coded by principal fibre orientation and binned into 0.2mm voxels. Details such as crossings in the centrum semiovale and fanning towards the cortical gyri can be observed, showcasing data quality.

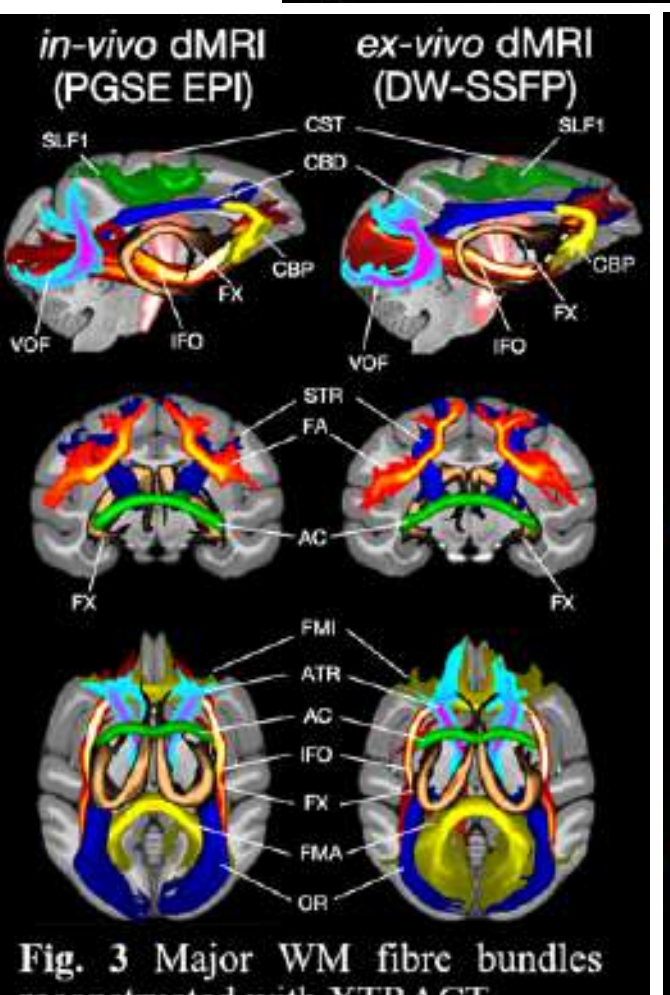

**Fig. 3** Major WM fibre bundles reconstructed with XTRACT.

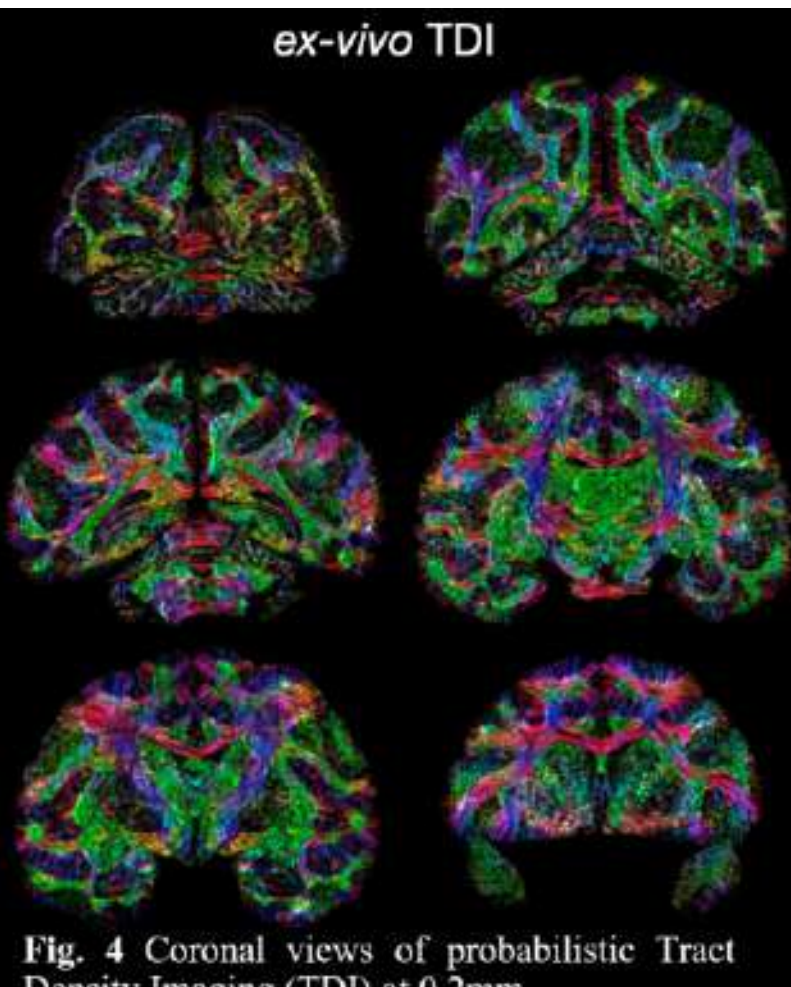

**Fig. 4** Coronal views of probabilistic Tract Density Imaging (TDI) at 0.2mm.

**Conclusion:** We have presented approaches for pushing the resolution of dMRI at 10.5T for the in-vivo and ex-vivo macaque brain. We anticipate that these developments will pave the way for human in-vivo and ex-vivo dMRI scanning at ultra-high field. The data will be publicly released and accompanied by PS-OCT microscopy and axonal tracing of the same brains.

**SBI Meets Tractography: A new approach for Bayesian inference in diffusion MRI models**

J.P. Manzano-Patrón[1*], Manuel Gloecker[2*], Cornelius Schröder[2], Jakob H. Macke[2,3*], Stamatios N. Sotiropoulos[1*]

*[1]Sir Peter Mansfield Imaging Centre, School of Medicine, University of Nottingham, UK. [2]MPI for Intelligent Systems, University of Tübingen, Germany. [3]Hertie Institute for AI in Brain Health, Tübingen, Germany. *Equal Contribution*

**INTRODUCTION:** Simulation-Based Inference (SBI) has emerged as a powerful framework for Bayesian inference. Neural networks are trained on in-silico simulations from a forward model, and subsequently applied to unseen data for rapidly estimating posterior distributions of the model parameters given the data (amortised inference, see Fig.1) [1]. We recently presented how SBI can be used for parametric spherical deconvolution from diffusion MRI (dMRI), for mapping uncertainty of fibre orientation estimates and performing probabilistic tractography [2]. However, inherent challenges remain: i) trained networks can be tied to the acquisition scheme and/or noise level/features used for training, ii) choosing the right model (including e.g. model complexity, number of modes, noise model, priors) can typically rely on heuristics/deterministic model selection [2,3]. Here, we propose a new transformer-based simulation-based model inference (SBMI) architecture [4,5] that addresses these two challenges, by allowing a) concurrent model selection and parameter inference over multiple model classes in a single framework, and b) flexible adaption to different acquisition schemes, priors, noise level and model configurations at inference time (i.e. post-training). We show results in the context of Bayesian inference for a multi-compartment model and probabilistic tractography, but the presented principles apply to any dMRI microstructure model.

**Fig.1 – Overview of inference paradigms in MCMC, SBI, and SBMI_transformer**

**METHODS:** We trained SBI networks on synthetic data using the multi-shell Ball&Sticks [6] as a forward model and acquisition schemes matched in length to the UK-Biobank dMRI protocol (105 volumes), as in [2], but with varying b-values (up to b=4000 s/mm$^2$) and gradient directions. We subsequently compared three approaches (Fig. 1) in their ability to resolve fibre crossings and the respective orientation uncertainty: random-walk MCMC, our previous SBI architecture [2] trained jointly on models with N=1,2,3 stick compartments, and the proposed SBMI_transformer, which was based on the simformer architecture [5]. Evaluations were first performed using UK Biobank-like data from [7]. For SBMI_transformer, model posterior probabilities were inferred jointly with parameter posteriors, preserving uncertainty across all plausible models. We assessed mean parameter estimates, uncertainty, and tractography by propagating posterior orientation samples to build spatial distributions of white matter (WM) pathways [8]. We finally tested the ability of SBMI_transformer in adapting to different acquisition schemes, by applying the network trained on UKB-like data to HCP-like data (i.e. considerably different b-vals, b-vecs and length of data compared to the training set).

**Fig.2 – Comparison between MCMC, previous SBI and new SBMI_transformer**

**Fig.3: Generalising to different acq. schemes – Non-HCP trained network applied to HCP-like data**

**RESULTS:** Compared to MCMC, both SBI approaches provide high agreement on mean estimates with MCMC (Fig.2A), demonstrating the feasibility of these frameworks. However, SBMI_transformer improves over previous SBI by achieving better estimation for the third orientations and their uncertainty, sharper contrast between WM and non-WM (Fig.2A), and greater coherence in the crossing fibres areas (Fig. 2B). These improvements are directly translated into a greater correlation with MCMC in the reconstructed tractography (Fig.2C) compared to SBI (0.92 vs 0.85 respectively). Importantly, we can use the same trained network for HCP-like data (Fig.3) showcasing the capability of SBMI_transformer to amortise now among any acquisition scheme, even when the number of volumes and q-space sampling differ substantially from those at training.

**CONCLUSION:** We introduced a transformer-based SBI architecture for Bayesian inference in diffusion MRI, that can inherently handle model selection, as well as parameter estimation and can generalise to unseen data from different-to-training acquisition schemes on inference time. Demonstrated on the Ball&Sticks model and probabilistic tractography, the principles are applicable for estimation and uncertainty mapping to any dMRI microstructure or orientation model.

# A Deep Diffusion Model Approach for Diffusion MRI White Matter Fiber Tractography

Yijie Li1, Wei Zhang1, Xi Zhu1, Ye Wu2, Yogesh Rathi3, Lauren J O'Donnell3, Fan Zhang1

1 University of Electronic Science and Technology of China, Chengdu, China
2 Nanjing University of Science and Technology, Nanjing, China
3 Harvard Medical School, Boston, USA

## *Introduction*

Complex fiber geometries like crossings, bottlenecks, and partial volume effects challenge white matter tractography. Traditional model-based approaches are computationally demanding and sensitive to noise. At the same time, most deep learning methods rely on fiber orientation distributions (FODs) approximations or discrete direction sampling, limiting their accuracy and continuity in streamline prediction[1, 2].

## *Method*

*Dataset*: We use dMRI data from the HCP-YA[3], with 8 subjects for training and 2 for validation. TractSeg provides reference tractography[4], yielding ~100k streamlines per subject. Evaluation is conducted on the ISMRM 2015 challenge dataset using Tractometer connectivity metrics[5, 6], with qualitative results shown for one randomly selected HCP subject.

*Method*: The pipeline normalizes diffusion MRI volumes using the b0 image and projects them into spherical harmonics (SH) (up to $l_{max} = 6$). From each voxel, a $3 \times 3 \times 3$ SH patch is extracted and processed by dual 3D CNNs to produce two embeddings: a temporal encoding $z_j$ and a spatial context vector $v_j$. The temporal embeddings are input to stacked RNN layers to model long-range dependencies and generate context vectors $c_j$. A 1D CNN-based U-Net diffusion model, tailored for sequential streamline data, then predicts the next direction via the conditional distribution $p(y_{t+1} \mid v_j, c_j)$.

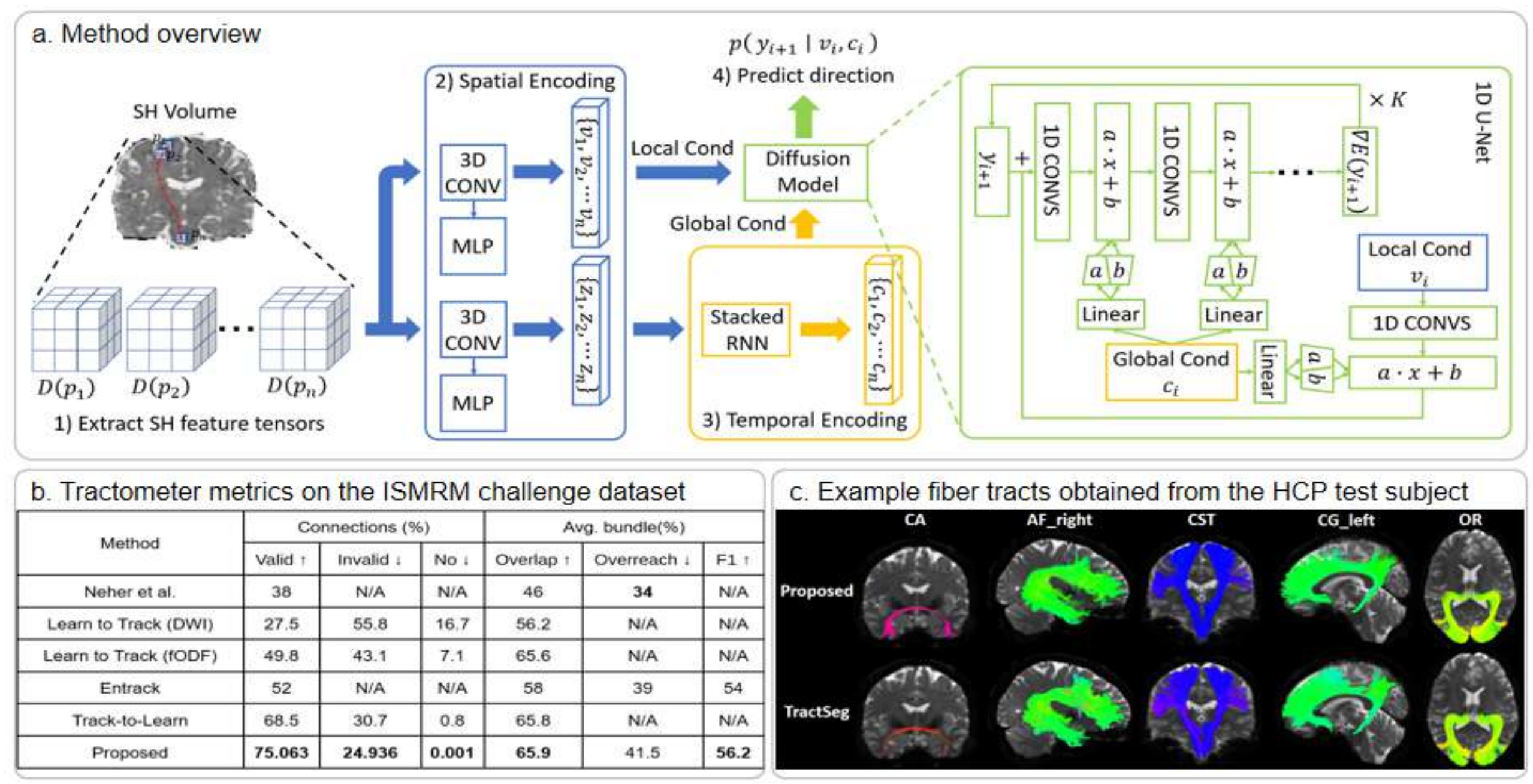

b. Tractometer metrics on the ISMRM challenge dataset

| Method | Connections (%) | | | Avg. bundle(%) | | |
|---|---|---|---|---|---|---|
| | Valid ↑ | Invalid ↓ | No ↓ | Overlap ↑ | Overreach ↓ | F1 ↑ |
| Neher et al. | 38 | N/A | N/A | 46 | **34** | N/A |
| Learn to Track (DWI) | 27.5 | 55.8 | 16.7 | 56.2 | N/A | N/A |
| Learn to Track (fODF) | 49.8 | 43.1 | 7.1 | 65.6 | N/A | N/A |
| Entrack | 52 | N/A | N/A | 58 | 39 | 54 |
| Track-to-Learn | 68.5 | 30.7 | 0.8 | 65.8 | N/A | N/A |
| Proposed | **75.063** | **24.936** | **0.001** | **65.9** | 41.5 | **56.2** |

Fig.1. The overview of our method and results.

## *Results*

Fig. 1b shows that our method achieves the highest valid connection rate and lowest invalid and missing connections, demonstrating strong accuracy and robustness. With over 60% overlap with ground truth bundles, it ensures broad anatomical coverage. Fig. 1c illustrates accurately reconstructed tracts from the HCP test set.

## *Conclusion*

We propose a diffusion model-based tractography framework that enhances the estimation of fiber orientations. The method exhibits robust and consistent performance across both phantom and in vivo datasets, demonstrating its generalization capability.

# Deconstructing DTI-ALPS:
## Clarifying the biological interpretation in aging and cerebral small vessel diseases

**Ami Tsuchida[1,2], Stanlislas Thoumyre[1,3], Quentin D'Acremont[1,2], Laurent Petit[1,4], Marc Joliot[1,4], Stephanie Debette[2,5]**

*1.Groupe d'Imagerie Neurofonctionnelle (GIN), IMN, UMR5293, U Bordeaux, France ; 2. Bordeaux Population Health (BPH), U1219, U Bordeaux, France, 3. Sherbrooke Connectivity Imaging Lab (SCIL), U Sherbrooke, Canada ; 4. IRP OpTeam, CNRS Biologie, France – U Sherbrooke, Canada ; 5. Institut du Cerveau (ICM) Paris, France*

**Introduction:** The glymphatic pathway helps maintain brain health by clearing metabolic waste, and its failure has been linked to age-related conditions, including cerebral small vessel disease (cSVD)[1]. Recently, a simple metric termed diffusion tensor imaging (DTI) analysis along the perivascular spaces (ALPS) has emerged as a promising non-invasive marker of glymphatic function, with reduction observed in cSVD and other age-related neurological conditions[2]. Although it has been used extensively as a measure of glymphatic clearance capacity, its biological specificity remains debated[3]. In particular, variability in crossing fibers across subjects and cSVD-related microstructural changes may confound ALPS measurements. For instance, cSVD-related fluid accumulation can reduce the index without altering the perivascular diffusivity the index is designed to measure. Here we propose a refined ALPS index (r-ALPS) that (1) minimizes the impact of crossing fibers and (2) incorporates free water correction (fwc) to better isolate perivascular diffusivity from extracellular fluid accumulation.

**Methods:** We computed both standard (std) and r-ALPS in a cohort of elderly volunteers aged >65 years (N=73, mean age 73.8 yrs), who either had minimal (CTL, n =29, Fazekas 0/1) or extensive cSVD (cSVD, n=44, Fazeka's 2/3). All subjects underwent an MRI session that included T1w, FLAIR, and multi-shell DWI (b-values = 300, 1000, and 2000 s/mm$^2$, 100 directions, 1.75 mm isotropic resolution) acquisitions. DWI data were preprocessed with TractFlow[4], followed by a standard DTI fitting using shells <1500 sec/mm$^2$ and fwc-DTI fitting using all shells. Figure 1 summarizes the std- (top) and r- (bottom) ALPS extraction. White matter hyperintensities (WMH) were segmented from T1w and FLAIR using SHIVA-WMH[5].

**Results:** The group comparison of std- and r-ALPS indices as well as free water fraction (FWF) in r-ALPS ROIs were analyzed with linear models, adjusting for age and sex. As shown in Figure 2, the std-ALPS was nominally lower in cSVD compared to CTL, but only significant on the right side ($p = 0.02$), which also showed significant effects of age ($p = 0.006$). The r-ALPS showed a similar pattern, with a significant group effect only on the right side ($p = 0.02$), but no longer with any age effects ($p > 0.05$). FWF was consistently elevated in cSVD relative to CTL (all $p < 0.01$), and with robust age effects ($p=0.005$ for the right side). In a separate analysis, FWF, but not ALPS indices, showed a highly significant association with WMH volume ($p<0.0001$, not shown).

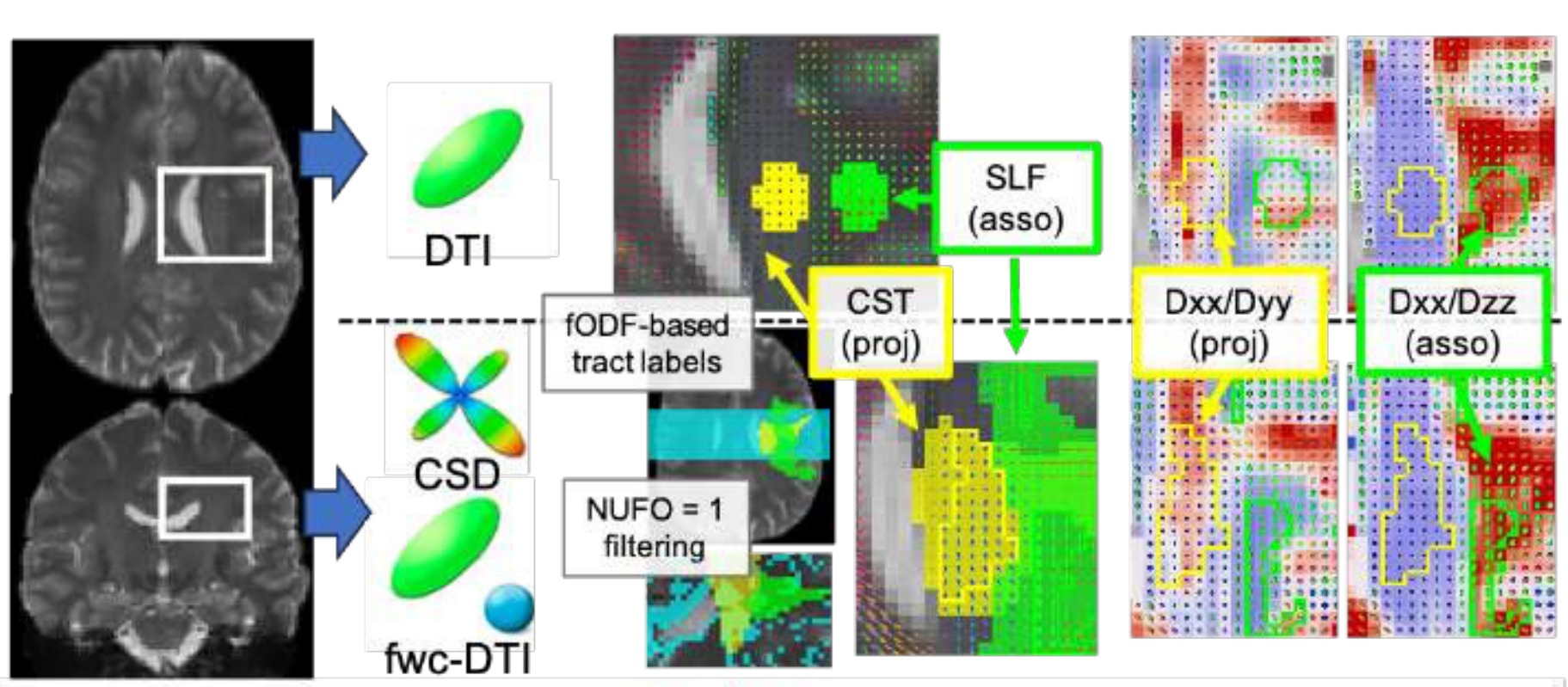

**Figure 1. Overview of standard (std-, *top*) and refined (r-, *bottom*) ALPS pipeline.** In std-ALPS pipeline, DWI (1) is fit with a DTI model (2), both to define ROIs in corticospinal tract (CST) and superior longitudinal fasciculus (SLF) (3) and to compute the ALPS index from radial diffusivity asymmetries (4). In r-ALPS, ROIs are refined using subject-specific tract labeling from fibre Orientation Distribution Function (fODF) and number of fibre orientation (NUFO) filtering (via constrained spherical deconvolution, CSD) to minimize crossing fiber effects. The ALPS index is then computed from fwc-DTI to isolate perivascular diffusivity.

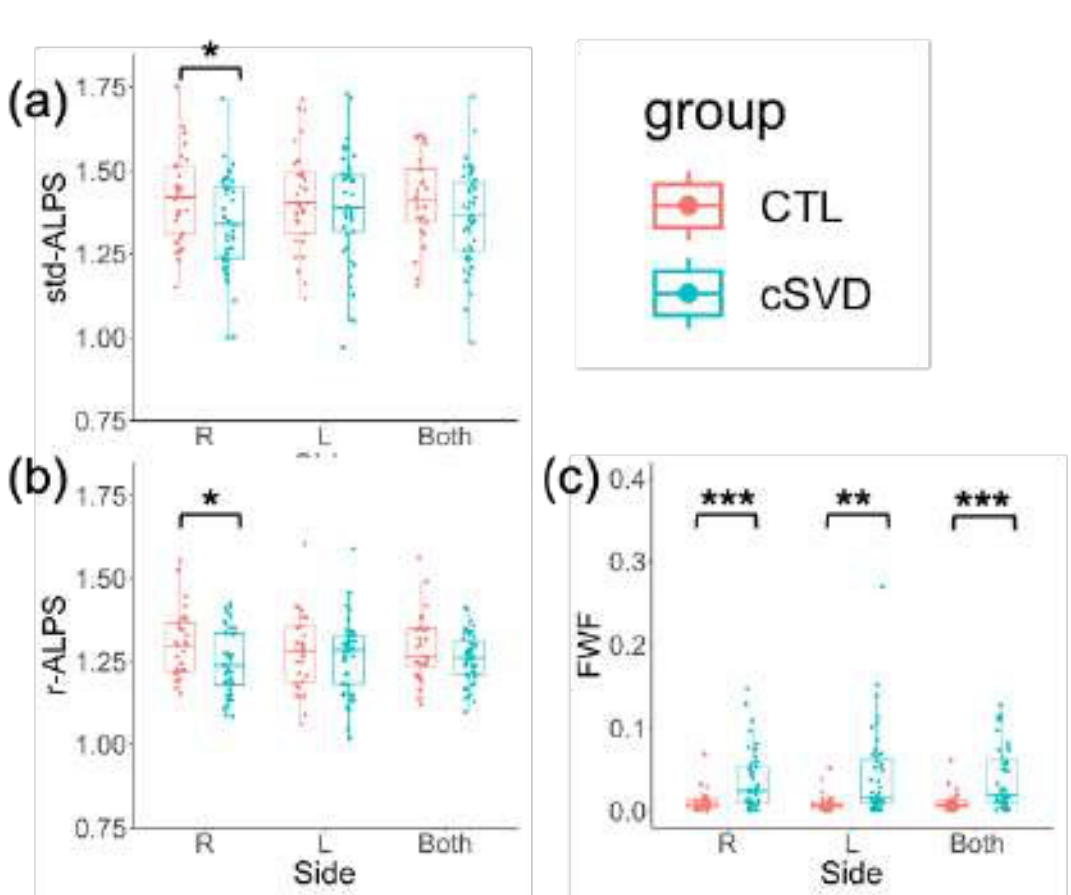

**Figure 2. Comparison of (a) std-ALPS, (b) r-ALPS, and (c) FWF between CTL and cSVD.** Both std- and r-ALPS were reduced in cSVD only on the right side (both $p = 0.02$), while FWF was consistently elevated (all $p < 0.001$) relative to CTL after controlling for age and sex in a linear model.

**Conclusion:** The FWF in ALPS regions indicates that age-related ALPS reduction is confounded by FWF increase with age, likely as a secondary effect of WMH. However, the significant group difference on the right hemisphere remained after free water correction. Future work is needed to validate the robustness of the cSVD-related reduction of the r-ALPS index in a larger sample and explore the spatial specificity of the radial diffusivity asymmetry found inside the ALPS regions.

# Supervised Learning for Tractogram Alignment

Gabriele Amorosino[1,3],Mattias P. Heinrich[2], Paolo Avesani[3]
*1. The University of Texas at Austin, Austin, TX, 2. University of Luebeck, Luebeck, Germany, 3. Fondazione Bruno Kessler, Trento, Italy*

**INTRODUCTION:** Precise tractogram alignment is essential for inter-subject comparison of white matter (WM) structures and clinical neuroimaging. Traditional methods based on volumetric registration often fail to capture WM fiber geometry. We propose DGTA, a deep learning-based approach that aligns tractograms directly using fiber information.

**METHODS:** DGTA represents tractograms as point clouds and uses a graph convolutional network (GCN) with loopy belief propagation (LBP) to learn displacement fields between fibers[1]. Three graph encoding strategies were evaluated: (i) pts (k-nearest neighbors), (ii) poly (fiber polylines), and (iii) trk (neighboring fibers). The poly strategy yielded the best results.

We used the TractoInferno dataset[2] (20 subjects, 29 bundles), splitting it into 16 training and 4 testing subjects. Each tractogram was subsampled to 1000 fibers. Pseudo-ground truth displacements were computed from bundle skeletons. The model was trained for 70 epochs with a learning rate decay from 0.01 to 0.001. Experiments included 5-fold cross-validation, ensuring robust generalization. Comparisons were made with ANTs SyN[3] using RMSE, Hausdorff distance (HD), and the Linear Assignment Problem Distance (LAPD) metrics.

**RESULTS:** DGTA outperformed ANTs SyN across all alignment metrics, especially in dense fiber regions. The poly encoding preserved fiber geometry more effectively than other strategies. DGTA reduced Bundle Minimum Distance (BMD)[4] relative to ACPC alignment and showed improved fiber-level alignment in bundles with complex geometry. LAPD analysis confirmed a major sensitivity to fiber-density rather than volumetric bundle convolution.

**CONCLUSION:** DGTA offers a fiber-informed, learning-based alternative to voxel-based alignment. Its ability to align dense and geometrically intricate regions more accurately makes it valuable for connectomics and clinical research.

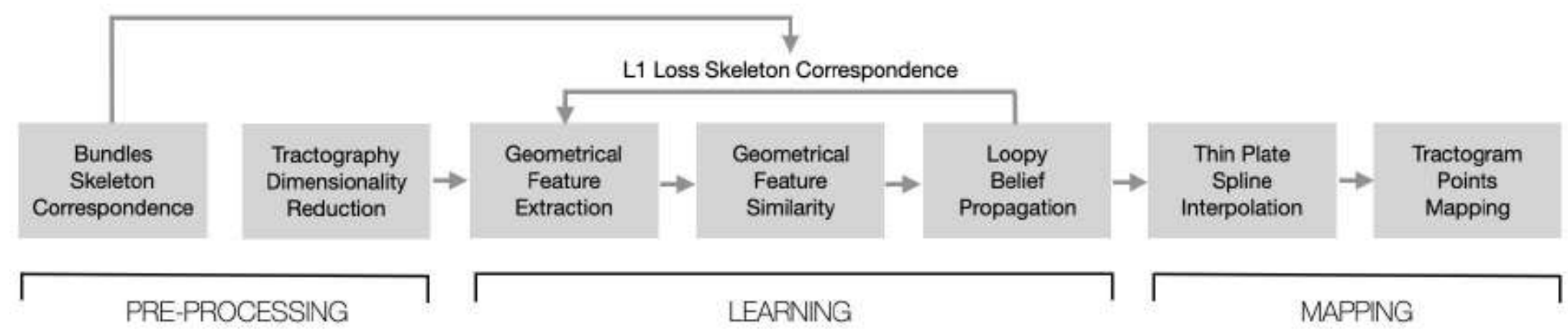

***Figure 1: DGTA Architecture.*** *The schema illustrates the components of the reference learning model based on point cloud registration. By contrast, are reported the extensions of the architecture to exploit the relational information encoded in the edges of the polylines, the digital representation of white matter fibers.*

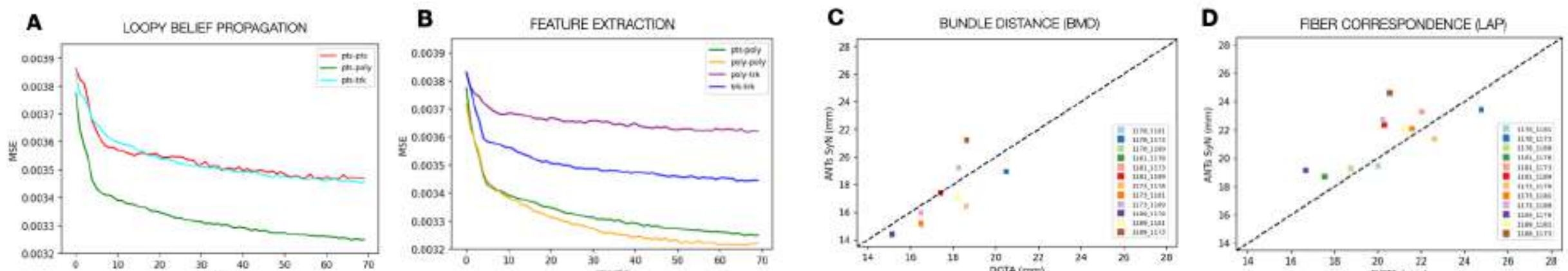

***Figure 2: Empirical results.*** *Left: Mean square error for (A) Loopy Belief Propagation models (pts-poly, pts-trk, pts-pts) and (B) Feature Extraction models (poly-poly, poly-trk, trk-trk) using pts-poly LBP. Right: Comparison of DGTA and ANTs in terms of (C) Bundle Minimum Distance (BMD) and (D) fiber correspondence distance (LAPD), computed with Hausdorff distance. Results averaged over all bundles across 12 test pairs.*

---

1 Hansen, L., & Heinrich, M. (2021). Deep learning based geometric registration for medical images
2 Poulin, P., et al. (2022). TractoInferno - A large-scale, open-source, multi-site database for machine learning dMRI tractography
3 Avants, B. B., et al. (2011). A Reproducible Evaluation of ANTs Similarity Metric Performance in Brain Image Registration
4 Garyfallidis, E., et al., (2012). QuickBundles, a Method for Tractography Simplification.

# DeepDisco: A Deep Learning Framework for Rapid Brain Connectivity Estimation

Anna Matsulevits[1,2], Thomas Tourdias[3,4], Michel Thiebaut de Schotten[1,2]

[1]Groupe d'Imagerie Neurofonctionnelle, Institut des Maladies Neurodégénératives 5293, Centre National de la Recherche Scientifique (CNRS), University of Bordeaux, 33076 Bordeaux, France
[2]Brain Connectivity and Behaviour Laboratory, Sorbonne Universities, 75006 Paris, France
[3]Centre Hospitalier Universitaire (CHU) de Bordeaux, Neuroimagerie Diagnostique et Thérapeutique, 33076 Bordeaux, France
[4]University Bordeaux, National Institute of Health and Medical Research (INSERM), Neurocentre Magendie, U1215, 33076 Bordeaux, France

<u>Introduction:</u> Network-based models of the brain have become central to understanding cognition, function, and disease[1,2]. Structural connectomes and Disconnection patterns offer valuable insights into behavioral mechanisms and disease biomarkers, yet practical barriers, particularly the reliance on resource-intensive tractography, limit their integration into large-scale or clinical pipelines[3–5].This limitation is especially salient as the field moves toward large lesion datasets and AI-driven modeling frameworks that demand accessible and modular tools[6].

<u>Methods</u>: Here, we introduce DeepDisco, a deep learning tool that rapidly estimates disconnection maps from binary lesion or region masks, bypassing traditional tractography while preserving anatomical fidelity. DeepDisco predicts four connectivity outputs: association, commissural, projection, and whole-brain disconnections (Figure 1). It is powered by a 3D U-Net trained on 5,332 lesion-disconnectome pairs serving as 'ground truth' generated with the BCBToolkit[7,8]. The model is optimized with a hybrid loss function (MSE + MAE) and supports real-time use, producing voxel-wise disconnection probability maps in under one second per lesion.

<u>Results:</u> Designed for scalability and ease of integration, DeepDisco offers a scriptable interface and a user-friendly GUI for batch processing, making it compatible with AI pipelines and large datasets. Empirical benchmarking against tractography-derived disconnectomes demonstrates high spatial correspondence and robustness across lesion types. When applied to post-stroke behavioral data, DeepDisco significantly improves long-term symptom prediction compared to atlas-based or unimodal models (Figure 2)[9].

<u>Conclusion:</u> In sum, DeepDisco provides a fast, accessible, and accurate solution to the longstanding bottleneck of disconnection mapping. Its open-source, cross-platform design enables wide adoption in both research and clinical contexts, supporting real-time applications, hypothesis testing, and population-level modeling.

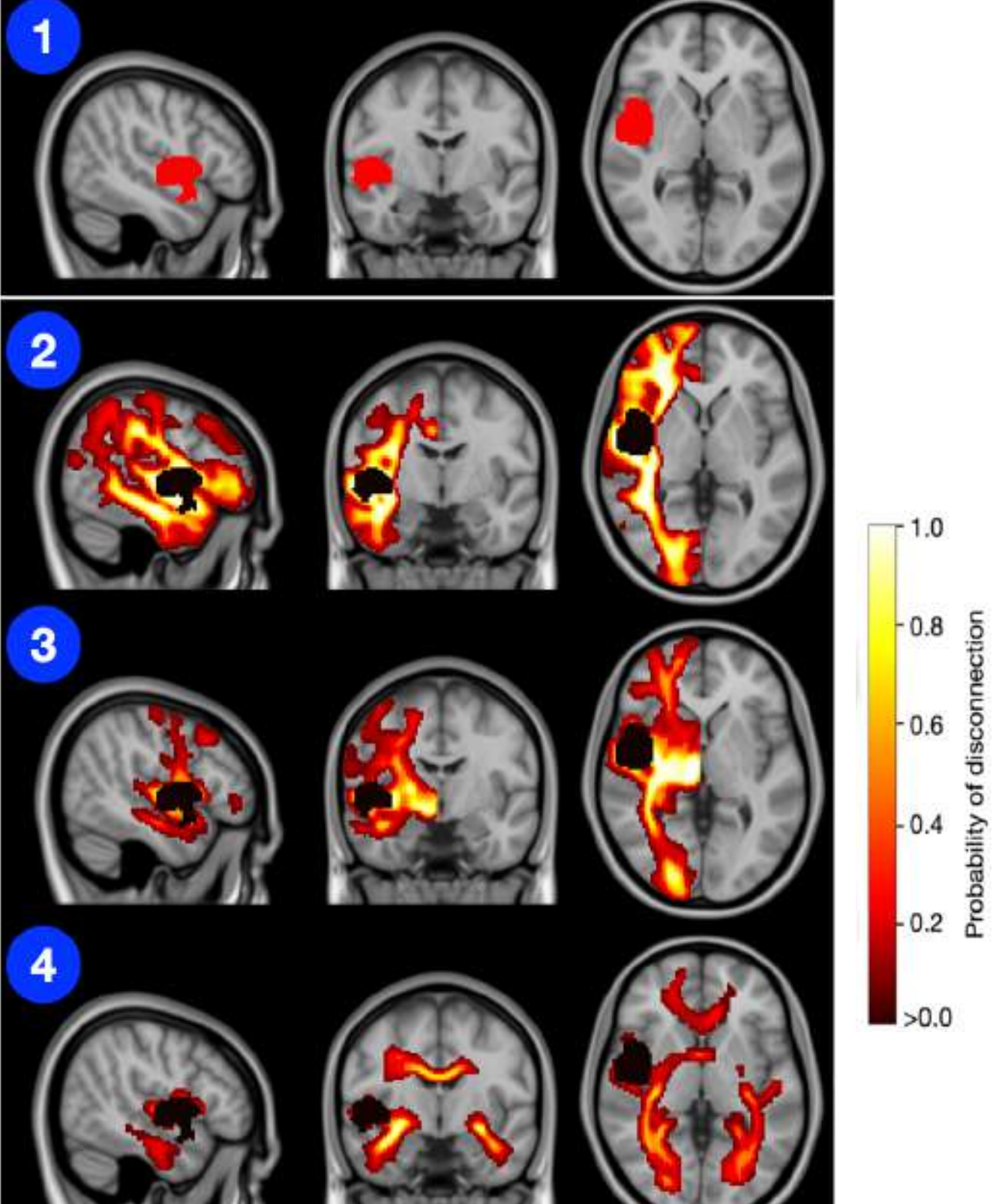

*Figure 1. The DeepDisco outputs that can be generated from the binary lesion or region input, marked in red (1). Panel (2) shows the output from the 'association fibre' model, panel (3) shows the output from the projection fibre model, and panel (4) shows the output from the commissural fibre model. While the input is binary, the output represents a probability of disconnections, with darker colors (red-black) representing a low probability and lighter colors (yellow-white) representing a high probability for a voxel to be disconnected*

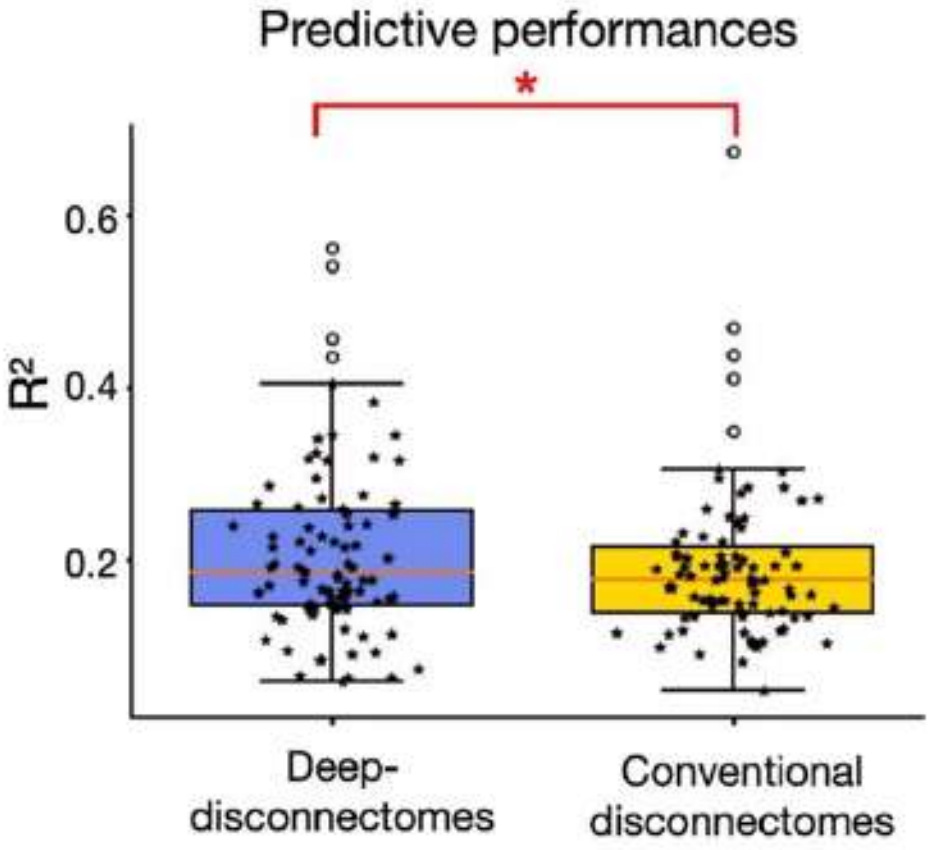

*Figure 2.* The boxplot shows all the *R*2 for the predictions for stroke survivors (dataset 3, *N* = 119) across *N* = 86 neuropsychological scores for the framework using deep-disconnectomes compared to the framework using disconnectomes obtained by BCBToolkit. The boxes represent the quartiles, the whiskers indicate the distribution, and the outliers are marked as dots. Inside the boxes, the median is visualized by the solid line. The *P*-value obtained from a paired *t*-test (2-tails): $*P < 0.01$ shows a significant difference ($t(85) = -1.663$, $P = 0.009$).

# White matter bundle segmentation with deformation features in glioma patients

Chiara Riccardi[1,2], Luca Zigiotto[1], Silvio Sarubbo[1], Paolo Avesani[1,2]
*[1] University of Trento, Centre for Mind/Brain Sciences (CIMeC), corso Bettini, 31, Rovereto, 38068, Italy*
*[2] Fondazione Bruno Kessler, Neuroinformatics Laboratory (NILab), Via Sommarive 18, Trento, 38123, Italy*

**INTRODUCTION**: Automated methods for white matter bundles virtual dissection are designed and validated on healthy individuals. The tacit assumption is to estimate the canonical spatial distribution of the connectivity structures. In the clinical context, where white matter bundles deviate from the norm, this assumption is no more compliant with the premises. Our contribution aims to propose a method that by design is conceived for clinical context.

**METHODS**: We trained a supervised geometric deep learning algorithm to perform the classification task, predicting if each streamline belongs or not to a specific bundle. Importantly, the model's input was not tractography expressed in spatial coordinates: each point P of a tractogram is encoded by 3 scalars, referred to as “deformation features”, that provide information about the possible tumor-displacement to the white matter fibers.

The deformation features are defined as: (i) the minimum distance of P from the skeleton[2] that is a representative streamline of the target bundle of the segmentation; (ii) the ratio between the tumor radius and the distance of P from the tumor’s center of mass; (iii) the absolute difference between the curvature of streamline in P and on the curvature of tumor surface. Taken together, deformation features are informative about the probability of the anatomy being modified by the tumor growth for each point of streamlines in the tractogram located in the area close to the lesion.

**RESULTS**: The preliminary results prove that our proposed method is effective in segmenting bundles when tumors is deforming the canonical anatomy, especially for major displacements. The empirical analysis is carried out using a dataset of tractograms with several hundreds of healthy individuals and the related major bundle annotations, a clinical dataset with several hundreds of tumor masks, and a simulator to generate plausible deformations combining the two datasets. In addition, we replicate the analysis with a state of the art method RecoBundlesX[3], as reported in Table 1.

| | inferior fronto-occipital fasciculus | arcuate fasciculus | frontal aslant tract | Inferior longitudinal fasciculus | pyramidal tract |
|---|---|---|---|---|---|
| Deformation Features | 0.84 ±0.08 | 0.72 ±0.07 | 0.82 ±0.07 | 0.76 ±0.05 | 0.87 ±0.07 |
| RecoBundlesX | 0.56 ±0.33 | 0.61 ±0.25 | 0.57 ±0.28 | 0.54 ±0.23 | 0.83 ±0.16 |

***Table 1:*** *The average and standard deviation of dice similarity coefficient, estimating coherence between tractography automatic segmentations with ground truth of proposed model and state of the art method RecobundleX, for 5 major bundles.*

**CONCLUSIONS**: We proposed a novel method for white matter bundles dissection in glioma patients. For the first time a computational learning model is conceived by design for the clinical context, where the canonical pathways of fibers deviate from the pattern of healthy individuals

**Identifying the Microstructural Neurobiological Signature of Brain Lesions and Disconnected Tissue Using the UK Biobank**

Anna Matsulevits[1,2], Oliver Parent[3], Thomas Tourdias[4,5], Michel Thiebaut de Schotten[1,2], Mallar Chakravarty[6]

[1]Groupe d'Imagerie Neurofonctionnelle, Institut des Maladies Neurodégénératives 5293, Centre National de la Recherche Scientifique (CNRS), University of Bordeaux, 33076 Bordeaux, France
[2]Brain Connectivity and Behaviour Laboratory, Sorbonne Universities, 75006 Paris, France
[3]Cerebral Imaging Centre, Douglas Mental Health UniversityInstitute, Verdun, Canada
4Centre Hospitalier Universitaire (CHU) de Bordeaux, Neuroimagerie Diagnostique et Thérapeutique, 33076 Bordeaux, France
[5]University Bordeaux, National Institute of Health and Medical Research (INSERM), Neurocentre Magendie, U1215, 33076 Bordeaux, France
[6]Department of Biological and Biomedical Engineering,McGill University, Montreal, Canada

Introduction: Understanding the microstructural consequences of brain lesions is critical for deciphering the mechanisms underlying neurological dysfunction and advancing therapeutic interventions. In this work, we investigated the microstructural and connectivity-based signatures of brain lesions across neurodegenerative pathologies using data from the UK Biobank.

Methods: We implemented a comprehensive and multimodal approach to characterize both the local and remote impact of brain lesions on tissue microstructure and connectivity. Our pipeline began with the selection of participants diagnosed with neurodegenerative diseases alongside matched healthy controls. From diffusion- and susceptibility-weighted imaging data, we derived biologically informative microstructural maps sensitive to fiber density, free water content, myelin integrity, and iron deposition (Figure 1). To contextualize these measures and enhance interpretability, we constructed voxel-wise normative models in the healthy cohort using Bayesian linear regression, adjusting for key covariates such as age and sex (Figure 2). Based on these models, we generated individual-level z-score maps, enabling the quantification of microstructural abnormality relative to normative expectations. For participants with pathology, we extracted lesion maps and applied a deep-disconnectome model to estimate personalized disconnectivity profiles, probabilistic maps reflecting white matter disconnection induced by focal damage[1]. These disconnectome maps were then spatially correlated with the z-scored microstructural abnormalities to reveal regionally specific patterns of remote degeneration. To explore disease-specific microstructural phenotypes, we will apply Uniform Manifold Approximation and Projection (UMAP) to the z-scored microstructural profiles[2]. This dimensionality reduction technique allows the construction of a morphospace where pathology-specific clusters can emerge, supporting the concept of microstructural "fingerprints" that differentiate diseases based on tissue properties and disconnection profiles. At the time of submission, data preprocessing and pipeline development are complete, and full-scale analyses are actively underway.

Results: Results will be available by the time of the conference, and are expected to yield insights into how structural brain injury alters tissue properties both locally and across distributed networks.

Conclusion: These findings have the potential to bridge the gap between connectomic disruption and microstructural pathology, offering new avenues for biomarker discovery and personalized interventions.

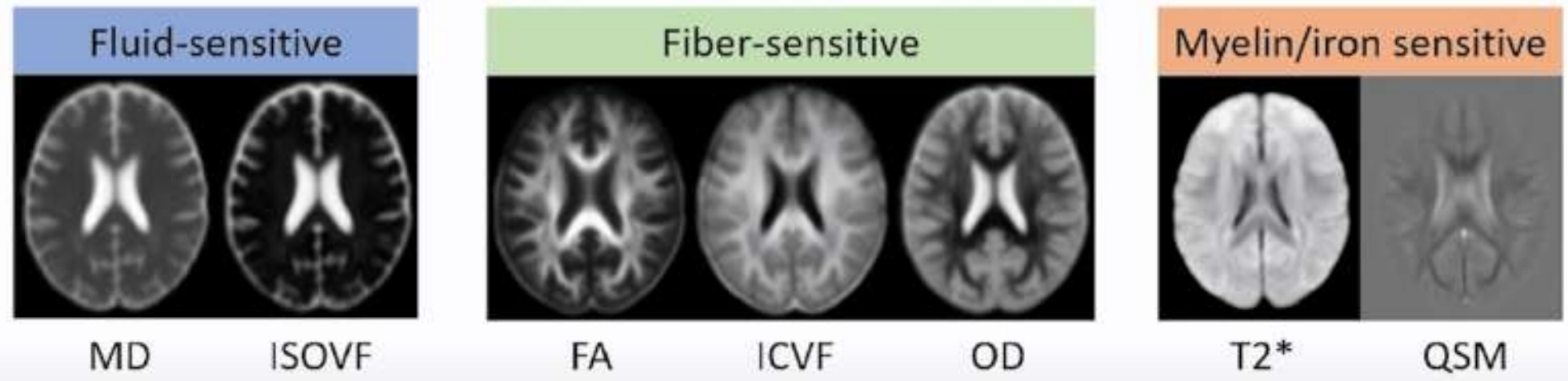

***Figure 1.*** *The microstructural markers used to construct microstructural signatures of brain lesions. MD = Mean Diffusivity, FA = Fractional Anisotropy, ICVF = Intra-Cellular Volume Fraction, ISOVF = Isotropic Volume Fraction, OD = Orientation Dispersion, QSM = Quantitative Susceptibility Mapping.*

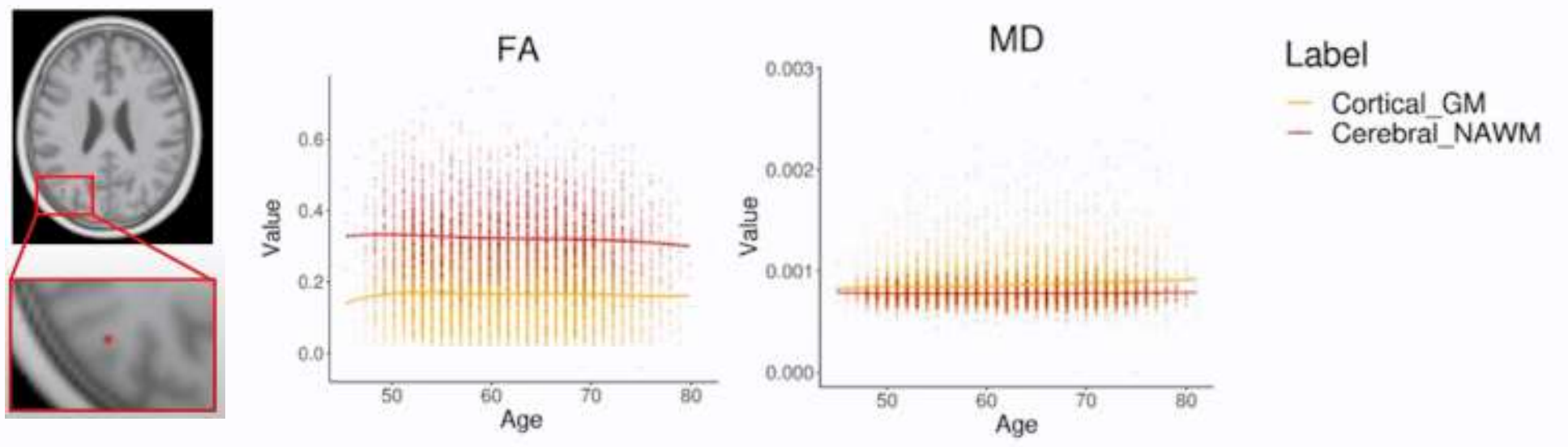

***Figure 2.*** *Visualization of the voxel-wise, tissue specific normative model obtained with the Bayesian linear regression (Micro ~ bs(age,3) + sex) and its output. Age and sex were used as covariates. GM = gray matter, NAWM = normal appearing white matter, FA = Fractional Anisotropy, MD = mean diffusivity.*

# Diffusion MRI tractography to reduce risks of postoperative neurological deficits: A systematic review and meta-analysis

Guido I. Guberman, MD, PhD[1], Guillaume Theaud, PhD[2], François Rheault, PhD[3], Joseph Y.-M. Yang, PhD[4-6], Maxime Descoteaux, PhD[3], Silvio Sarubbo, MD, PhD[7,8], Sami Obaid, MD, PhD, FRCSC[2,9]

1. Department of Neurology and Neurosurgery, Faculty of Medicine, McGill University, Montreal, Quebec, Canada
2. Neuroscience Research Axis, University of Montreal Hospital Research Center (CRCHUM), Montreal, Quebec, Canada
3. Computer Science Department, Université de Sherbrooke, Sherbrooke, Quebec, Canada
4. Department of Neurosurgery, Neuroscience Advanced Clinical Imaging Service, The Royal Children's Hospital, Melbourne, Australia
5. Neuroscience Research, Murdoch Children's Research Institute, Melbourne, Australia
6. Department of Pediatrics, The University of Melbourne, Melbourne, Australia
7. Center for Medical Sciences, Department of Cellular, Computational and Integrative Biology Center for Mind and Brain Sciences, University of Trento, Trento, Italy
8. Department of Neurosurgery, "S. Chiara" University-Hospital, Azienda Provinciale peri Servizi Sanitari, Trento, Italy
9. Department of Neuroscience, University of Montreal, Montreal, Quebec, Canada

Despite informing on the location of functionally relevant white matter tracts, diffusion MRI tractography is not routinely used to guide neurosurgical procedures. The potential of tractography to help avoid postoperative neurological deficits is not yet fully established. The objective of our study was to assess whether surgeries that incorporated tractography, either alone or in conjunction with other modalities, are associated with a lower risk of long-term postoperative neurological deficits in patients undergoing resective/ablative intracranial procedures.

We performed a systematic review with meta-analysis, searching through EMBASE and PubMed databases for all peer-reviewed articles published in English up until December 2024. Studies were included if they reported on intracranial resective or ablative surgeries, if they compared tractography-assisted against non-tractography-assisted approaches, and if they assessed new postoperative neurological deficits. No restrictions were placed on the age of patients. Studies were assessed for inclusion by two independent reviewers and disagreements were settled by a third. Data extraction was performed according to PRISMA guidelines, quality of studies was evaluated using the GRADE Framework, and risk of bias was assessed through a modified version of the Newcastle-Ottawa Quality Assessment Scale for cohort studies. Data were pooled using a random-effects model with a Mantel-Haenszel method for estimating risk ratios. The primary outcome consisted in any neurological deficits present at last follow-up (≥ 3 months).

Out of 5315 studies initially identified, eight were included after all stages of review, all of which consisted of resective surgeries. A meta-analysis of 629 patients revealed a 55% risk reduction of postoperative neurological deficits when incorporating tractography in the neurosurgical workflow. This benefit was still observed when assessing studies where tractography was exclusively used preoperatively. Further, the incorporation of tractography into intraoperative neuronavigation systems was associated with lower proportions of postoperative neurological deficits, compared to exclusively preoperative tractography. When considering the effect of additional brain mapping adjuncts, although a non-significant risk reduction was observed when excluding studies that incorporated other imaging modalities, results did show significant reductions in risk of postoperative neurological deficits after excluding studies that incorporated direct electrical stimulation. Finally, our results were also consistent after removal of the study with the largest weight, showing robustness to sample perturbations.

The addition of tractography is associated with a reduced risk of postoperative neurological deficits in intracranial resective surgeries. Tractography can complement gold standard brain mapping methods such as direct electrical stimulation during awake surgeries or serve as a helpful alternative when electrical stimulation is contraindicated.

**Title:** Surface-based Tractography uncovers 'What' and 'Where' Pathways in Prefrontal Cortex

**Authors:** Marco Bedini [1], Emanuele Olivetti [2,3], Paolo Avesani [2,3] & Daniel Baldauf [2]

**Affiliations:** 1. Institut de Neurosciences de la Timone (INT), Aix-Marseille University, Marseille, France; 2. Center for Mind/Brain Sciences (CIMeC), University of Trento, Trento, Italy; 3. NeuroInformatics Laboratory (NILab), Bruno Kessler Foundation (FBK), Trento, Italy

**Introduction:** The frontal eye field (FEF) and the inferior frontal junction (IFJ) are prefrontal regions that mediate top-down control. Accumulating evidence suggests a functional division of labor, with the FEF supporting spatial and the IFJ non-spatial processing[1]. We hypothesized that this dissociation is rooted in distinct structural connectivity profiles[2].

**Methods:** We localized the FEF and IFJ in standard space using an activation likelihood estimation (ALE) meta-analysis of fMRI studies that robustly engaged these regions[3]. Using 3T diffusion MRI data from the Human Connectome Project[4], we performed surface-based probabilistic tractography[5] on 56 unrelated subjects. For each subject, we seeded tractography from meta-analytically defined FEF and IFJ peaks and traced ipsilateral connections to dorsal and ventral visual stream regions, delineated on the native white matter surface[6] and parcellated using the multimodal Glasser atlas[7].

**Results:** The FEF showed higher structural connectivity likelihood with dorsal visual stream regions, particularly in the left hemisphere. Conversely, the IFJ demonstrated higher connectivity likelihood with ventral stream regions bilaterally. These patterns remained robust after controlling for Euclidean seed-to-target distance.

**Conclusions:** Our findings reveal distinct connectivity fingerprints for the FEF and IFJ, supporting their proposed specialization in spatial versus non-spatial processing. The results provide anatomical evidence that the dual-stream visual architecture extends into the lateral prefrontal cortex[9].

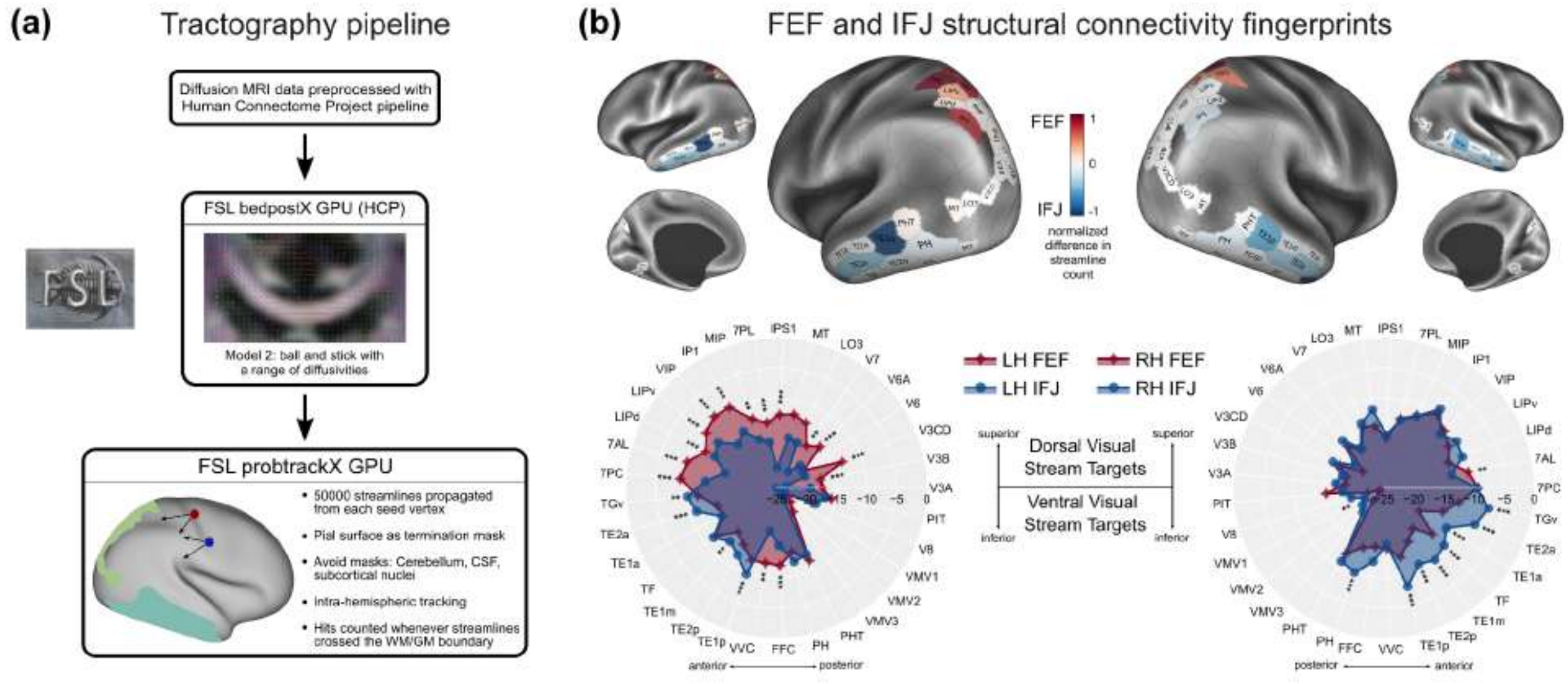

**Figure 1.** Panel A: Processing pipeline implemented with the FSL package[5]. Panel B: Summary of results. Asterisks indicate contrasts that remained significant after controlling for seed-to-target distance.

# GPU tractography: What can we learn from half a trillion streamlines?

Yanis Aeschlimann[1], Samuel Deslauriers-Gauthier[1], and Romain Veltz[1]

[1]Université Côte d'Azur, Inria, France

**Introduction:** Probabilistic fiber tractography is a stochastic process whose outcomes can be used to estimate the probability of connection between two brain regions. The most commonly used estimator for the probability of connection between regions $i$ and $j$ is given by $c_{ij}^{(n)}/n$, where $n$ represents the total number of seeds and $c_{ij}^{(n)}$ denotes the number of streamlines connecting the two regions. Despite its widespread use, the variance of this estimator is often overlooked. In this study, we leveraged a GPU-accelerated implementation of probabilistic tractography to generate a connectivity matrix from 500,000,000,000 streamlines. Using this large-scale dataset, we computed rigorous bounds on the confidence interval of the connection probability estimator.

**Methods:** When a streamline count connectivity matrix is normalized by the number of seeds, the entry in row $i$ and column $j$ can be interpreted as the probability of connecting regions $i$ and $j$ when generating a single streamline randomly. Each entry therefore follows a binomial distribution with parameters $n$ and $p_{ij} \in [0, 1]$ which corresponds to the true (unknown) probability of connection of the *tractography process* (distinct from the strength of brain connectivity [1]). The quantity $\hat{p}_{ij}^{(n)} = c_{ij}^{(n)}/n$ is an unbiased estimator of $p_{ij}$ whose variance is $\hat{\sigma}^2 = \hat{p}_{ij}^{(n)} \left(1 - \hat{p}_{ij}^{(n)}\right)/n$. Using the Clopper–Pearson method, the 95% confidence interval for the parameter $p_{ij}$ is computed. Probabilistic fiber tractography (step size 0.25 mm, maximum angle 30°) was performed on fiber orientation distribution function computed from the diffusion MRI data from subject 100206 of the Human Connectome Project. Our GPU accelerated implementation allowed us to generate a connectivity matrix from the Schaefer atlas [3] (400 regions) using $n = 0.5 \times 10^{12}$ streamlines and computed extremely precise bounds on the connection probability $p_{ij}$. For comparison, we also generated a connectivity matrix from the typically recommended $n = 2 \times 10^7$ streamlines.

**Results:** Figure 1 illustrates the ratio of the size of the confidence interval over the estimated probability and the histogram of the connection probabilities. For $n = 2 \times 10^7$, 33% of the values are above 1 (red region in the histogram), meaning that the uncertainty is at least as large as the estimate. This value drops to 1% (red and blue regions in the histogram) for $n = 0.5 \times 10^{12}$. The intra-hemispheric median ratios are 0.72 and 0.02 for $n = 2 \times 10^7$ and $n = 0.5 \times 10^{12}$, respectively. It highlights the high uncertainty of the estimated connectivity for $n = 2 \times 10^7$. In the histogram, the probabilities are in agreement only when the confidence interval is small compared to the estimated probability (in the red region).

**Conclusion:** GPU acceleration of tractography allows the generation of a large number of streamlines in a reasonable time. Our results show that many connectivity matrix entries are inaccurate with the current recommendation of $2 \times 10^7$ streamlines. Our conclusion differ from previous ones [2] because we focus on the accuracy of individuals probability of connection, rather than the global process of tractography. Our Python implementation is open-source and publicly available (`https://gitlab.inria.fr/cronos/software/tractography`).

**Acknowledgment:** We are grateful for the infrastructure of Slices RI (https://www.slices-ri.eu/) used in this study.

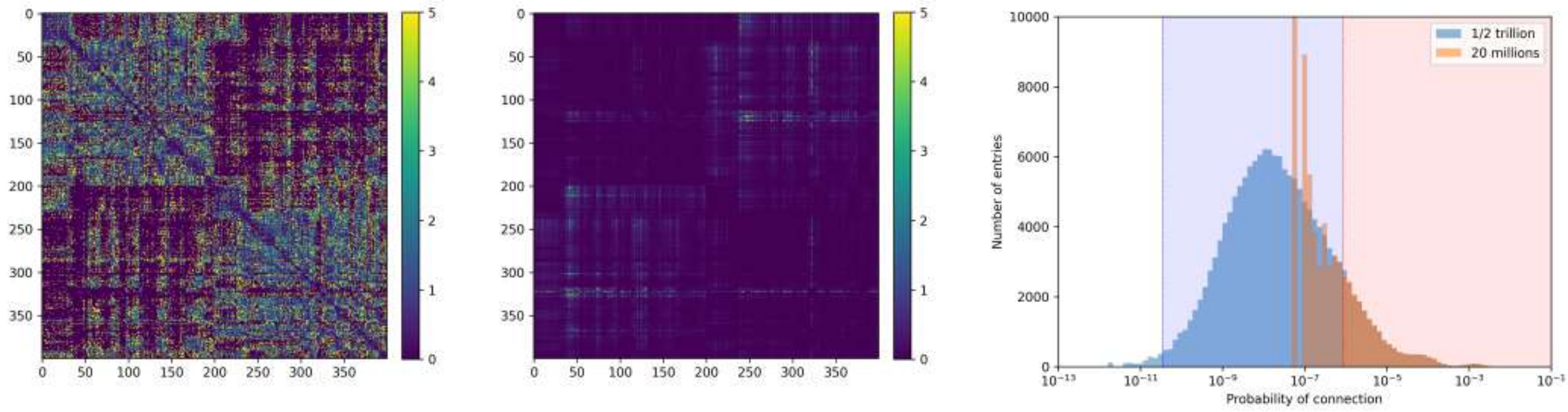

Figure 1: Ratio of the size of the confidence interval over the estimated probability $\hat{p}_{ij}^{(n)}$ for $n = 2 \times 10^7$ and $n = 0.5 \times 10^{12}$ streamlines. The histogram of the probabilities for 1/2 trillion and 20 million streamlines. The probabilities are reliable in the red regions for both estimates and only for $n = 0.5 \times 10^{12}$ in the blue region.

# A principled mathematical study of the limit of fiber tractography

Samuel Deslauriers-Gauthier[1] and Romain Veltz[1]

[1]Université Côte d'Azur, Inria, France

**Introduction:** Fiber tractography algorithms started as a numerical scheme to recover the trajectory of a streamline, parameterized by its arc length. These algorithms solved a well specified ordinary differential equation whose parameters were specified by diffusion tensor imaging (DTI) [2]. The limitation of DTI were quickly recognized and the numerical scheme was updated to make use of the more complex local information provided by the fiber orientation distribution function (fODF). This allowed researchers to overcome the limitations of DTI and explore the complex architecture of the white matter. However, this left fiber tractography algorithms in limbo with regards to the *mathematical problem* they numerical solve. As a direct consequence, many parameters of tractography algorithms (step size, maximum angle, number of seeds) are chosen arbitrarily with, at best, a qualitative description of their effect. In this work, we derive new tractography algorithms as the numerical approximation to the solution of an ordinary differential equation (deterministic) or stochastic differential equation (probabilistic). This principled approach not only unifies deterministic and probabilistic tractography but also overcomes many numerical limitations of current algorithms and opens the door to a deeper mathematical understanding of tractography. We believe we have identified the fundamental equation of tractography.

**Methods:** Deterministic fiber tractography is typically implemented as a two step process given by $\boldsymbol{x}_{i+1} = \boldsymbol{x}_i + \boldsymbol{u}_i \Delta t$ and $\boldsymbol{u}_{i+1} = g(\boldsymbol{x}_{i+1}, \boldsymbol{u}_i)$ where $\Delta t$ is the step size, $g$ is derived from the fODF, $\boldsymbol{x}_i \in \mathbb{R}^3$ and $\boldsymbol{u}_i \in \mathbb{S}^2$ are the location and direction of a particle at step $i$, respectively. We refer to this process as the Tractography Markov Chain (TMC).

In the limit $\Delta t \to 0$, the $\boldsymbol{x}$ update equation is differential $\dot{\boldsymbol{x}} = \boldsymbol{u}$ while the $\boldsymbol{u}$ update is algebraic $\boldsymbol{u} = g(\boldsymbol{x}, \boldsymbol{u})$ because it does not depend on $\Delta t$. We propose to move away from this differential–algebraic system of equations to a classic ordinary differential equation (ODE) formulation. For the probabilistic fiber tractography, the direction update is instead $\boldsymbol{u}_{i+1} \sim g(\boldsymbol{x}_{i+1}, \boldsymbol{u}_i)$. The limit $\Delta t \to 0$ is not very well-posed as the direction update dominates the $\boldsymbol{x}$ update.

**Results:** For the deterministic case, we are able to show that streamlines emerge naturally as the solution of the ODE $d\boldsymbol{x}_t = \boldsymbol{u}_t dt$ and $d\boldsymbol{u}_t = \nabla_{\boldsymbol{u}} \log f(\boldsymbol{x}_t, \boldsymbol{u}_t) dt$ where $f$ is the fODF. For the probabilistic case, the streamlines are the trajectories of particles satisfying $d\boldsymbol{x}_t = \boldsymbol{u}_t dt$ and $d\boldsymbol{u}_t = \gamma \nabla_{\boldsymbol{u}} \log f(\boldsymbol{x}_t, \boldsymbol{u}_t) dt + \sqrt{\gamma} dB_t^{\mathbb{S}^2}$ where $dB_t^{\mathbb{S}^2}$ is the spherical Brownian motion and $\gamma$ is related to the inverse maximal curvature. Given their similar form, these two equations can be combined into

$$
\begin{aligned}
d\boldsymbol{x}_t &= \boldsymbol{u}_t dt \\
d\boldsymbol{u}_t &= (1+\gamma) \nabla_{\boldsymbol{u}} \log f(\boldsymbol{x}_t, \boldsymbol{u}_t) dt + \sqrt{\gamma} dB_t^{\mathbb{S}^2}.
\end{aligned}
\tag{1}
$$

with $\gamma = 0$ corresponding to deterministic tractography and $\gamma > 0$ to probabilistic tractography. It is interesting to note that the solution of the above equation finds local maxima of $f$ which beautifully corresponds to previously used deterministic heuristic. Due to space limitations, we do not provide the proof here, but it is based on the infinitesimal generator [1] of a time re-scaling of the TMC and underlines the fundamental hypotheses underpinning the existence of a limit when $\Delta t \to 0$. The process solution of Eq. (1) solves the issues related to the parameter dependency of the TMC. We also studied the limit of the above process when $\gamma \to 0$ and discovered the links between the different forms of the tractography process (deterministic, probabilistic, etc).

An Euler scheme solving this system of equation is

$$
\begin{aligned}
\boldsymbol{x}_{i+1} &= \boldsymbol{x}_i + \boldsymbol{u}_i \Delta t \\
\boldsymbol{u}_{i+1} &= \mathrm{Exp}_{\boldsymbol{u}_i} \left( (1+\gamma) \nabla_{\boldsymbol{u}} \log f(\boldsymbol{x}_{i+1}, \boldsymbol{u}_i) \Delta t + \sqrt{\gamma \Delta t} B_i^{\mathbb{R}^2} \right)
\end{aligned}
$$

where Exp is the exponential map on the sphere and $B_i^{\mathbb{R}^2}$ is normally distributed. In previous probabilistic algorithms, reducing $\Delta t$ has the effect of amplifying noise whereas in the above numerical scheme converges to the solution of Eq. (1) as desired.

**Conclusion:** In this work, we have identified what we believe to be the fundamental equation of tractography, time will tell if we are correct. Our formulation is principled, numerically stable, can be studied mathematically, and opens the door to many radical new approaches for tractography. One example is to solve the Fokker–Planck equation associated to Eq. (1), obtaining the exact streamline density that would be obtained with an infinite number of seeds. Our implementation is open-source and publicly available (`gitlab.inria.fr/cronos/software/tractography`).

## Unbiased tractogram density optimisation for robust estimation of white matter connectivity differences

Philip Pruckner[1], Remika Mito[1,2], David Vaughan[1,2,3], Kurt Schilling[4], Victoria Morgan[4], Dario Englot[4], Robert Smith[1,2]
[1]The University of Melbourne, Australia; [2]The Florey Institute, Australia; [3]Austin Health, Australia; [4]Vanderbilt University, United States

**Introduction.** The prospect of quantitatively mapping longitudinal connectivity changes through diffusion magnetic resonance imaging (dMRI) holds significant promise for early diagnosis, disease monitoring, and personalized treatments. Streamline tractography is however subject to considerable methodological variance[1], making it difficult to reliably detect subtle biological effects. To address this problem, we here propose an *unbiased tractogram density optimisation* framework that dramatically reduces reconstruction variance by harnessing shared information between timepoints.

**Methods.** The human brain establishes its long-range projections predominately prenatally[2], so any subsequent developmental or pathological process can be postulated to operate within an overall fixed white matter scaffold. Leveraging this anatomical constraint, we here propose a novel analysis framework wherein the underlying streamline trajectories for a given subject are shared across timepoints, with only the densities ascribed to those streamlines determined based on session-specific image data. We here present two approaches for such unbiased optimisation of tractogram densities, both based on the widely adopted SIFT2 framework[3]. The first, *symmetric* unbiased optimisation, derives a session-average weighted tractogram which is then separately refined to fit session specific fibre densities. The second, *differential* unbiased optimisation provides a distinct methodological implementation specifically tailored to longitudinal analysis: instead of estimating streamline densities per session, it directly optimises a session-average weighted tractogram to fit *fibre density differences* between sessions. Figure 1 illustrates an exemplar output of differential optimisation—a differentially weighted tractogram derived from a synthetic kissing bundles phantom with simulated fibre loss in one bundle.

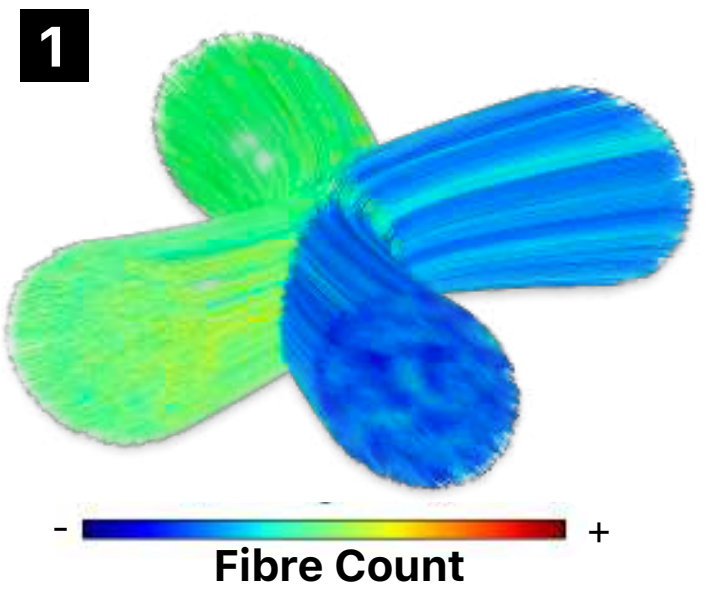

We demonstrate the benefits of unbiased tractogram optimisation to investigate longitudinal connectivity changes within a complex synthetic dMRI phantom, as well as in two *in vivo* datasets with a clear biological expectation of an effect. Synthetic dMRI data with a signal-to-noise ratio of 20 were derived from a modified DiSCo3 phantom[4], simulating a central lesion that causes a loss of fibres in a wide range of bundles. *In vivo* datasets comprised the Human Connectome Project's Scan-Rescan[5] *(n=44)* and a temporal lobe epilepsy surgery dataset *(n=52), with expectations of* no connectivity changes and pronounced connectivity decreases close to the resection[6] respectively. Results were benchmarked against cross-sectional reconstruction with and without SIFT2 optimisation. All connectivity estimates were quantified as Fibre Bundle Capacity[7], a measure of interregional information bandwidth.

**Results.** In synthetic phantoms, unbiased reconstruction resulted in significantly lower errors in longitudinal connectivity change quantification compared to cross-sectional tractogram reconstruction *(p<0.001, see Fig. A)*. In human scan-rescan data, cross-sectional reconstruction resulted in implausible increases and decreases; in contrast, unbiased connectivity quantification showed no relevant connectivity changes *(see Fig. B)*. Similar results were found in longitudinal analysis of surgical data, with unbiased reconstruction showing a more plausible pattern of effects, which were mainly restricted to the hemisphere ipsilateral to resection *(see Fig. C)*.

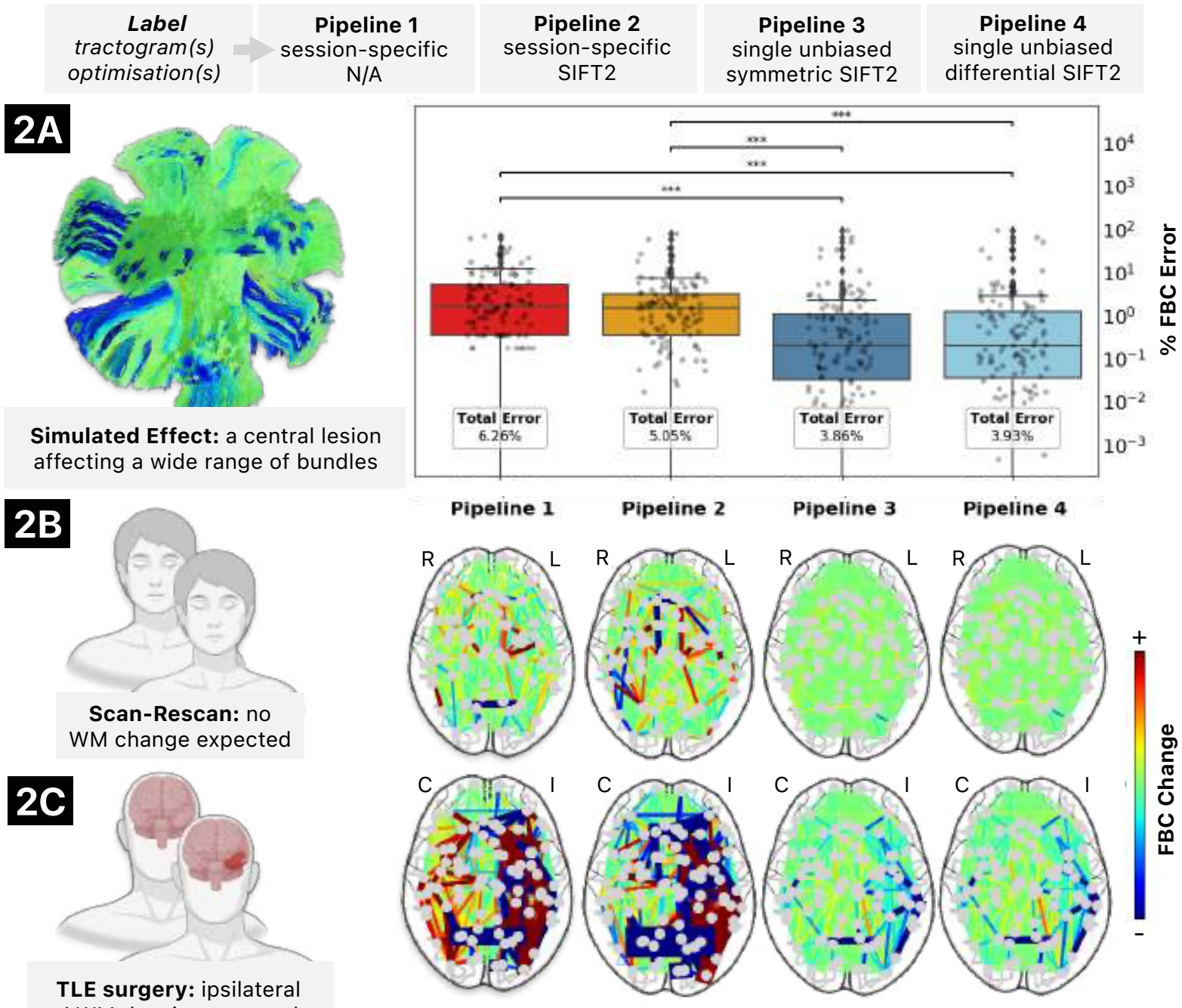

FBC = Fibre Bundle Capacity, L = left, R = right, I = ipsilateral, C = contralateral

**Conclusion**. Unbiased tractogram optimisation drastically reduces methodological imprecisions of estimating longitudinal connectivity changes, enabling robust quantification of white matter connectivity differences. This novel unbiased tractogram optimisation framework will especially benefit clinical studies, where inferences must be drawn from small cohorts or individual cases. Robust estimation of brain connectivity changes within an unbiased framework opens exciting development avenues for precision medicine applications of quantitative streamline tractography, bringing advanced diffusion imaging one step closer to clinical application.

# Intraoperative fast fibre tract segmentation in paediatric tumour patients

Dana Kanel[1], Fiona Young[1], Kiran K. Seunarine[1], Nikhita Nandi[1], Annemarie Knill[1,3], Enrico De Vita[1,3], Kshitij Mankad[4], Chris A. Clark[1], Kristian Aquilina[2], Jonathan D. Clayden[1]
[1]Developmental Imaging and Biophysics Section, UCL GOS Institute of Child Health, London WC1N 1EH
[2]Department of Neurosurgery, Great Ormond Street Hospital for Children, London WC1N 3JH
[3]MRI Physics Group, Radiology, Great Ormond Street Hospital for Children, London WC1N 3JH
[4]Department of Neuroradiology, Great Ormond Street Hospital for Children, London WC1N 3JH

**Introduction:** Segmenting white matter (WM) tracts is clinically useful for intraoperative surgical planning and navigation, as well as for relating post-operative outcomes with underlying structural connectivity. Streamline tractography is the current standard for reconstructing WM tracts, although it is limited by relatively lengthy data processing and lack of expertise in clinical settings [1]. Tractfinder [2,3] is an alternative method to tractography that requires minimal processing time and expertise. It uses a tract-specific orientation atlas, created using tractography, and tumour deformation modelling to enable fast fibre tract reconstruction in tumour patients. The current work extends the use of Tractfinder by applying it to the reconstruction of infratentorial tracts.
**Methods:** Intraoperative diffusion and T1-weighted MR images were collected from an initial set of 3 paediatric tumour patients. After minimal processing, Tractfinder was used to segment one cerebellar and four cortical WM tracts, bilaterally. Outputs were evaluated against probabilistic tractography and Tractseg (a semi-automated method) using bundle adjacency—a volume-based metric describing average distance of disagreement between bundles [4].
**Results:** Tractfinder successfully segmented WM tracts, including those in regions with large displacement. Outputs were well-aligned with tractography, with bundle adjacency <2mm for all comparisons, which is similar to or below typical inter-protocol variability [4]. Tractseg outputs had the same or higher bundle adjacency values as compared with Tractography, particularly in regions with large displacement. Figure 1 shows better performance of Tractfinder on reconstructing a cerebellar pathway.
**Conclusion**: Tractfinder works effectively with intraoperative diffusion data from paediatric tumour patients, providing a quick and accessible alternative to tractography. Implications of fast segmentation of cerebellar tracts include the prediction of mutism in pediatric tumour patients [5].

**Funding:** This work was funded by Children with Cancer UK (grant number 23-353). All research at Great Ormond Street Hospital NHS Foundation Trust and UCL Great Ormond Street Institute of Child Health is made possible by the NIHR Great Ormond Street Hospital Biomedical Research Centre. The views expressed are those of the author(s) and not necessarily those of the NHS, the NIHR or the Department of Health.

**Figure 1.** T1-weighted iMRI of paediatric tumour patient with automated Superior Cerebellar Peduncle segmentations

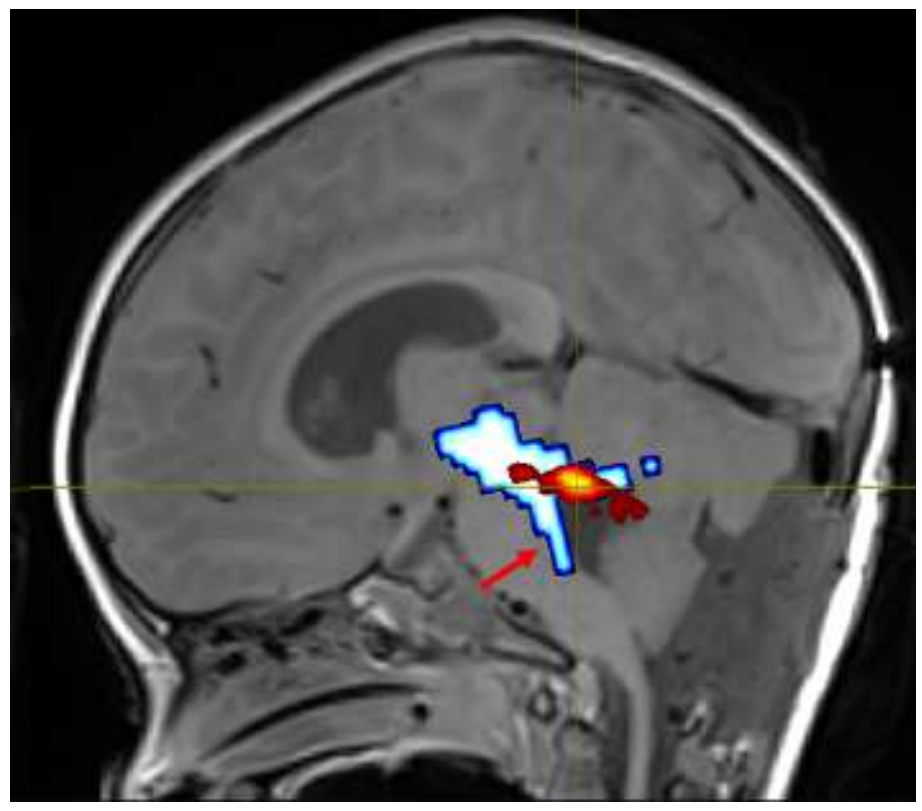

Segmentations created using Tractfinder (red) & Tractseg (blue). Figure indicates incorrect trajectory of Tractseg segmentation, incorporating erroneously identified fibres along the dorsal tegmentum (red arrow).

# Integrating normative and patient models of tractography for accurate prognosis in human glioblastoma

Joan Falcó-Roget[1,*], Gianpaolo Antonio Basile[2,†], Anna Janus[1,3,†], Sara Lillo[4,5], Letterio Salvatore Politi[6,7], Jan K. Argasinski[1,8], Alberto Cacciola[6,7,*]

[1]Computational Neuroscience Group, Sano Centre for Computational Medicine, Kraków, Poland. [2]Brain Mapping Lab, Department of Biomedical, Dental Sciences and Morphological and Functional Imaging, University of Messina, Messina, Italy. [3]Department of Neurophysiology and Chronobiology, Institute of Zoology and Biomedical Research, Faculty of Biology, Jagiellonian University, Krakow, Poland. [4]Radiation Oncology Unit, Clinical Department, National Center for Oncological Hadrontherapy (CNAO), 27100, Pavia, Italy. [5]Department of Internal Medicine and Medical Therapy, University of Pavia, 27100 Pavia, Italy. [6]Department of Biomedical Sciences, Humanitas University, Via Rita Levi Montalcini 4, Pieve Emanuele, 20072 Milan, Italy. [7]IRCCS Humanitas Research Hospital, Via Alessandro Manzoni 56, Rozzano, 20089 Milan, Italy. [8]Faculty of Physics, Astronomy and Applied Computer Science, Jagiellonian University, Krakow, Poland. †Equal contribution: Gianpaolo Antonio Basile, Anna Janus

***Introduction***: Since the advent of tractography, there has been great interest in its potential to improve brain tumor management [1-3]. Among the most aggressive and fatal are human glioblastomas (GBMs), with median survival rarely exceeding 1.5 years. Recent evidence reveals gliomas as active agents that establish synapses, promoting tumor growth [4]. This underscores the urgent need to map white matter-glioma interactions to predict tumor spreading and prognosis.
***Methods***: To mitigate the confounding effects of tumor-induced brain distortions [5,6], we leveraged normative tractography models derived from large healthy cohorts [7]. GBMs were embedded into these normative connectomes to identify the white matter scaffold structurally connected with each tumor [8]. We introduced a *Lesion-Tract Density Index* (L-TDI), based on average tract density, to quantify tumor-white matter involvement. This index was used to stratify and predict outcomes in two independent patient cohorts (N=367 and N=496), based on the distribution of L-TDI values.
***Results***: In both cohorts, overall survival rates ($p<0.01$, log-rank test) and times ($p<0.01$, Mann-Whitney U-test) significantly differed between high and low L-TDI groups across multiple stratification thresholds. Cox proportional hazard models confirmed the prognostic value of L-TDI when combined with clinical covariates, including age, methylation, extent of surgery, and cognitive performance. A logistic model based on the L-TDI predicted 12-month mortality with balanced accuracies of 0.68 and 0.65, and areas under the curve of 0.74 and 0.73 when training and testing in the two independent cohorts.
***Conclusions***: Normative tractography-derived biomarkers offer fast, robust, reproducible, and clinically meaningful insights that can enhance existing clinical workflows. While L-TDI offers a cost-effective and scalable strategy for predicting patients' prognosis and, potentially, for surgical and therapeutic planning, direct validation against patient-specific HARDI data remains essential to establish its precision, despite the higher acquisition costs.

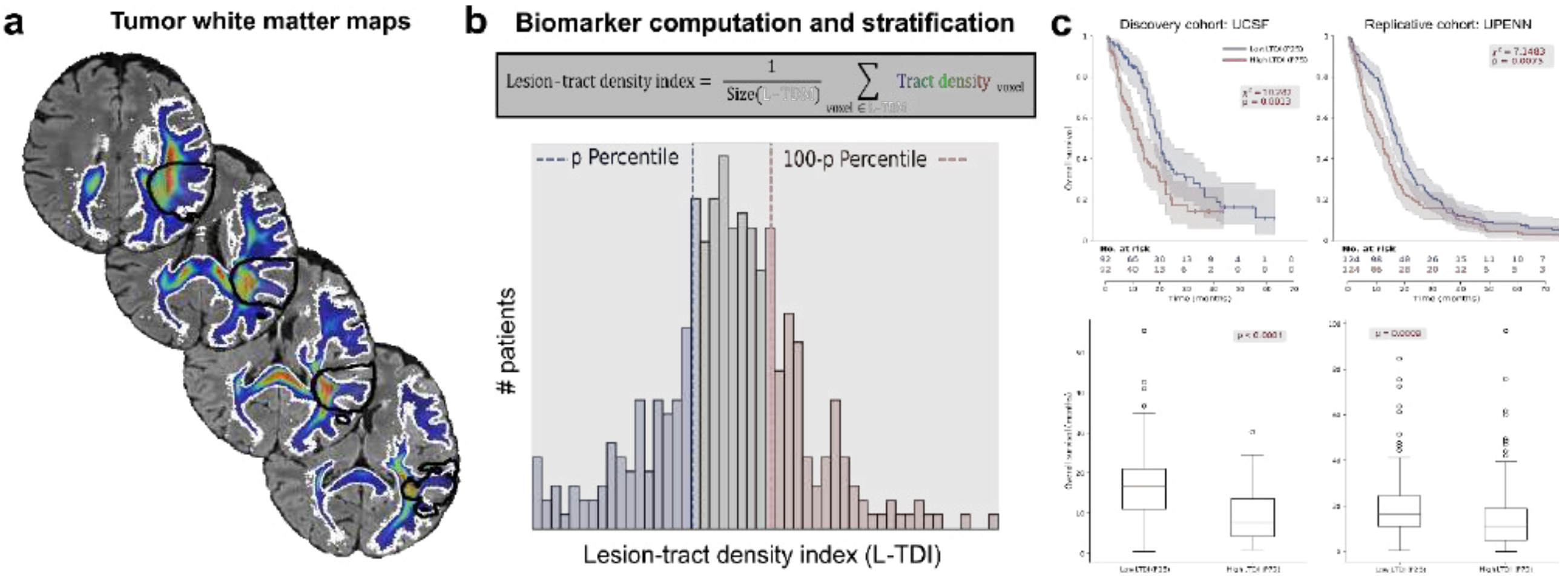

## Improving tractography reconstruction with asymmetric FOD tractography: preliminary evidence on the cortico-spinal tract

*Richard Stones and Flavio Dell'Acqua, King's College London*

*Introduction:* Tractography has become a crucial tool in neuroscience and neurosurgery for visualising the 3D structure of white matter pathways. The majority of spherical deconvolution (SD) approaches generate symmetric fibre orientation distributions (sFODs) to resolve crossing fibre configurations [1,2]. However, SD frameworks can be extended to reconstruct asymmetric FODs (aFODs) which can also represent more complex bending, branching and fanning fibre configurations [3,4,5]. Here we demonstrate the effect of aFODs on the anatomical representation of the cortico-spinal tract (CST) using whole brain deterministic tractography. The CST is particularly relevant because it is often challenging to reconstruct using standard deterministic algorithms which typically fail to capture the full extent of the bending and fanning of axon fibres into the lateral regions of the pre-central gyrus.

*Methods:* We compare tractography generated from two types of FOD shown in Figure 1: symmetric FODs generated using damped Richardson-Lucy SD [2], and asymmetric FODs generated using damped Richardson-Lucy SD including a graph-based regularisation term introduced in [5]. The same tractography algorithm is applied in both cases. We use a deterministic version of the tracking algorithm described in [3], where instead of probabilistically selecting the next stepping direction, we select the direction of maximum FOD amplitude within the current cone of propagation. We generated whole brain tractograms of each type for 10 HCP subjects [6] downsampled to 2.5mm resolution. Downsampling was performed to enable direct comparison with high resolution data, which will be used as reference in future analyses. The right CST was then dissected using MegaTrack [7]. A probability map was calculated for each FOD type showing the probability of finding the CST at each voxel across the 10 subject sample.

*Results:* Figure 2 shows the 3D probability maps for sFOD (blue) and aFOD (orange) tractography dissections of the CST. With aFOD tractography the lateral projections are more uniformly distributed across the pre-central gyrus and have higher probability compared to the sFOD version. In Figure 3 we show a comparison of CST dissections for two individual subjects. The individual dissections further demonstrate the more complete reconstruction of the CST when using aFODs, with missing anatomical features in the sFOD dissections. Overall, we find that the reconstructed CST dissections increase in volume by an average of 13.8% across the 10 subjects when using aFODs.

*Conclusion:* These results indicate that tractography using asymmetric FODs may enable more accurate anatomical representations of white matter pathways. While more testing is required, we think this is likely due to the ability of aFODs to represent more complex bending and fanning fibre configurations than sFODs, as well as to the smoothing present in the aFOD field introduced by the SD regularization term. Further work is ongoing to investigate the advantages of aFOD tractography on other white matter pathways, as well as fine tuning modelling and tracking parameters.

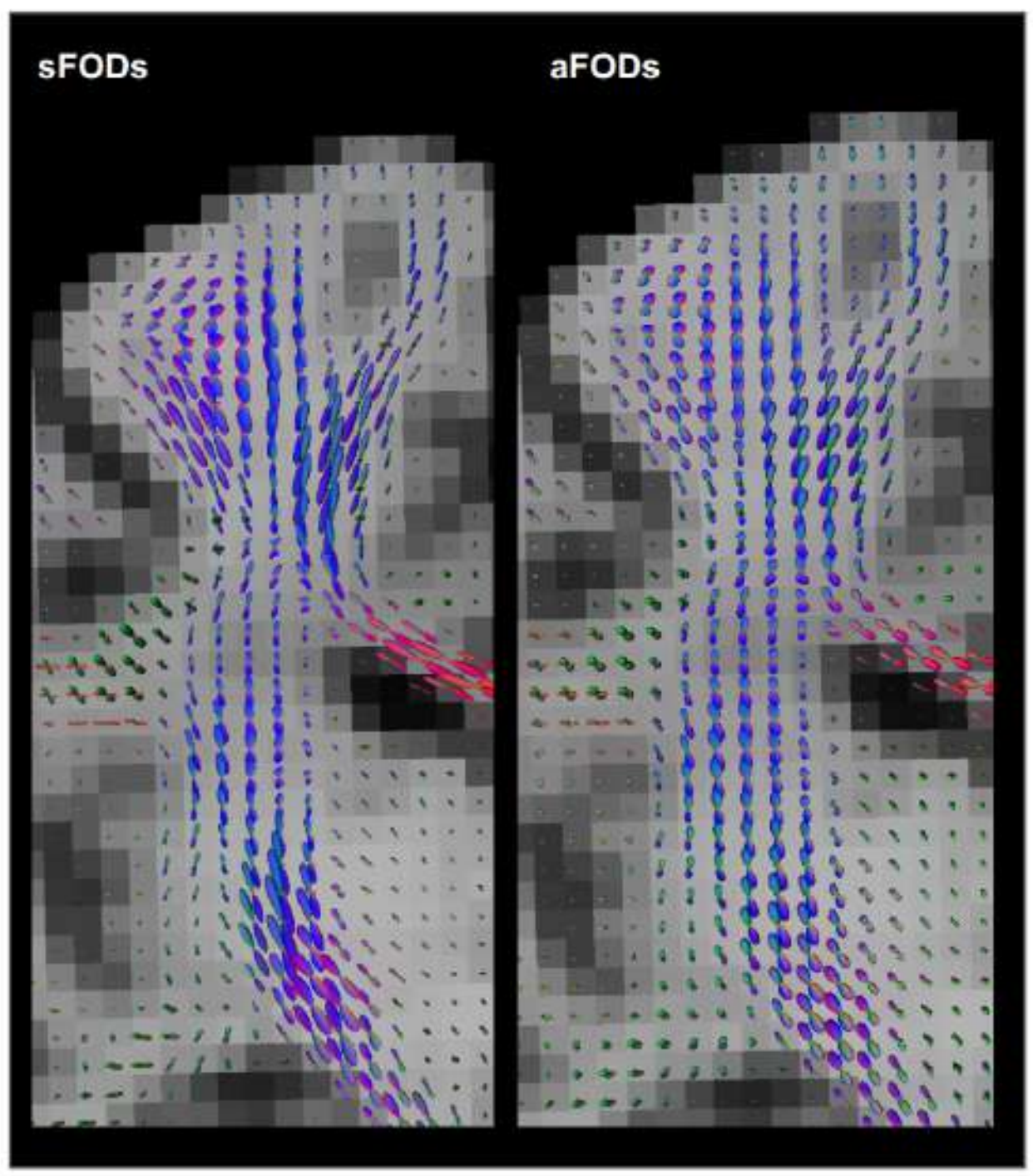

Figure 1: Comparison of symmetric FODs (left) and asymmetric FODs (right) in the CST.

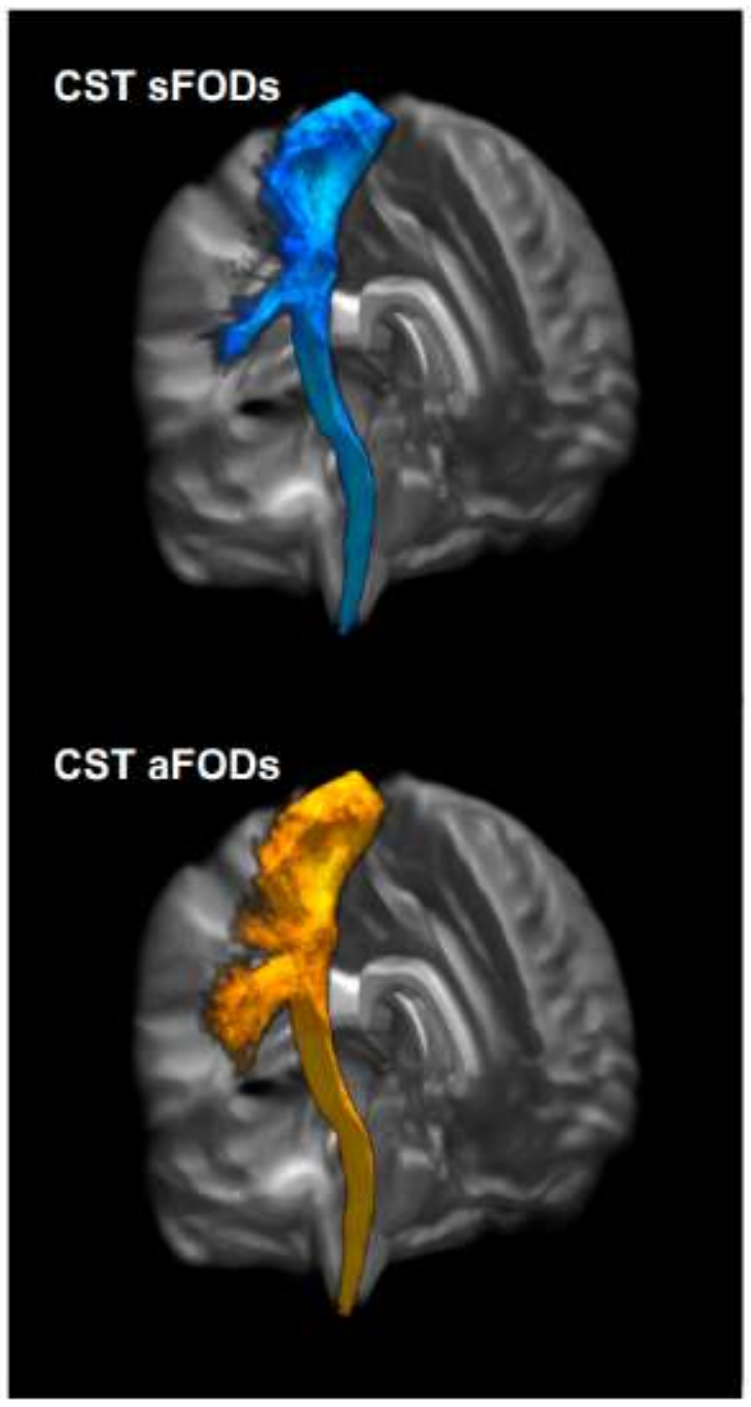

Figure 2: Probability maps of CST dissections from sFOD and aFOD tractography.

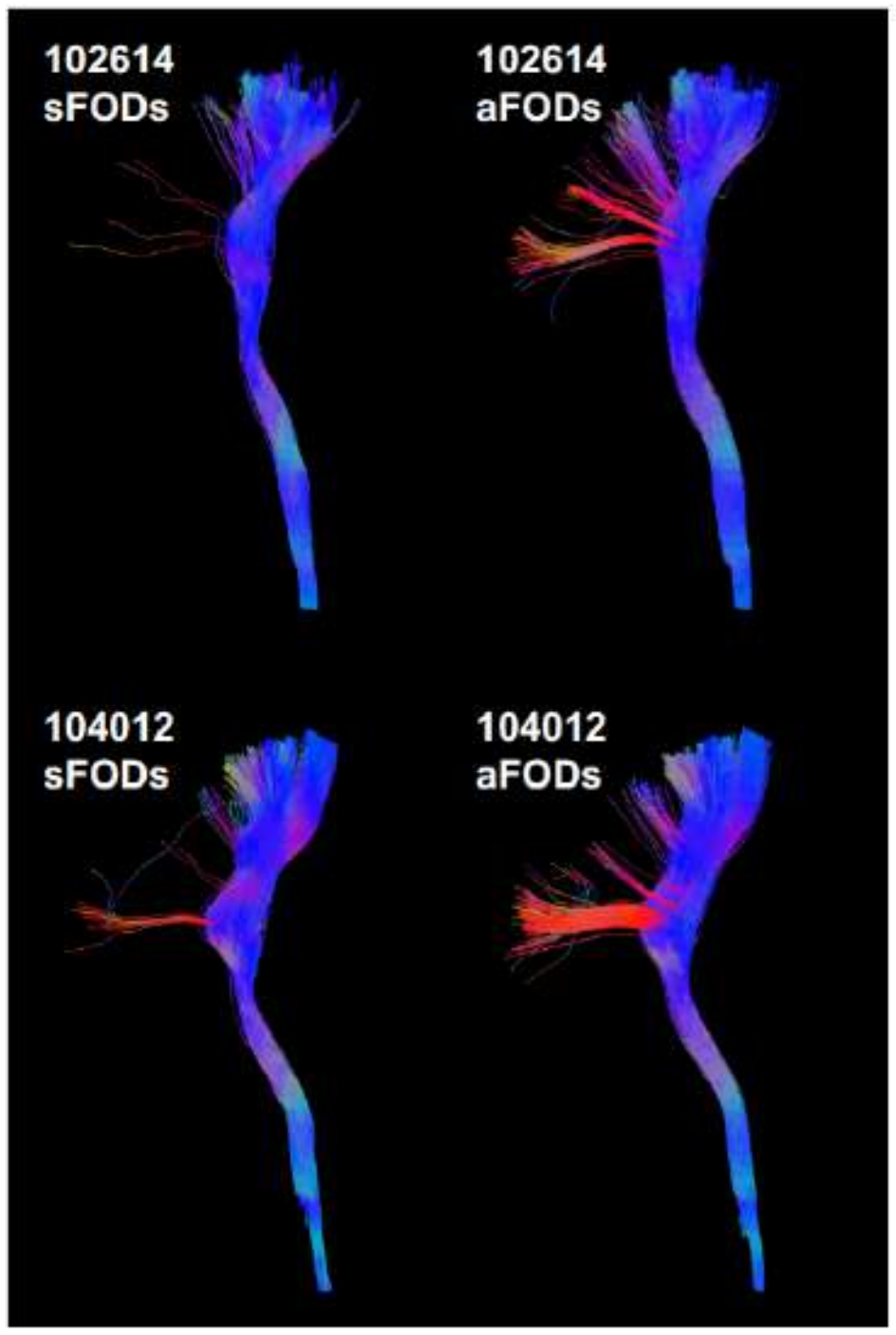

Figure 3: Comparison of individual CST dissections from sFOD and aFOD tractography for two HCP subjects.

# Tractography on Implicit Neural Representations of Diffusion MRI

Sanna Persson[1], Fabian Leander Sinzinger[1], Rodrigo Moreno[1]
[1]Department of Biomedical Engineering and Health Systems, KTH Royal Institute of Technology, Huddinge, Sweden

## I. Introduction

From a practical perspective, the acquisition of diffusion-weighted images is often constrained by scan time, resulting in a sparsely and non-uniformly sampled $Q$-space. To address the limitations of sparsely and non-uniformly sampled $Q$-space, which hinder accurate modelling and affect downstream tractography, we propose representing the raw DWI signal using *implicit neural representations* (INRs). Since INR-based DWI reconstructions only approximate the measured signal, it is important to assess their effect on practical use cases. In this work, we focus on streamline tractography derived from the reconstructed data.

## II. Methods

### A. Input encoding

In previous work on INRs, appropriate input encodings have been shown to improve learning (e.g., [1]). Following convention, we encode the spatial and b-value coordinates $(x, y, z, b)$ via Gaussian random features. For the $Q$-space coordinates $\mathbf{p} = (\alpha, \beta) \in \mathbb{S}^2$, we propose to encode them with real spherical harmonic basis functions up to degree $\ell_{\text{max}}$ at the respective position $\mathbf{p}$. We use even degrees only ($\ell = 0, 2, \dots$), enforcing antipodal symmetry.

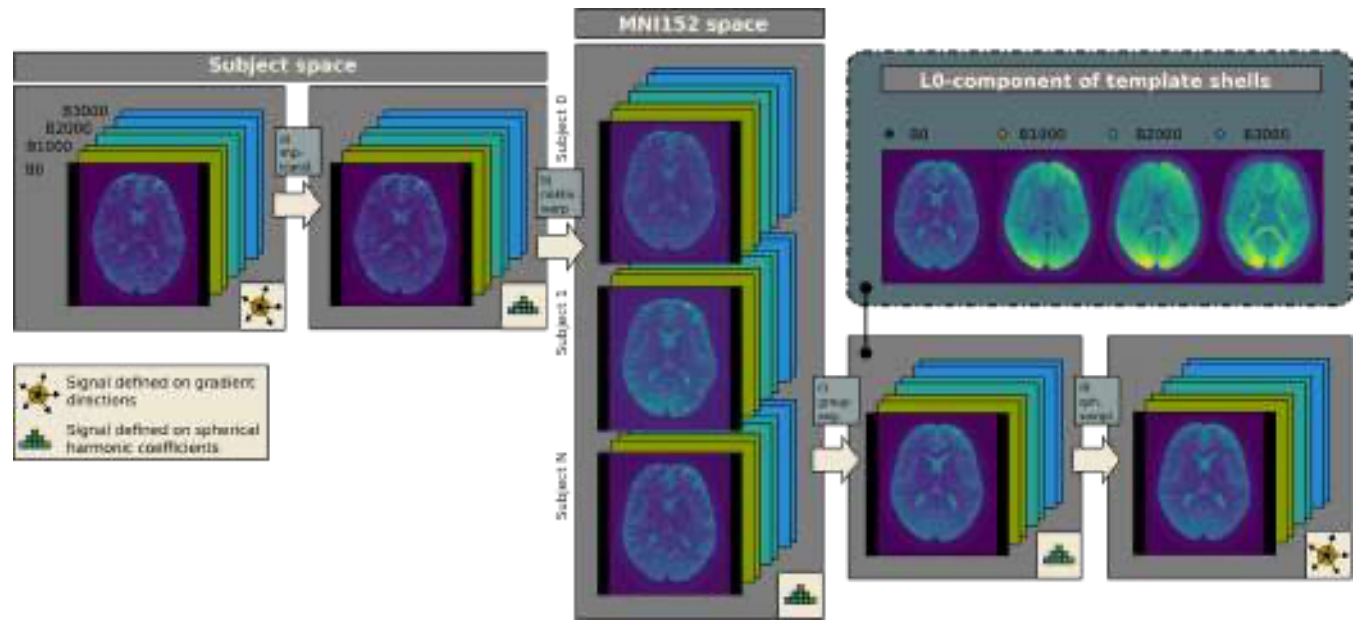

Fig. 1. a) The DWI signal is transformed into the spherical harmonics domain. b) Individual shells in spherical harmonics form are affine-transformed and nonlinearly warped to MNI152 space with deformations obtained from ANTS [2]. c) Spherical harmonics coefficients are averaged across 50 subjects. d) A DWI template is reconstructed by sampling the spherical harmonics representation.

### B. Implicit Neural Representations

We implement three different INR architectures and evaluate their performance on learning the representation of the full diffusion table. The base model consists of a three-layer perceptron with leaky ReLU activations. We further implement the SIREN [3] and WIRE [4] architectures for improved high-frequency representations.

### C. Dataset

We evaluate tractography generation from our INR-generated DWI surrogates on a multishell DWI template. The template was constructed from 50 HCP subjects by first applying a nonlinear transformation from subject space to MNI152, followed by averaging the separate shells in spherical harmonic coefficient form (cf. Figure 1).

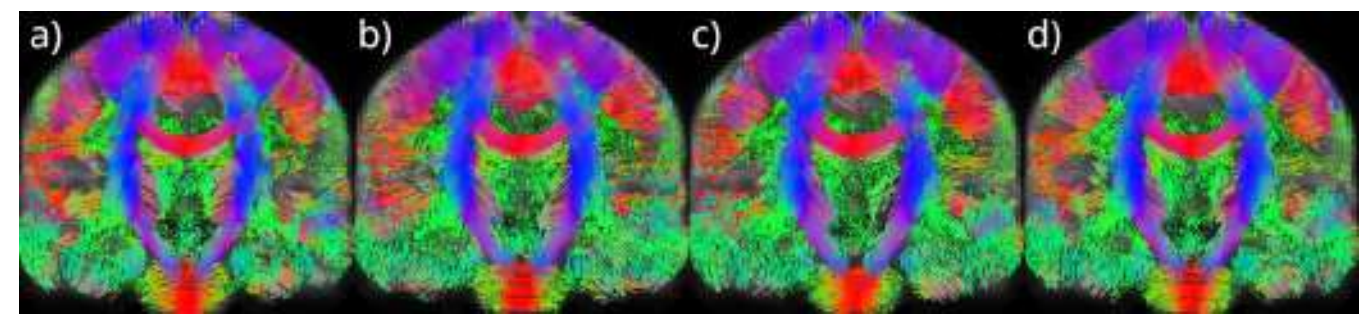

Fig. 2. Coronal slice showing derived streamlines from: (a) ground truth, (b) ReLU, (c) Siren, and (d) Wire INR architectures.

## III. Results

To investigate the validity of the proposed INR methods, we compare tractograms derived from the INR DWI with the tractogram obtained directly from the group template using *SD_STREAM* from MRtrix3[5]. Preliminary results are reported quantitatively in Table I and qualitatively in Figure 2. The Wire architecture generates the closest tractogram to the original one.

TABLE I
Average global streamline metrics per INR model

| Model | $\ell$ (mm) | $s$ (mm) | $\tau$ | $\kappa$ (rad) | $w_z$ | $c$ | $e$ |
|---|---|---|---|---|---|---|---|
| Original | 52.1 | 36.8 | 1.52 | 0.0161 | −0.0043 | 29.8 | 1.20 |
| ReLU | 51.1 | 40.0 | 1.30 | 0.0153 | −0.0040 | 34.0 | 1.14 |
| Siren | 48.4 | 38.3 | 1.29 | 0.0133 | −0.0025 | 32.9 | 1.13 |
| Wire | 51.1 | 37.9 | 1.42 | 0.0136 | −0.0014 | 31.6 | 1.17 |

**Symbol legend** $\ell$: streamline length, $s$: span (endpoint distance), $\tau$: tortuosity, $\kappa$: mean curvature, $w_z$: winding (z-axis), $c$: compactness ($s^2/\ell$), $e$: elongation ($\ell$/bbox diameter).

## IV. Conclusion

We explored the use of INRs for super resolution in the spatial and $Q$-space domains. Preliminary results show promise for $Q$-space interpolation from sparse data and for downstream applications such as tractography. One limitation of our current evaluation is that we derive tractograms by applying streamline tractography to a discretely sampled DWI representation. For future work, we aim to leverage the continuous pointwise signal representation offered by the learned INRs directly within the tractography process.

## Acknowledgments

We thank the National Academic Infrastructure for Supercomputing in Sweden (NAISS) for the use of Alvis. The project is partially funded by the Swedish Research Council through grant 2022-03389, Digital Futures and MedTechLabs.

## Title: Connectivity Patterns across Bipolar Disorder Stages: a Tractography-based Graph Analysis

**Authors:** Serena Capelli,[1]* Alberto Arrigoni,[1]* Salvatore Saluzzi,[2] Paolo Patani,[2] Stefano Martinelli,[2] Anna Caroli,[1] Simonetta Gerevini,[2] Annabella Di Giorgio.[2]

* These authors contributed equally to this work.

**Affiliations:** 1) Istituto di Ricerche Farmacologiche Mario Negri IRCCS, Italy; 2) ASST Papa Giovanni XXIII, Italy.

***Introduction.*** Diffusion MRI (DW-MRI) tractography is a powerful, non-invasive technique for mapping the brain connectivity and has become increasingly important for studying structural connectivity in psychiatric diseases.[1,2] In this study, we applied advanced DW-MRI processing through a pipeline integrating single-shell 3-tissue constrained spherical deconvolution (SS3T-CSD), anatomically constrained (ACT) probabilistic tractography, and graph connectivity analysis to a stratified cohort of individuals at different stages of Bipolar Disorder (BD), including those at familial risk.

***Methods.*** Subjects at risk of BD and patients with established BD were recruited from two clinical centers in Italy: ASST Papa Giovanni XXIII Hospital (Bergamo) and ASST Fatebenefratelli Sacco Hospital (Milan). Brain MRI scans were acquired prospectively between January and August 2024 at the Neuroradiology Unit of the Bergamo center using a 3T General Electric MRI scanner (Discovery MR750w; GE Healthcare, Chicago, IL).

Participants (*n* = 45) were classified according to Kupka and Hillegers staging:[3] Stage 0 (*n* = 7, having first-degree relative with BD, and not having psychiatric symptoms), Stage 1 (*n* = 6, having familial risk, and subthreshold symptoms or depression), Stage 2 (*n* = 8, BD patients with first hypo/manic episode), Stage 3 (*n* = 13, BD patients with recurrent mood episodes), and Stage 4 (*n* = 11, BD patients with chronic non-remissive course). Given their conceptual similarity and small sample sizes, Stage 0 and Stage 1 were combined into a single group (Stage 0-1: *n* = 13) for analysis.

The MRI acquisition protocol included a DW-MRI AP single-shell scan (b-values: 0 and 3000; number of directions: 45; voxel size: 2.4x2.4x2.4 mm), a b0 PA, and a T1-weighted axial scan (voxel size: 0.5x0.5x0.5 mm).

An in-house MRI processing pipeline was developed in Python (v3.7.10) utilizing the following packages: MRtrix3Tissue (v5.2.9), MRtrix3 (v3.0.4), FSL (v6.0.5), and FreeSurfer (v7.4.1). The pipeline included: correction of noise, artifacts, and distortions in the input scans; regional parcellation using FreeSurfer and the T1-weighted image; SS3T-CSD; probabilistic tractography via the iFOD2 algorithm; and connectivity analysis. Pre-processing also employed the Python tool DESIGNER-v2, incorporating denoising via MPPCA and Gibbs ringing correction (RPG) for partial Fourier images.

EPI distortion correction, eddy current and motion correction was achieved using *dwifslpreproc*. B1 inhomogeneity correction was computed during FOD normalization.

To further mitigate geometric distortions, SyN registration was performed between the topup-corrected image and the skull-stripped T1-weighted scan, with skull-stripping conducted using FreeSurfer's *mri_synthstrip*. After SS3T-CSD and iFOD2-ACT tractography (generating 5 million streamlines), the COMMIT2 algorithm was applied to reduce tractogram density based on diffusion signal fitting and anatomical coherence. The final tractogram was used to generate the connectome for graph analysis.

Brain connectivity measures were obtained using the *bctpy* Python package and compared across patient groups at different disease stages. For node-wise metrics, network-wide mean values were also calculated.

Group differences were assessed using the Kruskal-Wallis test, and pairwise comparisons with Wilcoxon rank-sum tests. Bonferroni correction was applied to adjust for multiple testing. Statistical significance was set at $p < 0.05$. Statistical analyses were performed using R (v4.3.2).

***Results.*** Mean nodal strength showed a significant group effect (Kruskal-Wallis, $p = 0.024$), with post hoc analysis indicating reduced strength in Stage 4 compared to Stage 2 (Wilcoxon, $p = 0.002$) and a trend toward significance when compared to Stage 3 ($p = 0.072$). Among node-wise metrics, nodal betweenness of the left posterior cingulate cortex showed (PCC) the most significant effect (Kruskal-Wallis, $p < 0.001$) that remained significant after Bonferroni correction.

***Conclusion.*** These results highlight the progressive structural alterations of the network in the stages of BD and the potential of DW-MRI-based biomarkers in its progression.

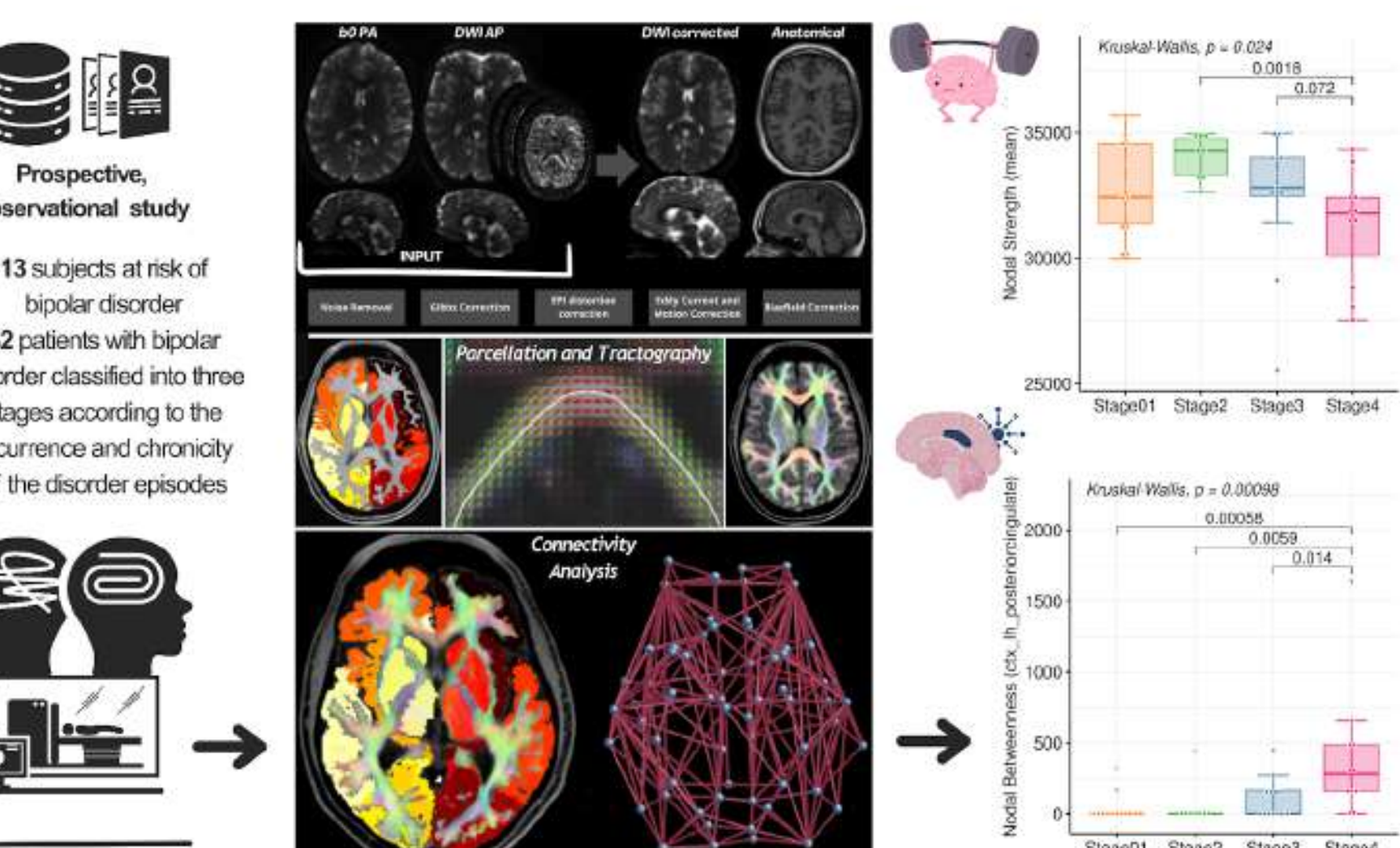

**Title:** Predicting Facial Nerve Condition and Functional Outcome in Cerebellopontine Angle Tumours Using MRI-Based Models

**Authors:** Alberto Arrigoni[1], Andrea Mangili[1], Serena Capelli[1], Giulio Pezzetti[2], Barbara Frigeni[2], Rachele Bivona[2], Giovanni Danesi[2], Anna Caroli[1], Simonetta Gerevini[2]

**Affiliations:** 1 Istituto di Ricerche Farmacologiche Mario Negri IRCCS, Italy; 2 ASST Papa Giovanni XXIII, Italy

***Introduction***. Cerebellopontine angle (CPA) tumors are typically benign but can exert a mass effect on adjacent structures, including the facial nerve (FN). Surgical resection is often indicated. However, the surgery carries a risk of iatrogenic nerve injury, potentially resulting in transient or permanent facial palsy [1]. This study investigates anatomical and diffusion-weighted MRI (DW-MRI) 's role in pre-surgical planning, first focusing on accurate reconstruction of the FN. Morphological biomarkers and microstructural features from DW-MRI are extracted using radiomic analysis and diffusion tensor imaging (DTI). These imaging-derived biomarkers are integrated with clinical variables to train machine learning models to predict the facial nerve's condition before surgery, the postoperative functional outcomes, and long-term follow-up status.

***Methods.*** Forty-seven patients with CPA who underwent surgery and pre-operative MRI examination were included in the study. The MRI acquisition protocol included a DW-MRI AP single-shell scan (b-values: 0 and 1500; 50 directions; voxel size: 2x2x2 mm;), a b0 PA, a post-contrast volumetric T1-w scan and a volumetric T2-w scan. An in-house MRI processing pipeline was developed in Python (v3.7.10) using the following tools: MRtrix3Tissue (v5.2.9) and FSL (v6.0.5). The pipeline included correction of noise, artifacts, and distortions in the input scans; lesion segmentation on the T1-weighted image; SS3T-CSD and probabilistic tractography via the iFOD2 algorithm by manually placing reconstruction seeds in anatomical references.

***Results.*** The pipeline successfully reconstructed the facial nerve in all patients, demonstrating the role of DW-MRI and tractography for CPA resection pre-surgical planning. The facial nerve exhibited significant microstructural alterations on the tumor-affected side, including increased fractional anisotropy (FA) and decreased mean diffusivity (MD) compared to the contralateral healthy side. Conventional prognostic indicators—including the blink reflex and Koos grading—showed positive correlations with diffusion metrics and lesion volume. Compound Muscle Action Potential (CMAP) correlated inversely with diffusion values and FN tract length. Postoperative and 1-year House-Brackmann (HB) grades of facial nerve function were significantly associated with lesion volume, FN length, and CMAP values. Machine learning models trained on all extracted features to predict nerve adherence to the tumor (FN integrity) and HB outcomes achieved accuracies exceeding 0.90, highlighting the potential of integrated radiomic-clinical approaches for prognostication.

***Conclusion.*** This study supports the utility of diffusion tractography in CPA resection pre-surgical planning and underscores the prognostic value of both morphological and diffusion-based MRI biomarkers. When combined with clinical and neurophysiological data, these features can improve the prediction of facial nerve integrity, immediate postoperative function, and long-term outcomes following CPA tumor resection.

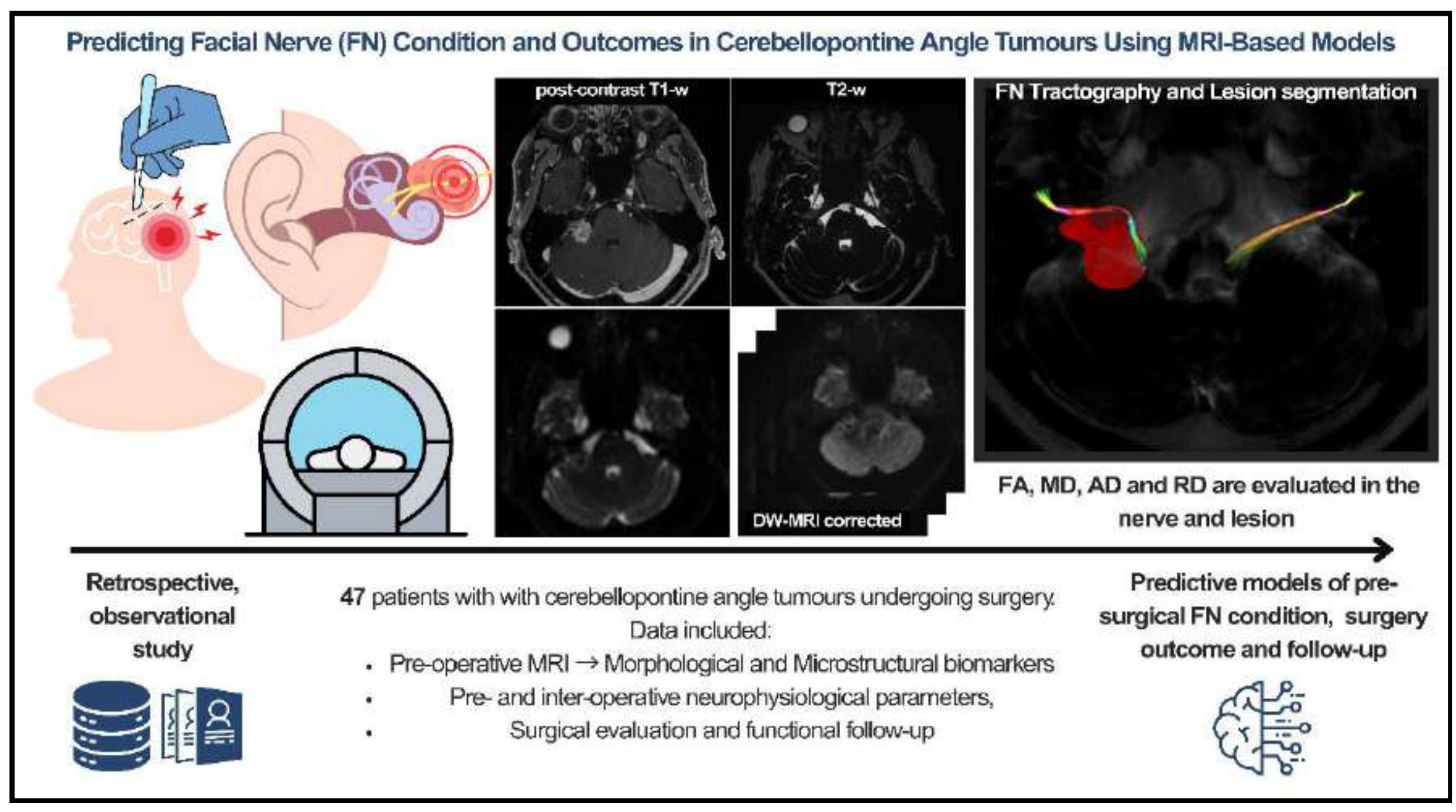

# Assessment of Different Tractography Methods for Superficial White Matter Reconstruction

**Xi Zhu[1], Xiaofan Wang[1], Yuqian Chen[2], Lauren J O'Donnell[2], Fan Zhang[1]**

[1] University of Electronic Science and Technology of China, Chengdu, China

[2] Brigham and Women's Hospital, Harvard Medical School, Boston, USA

## Introduction

The superficial white matter (SWM) contains short association fibers connecting nearby cortical regions [1]. These fibers are essential in neurodevelopment, aging, and various neuropsychiatric disorders [2]. In this study, we evaluate the performance of different tractography methods in the task of SWM fiber reconstruction. We adopt a filtering strategy proposed by Aydogan et.al. [3] to extract SWM fibers from whole brain tractography. We compare four distinct tractography methods: iFOD2 [4], surface-seeding-based iFOD2, PTT algorithm [5] and UKF [6].

## Method

We used MRI data from 10 HCP-YA [7] subjects for evaluation. Specifically, we create a grey–white matter interface as a seeding mask. For surface-based iFOD2, seed points were taken from the white matter surface (WSM) mesh. Finally, masks from the white–grey matter were used to constrain tractography. We employed three quantitative metrics to assess performance of different tractography methods: 1) Streamline Length; 2) Coverage: Percentage of WSM vertices considered "closest" to at least one end of one streamline; 3) Coverage bias [3]: Proportion of "covered" WSM vertices residing in gyri, corrected by proportion of all WSM vertices residing in gyri.

Table 1. Tractogram differences between different methods

| | iFOD2 | Surface-based iFOD2 | UKF | PTT |
|---|---|---|---|---|
| Mean Length (mm) | 23.35 | 23.20 | 25.47 | 22.99 |
| Coverage (%) | 81.89 | 83.25 | 68.09 | 85.87 |
| Coverage bias | 1.011 | 1.022 | 0.960 | 1.016 |

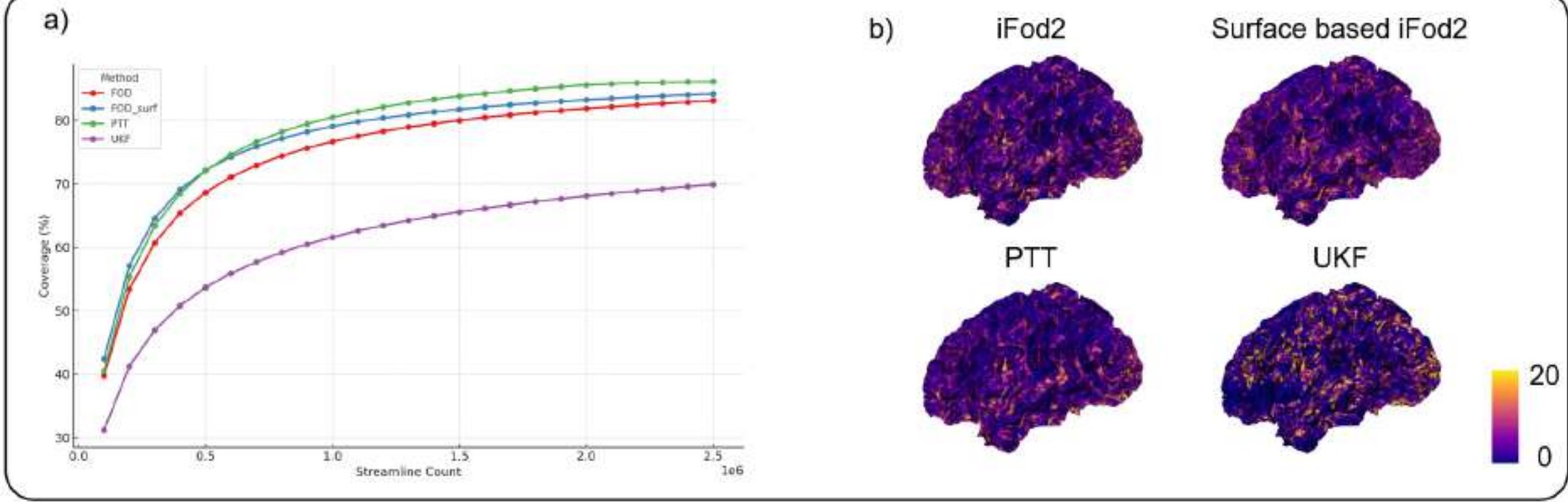

Fig.1. Results of different methods

## Results

The UKF tends to produce longer streamlines, while the PTT algorithm shows higher coverage at the same streamline count. Higher coverage bias indicates more streamlines terminating in gyral regions, consistent with U-fiber characteristics. Figure 1a shows how coverage increases with streamline count, and Figure 1b maps fiber density on the white matter surface.

## Conclusion

It is evident that the performance of different tractography methods in SWM reconstruction varies substantially.

**Structural connectivity-based individual parcellations using various tractography algorithms**

C. Langlet[1], D. Rivière[1], B. Herlin[1], I. Uszynski[1], C. Poupon[1], J.-F. Mangin[1]
*[1]Université Paris-Saclay, CEA, CNRS, Neurospin, Baobab, Saclay, France*

## Introduction

Mapping the human brain has been a long-standing goal of the neuroscience community as cerebral maps provide a spatial referential to conjointly study brain structure and function. One crucial application is the study of the connectome that needs parcellations to define structural pathways and functional interactions between regions of the brain. Such parcellations usually stem from atlases projected onto individuals via a cortical folding-based alignment, however this method fails to represent the anatomical peculiarities of individuals (Mangin, 2019). In this work, we generated individual parcellations using a data-driven algorithm based on the structural connectivity obtained from two tractography algorithms: FSL and MRtrix.

## Methods

We first computed FSL and MRtrix tractography data from the Human Connectome Project dataset to create complete connectivity matrices for all individuals. Using a group of 200 individuals and the Constellation software (Lefranc, 2016), we then generated average subdivisions of the Desikan atlas based on connectivity profiles at different resolution levels. These group subdivisions were then projected onto each individual of the dataset using their individual connectivity profiles. To create our final parcellations, we selected an adequate number of subdivisions based on cortical thickness and diffusion-based cortical microstructure features processed with Freesurfer (Glasser, 2013) and Ginkgo (Herlin, 2024) softwares.

## Results

We obtained group and individual parcellations for FSL and MRtrix tractography algorithms. At the whole brain level, our criterion - based on structural data - interestingly selected almost the same number of regions (399 for FSL and 400 for MRtrix). Additionally, for both FSL and MRtrix parcellations, individual parcels present a higher structural connectivity homogeneity than group parcels projected via a Freesurfer alignment.

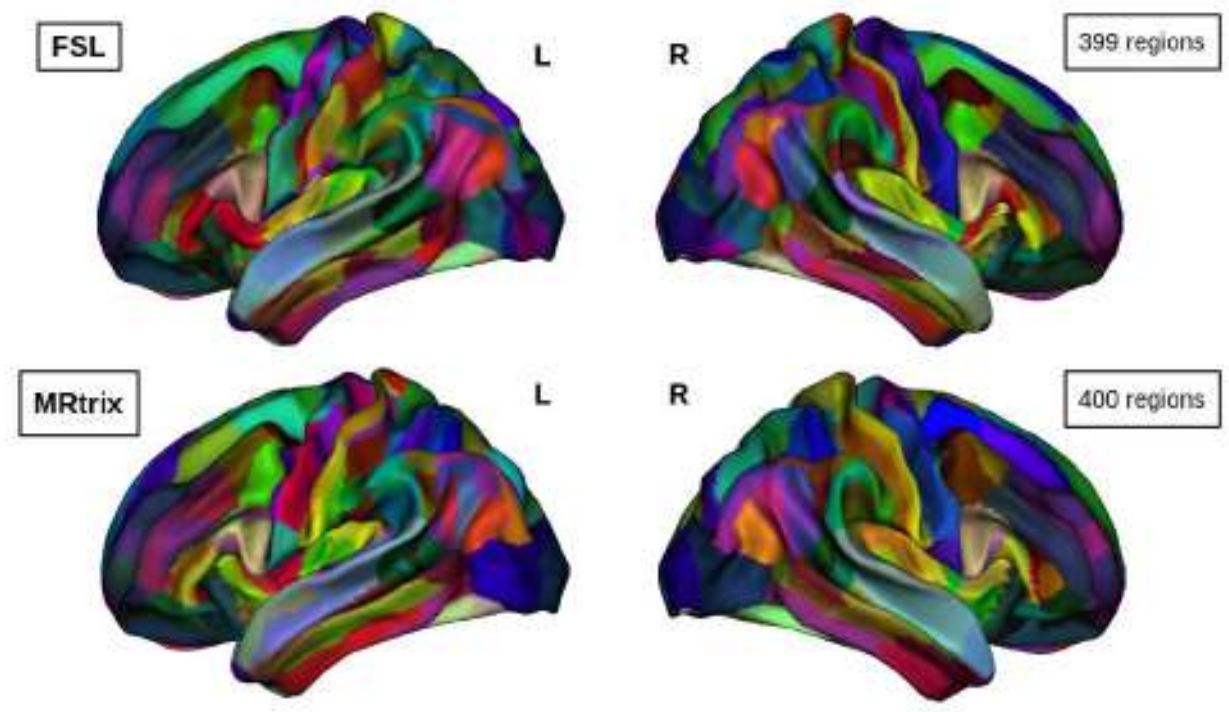

**Figure:** Group parcellations based on FSL (top) and MRtrix (bottom) tractographies

## Conclusions

This work paves the way towards the study of the human connectome by proposing coherent parcellations between individuals based on tractography data. Comparison of parcellations obtained with different tractography algorithms may deepen our understanding of the information encoded by each software and help us describe the human connectome more accurately.

# Using Large Language Models to Inform Tractography

*Authors: Elinor Thompson, Tiantian He, Anna Schroder, Ahmed Abdulaal, Daniel C. Alexander. Hawkes Institute, University College London, UK*

**Introduction** Prior knowledge on the existence and route of white matter pathways has been successfully used to constrain tractography for bundle reconstruction, leading to improved accuracy[1]. However, it is challenging to scale this approach to connectomics. Large language models (LLMs) provide a route to synthesise large amounts of textual information into quantitative outputs, and demonstrate promising results in neuroscience tasks due to their extensive training on the scientific literature[2]. Here, we develop and evaluate a pipeline using LLMs to provide quantitative priors for connectomics and demonstrate how this approach can be used to supplement microstructure-based tractography filtering approaches, by retaining connections whose presence is supported by the neuroanatomy literature.

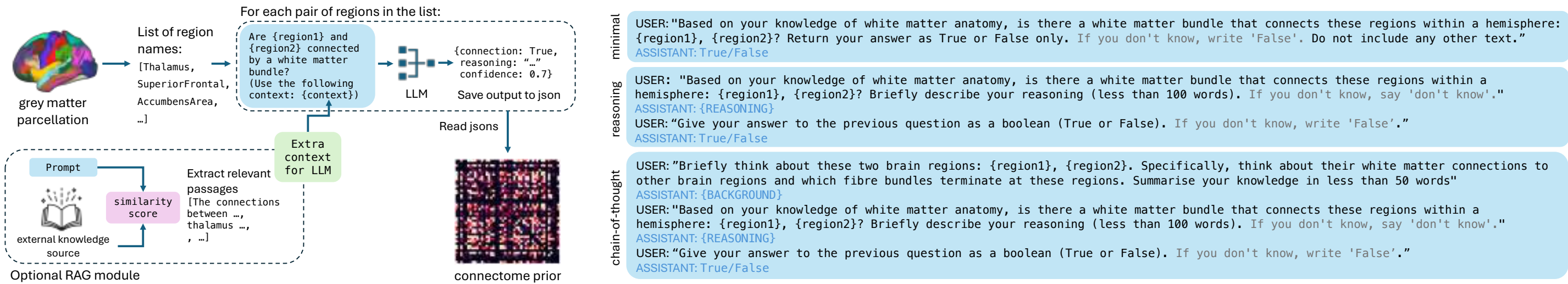

*Figure 1. Left: An overview of our pipeline for generating connectome priors from the LLM. Right: Overview of the three different prompting approaches: minimal, reasoning, and chain-of-thought (COT). The additional phrases encouraging the LLMs to express uncertainty are shown in grey.*

**Methods** The input to our pipeline is a list of brain regions from a parcellation. For each region pair, we query the existence of a white matter bundle connecting them. The outputs are saved to json files, from which the connectome priors can be retrieved (Fig. 1). We considered two parcellations, Desikan-Killiany and the Human Connectome Project's multi-modal parcellation (HCP-MMP)[3].

**Benchmarking the white matter knowledge of the LLMs:** We used a gold-standard tractography atlas[4] to generate an evaluation set, comprised of the 50 most strongly connected region pairs in the atlas and a random sample of 50 unconnected pairs. Different prompts and models were compared for their accuracy in classifying region pairs as connected or not connected. We also tested an approach where the LLM is explicitly encouraged to admit uncertainty (Fig. 1). For OpenAI models (gpt-4o and gpt4-turbo), we retrieved a model confidence score from the log-probs variable assigned to the Boolean output.

**Retrieval-Augmented Generation (RAG)** was used to integrate external knowledge sources into the prompt. Initial experiments showed that the LLM was lacking knowledge about regions in the HCP-MMP, so we used the supplementary information relating to the parcellation as a knowledge source[3]. The 80-page document was split into semantic chunks. We used BM25 keyword search to find three relevant text chunks relating to each region, which were summarised by the LLM and included in the prompt.

**Combining LLM-derived priors with Microstructure Informed Filtering:** Tractography filtering aims to remove false positive streamlines, but this can result in the loss of true positives.[5] We propose that LLM-derived priors can help to select neuroanatomically plausible connections to retain during filtering. We performed anatomically constrained tractography with MRtrix3[6] on 5 HCP subjects, generating 5 million streamlines for each. We then apply COMMIT2 filtering[7] but retain connections with high confidence scores from our LLM pipeline (confidence>0.5, minimal prompt, gpt4-turbo). We evaluate this approach by comparing the accuracy of a network diffusion model (NDM)[8] for modelling Alzheimer's pathology spread across three binary connectomes: unfiltered, COMMIT2-filtered, and LLM+COMMIT2-filtered. The target data for the model is an average tau-PET SUVR map from 155 amyloid- and tau-positive participants from the ADNI study, as described in[9]. Our hypothesis is that a more accurate connectome will provide a better model fit to the tau-PET data. We perform a null comparison for each connectome by repeating the process 1000 times with the same number of randomly chosen connections retained after COMMIT2 filtering.

## Results

| model | minimal prompt | | reasoning prompt | | chain-of-thought prompt | |
|---|---|---|---|---|---|---|
| | standard | uncertainty | standard | uncertainty | standard | uncertainty |
| claude3.5-sonnet | 0.76 ± 0.06 | 0.79 ± 0.03 | 0.80 ± 0.07 | **0.85 ± 0.03** | 0.66 ± 0.02 | 0.75 ± 0.01 |
| llama3-8b | 0.82 ± 0.01 | 0.83 ± 0.01 | 0.57 ± 0.01 | 0.84 ± 0.01 | 0.56 ± 0.01 | 0.50 ± 0.00 |
| gpt-4o | 0.78 ± 0.00 | **0.85 ± 0.02** | 0.65 ± 0.02 | 0.80 ± 0.02 | 0.68 ± 0.02 | 0.74 ± 0.04 |
| gpt-4-turbo | **0.89 ± 0.01** | **0.85 ± 0.04** | **0.82 ± 0.01** | 0.83 ± 0.02 | **0.89 ± 0.02** | **0.91 ± 0.02** |

*Table 1. Evaluation accuracy across models and prompts, calculated as agreement with a gold-standard tractography atlas across 100 region pairs from the Desikan-Killiany atlas. Results in bold show the best performing model for each prompt, and green indicates the best performing prompt for a given model. Mean±std across four repeats.*

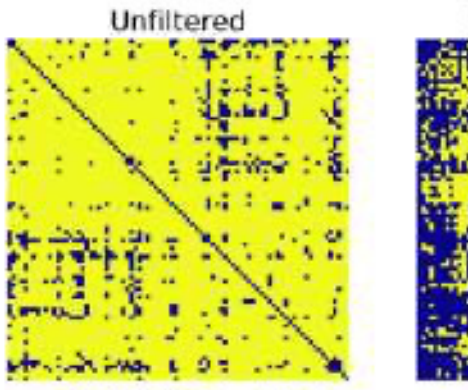

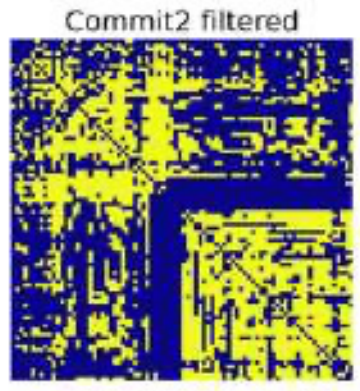

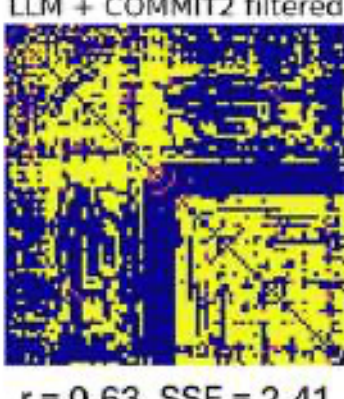

*Figure 2. Example results from LLM-augmented filtering. Connections retained using the LLM-priors are shown in pink on the right-hand plot. Model fit with the NDM is shown below: r = Pearson's correlation between tau-PET and model prediction, SSE = sum of squared errors.*

**Benchmarking performance:** OpenAI's gpt4-turbo model performed best across all prompt types. The most accurate was the uncertainty variant of the COT prompt (91%), although similar performance was attained with the minimal prompt. An additional stability analysis with ten repeats of the minimal prompt with gpt4-turbo showed consistent responses for 84 out of 100 region pairs, with an accuracy of 0.89±0.03 (range: 0.82-0.92). The LLMs had lower confidence scores for answers that contradicted the tractography atlas (independent t-test, minimal prompt: t=9.3, $p=1.3\times10^{-18}$, COT prompt: t=5.1, $p=4.9\times10^{-17}$).

**RAG:** For the HCP-MMP, providing contextual information about the parcellation increased the average accuracy in the evaluation set from 72±0.03% to 80±0.04% with the COT prompt (mean±std across four repeats). For the minimal prompt, accuracy slightly decreased with the additional context (from 0.73±0.04 to 0.70±0.02), due to an increase in false positives.

**LLM-Augmented filtering:** retaining connections selected by the LLM during filtering improved the fit of the pathology model for 4 of the 5 subjects, compared to filtering with COMMIT2 alone (mean±std improvement in $r^2$ = 20 ±24%). This improvement was significant compared to a null-distribution where the same number of randomly chosen connections were retained after filtering ($p<0.01$), indicating that the LLM-derived priors preserve valid connections that would otherwise be removed.

**Discussion & Conclusions** LLMs can provide accurate information about structural connectivity between grey matter regions, providing complementary information to tractography. A combination of a minimal prompt with the gpt-4-turbo model provides a good balance between accuracy and token efficiency, while other more detailed prompts can be used to probe the models' internal reasoning. We show that the models were more uncertain in their incorrect answers, which provides a route for uncertainty quantification in downstream tasks. We have also demonstrated how external knowledge can be used to improve accuracy in situations where the LLM's internal knowledge is lacking, although this is dependent on the prompting strategy. Our filtering approach showed promising results, with the retention of relatively few connections leading to a significant improvement in the accuracy of a pathology spreading model (Figure 2). Future work will extend this to more subjects and test with differing levels of regularisation in the COMMIT2 filtering.

# Towards Whole-Brain Tractography of the Mouse from Serial Optical Coherence Tomography

Charles Poirier[1], Frans Irgolitsch[2], Joël Lefebvre[3], Maxime Descoteaux[1]

[1]Université de Sherbrooke, QC, CA. [2]Polytechnique Montréal, QC, CA. [3]Université du Québec à Montréal, QC, CA.

**Context.** Optical coherence tomography (OCT) is an imaging modality relying on the intrinsic contrast of a sample. When applied to brain tissues, the OCT contrast is primarily driven by the myelin reflectivity [1]. As the OCT signal is acquired at multiple depths simultaneously, OCT is suitable for serial blockface histology [2] (SBH), which consists of imaging the surface of a sample before thinly slicing the top layer out, revealing a new surface imaged in turn. Due to its high resolution, on the order of microns, and its 3D nature, serial OCT (S-OCT) offers promise for studying long-range white matter (WM) connections at the microscopic scale. Fibre tractography has already been applied to polarization-sensitive OCT acquisitions [8, 9], where orientation information is readily available, but not to OCT alone. However, working directly on the OCT reflectivity signal has its advantages: the imaging setup is simpler, less expensive and the raw data is less noisy. A disadvantage of OCT, however, is its orientation dependence – while in-plane (lateral) WM fibres come out as brighter than grey matter (GM), out-of-plane (axial) fibres actually appear darker than GM [1]. In this work, we present the first results of fibre orientation density (FOD) estimation and fully-3D probabilistic fibre tractography of the mouse brain from S-OCT acquisitions.

**Methods.** **Data acquisition and reconstruction.** The fully-automated S-OCT system and the data used in this work have been described in a previous paper [4]. The system has an axial resolution of approximately 3.5 $\mu$m and lateral resolution of around 3 $\mu$m, thus resulting in nearly-isotropic micro-scale resolution for the whole mouse brain. A dissected mouse brain is first mounted onto a motorized stage, with the dorsal side of the brain facing towards the microscope objective. Then, volumetric tiles (covering a region of $750 \times 750 \times 1200$ $\mu$m$^3$) are acquired sequentially by translating the motorized stage until the entire blockface is imaged. When done, a slice of 200 $\mu$m thickness is removed from the sample by means of a vibratome. Then, the sample is moved up by 200 $\mu$m and the newly revealed blockface is imaged. To obtain a 3D reconstruction of the whole brain, the tiles for each depth are first stitched together into a 200 $\mu$m-thick volume, and these volumes are then stacked on top of each others, as in [3]. To reduce memory requirements, the 3D volume is downsampled to 10 $\mu$m isotropic resolution. **FOD estimation and tractography.** Histological FOD are estimated using the method from [6]. First, the image gradient is estimated using derivative of Gaussian filters with standard deviation (std) of 20 $\mu$m. Then, the principal direction of each voxel is estimated using structure tensor analysis with a Gaussian window with std of 50 $\mu$m. Finally, voxels are grouped into bigger, $60 \times 60 \times 60$ $\mu$m$^3$ voxels inside which a histogram of orientations (100 bins) is built from the principal directions. The result is finally projected onto a spherical harmonics basis with maximum order 8. Probabilistic tractography is performed using a step size of 30 $\mu$m and maximum angle of 20°. Seeding is done from manually-drawn regions of interest (ROI) targeting the corticospinal tract (CST) and the anterior commissures (AC). The entire brain mask is used as the tracking mask. Streamline of length lower than 0.5 mm are rejected. The resulting tractograms are further filtered using the seeding ROI as inclusion ROI [5].

**Results and discussion.** The reconstructed S-OCT volume (10 $\mu$m resolution) is shown in Figure 1a. The axial view shows the in-plane acquisition. Despite the variations in intensities between the stitched 200 $\mu$m-thick volumes, we report a good alignment between consecutive slices. Figure 1b shows the FOD fields estimated inside two ROIs. The choice for $60^3$ $\mu$m$^3$ voxels was determined empirically, as our experiments suggest that a smaller window results in spurious fibre orientations while a bigger window results in an increase in partial volume effects. As seen for the blue ROI, the estimated FOD follow the trajectories of the underlying WM fascicles. The sagittal view for the same region shows coherent vertical FOD in regions of the CST appearing dark (out-of-plane fibres). The changes in intensities between consecutive slices introduce a bias towards in-plane orientations, resulting in a lot of FOD being aligned with the axial plane. To mitigate this problem, we need to improve our 3D S-OCT reconstruction by implementing more advanced methods for compensating the signal attenuation and for blending consecutive slices, such as suggested in [3]. As seen in the blue ROI, the FOD estimated along the corpus callosum (CC) are oriented along the anterior-posterior axis rather than the expected ventral-dorsal axis. While this could be a consequence of the in-plane orientation bias, it could be that the weak signal in this region does not allow for correctly delineating the WM fascicles. This is not a problem for the IC fibres, as the contrast with the surrounding GM results have well defined borders. Figure 1b and Figure 1c show tractography results for the AC and CST, respectively. Both reconstructions are well aligned with the underlying anatomy. Moreover, the CST fibres are successfully tracked out-of-plane. However, no streamline actually reaches the motor cortex. Looking at the FOD, we see that no secondary directions are estimated along the CC, hence blocking the streamlines from crossing it. To solve this, we may need to develop a new method for estimating FOD, taking into account the orientation dependence of the OCT signal. Working closer to the acquisition resolution of 3 $\mu$m could also help, as more structures are visible at higher resolutions. As of now, there are also no constraints forcing the tractography to remain inside WM, as the whole brain mask is used as a tracking mask. The inclusion of anatomical priors [7] from brain tissue segmentation would be important to constrain streamlines to the WM and capture the true location of WM fascicles.

Figure 1: (a.) *Reconstructed 3D volume at 10 $\mu$m isotropic resolution (scale bar is 400 $\mu$m).* (b.) *Estimated FOD (60 $\mu$m) overlayed on the 10 $\mu$m volume.* (c.-d.) *Streamlines for the AC and CST, respectively.*

**Conclusion** In this work, we showed the first results of FOD estimation and fully-3D probabilistic fibre tractography of the mouse brain from S-OCT acquisitions. Despite many challenges, this work is a first step towards state-of-the-art tractography of microscopy S-OCT data. Future works will focus on improving the 3D reconstruction method, estimating FOD at resolution lower than 60 $\mu$m and including anatomical priors to constrain the tractography to the WM.

# Probing the clinical value of tractography reconstruction for glioma surgery

Ludovico Coletta [1,2], Luca Zigiotto [3,4,5], Paolo Avesani [1,2], Silvio Sarubbo [3,4]

1 Neuroinformatics Laboratory (NiLab), Bruno Kessler Foundation (FBK), Trento, Italy | 2 Center for Mind/Brain Sciences – CIMeC, University of Trento, Rovereto, Italy | 3 Department of Neurosurgery, "S. Chiara" Hospital, Azienda Provinciale per i Servizi Sanitari, Trento, Italy | 4 Center for Cellular, Computational and Integrative Biology, Center for Medical Sciences, University of Trento, Trento, Italy | 5 Department of Psychology, "S. Chiara" Hospital, Azienda Provinciale per i Servizi Sanitari, Trento, Italy

**Introduction.** Maximizing the extent of resection while at the same time optimizing functional preservation is the key goal of modern neurosurgery[1,2]. Recent evidence suggests that the disruption of the white matter scaffold of the human brain – as derived from tractography and non-invasive magnetic resonance imaging (MRI) data – is a robust predictor of lower survival rates and more pronounced functional damages in glioma patients[2.3]. Therefore, the identification and consequent physical preservation of specific portions of the white matter subtending a common function – so called white matter bundles – is of crucial importance. Within this context, the choice of tracking strategy used for mapping white matter bundles is critical. Especially in clinical populations, the reconstruction method can return radically different overviews into the intrinsic architecture of the brain, fundamentally shaping the subsequent downstream surgery planning. Here, we probed the specificity and sensitivity of deterministic and probabilistic tractography in capturing dysconnection-induced impairments as we all as their capacity in establishing a relationship between resection distance and functional preservation in glioma patients.

**Methods.** We investigated a cohort of 72 glioma patients (mean age= 53.1, std=14.1, 47 men, 57 high grade gliomas, 34 awake surgeries). Diffusion weighted MRI data for white matte bundles segmentation was obtained before surgery. After preprocessing[1], deterministic and probabilistic tracking[1] were performed on the MRI acquisition before surgery to extract five bundles of interest: the inferior longitudinal fasciculus (IFL), the frontal aslant tract (FAT), the inferior fronto-occipital fasciculus (IFOF), arcuate fasciculus (AF), and the superior longitudinal fasciculus (SLF). For each bundle and each tracking algorithm separately, we initially contrasted the subset of patients whose bundle was left intact with the patients with partial bundle resection during surgery. The stratification was performed based on overlap/no overlap between a volumetric representation of the bundle and the resection cavities obtained within 48h after surgery[1]. For AF, FAT, and SLF we tested whether patients with an intact bundle (either in the right or left hemisphere) showed better phonemic fluency scores than patients with partial resections pre surgery, 1 week and one month after surgery via Mann-Whitney U test for independent samples. Using the same non-parametric test, for AF, FAT, and SLF we tested whether patients with an intact bundle (left hemisphere only) showed better semantic fluency scores than patients with partial resections pre surgery, 1 week and one month after surgery. Moreover, for the subset of patients whose bundle was left intact, we tried to assess whether the minimal Euclidean distance between the resection cavity and the bundles was related to functional recovery/worsening via a correlational analysis. Similarly, for the subset of patients with partial bundle resection, we tried to assess whether the volume of resection was related was related to functional recovery/worsening.

**Results.** Across bundles and for both deterministic and probabilistic tractography, an overlap between resection cavities and the bundles was consistently associated with a worse functional outcome one month after surgery. Of note, this was the case when considering both right and left hemisphere lesions for phonemic fluency, as well as left lesions only with respect to semantic impairments. We found that minimal Euclidean distance between the resection cavity and the bundles was moderately related to the degree of functional recovery (the further away, the better). Intriguingly, a qualitative inspection of the results revealed the possibility of defining a safe-minimum distance zone that avoids surgery-induced functional impairment.

**Discussion.** Although preliminary, our analysis suggests that both deterministic and probabilistic tractography are sensitive enough to be effectively used intraoperatively. By increasing sample size and algorithmically define a safe threshold that avoids/minimizes the risk for functional impairment, we aim to corroborate this important set of findings.

**Pre- versus postoperative tractography in patients with (supra)sellar tumors: correlations to optic pathway deformations and vision**

Andrey Zhylka[1], Constanze Ramschütz[2], Lioba Grundl[2], Jan S. Kirschke[2], Claus Zimmer[2], Maria T. Berndt-Mück[2], Bernhard Meyer[2], Vicki M. Butenschoen[2], Nico Sollmann[3]

[1]UMC Utrecht, Utrecht, The Netherlands, [2]TUM University Hospital, Munich, Germany, [3]University Hospital Ulm, Ulm, Germany

**Introduction.** Visual impairments are the most common clinical symptoms of (supra)sellar masses (e.g. pituitary adenomas), with visual field (VF) deficits being a primary indication for surgical intervention in these patients(1,2). Diffusion tensor imaging (DTI) is a non-invasive MRI method that provides quantitative insights into white matter structure through measures such as fractional anisotropy (FA), which characterizes axonal architecture(3). Specifically, DTI allows to assess the extent of anterior optic pathway compression caused by (supra)sellar masses, giving clinicians and neurosurgeons detailed information on visual field compromise in relation to clinical findings and tumor extents(4-6).

However, the relationship between VF impairments and changes within the optic tract (OT) and optic radiation (OR) bundles is overlooked. In this study, we investigated whether FA of the OT and OR correlates with preoperative VF impairments and postoperative VF improvements in patients with pituitary adenomas.

**Methods.** ***Data*.** The MRI scans of 18 patients with pituitary adenomas were used. The dataset consisted of T1-weighted (voxel resolution $0.75x0.75x0.75mm^3$ isotropic) and diffusion scans (voxel resolution $2x2x2mm^3$ isotropic) that consisted of one volume corresponding to $b=0s/mm^2$ and 32 volumes corresponding to $b=1000s/mm^2$. Additionally, clinical records reporting the visual acuity (decimal scale) prior and after the surgery were available (a few days later and and about 3 months later). ***Preprocessing*.** All patients were scanned pre-operatively, 14 patients were also scanned post-operatively. Echo-planar imaging correction as well as correction for motion artifacts and eddy currents were performed using ExploreDTI(7). The diffusion MRI data were registered to corresponding T1-weighted scans. Fiber orientation distributions were estimated using the constrained spherical deconvolution algorithm (response function was based on the *tournier* algorithm) from MRtrix(8,9).***Tractography*.** The Fun-With-Tractography toolkit(10) was used in order to reconstruct bihemispheric OR and OT bundles. Bundle-specific reconstruction was chosen using the iFOD2 algorithm. Quality control of the volumes of interest near the optic chiasm was performed by a neuroradiologist with 2 years of experience in neuroimaging research and corrected manually (e.g., due to tumor-induced deformation).

***Analysis*.** Mean FA was computed per streamline and the median of the streamlines' FA values was assigned as the bundle FA for left and right OT and OR separately. Comparison was performed of the measurements obtained at the pre- and postoperative stages. Hausdorff distances were computed between the volume of interest estimated by FWT, which represented the expected location in case of tumor absence, and the volume of interest obtained after manual correction in order to estimate the level of deformation caused by the tumor. The resulting distance is further referred to as the deformation degree. The deformation degree was then correlated to the vision grade at the preoperative stage.

Linear regression and multiple correlation coefficients were used to analyze the relationship between vision grades and the FA change of the OR and OT bundles between pre- and post-operative stages.

**Results.** Vision grade showed stronger negative association with the degree of deformation for the left eye ($r = -0.36$, $p=0.1$) compared to the right eye ($r = -0.28$, $p=0.2$).The FA of the OT bundles was shown to have higher correlation with post-operative and follow-up vision grades compared to the OR bundles (Table 1).

| | Bundle | left vision (pre-op) | right vision (pre-op) | left vision (post-op) | right vision (post-op) | left vision (follow-up) | right vision (follow-up) |
|---|---|---|---|---|---|---|---|
| pre | OR | 0.50 | 0.43 | 0.41 | 0.49 | 0.43 | 0.49 |
| | OT | 0.54 | 0.53 | 0.70 | 0.67 | 0.72 | 0.51 |
| post | OR | | | 0.43 | 0.62 | 0.56 | 0.65 |
| | OT | | | 0.77 | 0.78 | 0.79 | 0.81 |

*Table 1 Multiple coefficient correlation between bundle FA and vision grades. The OT bundles show stronger association with the vision grades.*

Based on adjusted $R^2$ score, linear regression revealed inconsistent combinations of the bundles to be useful for interpretation of the vision grade change. For follow-up left vision grade — all bundles (adjusted $R^2 = 0.11$, $R^2 = 0.47$); follow-up right vision — left and right OT bundles (adjusted $R^2 = 0.2$, $R^2 = 0.36$); post-operative left vision grade — left and right OR bundles (adjusted $R^2 = 0.19$, $R^2 = 0.35$); post-operative right vision grade — left and right OT bundles (adjusted $R^2 = -0.02$, $R^2 = 0.18$). At the same time, weak association was observed between changes in bundle FA between pre- and post-operative scans and changes of vision grades, based on multiple correlation coefficient (Table 2).

| Bundle | left vision (change at post-op) | right vision (change at post-op) | left vision (change at follow-up) | right vision (change at follow-up) |
|---|---|---|---|---|
| OR | 0.53 | 0.04 | 0.32 | 0.24 |
| OT | 0.14 | 0.43 | 0.36 | 0.60 |

*Table 2 Multiple correlation coefficients between the pre-post operative FA difference and vision grade changes.*

**Conclusion.** We investigated correlation of the OT and OR bundles' FA measures with visual grades before and after surgical treatment in patients with pituitary adenomas. The results showed generally higher association between vision grades and the FA measures of the OT bundles. The OT bundles were also shown to be consistently important for explaining vision grade in most cases (except for the left eye at post-operative stage). Such observation is also supported by the observed correlation between pre-post operative FA differences and vision grade changes, which coincides with stronger compression of the left side, suggesting the need for shape analysis of the bundles. Overall, the results suggest that FA is a promising biomarker for estimating surgical outcome for adenoma patients.

**A comprehensive, high-resolution 7T atlas of structural brain connectivity in humans**

Hélène Lajous[1], Yasser Alemán-Gómez[1], Ileana Jelescu[1], Patric Hagmann[1]

[1] Department of Radiology, Lausanne University Hospital and University of Lausanne, Lausanne, Switzerland

**Introduction.**

Ultra-high-field diffusion magnetic resonance imaging (dMRI) enables unprecedented spatial resolution for probing the fine-scale architecture of human brain structural connectivity. Tractography derived from dMRI is a powerful method to associate biologically meaningful microstructural properties and conduction delays with specific cortico-cortical pathways and fiber-tract segments – thereby deepening our understanding of the anatomical foundations of cognition and behavior. Population-based brain atlases can enhance the interpretability of tractography results and support robust group-level inferences in structural connectomics. However, technical challenges and strong susceptibility artifacts at 7T have restricted the availability of large-scale diffusion datasets required to build such structural connectivity atlases so far. In this work, we introduce a pipeline that integrates advanced tractography techniques with robust and anatomically-informed streamline pruning strategies, applied to the 7T Human Connectome Project[1] (HCP) data. Our goal is to generate a high-resolution, neuroanatomically-grounded atlas of structural brain connectivity.

**Methods.**

***Data.*** This study included 142 subjects (90 females, mean age: 29.8 +/- 3.19 years) from the 7T HCP dataset[2].

***Tractography.*** We implemented an anatomically-constrained tractography[3] pipeline for deterministic whole-brain fiber tracking at 7T using MRtrix3[4]. Tracking parameters included a minimum and maximum streamline length of 15 mm and 150 mm, respectively, a step size of 0.5 mm, and a maximum curvature angle of 45° between steps. Ten million streamlines were seeded from the cortical gray-white matter interface, subcortical structures, deep gray matter nuclei, and the brain stem. An expert neuroradiologist qualitatively assessed the fiber orientation distribution functions computed via multi-shell, multi-tissue constrained spherical deconvolution[5], with particular attention to regions of complex fiber crossings. To leverage the high spatial resolution of 7T imaging (1.05 mm isotropic for the 7T HCP data, being an increase of 40% in volume compared to yet remarkable high-quality HCP data at 3T), we incorporated an open-source framework[6] that combines multiple fine-scale parcellations of the human brain substructures.

***Streamline filtering.*** The resulting tractograms were initially filtered using Spherical-deconvolution Informed Filtering of Tractograms, version 2[7] (SIFT2, MRtrix3) with a filtering threshold of 0.5. Anatomically-informed pruning followed, excluding implausible connections such as streamlines linking the thalamus to contralateral hemispheres. We implemented an innovative probabilistic approach for determining the behavior of white matter fiber bundles in subcortical structures such as the deep gray nuclei, the brain stem, and the cerebellum. According to expert's recommendation, we assumed that the probability $P_k$ for a streamline to cross any subcortical structure $k$ follows a monoexponential decay law: $P_k = \exp\left(-L_k * PV_{GM,k}^4\right)$, where: $L_k$ is the length of the streamline within the region of interest $k$, and $PV_{GM,k}$ is the mean partial volume of gray matter in $k$ obtained using FMRIB's Automated Segmentation Tool[8] (FAST).

***Connectivity analysis.*** We computed individual structural connectomes, scaling the connection strengths by the volume of the corresponding brain regions.

**Results.**

Although most parcellation algorithms have been trained on 3T datasets, Chimera[6] provided precise, fine-scale brain parcellation of 7T data. Moreover, the fiber orientation distribution functions were evaluated accurate, especially in regions where the main fiber bundles intersect, such as the corpus callosum, the corticospinal tract and the superior longitudinal fasciculus in the centrum semiovale, but also bend, for instance in the temporal lobe, near the uncinate fasciculus. Finally, we showed that anatomically-informed fiber pruning is crucial to handle dense connectomes at 7T without compromising biologically-meaningful connections.

**Conclusion.**

In this work, we present our methodology to build the first comprehensive, fine-scale atlas of the complex organization of supra- and infratentorial white matter in the human brain at 7T. Future steps will consist in exploring how anatomically-informed fiber pruning can enhance the reliability of tractography-based mapping of structural connectivity in the young adult.

## Improved Riemannian FOD averaging for fiber bundle priors incorporation in FOD-based tractography algorithms

G. Ville[1], E. Caruyer,[1*] J. Coloigner[1*]
[1]Univ Rennes, Inria, CNRS, Inserm, IRISA UMR 6074, Empenn ERL Rennes, U-1228, France
*These authors contributed equally to this work.

### INTRODUCTION

Guiding tractography algorithms with prior information from anatomy or microstructure can allow for better coverage of white matter bundles with streamlines, without increasing the number of false positives (improved sensitivity-specificity ratio). In [1], these priors are estimated from templates of streamlines and take the form of track orientation distributions (TOD) [2], which once incorporated voxel wise into fiber orientation distributions (FOD), help tractography algorithms to reconstruct the fiber bundles. In a similar approach [3], TOD are mixed with diffusion orientation distribution functions (ODF) thanks to a weighted averaging following a Riemannian framework [4]. The voxel-specific weights allow to adapt the importance given to priors to the fiber configuration and to the confidence in the ODF estimate. Following this work, we proposed an improved way to perform averaging between model and priors, which is also generalized to FOD.

### METHODS

The approach in [3] can be divided in three steps, that we also followed in our work: i) building for each bundle of interest an anatomical atlas from pre-segmented streamlines, ii) estimating the TOD-based priors from this atlas of streamlines and iii) incorporating the TOD into the ODF of the studied subject, once registered in the latter's space, by using the Riemannian weighted averaging framework mentioned above. The latter starts by computing the square root of the input distributions and has been devised for functions whose square integrates to 1 over the sphere [4].

While the averaging implementation from [3] only worked for ODF (diffusion-based), we extended it to include FOD (fiber-based). This required an adaptation, as (square roots of) FOD don't (squarely) integrate to 1 over the sphere, contrary to diffusion ODF. Thus, we normalized the square roots of FOD to make them squarely integrate to 1, by multiplying each of the spherical harmonics (SH) coefficients used to represent them by a constant factor. After averaging, we multiplied the output's SH coefficients by the inverse of this factor, to bring the distribution back to its initial density. Furthermore, we chose to use spherical designs [5] to compute a (point, value) representation of the FOD, useful to get their square root or their square after hand, instead of a simple grid sampling as it was done before. Spherical designs allow error-free replacement of integrals by sums; they are hence useful to find the representation by coefficients of the FOD after computing the square root or the square. Finally, another contribution was to constrain the computations within a white matter (WM) mask, to avoid falsely modifying gray matter (GM) and cerebrospinal fluid (CSF) distributions.

We tried our framework with 105 high resolution healthy subjects from the Human Connectome Project (HCP)[2]. They have a tractogram pre-segmented into fiber bundles, that we used as ground-truth. 100 subjects were used to build TOD-based priors for 6 bundles of interest. The 5 remaining subjects were used for testing. We first estimated FOD using MSMT-CSD [6]; that allowed us to create a WM mask from GM and CSF partial volume maps. We then incorporated the bundle priors into the FOD images thanks to our improved averaging framework. Afterward, we performed tractography with the iFOD2 algorithm [7], using as input both the classical FOD and the output of the averaging, that we call as in [1] enhanced FOD (E-FOD). We generated 1 million streamlines in these two experiments for equal comparison. The tractograms were then filtered with bundle extremities masks from TractSeg [8] to keep only the streamlines of interest for the bundle of study. Finally, we performed a clustering with one cluster using Quickbundles [9], to remove every streamline further than a bundle-specific threshold from the obtained centroid. That allowed us to remove some false positives. We compared this output to ground-truth using a weighted-Dice coefficient and the overlap score (which is the proportion of the ground-truth mask covered by streamlines).

**1** FOD image estimated with MSMT-CSD (left) and output E-FOD image after averaging with a cingulum TOD image using our framework (right) (Only a part of the image is shown). **2** Zoom on the voxel in the middle of the white square in Fig. 1: the FOD (left) is averaged with the TOD (middle) to get the E-FOD (right). **3** CST right estimated from a tractography using classical FOD (left) and E-FOD, namely guided by priors (middle), and associated ground-truth (right). **4** SLF_I right estimated from a tractography based on FOD (left) and E-FOD (right).

### RESULTS

Fig. 1 shows that we managed to generalize the Riemannian averaging framework to FOD, by preserving the input distributions' density while enhancing the bundle of study. Indeed, in the E-FOD image, the cingulum (CG) is more recognizable and covers a greater surface. The validity of the enhancing can be checked at the voxel scale with Fig. 2: the green peak, corresponding to CG, is more voluminous after priors incorporation, while the second peak is reduced.

Fig. 3 and 4 show that with our priors incorporation, tractography manages to better reconstruct the bundles, mostly at the bundles' extremities which become much better covered, even with a same input number of streamlines. For the right part of the corticospinal tract (CST right), the average overlap between the 5 subjects used for testing rises from 0.813 without priors to 0.947 with priors. For the right part of the superior longitudinal fascicle I (SLF_I right), it rises from 0.691 to 0.877. However, the average weighted-Dice score only goes from 0.648 to 0.651 for CST right and from 0.615 to 0.662 for SLF_I right. It wasn't possible to show all the results by lack of space, but we tried to display the most representative ones.

### DISCUSSION AND CONCLUSION

Thanks to priors incorporation, we managed to get a better overlap and spatial resolution at the bundle scale with the same number of streamlines, as did in [1] and [3], but using a weighted averaging framework generalized to FOD and technically improved (constrained by a WM mask and using spherical designs for sampling). However, most of the time, the weighted-Dice score increases very little or not at all. Fig. 3 gives us an explanation: incorporating the priors leads to estimate streamlines that were not present in the ground-truth, like in the fanning part of CST right. They are hence accounted for false positives and lower the weighted-Dice score. Conversely, the overlap sharply increases since it is not penalized by false positives. But the weighted-Dice may not be the most relevant metric, since it bases its computations on density maps, namely images showing the proportion of streamlines passing through each voxel. And streamline density is known to have its own biases in tractography. Our future plans include using Recobundles [10] to segment the output tractograms of our method into bundles instead of masks from TractSeg, and to add a streamline-based evaluation score. We also aim at evaluating our framework on clinical data and subjects with neurological disorders.

[2] https://www.humanconnectome.org/

# Multimodal Interactive White Matter Bundles Virtual Dissection

Garyfallidis E[1], Gor M[1], Koudoro S[1], Abouagour M[1], Vavassori L[2], Coletta L[2,3], Rheault F[5], Petit L[4], Sarubbo S[2], Avesani P[2,3]

*[1] Indiana University, Bloomington, IN, [2] University of Trento, Trento, Italy, [3] Fondazione Bruno Kessler, Trento, Italy, [4] GIN-IMN, Bordeaux, France, [5] University of Sherbrooke, Sherbrooke, QC*

**INTRODUCTION:** Virtual dissection of white matter bundles through tractography has become an essential tool in both modern neuroscience and clinical practice. Despite the emerging impact of automated methods, manual interactive identification of white matter pathways plays a key role in the investigation of less known bundles and in the clinical practice where fiber pathways deviate from canonical patterns.

Nevertheless manual virtual dissection suffers from two main drawbacks. First, the selection of relevant fibers is performed by manually sketching ROI on volumetric images, with voxel-based selection having relevant implications for the differentiation between plausible and implausible fibers. Second, the MR images do not easily support the identification of anatomical landmarks in the white matter.

**METHODS:** We propose a novel approach to white matter virtual dissection that combines two key innovative elements: (i) manual selection of streamlines driven by connectivity patterns[1] rather than volumetric waypoints, and (ii) multimodal imaging supporting the seamless integration of in-vivo and ex-vivo structural connectivity data[2] represented as polylines and textured meshes, respectively.

From a technological standpoint , the implementation of these features is carried out using the Free Unified Rendering in pYthon (FURY)[3]. This framework is supporting WebGPU, a new powerful and cross-platform API that provides a modern and efficient way to utilize the GPU for rendering and computational tasks, building on top of native APIs like Vulkan, Metal, and DirectX.

**RESULTS:** The manual virtual dissection driven by fiber connectivity patterns rather than ROIs has proven to achieve a more accurate characterization of white matter bundles according to the irregularity measure[4], especially in a clinical context where the canonical patterns of connectivity may deviate from healthy ones. In addition, the integration of textured mesh of white matter within the virtual dissection process provides further support to anatomical plausibility assessment.

**CONCLUSION:** The multimodal integration of white matter imaging of the human brain opens a new scenario for the manual identification of less known anatomical connectivity structures. Both perspectives, namely in-vivo and ex-vivo, can mutually contribute to a more accurate characterization of the bundles. Our advanced scientific visualization framework ensures a scalable and efficient across-platform for a wide user adoption.

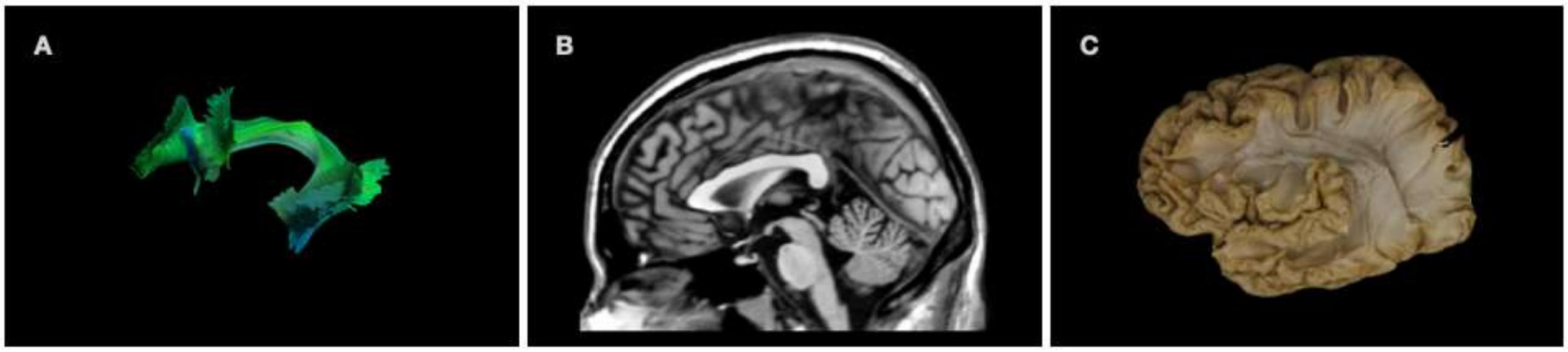

***Figure 1:** The multimodal integration for manual interactive bundle tractography dissection: (A) the digital representation of fibers as polylines as reconstructed by tractography, (B) the volumetric MR images and their DWI derivatives such as FA, (3) the mesh of white matter surface exposed after Klingler dissection and photogrammetric acquisition followed by 3D reconstruction.*

---

[1] Porro Munoz D, et al. (2015). Tractome, a visual data mining tool for brain connectivity analysis

[2] Vavassori L., et al., (2025). Brain Dissection Photogrammetry for Studying Human White Matter Connections

[3] Garyfallidis E., et al., (2021). FURY: Advanced Scientific Visualization

[4] Sarubbo S, et al. (2024). Changing the Paradigm for Tractography Segmentation in Neurosurgery

**Multi-compartment tractometry approach for white matter neuroinflammation investigation in late-life depression**

**Nathan Decaux[1], Gabriel Robert[1,2], Julie Coloigner[1]**

[1]Univ Rennes, INRIA, CNRS, INSERM, IRISA UMR 6074, Empenn ERL U 1228, 35000 Rennes, France.

[2]Adult University Psychiatry Department, Guillaume Régnier Hospital, Rennes, France.

## INTRODUCTION

Late-life depression (LLD) affects 7% of the population aged more than 60 years old. This prevalence is concerning, as LLD is not only an independent risk factor for mortality, but also a modifiable risk factor for dementia. The pathophysiology of LLD is multifactorial, involving inflammatory, degenerative, and vascular mechanisms. This complexity underscores the need for a deeper understanding of the underlying biological processes. Recent advances in diffusion imaging and tractography have led to the development of a novel framework—tractometry—for more precise assessment of white matter (WM) microstructure [1, 2]. In brief, these along-tract approaches generate a profile for each fiber bundle, mapping microstructural metrics onto a central streamline. Statistical analysis of these bundle profiles can provide a more specific and localized investigation than looking at a region of interest or tract-averaged measures. We hypothesize that a tract-based approach could yield novel biomarkers for LLD. Building on the work of Hédouin et al. [3], we performed an advanced microstructural compartment model (MCM) along fiber bundles, to derive more interpretable and sensitive metrics, of white matter integrity and inflammation that could not be obtained with a classical DTI model. We notably applied our framework on two metrics: free-water (FW) and fractional anisotropy (FA) between LLD and control groups.

## METHODS

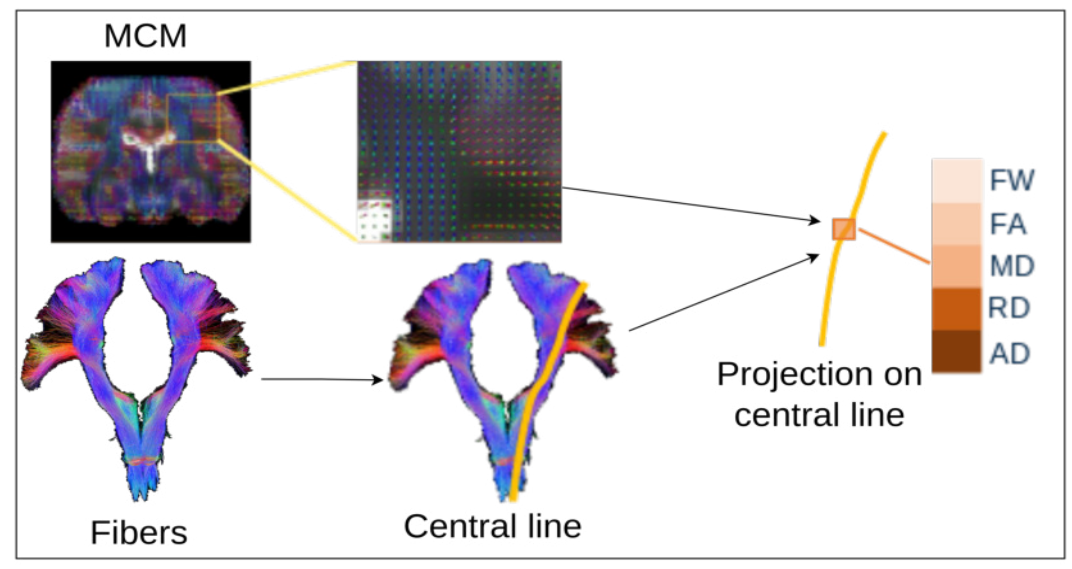

*Figure 1 : Illustration of the tractometry principle combining multi-compartment modeling (MCM) and tractography for microstructural analysis of white matter fiber bundles.*

We used the AMYNET database, comprising 31 patients with late-life depression (40 females, age 75.4 ± 6.2 years) and 21 healthy controls (16 females, age 74.8 ± 5.2 years), recruited from the French old-age psychiatry centers of Rennes and Tours, France. Diffusion MRI preprocessing followed the protocol established by Hédouin et al. [3], including denoising, Gibbs ringing correction, motion and eddy current correction, and bias field correction. T1-weighted images were used for anatomical reference and registration. Orientation distribution functions (ODFs) were estimated using the Multi-Shell Multi-Tissue Constrained Spherical Deconvolution (MSMT-CSD) algorithm implemented in MRtrix3 [7]. Tractography was performed using the iFOD2 algorithm with a target number of 1 million streamlines, seeded from the whole brain mask. Fiber bundle elements were extracted for each voxels following [9] approach and the MRTrix3 implementation. Main direction peaks were then derived from the fixels. A multi-compartment model (MCM) was also performed including one isotropic compartment representing free water, one iso-restricted compartment, and one to three anisotropic compartments, corresponding to distinct fiber populations. This number corresponds to detected peak directions of each ODF. From this model, we derived several metrics for each voxel including free water fraction (FW), iso-restricted water fraction (IRW), fractional anisotropy (FA), mean diffusivity (MD), axial diffusivity (AD), and radial diffusivity (RD) for each anisotropic compartment. White matter tracts were segmented using BundleSeg, an enhanced version of the RecoBundles algorithm [5] with a fixed search radius of 8mm. We performed the segmentation on each subject's tractogram using a white matter bundle atlas derived from the cohort of 105 annotated Human Connectome Project (HCP) data [6]. This template was constructed through an iterative registration process that aligned individual HCP subject data to a common template using a diffeomorphic approach, adapting the method proposed by [7] for diffusion tensor images. The resulting atlas contains 72 bundle models, each comprising streamlines from all atlas subjects registered to a standardized space. For complete methodological details regarding atlas creation, please refer to section 2.1 of [8]. We developed a tractometry approach that selects the anisotropic compartment most aligned with the local streamline direction at each point, rather than averaging all compartments as proposed in [3]. This preserves microstructural specificity in complex white matter regions. Microstructural metrics from the selected compartment were sampled at equidistant points along each bundle to create profiles for statistical comparison, as illustrated in Fig. 1. We performed a two-sample t-test for the FA and the FW between LDD patient and control groups with age, gender and site of MRI acquisition as covariates, correcting for multiple comparisons.

## RESULTS AND DISCUSSION

Among the 72 segmented bundles, six demonstrated significant group differences, as illustrated in Figure 2. The late-life depression (LLD) group exhibited elevated free water (FW) fractions in the left anterior thalamic radiation (ATR), left fronto-pontine tract (FPT), third segment of the corpus callosum (CC_3) and right thalamo-premotor tract (T-PREM) compared to healthy controls. The FW metric quantifies the relative fraction of freely diffusing water in the extracellular space, serving as a proxy for chronic low-grade neuroinflammation, which plays a pivotal role in the neuropathogenesis of various neurological and psychiatric disorders, including LLD. Consistent with previous research, we observed alterations in frontolimbic circuits implicated in LLD pathophysiology [3]. Additional findings such as decrease in FW of the right posterior parieto-occipital tract (POPT) and right inferior cerebellar peduncle (ICP) can likely be attributed to our multi-compartment modeling approach, which employs more realistic compartment estimations through fixel-based derivation. Furthermore, patients demonstrated significantly increased fractional anisotropy (FA) in the genu of the corpus callosum, while exhibiting decreased FA in the right inferior cerebellar peduncle—findings that diverge from previous studies. This discrepancy in results may be explained by our selection mechanism for identifying the anisotropic compartment most collinear with each tract.

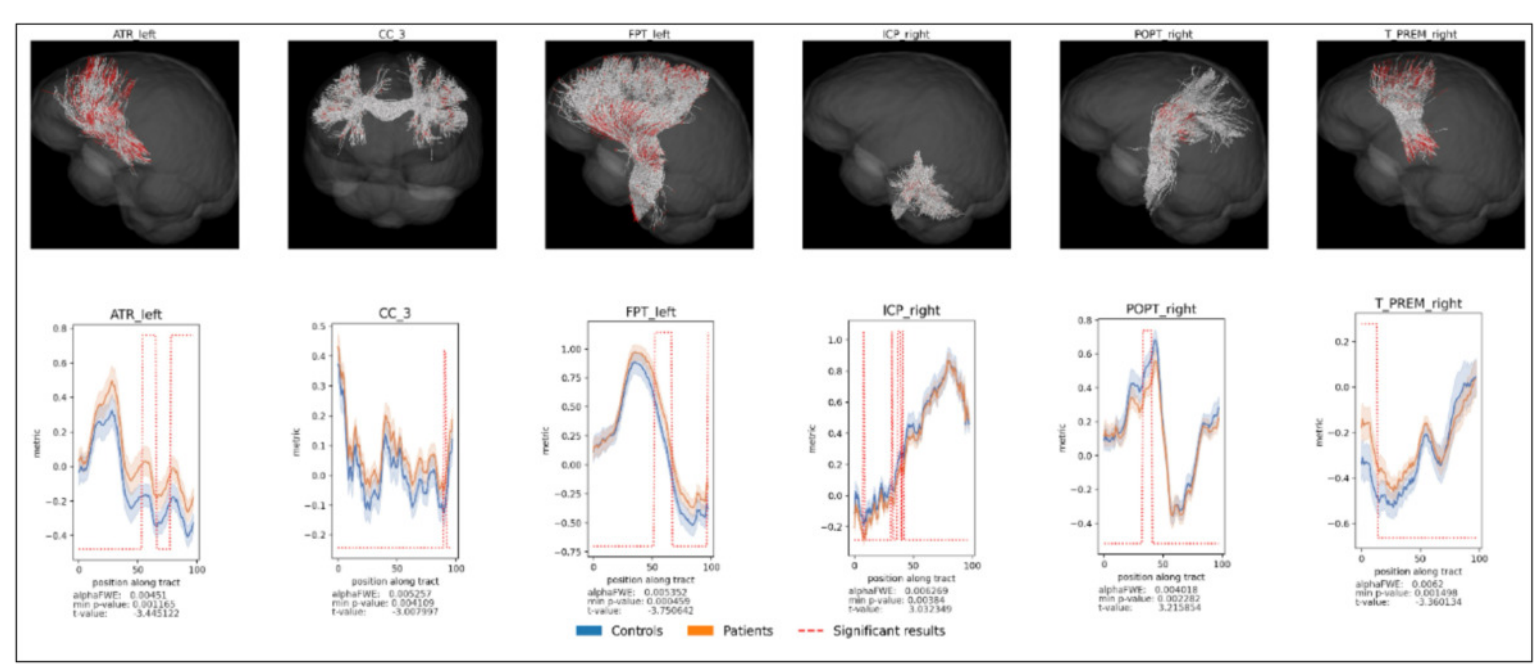

*Figure 2 : Tractometry analysis showing group differences in free water fraction (FW) between controls (blue) and patients (orange) across six white matter tracts. Top: 3D tract reconstructions with significant regions highlighted in red. Bottom: Along-tract profiles with statistical comparisons (red dashed boxes indicate significant differences, FWE-corrected $p < 0.05$).*

## DISCUSSION AND CONCLUSION

The difference in the neuroinflammation component value between LLD and HC in the ATR and T-PREM confirms previous results reporting an association between these tracts and serum inflammation in major depressive disorder. Moreover, those prefrontal corticostriatal loops are known to be involved in goal-oriented behavior, mediating motor behavior such as planning, learning and motor execution.

Our future plans include exploring new approaches for projecting microstructural metrics along fiber bundles which address cases where central line reduction is not appropriate, such as at fiber bundle extremities, with the aim of more precisely investigating altered regions in patient populations. Additionally, we plan to investigate correlations between these structural findings and functional data, such as actigraphy, to better understand the underlying mechanisms of apathy and late-life depression.

# An Atlas of the augmented Corticospinal tract

Guillaume Mahey[1], Aymeric Stamm[1], Bertrand Michel[2]

Nantes University[1], Centrale Nantes[2], Department of mathematics Jean Leray, UMR CNRS 6629

**Introduction** Diffusion MRI is a non-invasive imaging technique that has been proven useful in probing white matter tissue microstructure and structural connectivity. On one hand, diffusion images can be used to infer tissue microstructure using parametric models of water molecule diffusion [1]. On the other hand, diffusion images can also be used to infer structural connectivity, i.e. wiring of axons throughout the brain [8]. In terms of statistical analysis, previous work has mainly focused on tissue microstructure [2] or on structural connectivity [4]. Will some good practice focus on maximising robustness when inferring either microstructure or connectivity, there is still a high inter-subject variability. To overcome this, some works proposed to jointly analyse microstructure and connectivity by following the evolution of scalar microstructural parameters along streamlines.That being said, recent works have shown that inferring tissue information from a simple scalar diffusivity measure can be ill-posed [1]. Typically, two different microstructures can lead to the same scalar diffusivity measure. This statement pushed researchers to represent local diffusion with non-scalar measures [6]. For example, given the coarse spatial resolution of diffusion MRI, microstructure information can be provided in the form of mixtures of densities since multiple tissue populations are present in a voxel. In this article, we propose a comprehensive mathematical framework to manipulate what we call microstructure-augmented streamlines (MAS), which are 3-dimensional parameterised curves which carry local tissue organisation as a mixture of Gaussian densities. Using this framework, we propose an atlas of the augmented Corticospinal tract (CST) that preserves both global spatial information and local tissue microstructure.

**Methods** We propose a comprehensive framework at the intersection between functional data analysis and optimal transport (OT) [7, 3]. The former helps to deal with the functional nature of MAS while the latter provides dedicated spaces to analyse probability densities in general. To generate the microstructure-augmented streamlines, we first represent local diffusion by inferring a mixture of Gaussians at each voxel of the brain. Next, we trace 3D streamlines following local diffusion directions and we add to each point local microstructure information in the form of a Gaussian mixture obtained as an average from neighboring voxels using mixture-Wasserstein barycenters [5]. Mixture-Wasserstein barycenters are defined as Fréchet means of the mixture-Wasserstein distance. Equipped with a distance and Fréchet mean, we can generalize statistical methods originally conceived for multivariate data. In particular, clustering is useful when building an atlas of the CST because we need a mechanism that ensures that the atlas will preserve streamlines representative of each functional area of the motor cortex.

**Results** An important motivation for tractography is segmentation: given several observations of a particular tract, one may be interested in constructing an expected or average tract. In this work, we focus on constructing an average representation of the CST, which is typically endowed with a mixture of Gaussians. To achieve this, we first define a distance between MAS, taking into account structural connectivity via the Euclidean coordinates of the streamlines and microstructure via the Gaussian mixture densities. Our distance enables clustering among MAS, and in particular, the extraction of a representative set for each observed augmented tract. Next, we align these sets of representatives by applying a translation and rotation to each, in an optimal manner. Finally, we define our atlas as the average of the aligned representatives, summarizing the different observations of the CST.

**Conclusion** The proposed augmented atlas of the CST is motivated by the need to contrast a patient's brain against a reference atlas for anomaly detection. In the context of patients affected by traumatic brain injury (TBI), the CST is of particular importance as it is known to be the primary motor pathway in the human brain. We believe that the inter-subject variability of inferring, whether microstructure or connectivity, can be reduced by considering microstructure-augmented streamlines and analysing them through the combined spectrum of functional data analysis and optimal transport.In the future we aim to define a distance, followed by statistical tests, between augmented tractograms of patients, affected by TBI, and a healthy control atlas.

**Extracting tract-specific neurodegeneration by differentiating converging fibers using fixel-based analysis**

*Lloyd Plumart, Hinke N. Halbertsma, Mayra Bittencourt, Remco J. Renken, and Frans W. Cornelissen*

Laboratory for Experimental Ophthalmology, University Medical Center Groningen, Netherlands

## INTRODUCTION

A major challenge in tractography is analyzing complex white matter configurations, such as crossing and converging fibers[1]. If we could segment tracts inside regions where crossing or convergence occurs, we would also be able to identify tract-specific changes. Furthermore, these changes may reflect the underlying anatomy and pathologies more accurately. Crossing fibers can currently be disentangled using fixel-based analysis (FBA)[2]; this technique can be used to quantify white matter microstructure based on diffusion MRI data and provides insight into the anatomical properties of fiber populations within a voxel, called *fixels.*

However, thus far fixel-based metrics (e.g., fiber density and cross-section; FDC) cannot be attributed to distinct converging anatomical tracts. This becomes an issue when determining tract-specific changes in profiles, as the FDC values may reflect contributions from multiple converging tracts.

To tackle this challenge, we propose *tract-profile segmentation*, a method to segment fixel-based tract-profiles according to the respective contributions of different anatomical tracts along this profile. This approach aims to improve anatomical specificity when characterizing tract-level white matter changes. As a proof of concept, we applied this method to assess tract-specific FDC reductions in glaucoma—a disease causing neurodegeneration of the visual system[3].

## METHODS

We performed diffusion-weighted imaging (DWI) in 43 participants (26 controls, 17 glaucoma patients, FOV = $210 \times 210 \times 132$ mm$^3$, voxel size = $2.0 \times 2.0 \times 2.0$ mm$^3$, 66 slices, b-values = [1000, 2500] s/mm$^2$, gradient directions = [64, 64], $b_0$ images = 3) using the Siemens MAGNETOM Prisma 3T MRI scanner. We constructed a fixel-based profile of the optic radiations (OR) using probabilistic tractography following the standard MRtrix FBA pipeline. Subsequently, we estimated the contribution of two other visual white matter tracts that converge in the occipital lobe—the inferior fronto-occipital fasciculus (IFOF) and the forceps major (FM)—to each fixel in the OR profile. This enabled us to perform a segmentation of the FDC tract-profile based on the relative involvement of the two intervening tracts.

## RESULTS

Tract-profile segmentation revealed that portions of the OR profile contained major contributions from the IFOF and FM. A first manual segmentation based on visual inspection showed tract-specific FDC reductions present in the OR, IFOF, and FM in glaucoma compared to controls. We found relative FDC reductions that had strong contributions from the IFOF and FM, and to a lesser extent also the OR.

## CONCLUSIONS

Our tract-profile segmentation enhances the anatomical specificity of white matter analysis and may offer new insights into neurodegenerative processes affecting complex fiber architectures, as demonstrated in glaucoma. Our method overcomes a key limitation of fixel-based tract profiling by enabling tract-specific interpretation of microstructural changes in regions of fiber convergence.

---

[1] Klaus H. Maier-Hein et al. (2017), "The Challenge of Mapping the Human Connectome Based on Diffusion Tractography," https://doi.org/10.1038/s41467-017-01285-x.

[2] David A. Raffelt et al. (2017), "Investigating White Matter Fibre Density and Morphology Using Fixel-Based Analysis," https://doi.org/10.1016/j.neuroimage.2016.09.029.

[3] Shereif Haykal et al. (2019), "Fixel-Based Analysis of Visual Pathway White Matter in Primary Open-Angle Glaucoma," https://doi.org/10.1167/iovs.19-27447.

**Title**: *Quantifying the Impact of Probabilistic Streamline Turning Angle on Brainstem–Inclusive Whole-Brain Connectomes*

Monica Duran[1], Nicholas Cottam[1], Allison Block[1], Jose Guerrero-Gonzalez[1], Nagesh Adluru[1], Gregory Kirk[1], Brittany Travers[1].
[1]University of Wisconsin-Madison; Madison, WI, USA.

**Background** The brainstem hosts major neuromodulatory nuclei that are critical for maintaining consciousness, regulating arousal states, and integrating sensorimotor information. Its complex connections reach widespread cortical and sub-cortical territories, making it an important hub in the brain[1]. Because the brainstem is a challenging region to image, most diffusion MRI studies have omitted it from tractography analyses. Over the past decade a limited set of investigations has begun to retain the brainstem, yet methodological guidance remains sparse and key tracking parameters vary widely between reports. One such parameter is the maximum turning-angle: the largest angular deviation allowed between successive streamline segments. Published brainstem studies have used angles as low[2] as 4° and as high[3] as 90°, reflecting the trade-off between prematurely terminating genuine curved tracts (small angles) and admitting anatomically implausible trajectories (large angles). Although several groups have evaluated how other tracking settings affect connectome reliability [4-6], the specific influence of turning-angle, particularly in whole-brain connectomes that include the brainstem, has not been characterized. The objective of this study is to determine how commonly used turning-angles affect the reliability of global graph measures in pediatric brainstem–inclusive connectomes.

**Methods** Diffusion MRI (dMRI; multi-shell spin-echo EPI, 2.4 mm isotropic) and T1-weighted data were collected from 10 children (6–10 y) on a 3T GE MR750. The dMRI data were denoised and corrected for distortions and artifacts. Each dMRI series was up-sampled to 1 mm isotropic via the TiDi-Fused pipeline[7] to improve apparent resolution. Fiber orientation distribution maps were computed via constrained spherical deconvolution using MRtrix3[8]. For each subject, probabilistic tractography was performed using 3 different angles: 15°, 30° & 45° (angles of 4° and 90° were qualitatively screened and rejected beforehand). The following additional parameters remained fixed across all runs: dynamic seeding, 100 million streamlines, 1 mm step-size, 1 mm minimum length, and FOD cutoff 0.07. Given the probabilistic nature of the algorithm, three independent random-seed replicates were generated per subject and angle. The SIFT2 algorithm[9] was applied to minimize the effects of spurious streamlines. Each tractogram was parcellated into a 70-node network comprising 52 brain-stem regions (Brainstem Navigator Atlas[3]) and 18 limbic/cortical regions (FreeSurfer), producing 70 × 70 undirected adjacency matrices with edge weights quantified as Fiber Bundle Capacity (FBC), a quantitative proxy for axonal cross-section[10]. From every FBC-weighted matrix we extracted 6 global graph theory metrics: global efficiency, mean strength, mean clustering coefficient, mean betweenness centrality, mean shortest-path length, and modularity. Intraclass correlation coefficients, ICC (2,1), measured (1) replicate-to-replicate reliability within each angle and (2) cross-angle reproducibility of each metric.

**Results** *Within-angle* ICCs quantified replicate-to-replicate stability for each angle, a critical test because probabilistic tractography uses random seeding that can introduce run-to-run noise that can be exacerbated by curvature differences. At each angle, all 6 global graph theory metrics were highly consistent across the 3 replicates, with excellent ICCs ranging from 0.86 to 0.99. Differences between angles were marginal, suggesting that curvature settings might not impact the noise introduced from run-to-run. Given the strong consistency across replicates for all angles, we proceeded to test *cross-angle* ICCs with only one replicate per subject and angle, to assess the robustness of the graph theory metrics to angle differences. This analysis showed that most metrics are robust to curvature choice: efficiency and strength retained excellent agreement across angles (ICC ≈ 0.97), mean shortest path remained good (0.82), clustering and modularity were moderate–good (0.74–0.80), whereas betweenness was only moderate (0.61), suggesting its greater sensitivity to angle-dependent network changes. Overall, these results indicate that most global graph metrics are highly tolerant to the choice of curvature angle and exhibit strong replicate-to-replicate stability, meaning researchers can generate connectomes generated with 15–45° thresholds without jeopardizing reliability.

**Conclusion** All six global graph metrics displayed excellent replicate-to-replicate stability at 15°, 30°, and 45°, and their cross-angle agreement remained high, indicating that a single tractography run per subject is generally sufficient and that reliability is largely unaffected by which of these curvature thresholds is chosen. Limitations: (i) global metrics may obscure angle-dependent effects at specific nodes or connections, and (ii) we are still completing biological validation to confirm that the most reliable parameters also best reflect known neuroanatomy. Ongoing work will therefore assess node- and edge-level reliability, particularly within brainstem, cortical, and brainstem-to-cortex subnetworks, and integrate anatomical benchmarks to refine curvature-angle choices for targeted connectomics.

**Title: The Connectome Analysis for Pediatric Epilepsy Surgery (CAPES) Study: Leveraging Normative Disconnectome Mapping to Predict Seizure Outcomes**

Sudarsan Packirisamy[1]*, Cédérick Montplaisir[1]*, Kenza Sophie Lahlou[1], Tawfik Elsherbini[1], Roy Dudley[2], Jonathan Roth[3], Shimrit Uliel-Sibony[4], Avi Shariv[4], Mykhailo Lovga[5], Luke Tomycz[6], Cameron Elliott[7], Aris Hadjinicolau[8]†, Joseph Yuan-Mou Yang[9]†, Alexander G. Weil[10]†

**Affiliations:**

1. *Faculty of Medicine, Université de Montréal, Montreal, Quebec, Canada*
2. *Neurosurgery Division, Department of Surgery, Montreal Children's Hospital, Montréal, Québec, Canada*
3. *Department of Pediatric Neurosurgery, The Pediatric Brain Center, Dana Children's Hospital, Tel-Aviv Medical Center, Tel-Aviv University, Tel-Aviv, Israel*
4. *Pediatric Epilepsy Unit, Tel Aviv Medical Center, Tel Aviv University*
5. *Department of Pediatric Neurosurgery, Epilepsy Surgery and Neurophysiology, Western Ukraine Specialized Center, Lviv, Ukraine*
6. *Epilepsy Institute of New Jersey, Jersey City, New Jersey, United States*
7. *Division of Neurosurgery, University of Alberta, Edmonton, Alberta, Canada*
8. *Division of Neurology, Department of Pediatrics, Sainte-Justine University Hospital Centre, Montréal, Québec, Canada*
9. *Neuroscience Advanced Clinical Imaging Service (NACIS), Department of Neurosurgery, The Royal Children's Hospital, Melbourne, Australia*
10. *Division of Neurosurgery, Department of Surgery, Sainte-Justine Hospital Centre, Montréal, Québec, Canada*

**Equal Contribution, † Co-senior authorship*

**Introduction:** Surgical intervention for drug-resistant epilepsy achieves seizure freedom in approximately 60% of temporal lobe epilepsy (TLE) cases and around 25-40% of extratemporal lobe epilepsy.[1] In adults, improved seizure outcomes in TLE have been linked to resections that precisely target the mesial temporal lobe structures, and/or disconnecting the temporal outflow white matter tracts.[2,3] However, pediatric epilepsy differs markedly from adult epilepsy, with a higher incidence of cortical malformations and extratemporal foci, making predictors of surgical success less well defined. Normative disconnectome is a diffusion MRI tractography based technique that estimates the likely white matter disconnections caused by a brain lesion or surgical resection by mapping it onto a population-based tractography atlas derived from healthy individuals.[4]

This study aims to improve understanding of surgical outcome predictors in pediatric drug-resistant epilepsy and to identify white matter pathways whose disruption may influence postoperative seizure recurrence using a normative disconnectome approach. The findings will inform pre-surgical planning and support more accurate prognostic modeling in children.

**Methods:** This multicenter international retrospective study includes pediatric patients who underwent frontal, temporal, or insular lobe epilepsy surgery. Pre- and postoperative T1-weighted MR images are used to manually delineate resection cavities using ITK-SNAP.[5] These lesion masks are then non-linearly registered to an age-specific standard brain template in Montreal Neurological Institution (MNI) space using Advanced Normalization Tools (ANTs)[6], enabling accurate cross-subject comparison. To identify regions associated with seizure outcomes, each lesion mask is weighted by postoperative seizure status to generate a probabilistic z-score map identifying regions associated with better seizure control.

In parallel, normative disconnectome analysis was performed using high-quality pediatric diffusion MRI data from the Human Connectome Project–Development (HCP-D 2.0 release; *DOI: 10.15154/1520708*).[7] A group-averaged diffusion-weighted template was created using ANTs, and a normative structural connectome was generated via probabilistic tractography with constrained spherical deconvolution (CSD) in MRtrix3.[8] For each patient, the lesion mask was projected onto the normative connectome to estimate likely white matter disconnections.

Finally, linear regression models were used to assess associations between specific white matter tract disconnections and seizure outcomes, providing insight into structural pathways that may influence surgical success.

**Results:** We are currently acquiring imaging and clinical data from six centers across three continents and segmenting pre- and postoperative MRIs. Outcome data are expected before October 2025. Figure 1 summarises the proposed imaging processing pipeline, using a case example. The case is an 8-year-old boy with drug-resistant epilepsy referrable to a right frontal polymicrogyria who underwent an anterior frontal lobectomy. He had seizure recurrence 3 years following surgery.

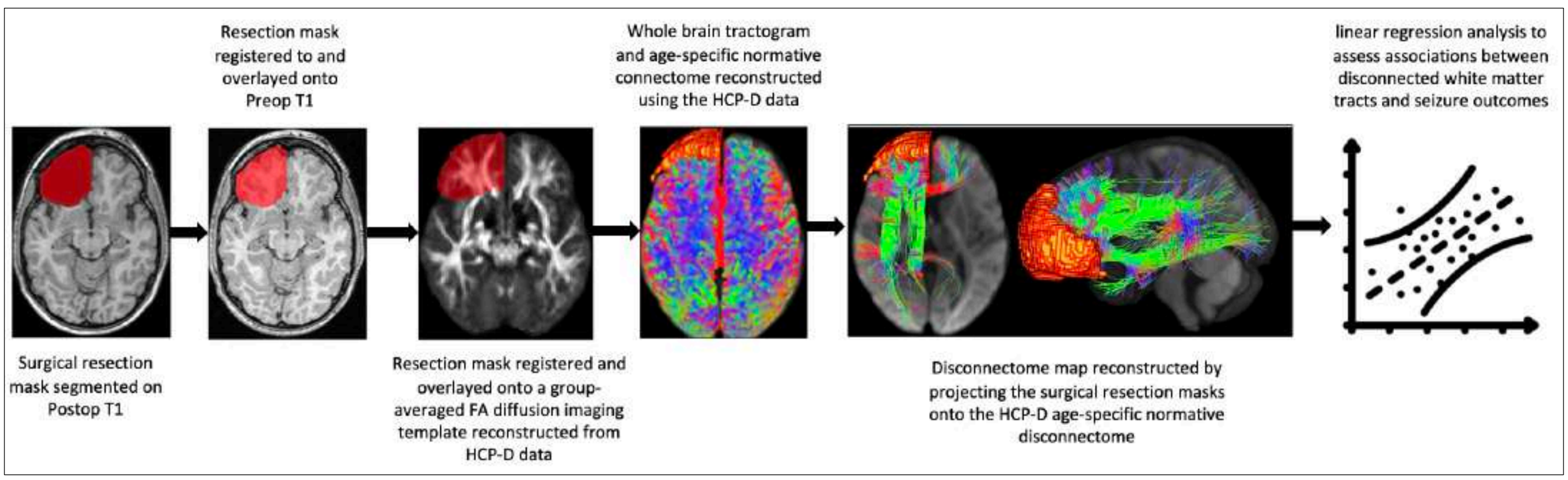

*Figure 1. Summary of the proposed normative disconnectome processing and analysis pipeline in this study*

**Conclusion:** Seizure outcome prediction remains a critical yet challenging goal in children undergoing lesionectomy for focal drug-resistant epilepsy, as standard clinical variables often fall short. This study will demonstrate the potential of a normative disconnectome approach to identify patients at risk of postoperative seizure recurrence. By leveraging publicly available, high-quality tractography datasets, this method offers a practical alternative to acquiring high-resolution diffusion MRI during presurgical workup-an approach that is often technically and logistically infeasible or challenging in routine pediatric epilepsy care.

# SWM bundles segmentation using streamlines and voxel information in VAE latent space

S. Navarrete[1], J. Molina[1], N. Vidal[1], C. Hernández[1], P. Guevara[1]
[1]Faculty of Engineering, Universidad de Concepción, Chile

## Introduction

Superficial white matter (SWM) refers to the layer of white matter located immediately beneath the cortical gray matter [1]. It is composed of short-range association fibers that follow the contours of sulcal valleys [2]. Despite the relevance of SWM fibers, research on them remains limited compared to that on deep white matter (DWM), largely due to the limitations of dMRI tractography and the lack of consensus regarding their definition [1,3]. Most automatic fiber bundle segmentation methods are designed for DWM. The streamline segmentation based on Euclidean distance (SSBED) [4] utilizes an Euclidean distance-based measure and anatomical atlases to guide the process. Multiple deep learning models have been proposed for segmenting fiber bundles [5], which, in general, can be categorized into voxel-based [6] and streamline-based methods [7]. In this work, we design a variational autoencoder (VAE) that takes as input both the streamline geometry and voxel-wise features sampled along the fiber trajectories. The latent space learned by the model is then used to perform the segmentation of SWM bundles based on a bundle atlas [8].

## Methods

We used test-retest acquisitions of the HCP S1200 preprocessed dataset to generate whole-brain probabilistic tractography using MRtrix3 software. A nonlinear warp was applied to register all data to the MNI152 standard space. The model uses a convolutional VAE with a structure similar to FINTA [7], but with a modified input. Rather than relying solely on the streamlines, the model incorporates additional dimensions containing voxel-wise values sampled along the fiber paths. These values include the intensity from the T1w, FA, and the first two peaks of the fODF. The model was trained using 16,000,000 streamlines, and the corresponding voxel-wise features of each streamline obtained from 24 subjects. Once the model was trained, both a modified SWM bundle atlas [8] and the new tractogram to be segmented are projected into the model's latent space. A Radius Neighbors Classifier (RNC) is then applied in the latent space to assign labels to the streamlines of the new tractogram. To optimize the RNC radius of each label, we used as reference the SWM fascicles from six subjects in the HCP database, segmented using the SSBED method [4] and filtered using the Convex Hull [10].

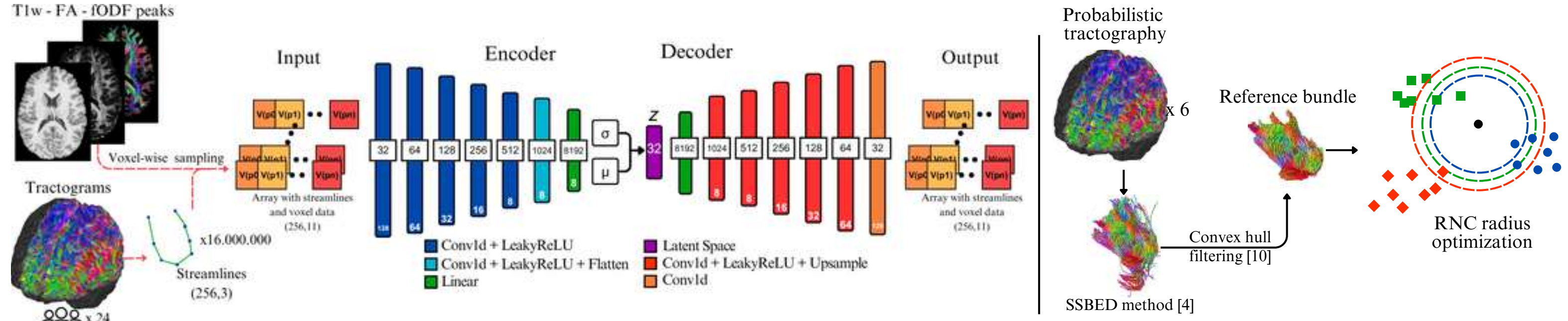

Fig. 1. Overview of the proposed model. Left: model architecture and training methodology. Right: radius optimization process.s.

## Results

The proposed segmentation method and the SSBED method [4] were applied to 10 test-retest subjects. When evaluating the 525 bundles from the atlas, and averaging across subjects, the proposed method showed better results in 300 bundles for the bundle adjacency (BA) streamline metric, 385 for BA voxel, and 141 for the Dice score. When averaging across all bundles, the proposed method achieved a lower BA streamline value 3.96 mm vs. 4.00 mm and a lower BA voxel value 0.55 mm vs. 0.61 mm, indicating better spatial alignment between test and retest segmentation. For the Dice score, where higher values reflect better overlap, the proposed method obtained of 0.62 compared to 0.65 with SSBED. The proposed method showed better performance in small, compact U-shaped bundles, reducing the inclusion of noisy fibers, while it was less effective in segmenting larger bundles that connect distant cortical regions.

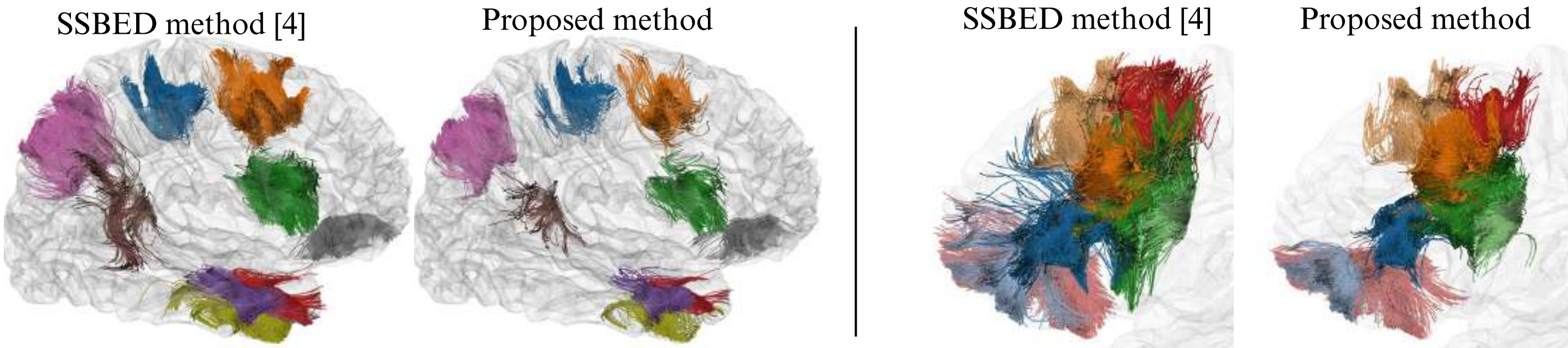

Fig. 2. Visual comparison of the segmentation from both methods. Left: examples of large bundles connecting multiple cortical regions. Right: examples of smaller, more compact U-shaped bundles.

## Conclusion

We proposed a method for SWM bundle segmentation using a variational autoencoder that integrates both streamline coordinates and voxel-wise features. The model achieved accurate segmentation of short, compact U-shaped bundles. Increasing the amount of training data may enhance its ability to segment larger SWM tracts more effectively.

# Nevrolens 2.0: Augmented Reality Atlas for Cross-Species Neuroanatomical Understanding

Thanh P. Doan [1,2] *, Eirin Svinsås [3], Thadshajini Paramsothy [3], Mikhail Fominykh [4], Ekaterina Prasolova-Førland [4]

1. Department of Neuromedicine and Movement Science, NTNU, Trondheim, Norway. 2. St.Olav's University Hospital, Trondheim, Norway. 3. Department of Computer Science, NTNU, Trondheim, Norway. 4. Department of Education and Lifelong Learning, NTNU, Trondheim, Norway.
* corresponding author: thanh.p.doan@ntnu.no

**Introduction:** Despite major advances in diffusion MRI, tractography in humans remains an indirect method, lacking direct anchoring to ground-truth axonal connectivity. In contrast, animal studies using in vivo neural tracers still offer unmatched resolution at the micro- and mesoscale. We previously used such methods to revise the canonical model of medial temporal lobe (MTL) connectivity in rodents, revealing convergent input streams into the lateral entorhinal cortex [1]. This inspired human studies leveraging tractography to uncover similar organizational principles [2]. However, the gap between animal microscale anatomy and human tract-based models persists—slowing translational progress in clinically relevant connectomics. One promising strategy involves embedding validated animal neuroanatomy into human frameworks using mixed reality, enabling anatomically grounded, visually intuitive platforms to bridge this divide [3].

**Methods:** We introduce Nevrolens 2.0, a novel augmented reality (AR) application developed for Apple Vision Pro that enables real-time morphing between rat and human brain structures. Built on the original Nevrolens system [4], this version includes 222 structures from the Waxholm Space rat atlas [5] and a homologous subset of human anatomy based on the CerebrA/Mindboggle-101 atlas registered to MNI-ICBM152 template space [6]. Users can manipulate, align, and morph between species with anatomically constrained 3D models. Designed through an iterative, user-centered process, the platform was evaluated by anatomists, clinicians, and neuroscientists [7]. While tractography data is not currently embedded, the architecture is built to support future integration of other modalities, such as rodent tracer-based and human diffusion-derived pathways—potentially through collaborations with the IST community.

**Results:** Expert reviewers rated the platform highly in usability, anatomical fidelity, and translational value. Key features include: Interactive structure morphing with transparency modulation, multiplanar dissection views, multi-user collaboration with synchronized model interaction and voice chat, and educational annotations and quizzes. These tools foster a shared spatial language between systems neuroscience and clinical neuroimaging—paving the way for pluridisciplinary and cross-species connectivity collaborations.

**Conclusion:** Nevrolens 2.0 offers a foundational AR-based framework for cross-species translation. It brings together the perspectives of fundamental anatomists, computer scientists, and practicing clinicians to support a new generation of anatomically informed, mixed-reality tools. With future integration of white matter tracts, this platform could help ground tractography in validated circuit anatomy—bridging model systems and human imaging with direct relevance for IST collaborators.

## High-Resolution Tractography Shows Later Maturation of Superficial White Matter Across the Lifespan

J. Urbina-Alarcón[1], J. A. Acosta-Franco[1], C Beaulieu[1,2]

[1]Biomedical Engineering and [2]Radiology & Diagnostic Imaging, University of Alberta, Edmonton, Alberta, Canada

**Introduction:** Superficial white matter (SWM) is located just beneath the cerebral cortex and is comprised of short-range association fibers (including U-fibers) with axons that support local cortico-cortical communication. Relative to deep white matter (DWM), SWM changes with 'healthy' development and aging has been understudied with diffusion MRI (dMRI), partly due to its thin size adjacent to cortex and its complex trajectories (e.g. curving, crossing, fanning) [1,2]. With a focus on diffusion tensor imaging (DTI) parameters such as fractional anisotropy (FA) and mean diffusivity (MD), SWM in children and adolescents has shown greater FA and lower MD changing linearly with age [3,4]. Extending the age range of neurodevelopment studies to 30 years has identified non-linear increases of FA and decreases of MD that level off in SWM [5,6]. Aging studies of SWM from young adults to the elderly have shown the opposite with decreases of FA and increases of MD with age [7,8], including the only study with multi-shell constrained spherical deconvolution (CSD) tractography to delineate the SWM [9]. However, there have been no dMRI studies of SWM over the 'lifespan'. Further, all the above studies used low spatial resolution (2.0 to 2.6 mm isotropic). The purpose here is to use 1.5 mm isotropic, multi-shell CSD tractography to identify changes of DTI parameters in SWM, relative to DWM, over the lifespan (5-74 years, n=185).

**Methods:** Whole-brain dMRI was acquired on a 3T Siemens Prisma with a 64-channel RF coil from 185 healthy participants (99 females, 5-74 years) using single-shot EPI: 90 axial slices at 1.5 mm thickness, in-plane resolution 1.5x1.5 $mm^2$ zero-filled to 0.75x0.75 $mm^2$, TR/TE=4700/64 ms, 6 b=0, 30 b=1000 s/$mm^2$, 30 b=2000 s/$mm^2$; and scan time 5:59 min. Tissue segmentation into cortex, deep gray matter, WM, and cerebrospinal fluid was performed in native dMRI space using the bias-corrected mean b0 and SynthSeg 2.0. Tractography was performed in MRtrix3 using CSD (FOD cutoff=0.05, angle=45°, min/max length=1.5/50 mm, 200k streamlines) with seeds placed in voxels within cortex immediately adjacent to the underlying WM. Streamlines that did not start and end within the seeding mask were excluded, and outliers extending into DWM were removed using clustering algorithm DBSCAN. The resulting tracts were binarized to define the SWM mask, and the DWM mask was defined as its complement within total WM. DTI metrics were computed with the b1000 shell using Dipy v1.7.0. DTI metrics and volumes were averaged across all SWM and DWM separately and assessed for sex differences. DTI-vs-age trajectories were modeled using a Poisson fit [10] to capture their asymmetrical shape.

**Results:** The proposed method well identified the SWM adjacent to cortical tissue (**Figure A** of 31-year-old male), consistently across the lifespan. Volumes showed sex differences with males larger (SWM 336±32 $cm^3$; DWM 138±29 $cm^3$) than females (SWM 306±21 $cm^3$; DWM 119±18 $cm^3$), both indicating that SWM is ~71% of the total WM volume. In contrast, there were no significant sex differences in FA or MD (data not shown). FA and MD versus age showed significant Poisson fits across the lifespan (SWM in orange, and DWM in blue, **Figure B**). The FA curves were consistently separated across the lifespan with DWM FA greater than SWM FA as expected. Interestingly, the MD curves were quite similar in childhood/adolescence and then diverged in adulthood and aging with SWM MD becoming much lower than DWM MD. FA peaked at 0.38 for SWM at 32 years and at 0.49 for DWM at 25 years. MD reached minimum values of 0.75x$10^{-3}$ $mm^2$/s for SWM at 42 years and 0.80x$10^{-3}$ $mm^2$/s for DWM at 30 years.

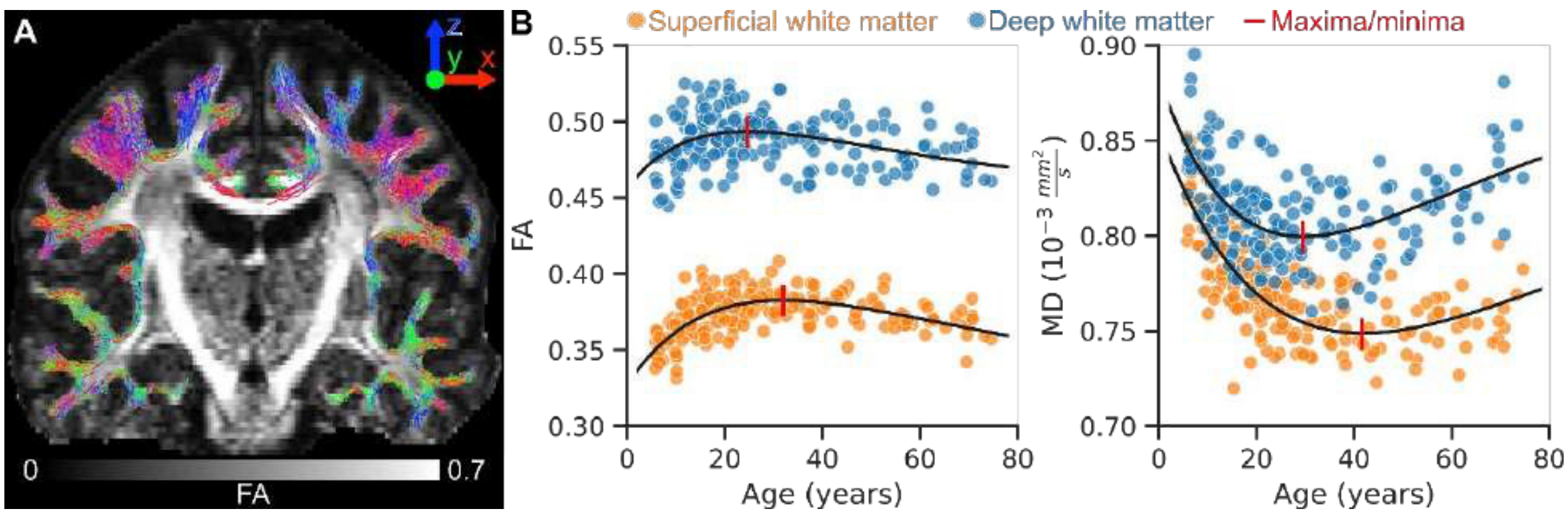

**Discussion:** Using higher spatial resolution (1.5 mm iso) and CSD tractography, the 'lifespan' curves for SWM were consistent with earlier reports either in neurodevelopment with higher FA/lower MD with age [3-6] or in aging with lower FA/higher MD with age [7-9]. However, the 'lifespan' age coverage here enables the assessment of the maxima/minima of FA/MD in SWM and DWM which occurred between 25 and 42 years, consistent with prior DTI findings for major WM tracts [10]. Notably, SWM reached peak FA and minimum MD later than DWM by 7 and 12 years, respectively, possibly due to prolonged myelination in SWM [1] as well as a delayed onset of age-related decline relative to DWM. These DTI metric delays in SWM relative to DWM have not been appreciated in the prior development/aging studies. A limitation here was the aggregation of all SWM, but future work will parcellate the SWM into lobes and specific gyral connections to examine regional superficial white matter differences of maturation and degradation with age.

## Structural Connectivity Mapping of the Central Amygdala

Vinod Kumar[1], Wenfei Han[1], Ivan de Araujo[1]
[1]Max Planck institute for Biological Cybernetics, Tuebingen, Germany

### Introduction

The central amygdala (CeA) is a key hub for integrating emotion, interoception, and autonomic responses, yet its structural connectivity remains largely unmapped. To address this, we investigated the structural connectivity of the CeA in both humans (using high-resolution diffusion MRI) and mice (using the Allen Mouse Connectivity Atlas), aiming to reveal its subcortical and cortical projection patterns and functional integration across domains.

### Methods

Diffusion MRI data from 730 subjects in the Human Connectome Project (Van Essen et al., 2012) were processed using multi-shell modeling via BEDPOST (Jbabdi et al., 2012; Jenkinson et al., 2012) and probabilistic tractography with PROBTRACKX (Behrens et al., 2007). The central amygdala (CeA) was defined using the CIT168 Subcortical Atlas (Pauli et al., 2018) and used as the seed region. Group-level connectivity probability maps were anatomically assigned using the HCP XTRACT Atlas, Harvard–Oxford Atlas, and Brainstem Navigator Atlas.
The Allen Mouse Connectivity Atlas provides voxel-wise efferent projections from the CeA using anterograde tracers (Kuan et al., 2015). The analysis combines the top 10 experiments (by injection volume) to delineate 3D projection density maps from the CeA to different brain regions.

### Results

The CeA exhibited widespread structural connectivity across cortical, subcortical, brainstem, and cerebellar regions, supporting its integrative role in emotion, memory, and autonomic regulation. Key pathways included the fornix, optic radiation, and fronto-occipital fasciculus. The strongest cortical associations were found in the subcallosal, cingulate, and parahippocampal areas, while subcortical targets included the thalamus, hippocampus, and amygdala. Brainstem nuclei such as the PCRtA, LPB, DR, and PAG showed high connectivity, highlighting roles in arousal and visceral control. The structural connectivity maps in the mouse also show widespread projection patterns to different cortical and subcortical areas.

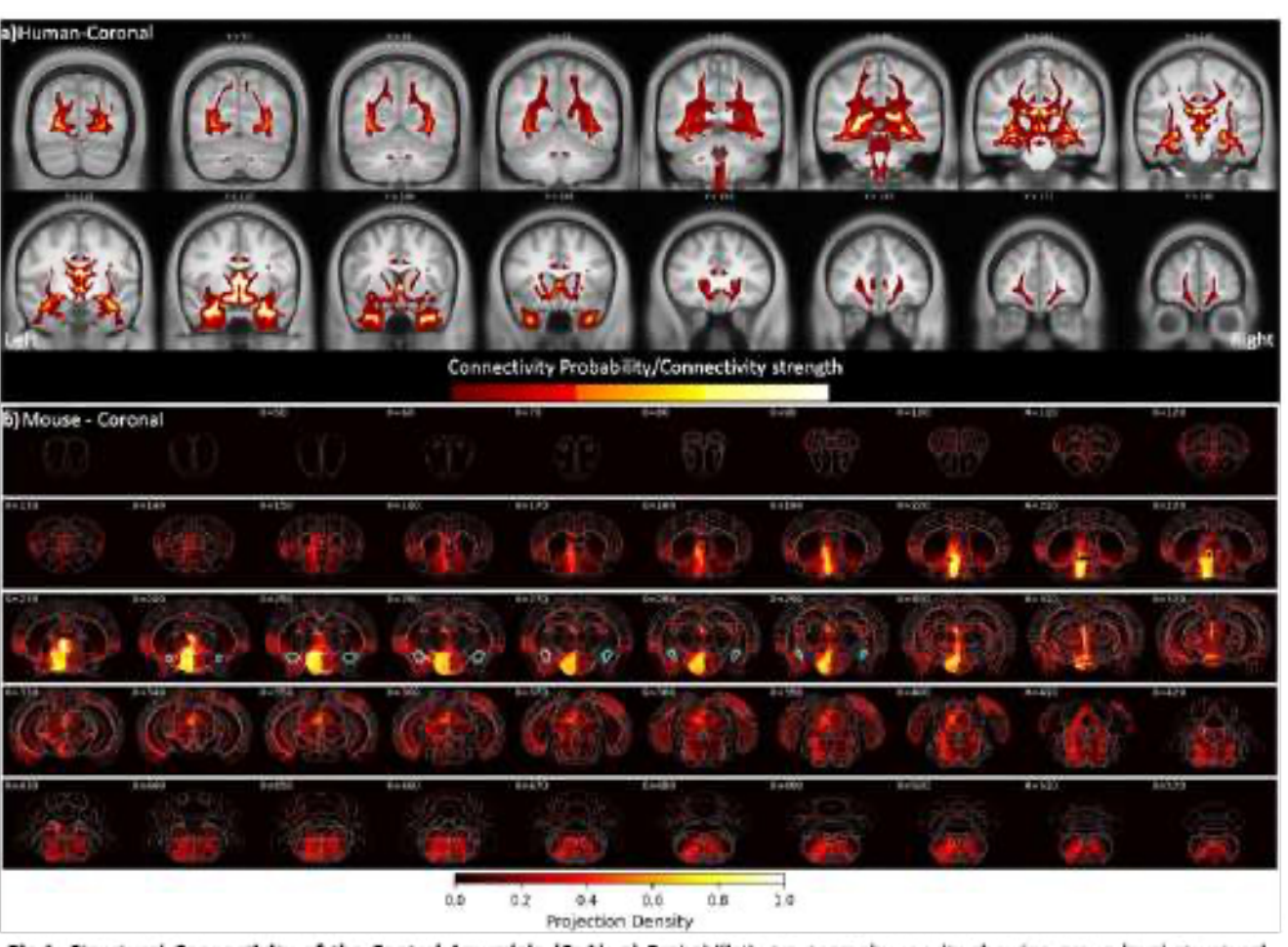

**Fig 1: Structural Connectivity of the Central Amygdala (CeA): a)** Probabilistic tractography results showing group-level structural connectivity maps of the left and right CeA to the ipsilateral hemisphere in 730 subjects. b) Coronal slice view showing projection density from the central amygdala (CeA), highlighting efferent connectivity patterns across brain regions.

### Conclusion

Our findings demonstrate that the CeA exhibits widespread structural connectivity with cortical, subcortical, and brainstem regions, consistent with its role as an integrative hub for emotion, interoception, and autonomic regulation. Robust connections to the PAG, PBN, NTS, DMV, and LC highlight its involvement in visceral pain and gut–brain communication. In the mouse brain, anterograde tracer data similarly reveal broad projection patterns, underscoring the CeA’s engagement in diverse brain–body functions. While diffusion tractography provides valuable macroscale insights, it remains an indirect measure of water diffusion along fiber pathways, which does not reflect actual axonal projections with the accuracy of tracer studies and is therefore susceptible to false positives and anatomical misinterpretation.

## Post-operative Clinical Outcome Is Not Correlated to Fronto-Striatal Tract Involvement by Diffuse Gliomas of the Supplementary Motor Area: Preliminary Results

Jahard Aliaga-Arias[1], Jose Pedro Lavrador[1], Richard Gullan[1], Ranjeev Bhangoo[1], Keyumars Ashkan[1], Flavio Dell'Acqua[2] and Francesco Vergani[1]

[1] Department of Neurosurgery, King's College Hospital NHS Foundation Trust, London, United Kingdom

[2] Department of Neuroimaging, Institute of Psychiatry, Psychology and Neuroscience, King's College London, London, United Kingdom

**Introduction. In** a recent study we reported the association of the interhemispheric proportions of the Frontal Aslant Tract volumes with the development of a verbal Supplementary Motor Area (SMA) syndrome after resection of diffuse gliomas (DGs) located in this brain region.[1] Following these results we now extended the analysis to another related tract: the Fronto-Striatal-tract (FST) portion connecting to the SMA We explored its role in the clinical outcome after surgery of SMA tumours.[2]

**Methods.** A retrospective single centre analysis was performed of a consecutive series of patients that underwent resection of diffuse gliomas in the SMA, over the previous 4 years. Clinical and radiological data were collected from electronic record systems. Preoperative diffusion weighted imaging data was acquired with 60 DW directions at b=1500s/mm$^2$ and 2.5 mm isotropic voxel on a 1.5 T MRI system. Data was pre-processed and probabilistic Spherical Deconvolution Tractography was performed using StarTrack. The Fronto-Striatal tract (FST) of both hemispheres was virtually dissected using TrackVis, with ROIs located on the SMA and caudate nuclei.[3] Mean volume, length, hindrance modulated orientational anisotropy (HMOA) and mean apparent diffusion coefficient (ADC) of the tracts were extracted. FST data were analysed for associations with functional outcomes after surgery.

**Results.** Of N=25 cases included in the study, n= 22 had FST identifiable on both hemispheres preoperatively, (n=2 had no tract reconstructed on the tumour side, and n=1 had no clear tract identifiable on either hemisphere. As expected, the reconstructed FSTs on the tumour side were significantly smaller in volume (median volume 9.37 cm$^3$ vs 5.38 cm$^3$, p= 0.001) and more elongated than the contralateral (median length 48.78 mm vs 42.63 mm, p= 0.033). The FST on the tumour side had higher ADC (median 0.73 vs 0.67, p< 0.001), and the HMOA was smaller (close to statistical significancy) compared to the contralateral side (median 0.009 vs 0.010, p= 0.051). N=12 cases developed an SMA syndrome, of which n=8 with motor features and n=6 with verbal features. No correlations were identified among any of the FST values extracted and pathological features of the tumour (WHO grade and IDH mutation status) or the onset of SMA syndrome. A further analysis of the interhemispheric lateralisation of the volumes and diffusion metrics of the tracts did not identified associations with the clinical outcome.

**Conclusions.** Our probabilistic SD tractography approach reliably delineates the FST in the context of diffuse gliomas invading the SMA and identify microstructural alterations caused by the tumour. No associations of preoperative FST features with clinical outcome after surgical tumour resection were identified in this study, suggesting a bilateral parallel function able to compensate the resection of the tract on a single hemisphere. Further studies with a larger cohort or with postoperative tractography analysis are currently on going to fully confirm the lack of contribution of the FST to the clinical outcome in surgery for gliomas of the SMA.

# **TractoSearch**: a Faster Streamline Search for Scalable Tractography Analysis

Etienne St-Onge[1]

[1]Department of Computer Science and Engineering, Université du Québec en Outaouais

**1. Introduction:** Tractography allows the in-vivo study of white matter pathways and brain connectivity. However, large tractograms, composed of millions of streamlines, pose significant computational challenges for subsequent tasks such as filtering, segmentation, clustering, and outlier detection. To reduce this complexity, existing methods often rely on approximation or non-deterministic strategies, which can compromise connectivity analyses accuracy and reproducibility.

Recently, in *Fast Streamline Search* (FSS) [St-Onge *et al.* 2022], we introduced a hierarchical search algorithm to efficiently perform exact proximity queries on tractography streamlines using the *average point-wise Euclidean distance*, which is equivalent to the $L^{2,1}$ mixed norm. This metric is commonly used to identify similar streamlines [O'Donnell *et al.* 2007; Olivetti *et al.* 2017; Garyfallidis *et al.* 2018]. FSS significantly improves upon the naïve pairwise comparison approach, which has quadratic time complexity $O(N^2)$, by reducing it to $O(N \log(N))$ through a hierarchical structure. While more precise and efficient than previous methods, FSS still involves redundant computations and increased memory usage, limiting its scalability for large tractograms.

**2. Methods:** To address these limitations, we developed TractoSearch, an open-source Python multithreaded framework, optimized for exact nearest neighbors and radius searches. We extended the standard k-d tree, typically used for point cloud data with Euclidean distance ($L^2$ norm), to support arbitrary $L^{p,q}$ Minkowski mixed norms. This enhancement enables both efficient indexing and exact querying of tractography streamlines. The generalization is based on an upper bound derived from Hölder's inequality [Albuquerque *et al.* 2017], using a scaling factor of $D^{(1/q-1/p)}$; where D is the space's dimensionality, subject to $1 \leq p \leq q < \infty$. For two streamlines ($S$ and $S'$) represented as an array sequences of 3D points, using the $L^{2,1}$ norm: $\|S - S'\|_2 \leq \|S - S'\|_{2,1} \leq 3^{(1-½)} \|S - S'\|_1$.

The resulting C++ k-d tree implementation builds upon the nanoflann library [Blanco et al., 2014] and offers Python bindings compatible with Scikit-Learn, facilitating integration with machine learning methods, some of which are described below.

Evaluation: Computation time was evaluated using 42 full-brain probabilistic tractograms from the Human Connectome Project (HCP), each containing 2.5 million streamlines ranging from 40*mm* to 250*mm* in length, and a white matter atlas containing 415k streamlines. Atlas bundles identification and adjacency matrices were computed using a radius search of 8*mm* (average distance along streamlines). Performance benchmarks were done using an Intel i7-14700K processor at 5.5Ghz, only using performance-cores, and 32 GB of RAM.

**3. Results & discussion:** TractoSearch k-d tree construction took 1.5 seconds, and 145 seconds for identification of the nearest atlas bundle, reduced to 22 seconds using 8 CPU cores (Figure #1). Constructing the adjacency matrix required 72 minutes on a single core and 11 minutes with 8 cores. In contrast, this operation was not feasible with the original FSS implementation due to excessive memory use, even on a system with 128 GB of RAM. By comparison, a naïve brute-force approach is estimated to require 4 days for nearest atlas segmentation and up to 22 days for adjacency matrix construction (approximated from a subset).

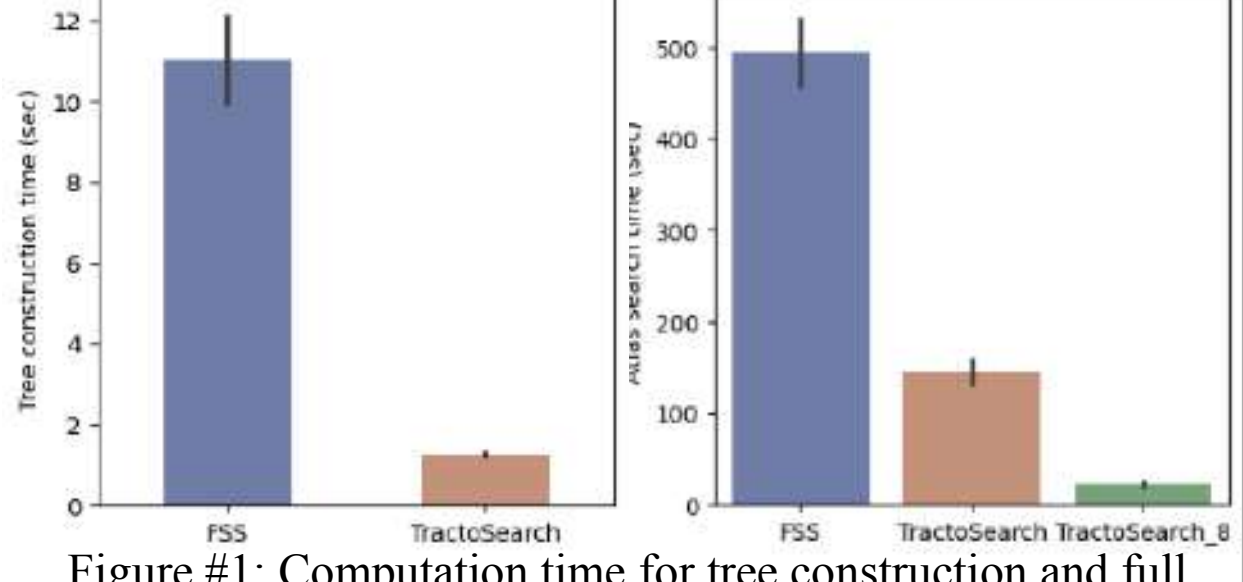

Figure #1: Computation time for tree construction and full brain segmentation in 33 bundles using KNN search.

The construction of this adjacency matrix (a neighborhood graph) allows for capturing local streamline relationships and serves as a foundation for a variety of tractography analyses, including density estimation, outlier filtering, and clustering. As shown in Figure 2, local streamline density can be used to: a) group similar streamlines, simplifying subsequent analysis; b) filter spurious fibers using fixed or statistically inferred thresholds; (c) analyze spatial streamline distributions for improved multi-subject comparisons.

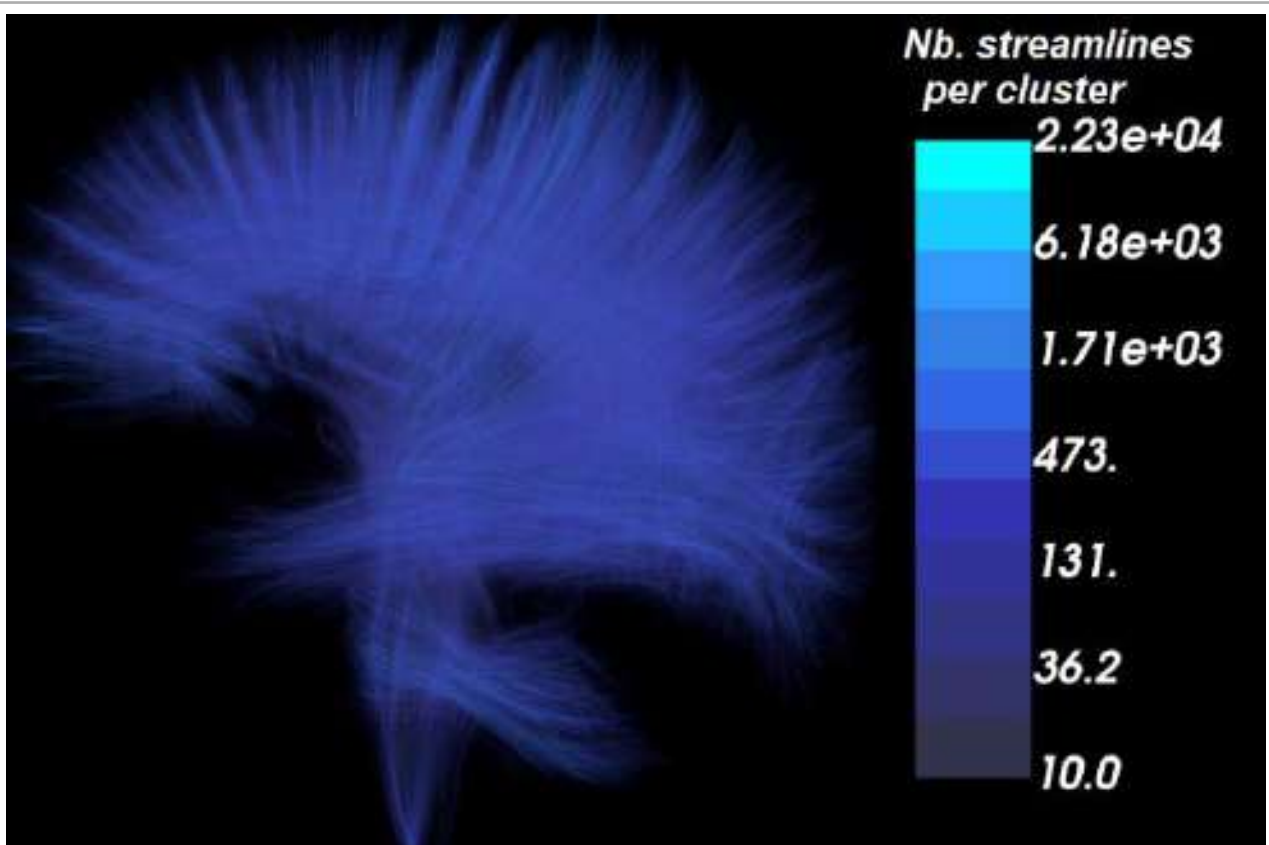

Figure #2: Streamlines clustered with an 8 mm radius; groups with fewer than 10 streamlines were filtered out. Clusters are shown as mean streamlines, color-coded by local density.

**4. Conclusion:** TractoSearch offers a fast and exact streamline distances computation using any $L^{p,q}$ mixed norms. It enables the construction of sparse neighborhood graphs with large tractograms, by avoiding redundant computation and memory overhead of a two-stage approach. This enables scalable and accurate tractography analysis and downstream tasks that require bundle segmentation, clustering, or density measures.

**Title: Amount of white matter activation and microstructures explain depression recovery in subcallosal cingulate deep brain stimulation**

**Authors: Ha Neul Song, Ki Sueng Choi, Helen S. Mayberg**

*Introduction*

Depression is increasingly understood as a disorder involving dysfunction across distributed brain networks rather than a localized pathology. SCC-DBS is thought to exert its effects by modulating interconnected systems, including the salience, default mode, and limbic networks, through stimulation of three major white matter (WM) tracts connected to the SCC: the cingulum bundle (CB), forceps minor (FM), and subcomponents of the uncinate fasciculus (UF). Evidence from tractography studies suggests that effective SCC-DBS engages projections from the SCC toward the rostral anterior and midcingulate, medial prefrontal, insular, and hippocampal regions, reinforcing the importance of circuit-level modulation via three WM tracts. These findings have contributed to a shift from anatomical landmark-based toward connectome-based targeting that emphasizes patient-specific connectivity profiles. To date, most neuroimaging studies and surgical planning of SCC-DBS have focused on large-scale connectivity patterns, with less attention given to WM microstructure, as inferred from diffusion-weighted imaging (DWI). This study investigated whether stimulated WM tracts and baseline microstructures are related to clinical outcomes following SCC-DBS.

*Methods*

A total of 52 SCC-DBS patients underwent bilateral SCC DBS surgery and had available weekly clinical data (HDRS17) for 24 weeks. Structural T1w and DWI were collected using a following parameters: 3D MPRAGE sequence, sagittal orientation, $1.0 \times 1.0 \times 1.0$ mm³ resolution, TR = 2,600 ms, TI = 900 ms, TE = 3.02 ms, flip angle = 8°, and 60 directions with five b0s images, b=1000 (n=37) and 1200 (n=15) s/mm², $2 \times 2 \times 2$ mm³ resolution, reversal-phase encoding scans. Standard DWI preprocessing (denoise, eddy current and motion correction) was applied. Postoperative CT scans ($0.46 \times 0.46 \times 0.65$ mm³ resolution) were coregistered to preoperative T1w images using Advanced Normalization Tools.[10] DBS electrode localization was performed using TRAC/CORE and PaCER toolboxes in Lead DBS. Weekly DBS settings (active contact configuration and amplitude) were used to estimate bilateral volumes of tissue activated (VTAs) in template space. Two distinct WM measures were evaluated: stimulation extent within major WM tracts and baseline microstructural integrity estimated by fractional anisotropy (FA) using Linear mixed-effects models to examine tract-specific joint influence of WM features on clinical improvement.

*Results*

We found a significant association between the stimulated WM proportion and clinical improvements in bilateral mid-CB and left FM (Figure 1). Higher stimulated WM proportions are associated with greater clinical improvement. Although no significant main effects were identified between baseline FA and clinical improvements, interaction effects between stimulated WM proportion and FA were found in left mid-CB and left FM. These interactions exhibited opposing directions: the mid-CBL showed a positive interaction, indicating the greater stimulation effect on clinical improvement in individuals with higher baseline FA, whereas the L-FM showed a negative interaction, suggesting the greater stimulation effect in those with lower FA.

*Conclusions*

Our findings suggest the importance of WM microstructural integrities in the therapeutic response of SCC DBS. Combining stimulation proportion and WM integrity will refine the connectomic SCC DBS targeting and allow us to optimize stimulation parameters in SCC DBS.

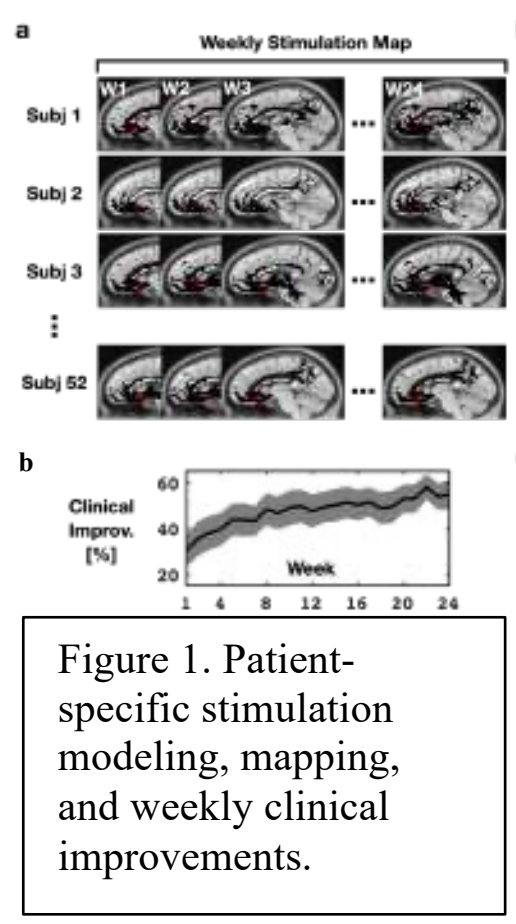

Figure 1. Patient-specific stimulation modeling, mapping, and weekly clinical improvements.

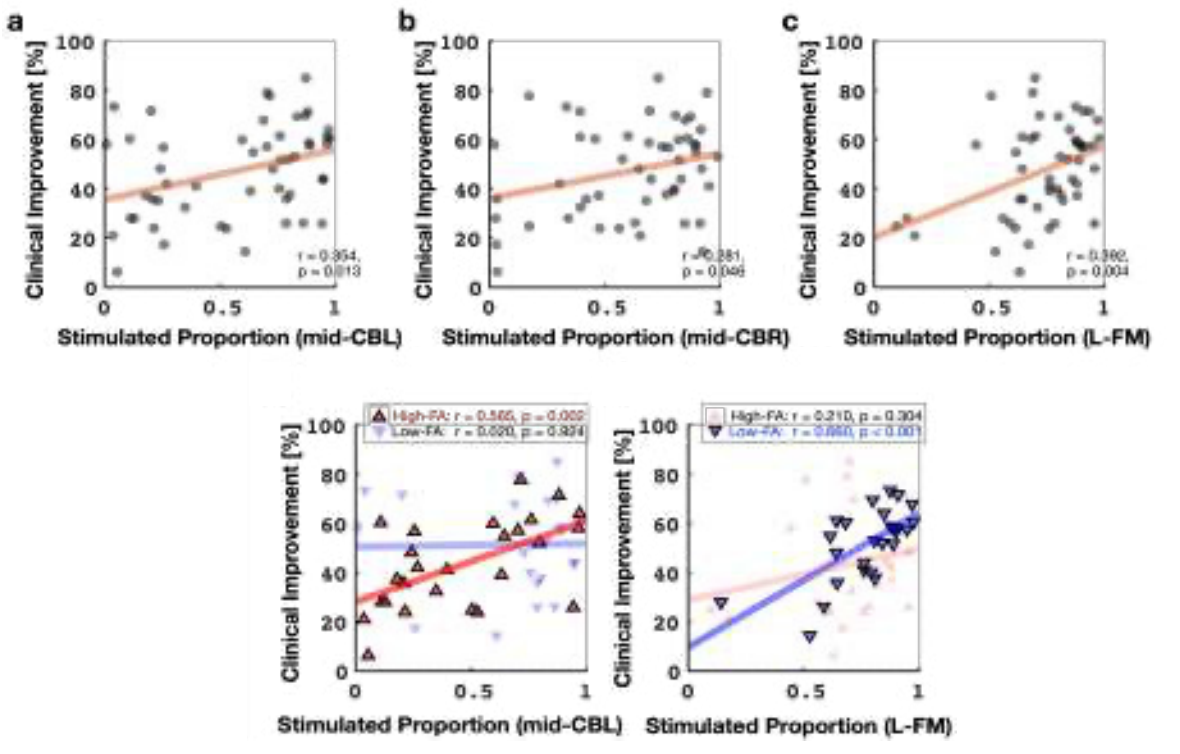

Figure 2. Significant linear association between stimulated proportion and clinical improvements: in a) left mid-CB, b) right mid-CB, and c) left FM. Significant interaction between stimulated proportion and baseline WM microstructures (FA) in d) left mid-CB and e) left FM

# Generation of synthetic data for validating tractography-based cortical parcellation and fiber clustering algorithms

Elida Poo[1], Joaquín Molina[1], Jean-François Mangin[2], Cecilia Hernández[1], Pamela Guevara[1]
[1] Faculty of Engineering, Universidad de Concepción, Concepción, Chile
[2] CEA, CNRS, Baobab, Neurospin, Université Paris-Saclay, Gif-sur-Yvette, France,

**Introduction.** Diffusion Magnetic Resonance Imaging (dMRI) tractography [1] has enabled the study of white matter connectivity and the development of automated methods for both fiber bundle clustering and segmentation and cortical parcellation. However, objective validation of such methods is limited by the lack of anatomical ground truth. We present two tools to address this gap: PhyberSIM [2], a white matter fiber bundle simulator, and a synthetic data generator that produces random cortical parcellations based and their connections [3], to validate tractography-based cortical parcellation (TBCP) methods.

**Materials and Methods.** PhyberSIM generates fiber bundles using a tubular model parameterized by a centroid, selected from the tractogram of a subject, and five radii along the bundle trajectory, used to generate fibers based on spline curves (Fig. 1). It was validated using bundles from a Deep White Matter (DWM) atlas [4] and employed to evaluate two fiber clustering algorithms, QuickBundles (QB) [5] and FFClust [6] through five classic clustering metrics. The cortical parcellation data simulator creates synthetic data consisting of a geodesic distance-based cortical parcellation based on the cortical mesh of a subject and the connections between the generated parcels. For the simulation of each bundle, it defines a centroid based on the subject's tractogram and adapted end points to fit the shape of the connecting parcels. We generated a database of 20 subjects with 150 parcels per hemisphere that was used to validate and improve a TBCP algorithm based on a two-level fiber clustering [7] (Fig. 4).

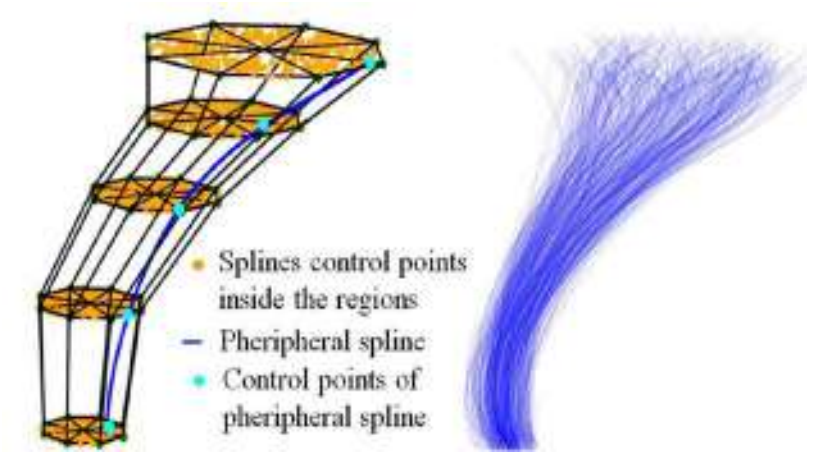

Fig. 1: Schematic of a simulated bundle generated with PhyberSIM (tubular spline model) [2].

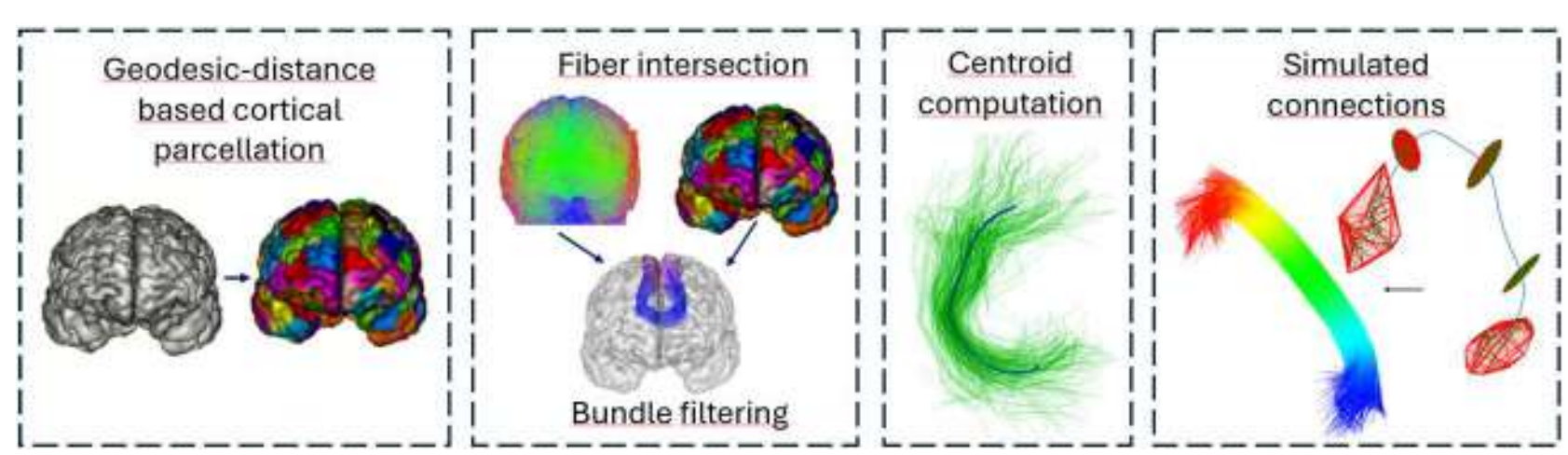

Fig. 2: Schematic of the generator of a random cortical parcellation with its connections for tractography-based cortical parcellation validation [3].

**Results.** PhyberSIM achieved an average 76.5% overlap with DWM atlas bundles and revealed a good performance for both fiber clustering algorithms, maintaining robustness to different number of simulated bundles and input order permutation. It also detected differences between clustering methods: FFClust tends to over-segment, while QB tends to merge close bundles. Regarding the cortical parcellation data, the tested CBCP algorithm could be improved and for its best parameter configuration, detected 118 parcels for the left hemisphere and 120 parcels for the right hemisphere, based on a DICE > 0.5, with a mean DICE of 0.61.

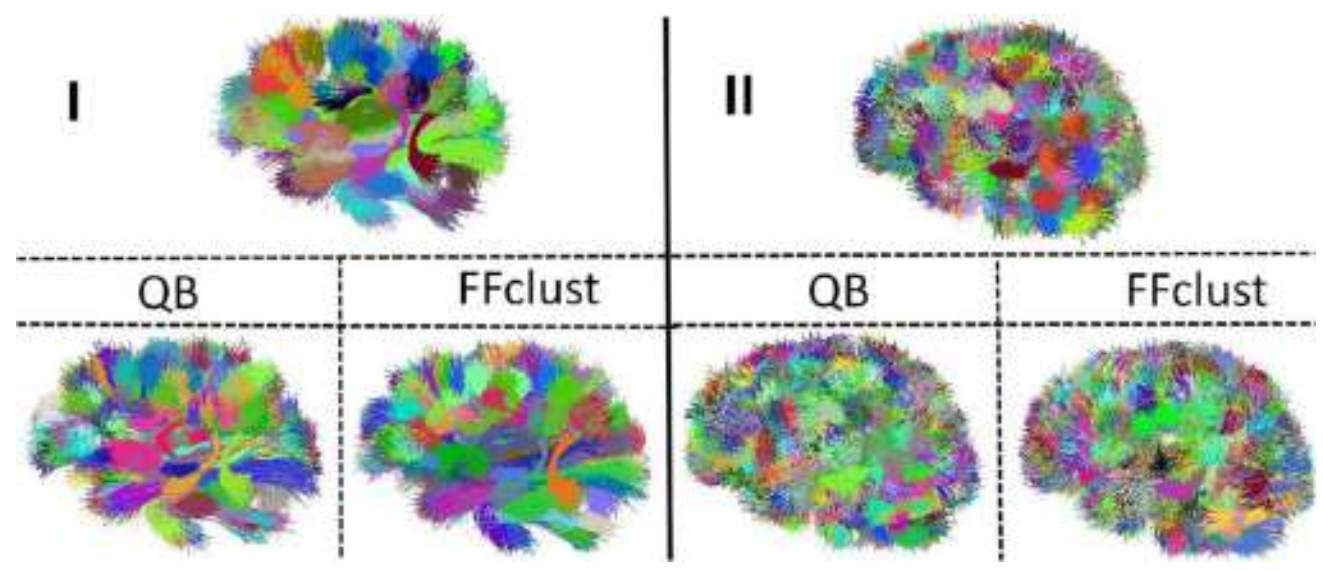

Fig. 3: QB and FFClust results for simulated tractography datasets with 100 bundles (I) and 500 bundles (II) for a distance of 10 mm.

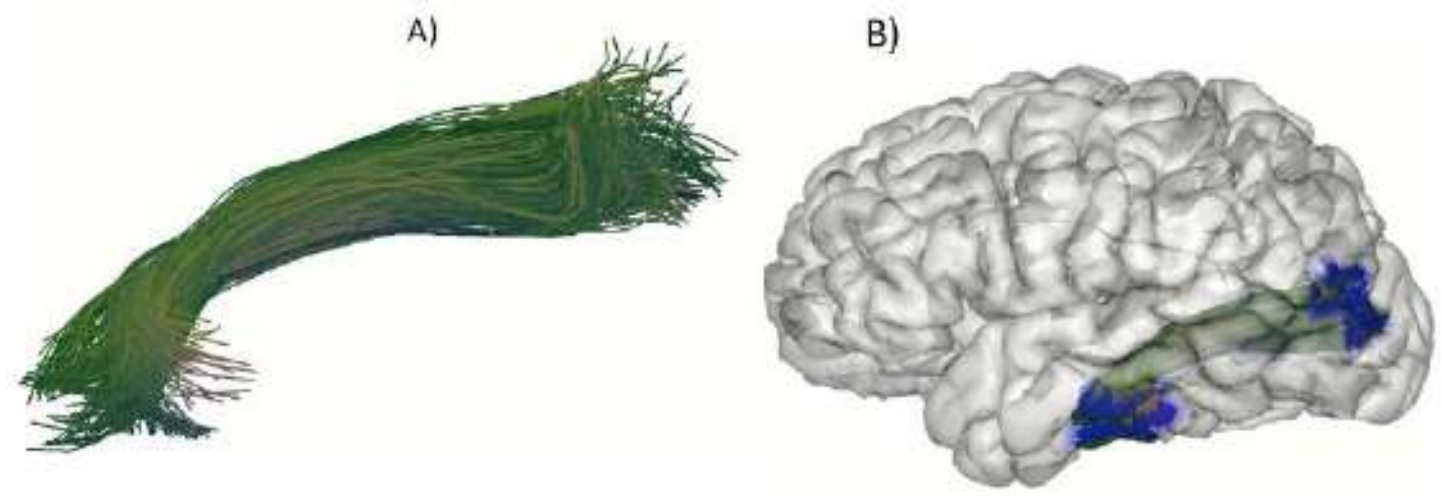

Fig. 4: Example of a simulated bundle generated from two cortical parcels. A) The simulated bundle. B) The simulated bundle (green) connecting a pair of parcels (blue) on the cortical mesh.spline curves.

**Conclusions.** Together, PhyberSIM and the cortical parcellation simulator offer a comprehensive, controlled validation framework for fiber clustering and tractography-based cortical parcellations algorithms. These tools enable reproducible experiments with known ground truth, providing valuable resources for the dMRI tractography community. Further improvements, such as added noise and flexibility on bundle complexity and configuration are planned for future work.

# Principal Component Analysis of Diffusion MRI and Magnetization Transfer Metrics Reveals Distinct Lesion Microstructure in Multiple Sclerosis

E. Hernandez-Gutierrez[1], R. Coronado-Leija[3], M. Edde[1], M. Descoteaux[1,3], M. Dumont[2], JC. Houde[2], M. Barakovic[5], S. Magon[5], A. Ramirez-Manzanares[4]

[1]Sherbrooke Connectivity Imaging Lab (SCIL), Department of Computer Science, University of Sherbrooke, Sherbrooke, QC, Canada; [2]Imeka Solutions Inc., Sherbrooke, QC, Canada; [3]Bernard and Irene Schwartz Center for Biomedical Imaging, Department of Radiology, New York University School of Medicine (NYU), New York, NY, United States; [4]Centro de Investigación en Matemáticas A.C. (CIMAT), Department of Computer Science, Guanajuato, GTO, Mexico; [5]Pharma Research and Early Development, Neuroscience and Rare Diseases Roche Innovation Center Basel, F. Hoffmann-La Roche Ltd., Basel, Switzerland

**INTRODUCTION:** Multiple sclerosis (MS) lesions exhibit heterogeneous microstructural changes that are not fully captured by individual imaging metrics. Advanced diffusion MRI (dMRI) and quantitative magnetization transfer (MT) imaging provide complementary information on tissue integrity and demyelination [1]. However, the high dimensionality of these data complicates lesion characterization [2]. Visualizing data from MS lesion studies presents significant challenges due to the inherently multidimensional nature of the data. MS research typically involves large patient cohorts, with multiple scans conducted per patient. Each scan reveals several lesions, each characterized by multiple metrics. The complexity further increases when considering specific WM tracts, adding an additional dimension through the subdivision of tracts into multiple sections. We applied principal component analysis (PCA) to reduce dimensionality and identify differences in lesioned tissue across a cohort of MS patients.

**METHODS:** Twenty MS patients and twenty-six healthy controls (HC) underwent five longitudinal MRI sessions each, including multi-shell dMRI, MT and T2-weighted fluid-attenuated inversion recovery (FLAIR) [3]. Each patient scan was pre-processed with *Tractoflow* [4], which included constrained spherical deconvolution (CSD) [5] and particle filtering tractography [6] to generate a tractogram. Tractograms were automatically segmented into major bundles using RecoBundlesX (`github.com/scilus/rbx_flow`). The multi-tensor model [7], with up to 3 tensors per voxel, was fitted to pre-processed images using the MRDS framework [8] to obtain multi-tensor fixel-based metrics (fixel-AD, fixel-RD, fixel-MD, fixel-FA), including isotropic volume fraction (ISOVF). The number of tensors per voxel was determined using track orientation density imaging [9]. As fixel-based metrics provide multiple values per voxel (one for each tensor), we generated, for each segmented bundle, a metric map in which only the metric value from the tensor most closely aligned with the local bundle direction was assigned to the voxel. The *ihmt_flow* (`github.com/scilus/ihmt_flow`) was employed to extract MTsat and MTR metrics registered to diffusion space. Lesions were segmented from T2-FLAIR images by *NeuroX* (`neurorx.com`) and labeled individually with the *Scilpy* tools (`github.com/scilus/scilpy`). For each lesion in each session of each patient, we extracted the median value of each metric, resulting in a lesion-by-metric matrix (N lesions × 4 metrics: fixel-AD, fixel-RD, ISOVF, MTsat), where N is the total count of lesions across the whole cohort. For fixel-based metrics, computed per-bundle metric maps were used to extract the median value for each lesion. Finally, PCA was performed to reduce the data to three principal components (PCs), followed by k-means clustering (k=3) to identify lesion subgroups.

**RESULTS:** The first three PCs explained 87.9% of the variance. Feature loadings indicated that PC1 was primarily driven by MTsat (0.63) and negatively by fixel-RD (-0.54), while PC2 was dominated by fixel-AD (0.72). K-means clustering revealed three lesion groups; a cluster containing degenerated tensor metrics was excluded. Figure 1 illustrates the spatial distribution and microstructural characteristics of the identified lesion clusters. Both clusters exhibited the classic pattern of demyelination with increased fixel-RD and ISOVF, and reduced MTsat [1] compared with HC. However, lesions in cluster 2 demonstrated significantly higher MTsat ($3.07 \pm 0.44$; $p = 0.23e\text{-}5$) and lower ISOVF compared to cluster 1 ($0.21 \pm 0.05$; $p = 0.42e\text{-}4$). Multi-tensor fixel-based metrics revealed that Cluster 1 and 2 lesions maintained higher fixel-AD ($1.49 \pm 0.2$ ms/$\mu$m$^2$; $p = 1.34e\text{-}2$) compared to HC.

**CONCLUSION:** PCA of combined diffusion and MTsat metrics effectively distinguishes microstructural clusters in MS lesion, supported recent applications of unsupervised machine learning in MS [2]. The approach has limitations to differentiate between inflammation subtypes or between demyelination and inflammation. Unexpected higher fixel-AD values in Clusters 1 and 2 compared with HC maybe due to contamination of the fixel-based metrics by isotropic contribution, as reported before [10]. Overall results suggest that, even among lesions with broadly comparable profiles, subtle but statistically significant microstructural differences exist, potentially reflecting varying degrees of demyelination and tissue damage.

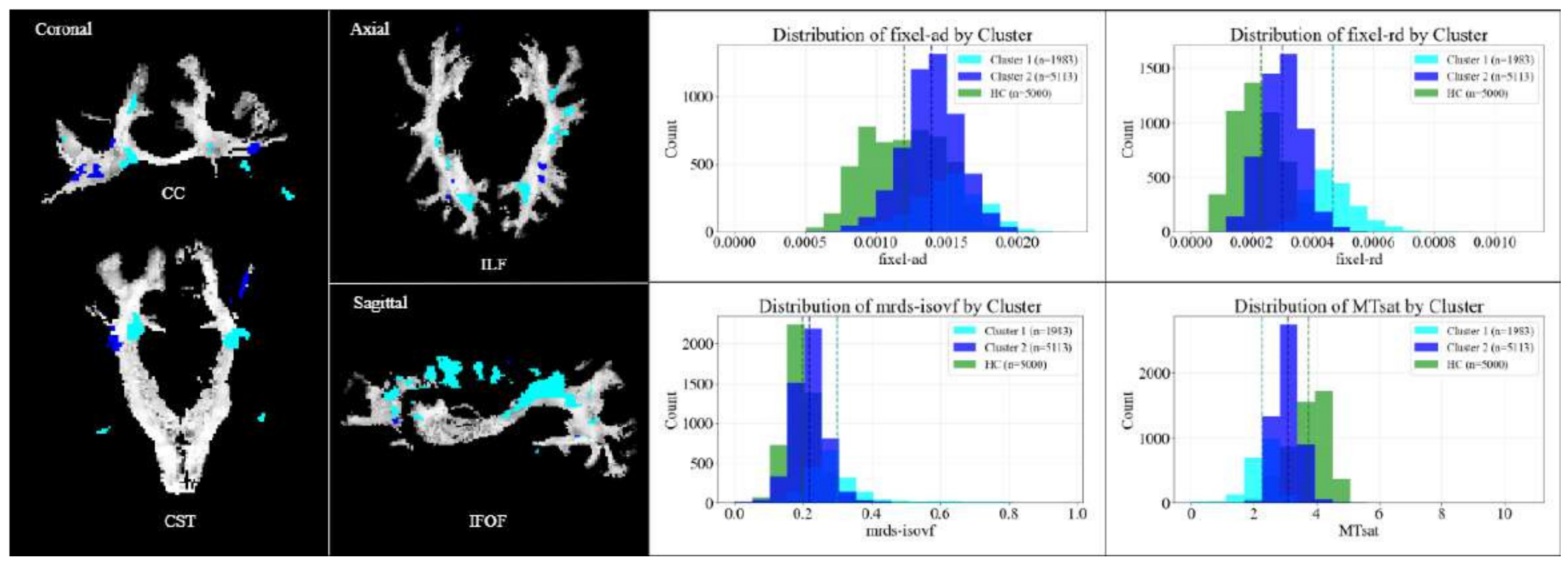

Figure 1: Spatial distribution and microstructural characterization of MS lesion clusters identified through PCA analysis. (left) Axial, coronal, and sagittal views of a single patient showing lesions colored according to cluster assignment (Cluster 1: cyan; Cluster 2: discarded; Cluster 3: blue) in four major bundles (CC, CST, ILF, IFOF). Lesions are overlaid on per-bundle fixel-FA map. (right) Histograms comparing the distribution of key metrics between clusters: fixel-AD (axial diffusivity), fixel-RD (radial diffusivity), isotropic volume fraction (ISOVF), and MTsat. Cluster 2 (cyan) exhibits higher MTsat values and lower fixel-RD compared to Cluster 1 (blue), indicating better preserved myeline. Clusters 1 and 2 demonstrates characteristic pattern of demyelination compared with healthy controls (HC: green). All metric distributions show statistically significant differences between clusters ($p < 0.001$).

# Hybridization Strategies for Robust Brain Tractography

Jesús Martínez-Miranda,*[,1] Gabriel Girard,[2] and Alonso Ramírez-Manzanares.[1]

[1] *Department of Computer Science, Centro de Investigación en Matemáticas A.C., Guanajuato, México.*

[2] *Department of Computer Science, Université de Sherbrooke, Sherbrooke, Canada.*

*jesus.martinez@cimat.mx

**Introduction.** Tractography is a technique to estimate connectivity pathways between brain regions using diffusion-weighted magnetic resonance imaging. Despite advances in tractography, existing methods continue to face challenges in reliably reconstructing different bundles in the brain [1].

**Methods.** As in many problem-solving scenarios, the hybridization from the best methods allows us to develop a novel approach that takes advantage of their most robust characteristics. Parallel Transport Tractography (PTT) [2] is a streamline propagation approach known for producing geometrically smooth curves through its framework with a local spatial support. Particle Filtering Tractography (PFT) [3] reduces the number of streamlines that prematurely terminate in white matter or cerebrospinal fluid by using maps of anatomical information. PFT uses a trajectory correction (backtracking) allowing streamlines to reach gray matter. Flocking Tractography (FT) [4] generates streamlines using the collective behavior of a group of particles that communicate and correct their trajectories through spatial information exchange. In this work, we implemented FT within the DIPY software library [5]. The FT's new implementation allows comparing and contrasting the performance of the three state-of-the-art methods above, as well as hybridize them, combining PTT and PFT algorithms (PTT-PFT), and FT and PTT algorithms (FT-PFT). The two hybrid methods follow their original direction selection strategies, while also using the PFT backtracking correction-strategy when a tracking problem prevents the particles to continue moving, for instance when a streamline ends in CSF. To evaluate their performances, we used the synthetic phantoms from the DiSCo Challenge 2021 (DiSCo) [6] and the ISMRM Tractography Challenge 2015 (ISMRM) [1]. For the DiSCo phantom, signal noise was added with SNR of 5, and a single b-shell of 3094 with only 30 b-vectors was used to simulate clinical conditions. For each phantom, the official scoring metrics from each challenge are used as follows. For the DiSCo data, we compute the Pearson correlation coefficient ($r$) between the estimated and ground-truth (GT) connectivity matrices [6]. For the ISMRM data, we compute the percentage of valid connections (VC), the average percentage of volumetric overlap (OL) and overreach (OR) per bundle [1]. Moreover, for both phantoms, we evaluated the volumetric Dice coefficient over the bundles [7]. All methods used the same Fiber Orientation Distributions (FODs) [5]. Performances of the DIPY probabilistic tractography method [5] are shown as reference.

**Results and Conclusion.** Tables 1 and 2 report the quantitative results (best scores highlighted in bold) on the DiSCo and ISMRM dataset, respectively. Figure 1 shows examples of the trajectories recovered for the Corticospinal Tract (CST) bundle of the ISMRM dataset. Both the Figure and Tables show that the hybridized proposals recover a greater portion of GT volume than the standalone methods. The PFT backtracking improves valid connections and overlap with the GT, but also increases overreach. This suggests a trade-off between coverage and precision, which future work will aim to optimize. The hybrid PTT-PFT method stands out in particular, as it achieves the best correlation performance in DiSCo, and valid connections and volumetric overlap in ISMRM.

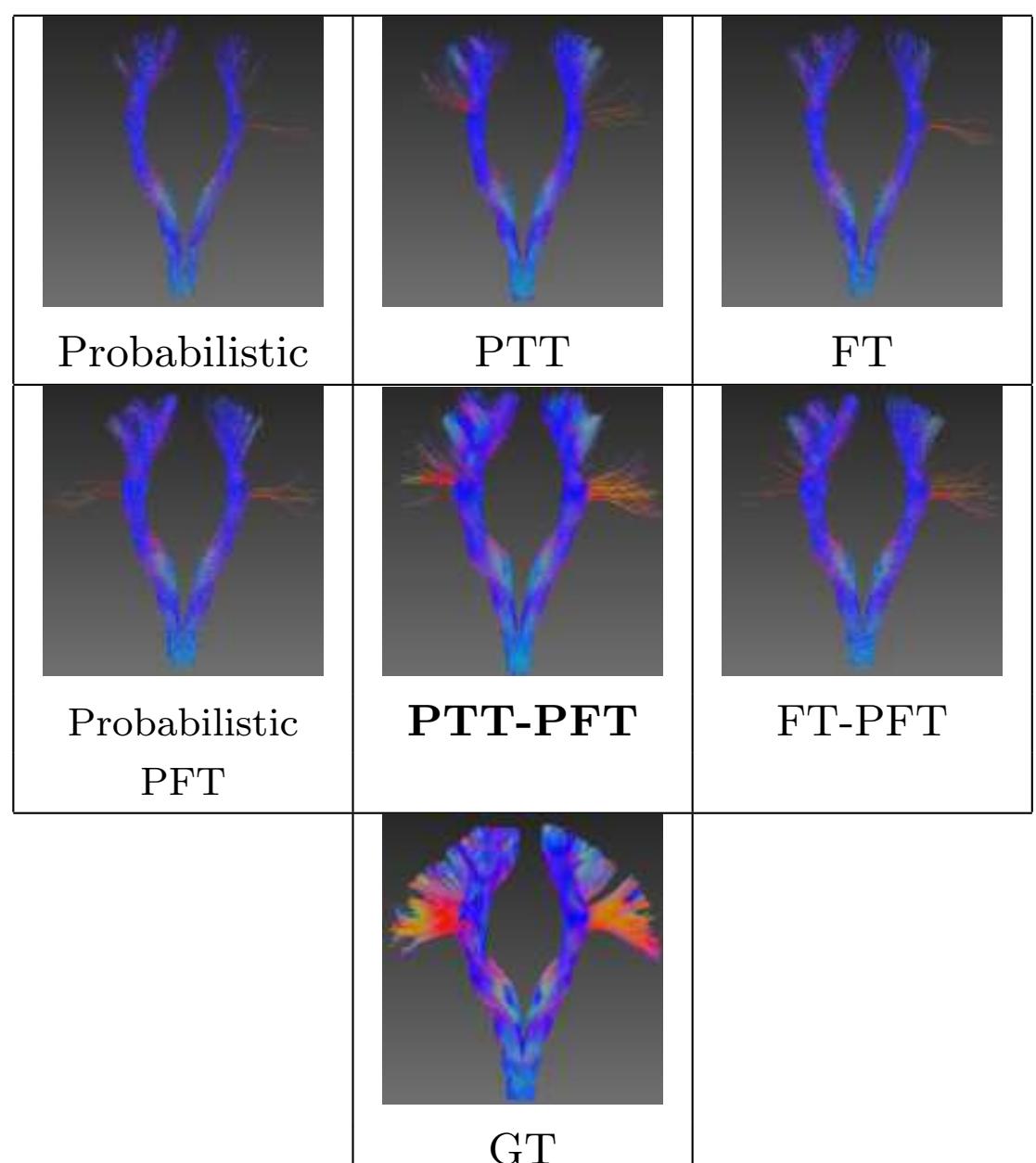

Figure 1: ISMRM dataset CST from different tractography methods. Individual methods on top, hybrid methods in the middle. From Tables 1 and 2, PTT-PFT performs best.

**Table 1:** Tractogram correlation with the GT connectivity matrix of the DiSCo dataset. Best values in bold font.

| Method | $r$ | Dice |
|---|---|---|
| Probabilistic | 0.818 | 0.636 |
| Probabilistic-PFT | 0.835 | 0.619 |
| PTT | 0.850 | **0.648** |
| PTT-PFT | **0.856** | 0.603 |
| FT | 0.846 | 0.642 |
| FT-PFT | 0.850 | 0.601 |

**Table 2:** Valid Connection (VC), Volumetric overlap (OL), Overreach (OR) and volumetric Dice of the different methods on the ISMRM dataset. Best values in bold font.

| Method | VC (%) | OL (%) | OR (%) | Dice |
|---|---|---|---|---|
| Probabilistic | 54.89 | 51.74 | **24.42** | 0.593 |
| Probabilistic-PFT | 61.89 | 69.22 | 34.23 | **0.658** |
| PTT | 64.47 | 59.62 | 28.59 | 0.605 |
| PTT-PFT | **65.61** | **72.33** | 45.06 | 0.556 |
| FT | 56.11 | 54.98 | 25.64 | 0.598 |
| FT-PFT | 63.15 | 67.20 | 31.98 | 0.642 |

## Clinical-ComBAT: Towards Flexible Harmonization of White Matter measures in Clinical Diffusion MRI

Manon Edde[1], Gabriel Girard[1], Felix Dumais[1], Yoan David[1], Maxime Descoteaux [1,2], Pierre-Marc Jodoin [1,2]
[1] Université de Sherbrooke, QC, Canada, [2] Imeka Solutions inc, QC, Canada.

**INTRODUCTION.** ComBAT[1] is a widely used neuroimaging harmonization method that corrects scanner-related effects while preserving biological signals. While it's effective in a controlled research context, its assumptions (linear covariate effects, homogeneous populations, and full data access) **limit its use in clinical practice**. In real-life situations involving diverse patients, continuous data collection and continually evolving imaging centers, the usual approach of aligning data to an average site becomes inappropriate. Furthermore, ComBAT's linear model struggles to handle nonlinear patterns, such as age-related changes in diffusion imaging-derived measures. This can introduce residual bias, which is particularly problematic when harmonization is used to support normative modelling, diagnosis or the monitoring of disease progression.

**METHOD.** We introduce **Clinical-ComBAT**, a novel harmonization method specifically designed to meet the constraints encountered in real-life clinical environments. It incorporates three key innovations: (1) **site-specific harmonization**, where each site is aligned to a fixed reference dataset, allowing new clinical sites and patient data to be added incrementally. Clinical-ComBAT uses informed priors from the reference site to enhance the robustness of parameter estimation, particularly variances and slopes, enabling good performance to be maintained even with a low number of subjects; (2) **polynomial modeling** of covariates such as age, to better capture the non-linear relationships between these variables and imaging measures; (3) a **robust variance estimation** framework adapted to limited sample sizes, as well as an index of **harmonization quality** using the Bhattacharyya distance (BD).

Clinical-ComBAT was evaluated on Diffusion Weighted Images (DWI) from The National Institute of Mental Health[2] (NIMH, n=119), and the Cambridge Centre for Ageing Neuroscience[3,4] (CamCAN, n=441). All datasets were processed with the TractoFlow[5] pipeline. DTI scalar maps, Fractional Anisotropy (FA) and Mean Diffusivity (MD), were derived from b-values <1200 s/mm². The apparent fiber density map (AFD) was obtained from the fiber Orientation Distribution Function generated using a spherical harmonics order of 8 and a standardized response function[6]. All measurements were registered in MNI space, and the average scalar values for **42 white matter bundles** were extracted using the streamline density map and the **IIT Human Brain Atlas**[7] (v5.0).

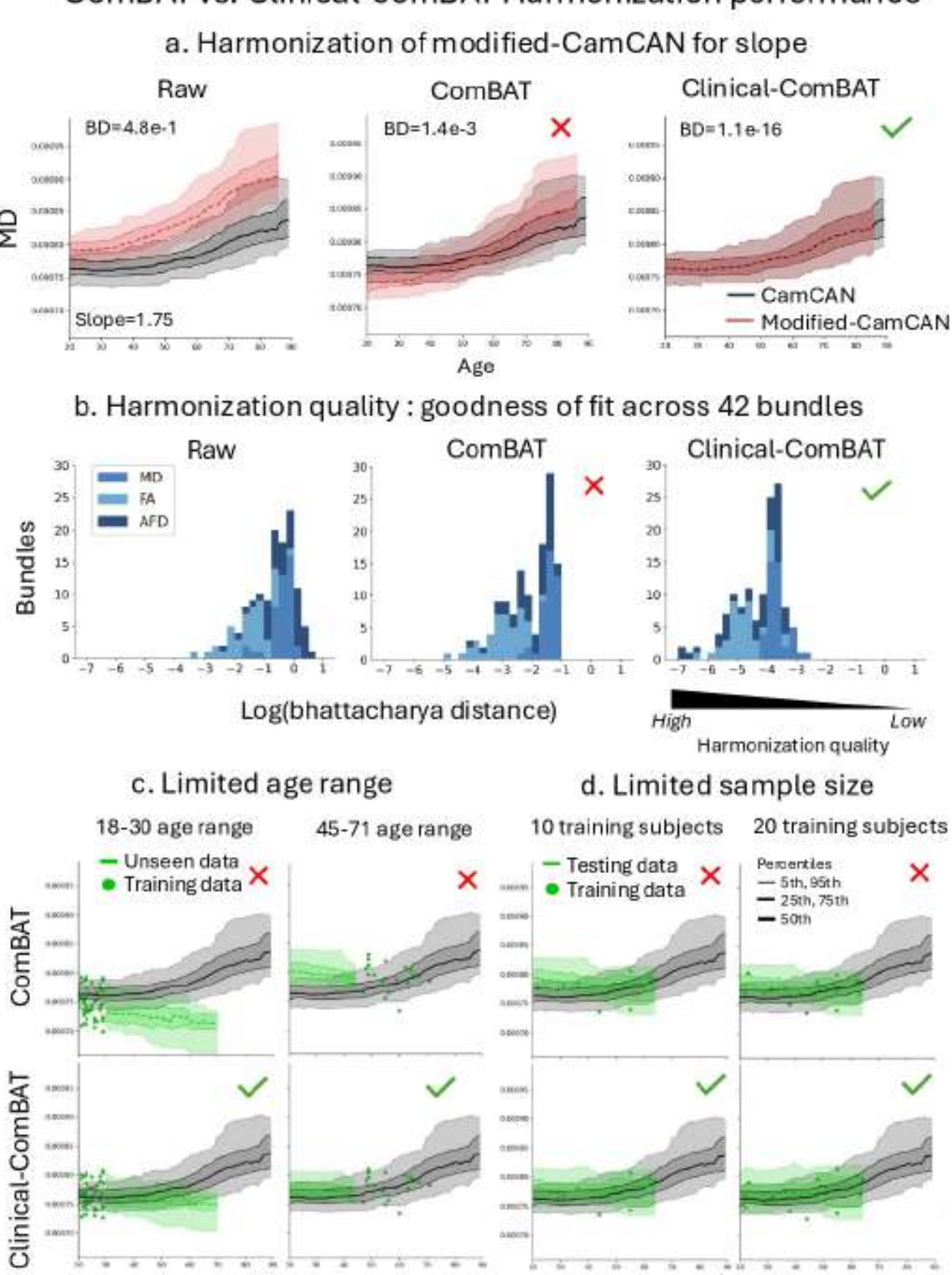

**Figure 1.** Harmonization results for DWI-metrics using Clinical-ComBAT and ComBAT.

We compare the harmonization performance of **Clinical-ComBAT versus ComBAT** using the Bhattacharya distance (BD). The NIMH was used as a **moving site**, and the CamCAN dataset is used as a fixed **reference site**. The moving site datasets have been split into training and testing datasets.

**RESULTS.** Using a modified version of the CamCAN data, Figure 1(a) shows that compared with the ComBAT method, Clinical-ComBAT corrects additive, multiplicative, and slope biases **more accurately without causing data overfitting**. Clinical-ComBAT also demonstrates a **more accurate alignment** of the moving site distribution with that of the reference site. This is evident from the consistent improvement in harmonization quality observed across all bundles and scalar maps (BD decrease, Fig. 1b). Clinical-ComBAT also stands out for **its ability to generalize harmonizations** to new subjects (green percentiles) and age groups that were not observed during the training subjects (green dots) (Fig. 1c). Finally, this method remains **robust** even when the number of healthy subjects available per site is low (Fig. 1d).

**CONCLUSION.** Clinical-ComBAT tackles the limitations of traditional harmonization methods by adapting to the dynamic and heterogeneous nature of clinical imaging workflows. By decoupling harmonization from centralized cohort structures and adopting non-linear modelling, it enables **scalable and reproducible harmonization** tailored to clinical translation. This approach paves the way for robust multi-site integration of diffusion MRI into routine care, ultimately improving early detection and follow-up of brain diseases.

**Title: Variability of white matter activation pathways for connectomic targeting in subcallosal cingulate deep brain stimulation**

**Authors: Ha Neul Song, Helen S. Mayberg, Ki Sueng Choi**

*Introduction*

The subcallosal cingulate deep brain stimulation (SCC DBS) has shown efficacy in treating treatment-resistant depression (TRD). Recent development of connectomic targeting has shifted the focus from a focal target to a multi-node network target within the SCC. The SCC is interconnected with other brain regions through white matter (WM) bundles, and recent studies reported that stimulation of all connections is necessary for a clinical response. Connectome-based surgical targeting of SCC DBS relies on individual structural connectivity analysis to maximize the activation of critical WM bundles: cingulum bundle, forceps minor, and subcortical junction. However, normative connectome data-based individual patient target selection was suggested due to the limitations of the clinical environment in collecting high spatial and angular diffusion-weighted data, and the convenience of selecting a personalized SCC target, necessitating defining individual diffusion data for each case. This study explored inter- and intra-subject variabilities of WM activation pathways in SCC targets using 100 subjects' human connectome data to investigate connectome-based targeting accuracy.

*Methods*

Whole brain tractography was performed using FSL probtractx2 toolbox on human connectome data (HCP; n=1000) using a probabilistic stimulation map seed (Figure 1, Left: x = -9, y = 28, z = -6, Right: x = 7, y = 29, z = -4, radius = 3 mm), which is derived from bilateral volumes of tissue activated (VTAs) of 47 TRD patients with given patient-specific stimulation settings (amplitude and contact configuration) and the difference in patients' Hamilton Depression Rating Scale (HAMD-17) using Lead DBS software. Correlation coefficient values were used to measure the inter-subject similarity in each critical WM bundle. Furthermore, intra-subject similarity (spatial similarity) was measured in each critical WM bundle while the seed was moved in (1) superior-inferior, (2) anterior-posterior, and (3) medial-lateral directions.

*Results*

We found low inter-subject similarity in the cingulum bundle compared to others ($p < 0.001$). Spatial similarity analysis demonstrated a low similarity when the seed was moved in the medial-lateral axis (0.56 ± 0.14), whereas high spatial similarity was observed with movement in the anterior-posterior axis (0.84 ± 0.11; $p < 0.001$, Figure 2).

*Conclusions*

Our findings support the importance of patient-specific targeting using individual connectivity profiles. Identifying SCC targets using normative data may compromise treatment efficacy due to low inter-subject similarity. Spatial similarity results emphasize the necessity of delivering a precise target identification due to the large variability of WM activation pathways in medial-lateral directions.

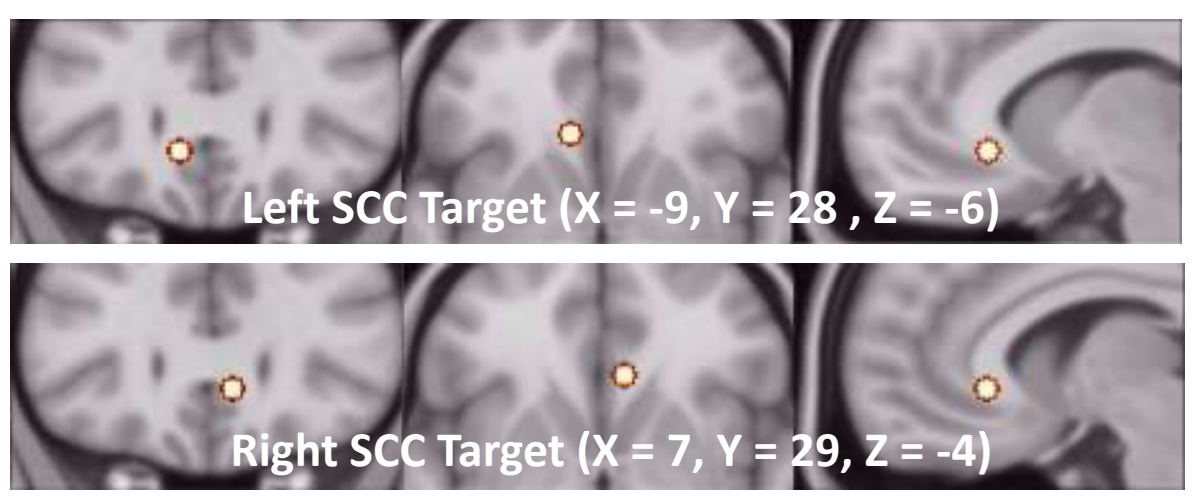

*Figure 1. Left and Right SCC seeds. Seeds were derived from 47 TRD patients who underwent bilateral SCC DBS.*

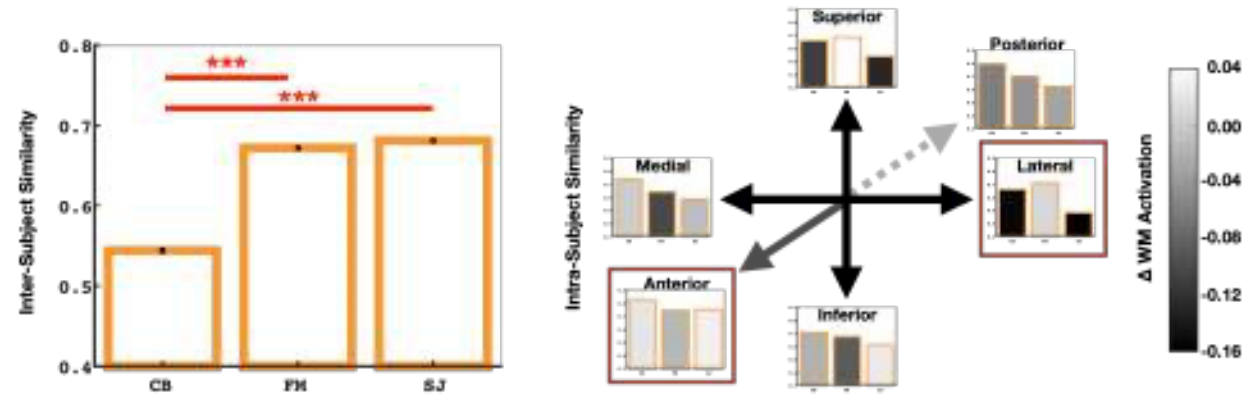

*Figure 2. Inter- and Intra-subject similarity. The cingulum bundle shows lower similarity compared to others in intersubject similarity analysis. Intra-subject similarity (spatial similarity) shows vulnerability to seed location in the medial-lateral direction, while robustness in the anterior-posterior direction.*